\documentclass[a4paper,fleqn,usenatbib]{mnras}

\usepackage{txfonts,pdflscape,longtable}

\usepackage[T1]{fontenc}
\usepackage{ae,aecompl}

\usepackage{graphicx}	
\usepackage{amssymb}	
\usepackage{float}
\usepackage{color}
\usepackage{setspace}

\title[Most massive stars in R136]{The R136 star cluster dissected with {\it Hubble Space Telescope}/STIS. II. Physical properties of the most massive stars in R136}

\author[J. M. Bestenlehner et al.]{
Joachim\,M.\,Bestenlehner,$^{1}$\thanks{E-mail: j.m.bestenlehner@sheffield.ac.uk}
Paul\,A.\,Crowther,$^{1}$
Saida M. Caballero-Nieves,$^{1,2}$
\newauthor Fabian R. N. Schneider,$^{3,4}$
Sergio Sim\'{o}n-D\'{i}az,$^{5,6}$
Sarah A. Brands,$^{7}$
Alex de Koter,$^{7,8}$
\newauthor G\"otz Gr\"afener,$^{9}$
Artemio Herrero,$^{5,6}$
Norbert Langer,$^{9}$
Daniel J. Lennon,$^{5,6}$
\newauthor Jesus Ma\'{i}z Apell\'{a}niz,$^{10}$
Joachim Puls,$^{11}$
and Jorick S. Vink$^{12}$
\\
$^{1}$Department of Physics \& Astronomy, Hounsfield Road, University of Sheffield, Sheffield, S3 7RH, UK\\
$^{2}$Department of Aerospace, Physics and Space Sciences, Florida Institute of Technology, 150 W. University Boulevard, Melbourne, \\~~FL 32901, USA\\
$^{3}$Zentrum f\"ur Astronomie der Universit\"at Heidelberg, Astronomisches Rechen-Institut, M\"onchhofstr. 12-14, 69120 Heidelberg, Germany\\
$^{4}$Heidelberger Institut f\"ur Theoretische Studien, Schloss-Wolfsbrunnenweg 35, 69118 Heidelberg, Germany\\
$^{5}$Instituto de Astrof\'isica de Canarias, E-38200 La Laguna, Tenerife, Spain\\ 
$^{6}$Departamento de Astrof\'isica, Universidad de La Laguna, E-38205 La Laguna, Tenerife, Spain\\
$^{7}$Anton Pannenkoek Institute for Astronomy, University of Amsterdam, 1090 GE Amsterdam, The Netherlands\\
$^{8}$Institute of Astrophysics, KU Leuven, Celestijnenlaan 200D, 3001 Leuven, Belgium\\
$^{9}$Argelander-Institut f\"{u}r Astronomie der Universit\"{a}t Bonn, Auf dem H\"{u}gel 71, 53121 Bonn, Germany\\
$^{10}$Centro de Astrobiolog\'ia, CSIC-INTA, Campus ESAC, Camino bajo del castillo s/n, E-28 692 Villanueva de la Ca\~nada, Madrid, Spain\\
$^{11}$LMU  Munich, Universit\"ats-Sternwarte, Scheinerstrasse 1, 81679 M\"unchen, Germany\\
$^{12}$Armagh Observatory and Planetarium, College Hill, Armagh BT61 9DG, UK
}

\date{Accepted XXX. Received YYY; in original form ZZZ}

\pubyear{2020}

\begin{document}
\label{firstpage}
\pagerange{\pageref{firstpage}--\pageref{lastpage}}
\maketitle

\begin{abstract}
We present an optical analysis of 55 members of R136, the central cluster in the Tarantula Nebula of the Large Magellanic Cloud. Our sample was observed with STIS aboard the Hubble Space Telescope, is complete down to about 40\,$M_{\odot}$, and includes 7 very massive stars with masses over 100\,$M_{\odot}$. We performed a spectroscopic analysis to derive their physical properties. Using evolutionary models we find that the initial mass function (IMF) of massive stars in R136 is suggestive of being top-heavy with a power-law exponent $\gamma \approx 2 \pm 0.3$, but steeper exponents cannot be excluded. The age of R136 lies between 1 and 2\,Myr with a median age of around 1.6\,Myr. Stars more luminous than $\log L/L_{\odot} = 6.3$ are helium enriched and their evolution is dominated by mass loss, but rotational mixing or some other form of mixing could be still required to explain the helium composition at the surface. 
Stars more massive than 40\,$M_{\odot}$ have larger spectroscopic than evolutionary masses. The slope of the wind-luminosity relation assuming unclumped stellar winds is $2.41\pm0.13$ which is steeper than usually obtained ($\sim 1.8$). The ionising ($\log Q_0\,[{\rm ph/s}] = 51.4$) and mechanical ($\log L_{\rm SW}\,[{\rm erg/s}] = 39.1$) output of R136 is dominated by the most massive stars ($>100\,M_{\odot}$). R136 contributes around a quarter of the ionising flux and around a fifth of the mechanical feedback to the overall budget of the Tarantula Nebula. For a census of massive stars of the Tarantula Nebula region we combined our results with the VLT-FLAMES Tarantula Survey plus other spectroscopic studies. We observe a lack of evolved Wolf-Rayet stars and luminous blue and red supergiants.


\end{abstract}




\section{Introduction}

The evolution of massive stars is still insufficiently understood, owing to uncertainties in nuclear reaction rates, stellar structure, internal mixing processes and mass-loss properties \citep{langer2012}. The uncertainties increase with stellar mass \citep[e.g.][]{martins2013}. Binary and higher-order multiple systems magnify the complexity and add additional evolutionary channels \citep[e.g.][]{eldridge2017}. The small number of massive and very massive stars raises the challenges to better understand the nature of such rare but important objects in vigorously star forming galaxies. Metal-poor very massive stars \citep[VMS $> 100\,M_{\odot}$][]{vink2015:IAU} are believed to be progenitors of gamma-ray bursts and pair-instability supernovae and produce more metals than the entire stellar mass function below \citep[e.g.][]{langer2009, kozyreva2014}. 
In addition, with their strong outflows and high ionising fluxes they dominate and shape the evolution of galaxies and are the main indicator of the star forming rate in galaxies. \cite{doran2013} confirmed the importance of VMS in the ionising budget of young starburst regions like 30 Doradus providing evidence that those stars contribute vitally to the ionisation and shaping of the interstellar environment in their host galaxies.

Based on ultraviolet (UV) and optical spectroscopy with the Hubble Space Telescope (HST), \cite{deKoter1997, deKoter1998} identified stars for the first time with initial masses exceeding $100\,M_{\odot}$ in the cluster R136 at the centre of NGC\,2070 in the Large Magellanic Cloud (LMC). 
Based on spectral type calibrations and optical HST data \cite{massey1998} suggested that more than 10 VMS are located in R136. \cite{crowther2010} identified stars with initial masses up to 320\,$M_{\odot}$ within the cluster core albeit with large uncertainties in the mass estimate. The existence of such VMS challenges the canonical upper mass limit of $150\,M_{\odot}$ proposed by \cite{figer2005}, and bring them into the predicted initial stellar mass range of pair-instability supernovae of 140 to 260\,$M_{\odot}$ at low metallicity \citep{heger2002, langer2007}. The finding of stars with initial masses in excess of 150\,$M_{\odot}$ is supported by \cite{bestenlehner2011}, \cite{hainich2014} \& \cite{bestenlehner2014} based on spectroscopic analysis, and by \cite{tehrani2019} through dynamical and spectroscopic analysis of Mk\,34, the most massive binary star known today ($139^{+21}_{-18}\,M_{\odot}  + 127^{+17}_{-17}\,M_{\odot}$). Those VMS might be formed in a similar way to low mass stars \citep{krumholz2015} or via stellar merger \citep{banerjee2012b}. The latter formation channel may result in an apparent age younger than the lower mass cluster members \citep{schneider2014b}.

The VLT-FLAMES Tarantula Survey \citep[VFTS,][]{evans2011} is the largest spectroscopic survey of massive stars today. They obtained multi-epoch spectra of over 800 O, B, and Wolf-Rayet (WR) stars in the Tarantula Nebula covering NGC\,2060 and NGC\,2070 of the 30 Doradus region, but excluded the core of R136 due to crowding. 

\cite{sabin2014, sabin2017} undertook a spectroscopic analysis of VFTS O dwarfs. 
Most O dwarfs in the Tarantula Nebula have weak winds and show a large dispersion in the wind-luminosity relation (WLR) for stars more luminous than $\log L/L_{\odot} > 5.1$. The mass discrepancy between evolutionary and spectroscopic masses is rather small, but evolutionary models systematically predict slightly large surface gravities \citep{sabin2017}. VFTS O giants and supergiants were spectroscopically analysed by \cite{ramirez2017}. The WLR agreed with the theoretical prediction by \cite{vink2000, vink2001}. 
5 stars are helium enriched and show only modest projected rotational velocities ($\varv \sin i$) that are not in agreement with the prediction of rotational mixing in main-sequence single star stellar structure calculations \citep{brott2011, koehler2015}. 

In addition, by studying VFTS luminous O, Of/WN and WNh stars \cite{bestenlehner2014} found no correlation between projected rotational velocity and helium composition ($Y$) at the stellar surface, but they discovered a strong correlation of $Y$ with mass-loss rate ($\dot{M}$) over stellar mass ($M_{\star}$) for $\log \dot{M}/M_{\star} > -6.5$. This suggests that rotational mixing is a relatively unimportant factor in helium enhancement for these stars with strong winds, but that shedding the stellar envelopes through mass loss might be the key process in chemically enriching the stellar surface with nucleosynthesis products of these very massive stars. Also for OB stars in the Small Magellanic Cloud \cite{ramachandran2019} did not find evidence for a correlation of chemical mixing with rapid rotation.

Based on pre-main-sequence stars and tracks \cite{cignoni2015} report that the star formation rate in the NGC\,2070 complex peaked between 1 and 3 Myr ago. \cite{deKoter1997, deKoter1998} and \cite{massey1998} estimated an age around 2 Myr based on the brightest stars within and in close proximity to the dense central cluster R136. \cite{crowther2016} inferred a median age of $\sim$1.6 Myr from UV calibration for the central stellar population of R136. Based on stars in the periphery of R136 \cite{massey1998} found that the slope of the initial mass function (IMF) is consistent with a Salpeter IMF \citep{salpeter1955}.

\cite{schneider2018b} combined the OB star results from VFTS and studied the massive star formation in 30 Doradus. They found that massive stars with all masses and ages are scattered throughout 30 Doradus. This suggests that they are not only formed in the dense stellar populations NGC\,2070 or NGC\,2060 but also in relative isolation in the field \citep{bressert2012}. The formation of massive stars swiftly increased around 8 Myr ago by forming stars in the field which continued inside NGC\,2060 (5.7 Myr ago) and NGC\,2070 (3.6 Myr ago) with a declining star formation rate in the last 1 Myr. R136 formed last in the centre of NGC\,2070. The IMF of 30 Doradus without R136 is densely populated up to 200\,$M_{\odot}$, with a shallower power-law exponent of $1.90^{+0.37}_{-0.26}$ for stars more massive than 15\,$M_{\odot}$ predicting more massive stars than inferred using the standard \cite{salpeter1955} $2.35$ value \citep{schneider2018}.  

The star cluster R136 in the centre of NGC\,2070 had been excluded by the VFTS because of crowding. 
To add the missing mosaic and enable study of the entire massive star population up to 300\,$M_{\odot}$ in 30 Doradus \citet[Paper\,I]{crowther2016} observed the cluster R136 in the optical and ultraviolet with the instrument STIS aboard the Hubble Space Telescope (HST). Paper\,I provides a far-UV spectroscopic census of R136 and studied the origin of He\,{\sc ii} $\lambda1640$ in young star clusters. The current Paper\,II undertakes an optical spectroscopic analysis to aim for the physical properties of most massive stars in R136 using consistent optical diagnostics and spectroscopic tools to VFTS. The third paper (Caballero-Nieves et al.~in prep., Paper\,III) focuses on the blue optical observations deriving spectral types and investigating multiplicity and rotational properties of the stellar content of R136. A future study of this series will explore the ultraviolet properties of these stars with attention to the stellar wind parameters and investigate systematics between UV+optical and optical-only spectroscopic analyses (Brands et al.~in prep., Paper\,IV).

This paper is structured as follows. In Sect.\,\ref{s:obs} we summarise the spectroscopic and photometric data used in this work. Our spectroscopic and error analysis are described in Sect.\,\ref{s:spec_ana}. We present the results (Sect.\,\ref{s:results}) in the context of the Hertzspung-Russell diagram (Sect.\,\ref{s:hrd}), their stellar masses and ages (Sect.\,\ref{s:masses_ages}), and their wind-momentum -- luminosity relation (Sect.\,\ref{s:winds}). In Sect.\,\ref{s:disc} we discuss the surface helium composition of our sample (Sect.\,\ref{s:he-abund}) and place our results in the context of cluster age and initial mass function of R136 (Sect.\,\ref{s:r136_age} and \ref{s:r136_imf}), ionising and mechanical feedback of R136 (Sect.\,\ref{s:io_mech}) and put R136 in the wider context as a stellar population within the Tarantula Nebula (Sect.\,\ref{s:vfts}). We conclude in Sect.\,\ref{s:con}.


\section{Observational data}
\label{s:obs}

\subsection{Spectroscopic data}
The current study makes use of the blue-optical HST-STIS/G430M and $\mathrm{H}{\alpha}$ HST-STIS/G750M observations described in Paper\,I. They cover a wavelength range from $\lambda 3793-4849$\,{\AA} with a resolving power of $\sim$\,7700 at $\lambda 4400$\,{\AA} (HST-STIS/G430M) and $\lambda 6482-7054$ with a resolving power of $\sim$\,6000 at $\mathrm{H}{\alpha}$ (HST-STIS/G750M), respectively. 

For the spectroscopic analysis (Sect.\,\ref{s:spec_ana}) we require rectified and radial velocity corrected spectra. Based on the spectral classification for Paper\,III we selected a synthetic template spectrum for each target, which is used as a reference spectrum to normalise and correct for radial velocity shifts. A single spectrum was created by stitching several grating settings together. Stars with $\log L/L_{\odot} < 5.3$ have low S/N spectra and the uncertainties of the stellar parameters are systematically larger. Less luminous but cooler stars have stronger He\,{\sc i} lines and we were still able to derive reasonable effective temperatures and luminosities. 

We do not consider spectra with a S/N below 5 per resolution element and removed the blended object 118 from \citet[hereafter H118]{hunter1995}, which was heavily contaminated by bright nearby sources, and spectroscopic double line binaries (SB2, H42 and H77). Our sample of 55 stars consists of 22 apparent single stars, 7 potential spectroscopic binaries (SB1/SB2), 19 stars with low S/N, and 7 stars which show to some extend cross-contamination due to crowding. The spectral types are taken from Paper\,III. The spectral resolution of the G430M grating is higher than the typical value used for spectral classification. The spectral classification was performed on a degraded resolution but with an improved S/N. More details on spectral classification and observational properties of the sample will be discussed in Paper\,III.


\subsection{Photometric data}
\begin{figure}
\begin{center}
\resizebox{\hsize}{!}{\includegraphics{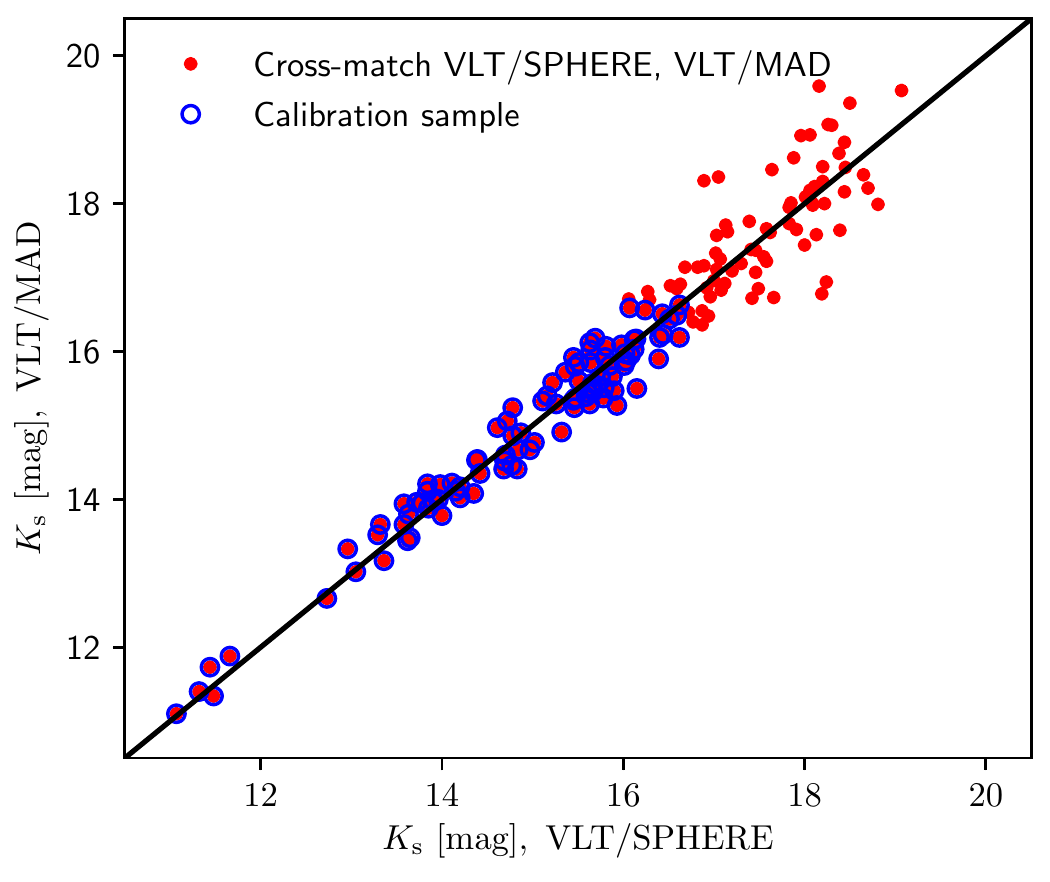}}
\end{center}
\caption{VLT/MAD \citep{campbell2010} versus VLT/SPHERE \citep{khorrami2017} $K_{\rm s}$-band photometry. Stars brighter than 16.7\,mag were used to determine the offset of $0.001$\,mag.}
\label{f:sphere-mad}
\end{figure}
We derived the stellar luminosity by modelling the stellar spectral energy distribution (SED) including the interstellar extinction (Sect.\,\ref{s:lum_red}). The following photometric data were used: optical HST/WFC3 F438W ($B$-band) and F555W ($V$-band) from \cite{deMarchi2011} and near-infrared (near-IR) $K_{\rm s}$-band photometry from \cite{khorrami2017}. For stars where the HST/WFC3 data were incomplete, the HST/WFPC2 F336W ($U$-band) and F555W ($V$-band) photometric data were used from \cite{hunter1995}. We applied the magnitudes offset to WFPC2 photometry, which were obtained by \cite{crowther2016} to the F336W ($\Delta \mathrm{mag}= -0.15$) and F555W ($\Delta \mathrm{mag}= -0.17$) magnitudes.

\cite{khorrami2017} adopted the instrumental zero-points and verified the VLT/SPHERE $K_{\rm s}$-band photometry with the $K$-band fluxes/magnitudes of 6 stars from \cite{crowther2010}, which were inferred from flux calibrated VLT/SINFONI spectra or from VLT/MAD $K_{\rm s}$ photometry \citep{campbell2010}. They were not able to verify the $J$-band photometry because the VLT/MAD observations used by \cite{campbell2010} were taken in the $H$ and $K_{\rm s}$-band. Therefore, we did not use the VLT/SPHERE $J$-band data in this study. 

The derived $K$-band magnitudes on the basis of flux calibrated spectra are uncertain because they are not direct measurements of the $K$-band flux. Therefore, we cross-matched the \cite{khorrami2017} $K_{\rm s}$-band catalogue with the VLT/MAD $K_{\rm s}$-band catalogue from \cite{campbell2010}. We recognised a slight disagreement between both coordinate systems with increasing distance from the frame centre of the VLT/SPHERE catalogue. We rotated the VLT/SPHERE coordinate system to match the VLT/MAD and both catalogues agree within 0.1 arcsec. In cases where we found multiple cross-matches, we selected the match with the smallest absolute K-band difference in the VLT/SPHERE and VLT/MAD photometry. Up to 16.7 magnitude the standard deviation between the two photometric catalogues was nearly constant, but suddenly increased for fainter objects (Fig.\,\ref{f:sphere-mad}). Therefore, we calculated the mean offset for stars brighter than 16.7\,mag in both catalogues. 
In Fig.\,\ref{f:sphere-mad} we show the cross-correlation between the two photometric catalogues. The mean offset is $0.001$\,mag and is negligible compared to the photometric errors.

\section{Spectroscopic analysis}
\label{s:spec_ana}
In total we analysed 55 stars and derived their stellar parameters. Paper\,I has observed R136 in the ultraviolet (UV) permitting measurements of terminal velocities ($\varv_{\infty}$) from P-Cygni resonance lines. Our sample of 55 stars comprises 50 targets that are in common with Paper\,I, as well as 5 additional targets (H120, H129, H139, H159, H162) which lie beyond the MAMA detector and have no determined $\varv_{\infty}$. Terminal velocities of O stars can only be measured in the UV and are essential for calculating the mass-loss rates of our sample (Sect.\ref{s:scaling-mdot}). The UV wavelength range is crucial to constrain additional wind parameters such as velocity law exponent $\beta$ and volume filling factor $f_V$ of O stars (Paper\,IV). 

We derive the line broadening parameters (Sect.\ref{s:line_broad}) as a required input for the spectroscopic analysis, which is followed by a discussion on mass-loss rate scaling relations (Sect.\,\ref{s:scaling-mdot}). In Sect.\,\ref{s:fw} we describe our O star analysis, which employs the stellar atmosphere code FASTWIND \citep{santolaya1997, puls2005, rivero2012a}. We chose FASTWIND to analyse as many stars as possible to be comparable with the results from the VFTS \citep{evans2011} as most O stars were analysed with FASTWIND: O dwarfs \citep{sabin2014, sabin2017}, O giants and O supergiants \citep{ramirez2017} and their nitrogen abundances \citep{grin2017}. 
The three WN5h stars in our sample are analysed with the stellar atmosphere code CMFGEN \citep[Sect.~\ref{s:cmfgen},][]{hillier1998}. The bolometric luminosity is determined by matching photometric data with reddened theoretical spectral energy distributions (Sect.~\ref{s:lum_red}). In Sect.~\ref{s:FW_CMFGEN} we compare three transition objects, which are analysed with FASTWIND and CMFGEN, to estimate the systematics between both methods and codes. 

FASTWIND and CMFGEN are non-LTE stellar atmosphere codes which consider spherical geometry, though take different approaches to the treatment of metal line blanketing. While CMFGEN considers all lines explicitly, FASTWIND uses an approximated approach. FASTWIND only calculates in detail the elements employed for the spectroscopic analysis and uses averages for background line opacity and emissivity. This approximation is most suitable for stars with optically thin winds and reproduces the line-blanketing in a realistic way. 

\subsection{Line broadening\label{s:line_broad}}

To obtain the stellar spectral line broadening of the O-type stars we use IACOB-BROAD \citep{simon-diaz2014} to derive the projected rotational velocity ($\varv \sin i$) and macro-turbulent velocity ($\varv_{\rm mac}$). IACOB-BROAD is an interactive analysis tool, that combines Fourier transformation and goodness-of-fit methods. The code is publicly available and is written in the interactive data language (IDL). A careful line broadening determination is, e.g., crucial to accurately derive surface gravities. An underestimation of the line broadening results in an over estimation of the surface gravity ($\log g$) and vice versa. 	

The signal to noise (S/N) ratio of our spectra was too low for the faint metal lines to be used, which provide more accurate line broadening parameters. Instead our line broadening analysis relied on He\,{\sc i} $\lambda \,4026$ and $4471$ and He\,{\sc ii} $\lambda \,4200$ and $4542$, but He\,{\sc i} $\lambda \,4026$ coincides with He\,{\sc ii} $\lambda \,4026$ and could only be used for stars with $\lesssim 35\,000$\,K. The quality of the spectra did not allow us to disentangle $\varv \sin i$ and $\varv_{\rm mac}$ broadening profiles. In addition, the He lines are also broadened by the Stark effect. He\,{\sc i} lines disappear at temperatures around 45\,000\,K. He\,{\sc ii} lines could be used at temperatures above 40\,000\,K, but their line strength was only considerably above the noise level at temperatures $\gtrsim 50\,000$\,K. The low S/N ratio made it difficult to find the first order minimum when the Fourier transformation method of IACOB-BROAD was applied. Therefore, we used the goodness-of-fit assuming all broadening is produced by rotation and estimated a combined broadening $\varv \sin i_{\rm max}$ with $\varv_{\rm mac} = 0$\,km/s so that our values are an upper limit to the actual $\varv _{\rm rot}$. In our spectroscopic analysis we also considered a broadening profile as if the line was only broadened by rotation. The adopted line broadening is listed in Table\,\ref{t:sp}.

We could not use IACOB-BROAD for the three WNh stars because their emission lines are formed in the stellar wind above the hydrostatic layers. This does not only add an additional broadening due to the velocity gradient but also the measured $\varv \sin i$ is lower as a result of the differential rotation. To account for broadening caused by the wind velocity law we convolved the synthetic spectrum with a rotation profile to match the line broadening of the observations (Fig.\,\ref{f:r136a1} to \ref{f:r136a3}). We used N\,{\sc v} at $\lambda$4604 and 4620 which have the closest line-forming regions to the hydrostatic layers for the given wavelength range. Like for the O stars the macro-turbulent velocity was set to zero assuming all broadening is produced by rotation. We estimated an upper $\varv \sin i$ limit of not more than 100\,km/s for R136a2 and R136a3. The projected rotational velocity of R136a1 is somewhat higher and between 130 and 150\,km/s. Taking differential rotation into account we adopted a $\varv \sin i_{\rm max} = 190$\,km/s for R136a1 and 150\,km/s for R136a2 and R136a3 (Table\,\ref{t:sp}).  


\subsection{Wind strength Q versus transformed mass-loss rate $\dot{M}_{\mathrm{t}}$ \label{s:scaling-mdot}}
The strength of emission features scales not only with the mass-loss rate but also with the volume-filling factor ($f_{\rm v}$), terminal velocity and radius of the star which are compressed into one parameter to reduce the effort when computing model grids. There are two ways to spectroscopically quantify mass-loss rates of hot massive stars using scaling relations. Stars with optically thin winds (OB stars) the wind strength parameter $Q$ is usually applied \citep{puls1996, sabin2014, sabin2017, holgado2018}, where $Q$ is proportional to the integrated optical depth over the resonance zone:
\begin{equation}
Q=\frac{\dot{M}\,[M_{\odot}\mathrm{yr}^{-1}]/\sqrt{f_{\rm v}}}{(R\,[R_{\odot}] \varv_{\infty}\,[\mathrm{kms^{-1}}])^{3/2}}.
\end{equation}
and $f_V$ has been set to unity (see Sect.~\ref{s:winds}). For optically thick conditions, there is an additional dependence on $\varv_{\infty}$ \citep[e.g.][]{puls1996}. The transformed radius \citep{schmutz1989,graefener2002,hamann2004} or the equivalent approach of the transformed mass-loss rate \citep[$\dot{M}_{\rm t}$,][]{bestenlehner2014} is usually used for optically thick winds (WR stars), where the line equivalent width is preserved: 
\begin{eqnarray}
  \log(\dot{M}) = \log(\dot{M}_{\rm t}) + 0.5\log(f_{\rm v}) + \log \left(\frac{\varv_{\infty}}{1000\,\mathrm{km\,s}^{-1}} \right) \nonumber \\ 
  +~0.75\log \left(\frac{L}{10^6L_{\odot}} \right).
\end{eqnarray}
Both scaling relations are equivalent except for the exponent of the $\varv_{\infty}$ dependence, $\dot{M} \varpropto \varv_{\infty}^{3/2}$ (wind strength $Q$) and  $\dot{M} \varpropto \varv_{\infty}$ (transformed mass-loss rate $\dot{M}_{\mathrm{t}}$). In our study we compared both scaling relations and find that optically thin winds are preferably scaled with the wind strength parameter while optically thick winds are better scaled with the transformed mass-loss rate. If $\varv_{\infty}$ in the model has a reasonable value, the differences between both scaling relations are small. However, if the line is in emission and the terminal velocity of the synthetic spectrum is too high, the line centre is fitted well with the $Q$ scaling relation, but the synthetic spectrum shows extended wings. This overestimates the actual mass-loss rate. By fitting Balmer lines in absorption a degeneracy between $\log g$ and $\dot{M}$ can occur using the $\dot{M}_{\mathrm{t}}$ scaling relation. We used the $\dot{M}_{\mathrm{t}}$ scaling relation for R136a1,\,a2,\,a3,\,a5,\,b and H36 and $Q$ for the remaining O stars in our sample.

\subsection{FASTWIND analysis\label{s:fw}}

The majority of our targets (52 out 55) were analysed with the stellar atmosphere and radiative transfer code FASTWIND \citep{santolaya1997, puls2005, rivero2012a} including nitrogen as an explicit element \citep{rivero2011, rivero2012a, rivero2012b}. The three WN5h stars in the core of R136 (R136a1,\,a2,\,a3) have such strong stellar winds that they could not be analysed with FASTWIND.

The stellar parameters were determined using the automated spectroscopic analysis tool IACOB-Grid Based Automatic Tool \citep[IACOB-GBAT,][]{simon-diaz2011, holgado2018}. IACOB-GBAT uses a $\chi^2$ algorithm to match the observed line profiles (here: H\,{\sc i}, He\,{\sc i} and He\,{\sc ii}) with a grid of pre-computed FASTWIND synthetic spectra. Typically in normal $\chi^2$ deep lines with many wavelength points dominate over narrow lines. To avoid these issues as much as possible an optimized (iterative) strategy was incorporated in IACOB-GBAT which is described in Appendix A of \cite{holgado2018}. Details and full description of the grid are given in \cite{sabin2014}. At an effective temperature at optical depth $\tau = 2/3$ ($T_{\rm eff}$) of about $\gtrsim$45\,000K, He\,{\sc i} becomes weak or disappears and temperature determination using the He\,{\sc i-ii} ionisation balance is not possible. Therefore we recomputed the grid from \cite{sabin2014} for $T_{\rm eff}$ greater than 30\,000\,K and added nitrogen as an explicit element. The grid has a half-solar metallicity with respect to \cite{asplund2005}, except for CNO abundances. For those we adopted values relative to hydrogen ($\epsilon_{\rm x} = \log(n_{\rm x} /n_{\rm H}) + 12$) according to \cite{korn2002} ($\epsilon_{\rm C} = 8.06$, $\epsilon_{\rm N} = 7.01$, $\epsilon_{\rm O} = 8.37$). To match the observed nitrogen line intensity of the WNh stars we used a N-abundance of $\epsilon_{\rm N} = 8.5$ (Sect.\ref{s:cmfgen}). Therefore, for CNO-processed atmospheres we set the nitrogen abundances to 8.5 (factor $\sim$30 enhancement) and an intermediate enrichment of 8.2 (factor $\sim$15 enhancement), with carbon and oxygen been reduced accordingly.

Based on photo-ionisation nebular models \cite{pellegrini2011} derived a N-abundance of $\epsilon_{\rm N} = 7.09$ for 30\,Dor, with C-abundance not measured. The evolutionary models used in this study \citep{brott2011, koehler2015} adopted a nitrogen base line of $\epsilon_{\rm N} = 6.90$, which represents the LMC average \citep{hunter2007, brott2011}. The value of \cite{korn2002} lies somewhat in between and has been therefore chosen. However, the actual N-abundance of our objects is not determined in this study, but we provide an indication, if N is enriched at the surface for stars hotter than $\sim 45\,000$K. The assumed N-abundance can affect the temperature determination for those stars, where $T_{\rm eff}$ is based on the ionisation balance of N\,{\sc iv} and N\,{\sc v}. The degeneracy between $T_{\rm eff}$ and N-abundances is discussed in Sect.\,\ref{s:FW_CMFGEN}.

The stellar parameters were derived using the following spectral lines: Balmer H${\alpha}$ and H${\gamma-\epsilon}$, He\,{\sc i} $\lambda$4026, 4121, 4144, 4388, 4471 and 4713, He\,{\sc ii} $\lambda$4026, 4200, 4542, 4686, N\,{\sc iii} $\lambda$4634 and 4641, N\,{\sc iv} $\lambda$4058 and 6381 and N\,{\sc v} $\lambda$4604 and 4620 (Fig.\,\ref{f:h35}). We applied IACOB-GBAT to all O-type stars in our sample and all stellar parameters were set free ($T_{\rm eff}$, $\log g$, $Q$, velocity law exponent $\beta$, helium mass fraction $Y$ and micro-turbulent velocity $\varv_{\rm mic}$). The micro-turbulent velocity was treated depth independent and homogeneous winds were assumed with $f_{\rm v} = 1.0$. IACOB-GBAT aims at a global optimisation, uses all lines in parallel to derive all parameters in parallel and takes into account correlations between the various parameters. The largest weight on $T_{\rm eff}$ is from the ionisation balance of the He\,{\sc i} and {\sc ii} while $\log g$ is mainly constrained by H${\gamma,\delta,\epsilon}$ and $Q$ by H${\alpha}$ and He\,{\sc ii} $\lambda$4686 assuming a typical velocity law exponent $\beta$ for a given luminosity class (Sect.\,\ref{s:beta-mdot}). Using $Q$ mass-loss rates are calculated with the wind strength scaling relation (Sect.\,\ref{s:scaling-mdot}) and terminal velocities from Paper\,I, if available (Table\,\ref{t:sp}). Stellar radii ($R_{\rm eff}$) were calculated with $L= 4 \pi \sigma R^2 T_{\rm eff}^4$ and Stefan-Boltzmann constant $\sigma$ to scale $\dot{M}$ with $Q$. He-abundances are determined by the line ratio of hydrogen and helium lines, while $\varv_{\rm mic}$ is constrained by the line strength of He lines. This results in a degeneracy between $Y$ and $\varv_{\rm mic}$ (Sect.\,\ref{s:y-mic}).

For stars hotter than $\gtrsim$40\,000\,K we adjusted the temperature by eye using the ionisation balance of the nitrogen lines  and adjusting the value of $\log g$. We ran IACOB-GBAT with a fixed temperature (Fig.\ref{f:chi_square_hot}). If $\log g$ was different to the assumed value, we checked the nitrogen ionisation balance again and re-ran IACOB-GBAT. We iterated until the temperature and surface gravity converged.

IACOB-GBAT provides the uncertainties associated to each parameter and takes into account correlations between the various parameters. In some cases the S/N ratio of the spectrum was so low, that all stellar parameters appeared to be degenerate (Fig.\,\ref{f:chi_square}). Even though the $\chi^2$ distribution was not completely flat, we were unable to derive an error. In such cases we set free only two parameters at a time $T_{\rm eff}$ and $\log g$, $Q$ and $\beta$ and, $Y$ and $\varv_{\rm mic}$ while the others were fixed. In this way the $\chi^2$ distribution was better characterized and errors could be to some extent estimated. Such stars are labelled as low S/N objects. The lower bound of $Y$ was considerably below the physical limit $\sim$\,0.25 for some stars. In these cases we truncated the lower error such that the lower limit was not below 0.2.

The results with their uncertainties ($1 \sigma$) are given in Table\,\ref{t:sp}.


\subsubsection{Degeneracy of $\beta$-type velocity law and mass-loss rate\label{s:beta-mdot}}
The velocity field in the stellar atmosphere codes is parametrised by a $\beta$-type velocity law. Smaller values of $\beta$ correspond to larger velocity gradients ($\mathrm{d} \varv/\mathrm{d} r$) in the inner and lower $\mathrm{d} \varv/\mathrm{d} r$ in the outer wind and vice versa. For example, larger values of $\beta$ lead to a denser wind in the onset region of the flow and result in a lower mass-loss rate estimate. The $\beta$-type velocity law and mass-loss rate are degenerate in the absence of necessary diagnostics and/or too weak stellar winds. The stellar spectrum can be matched with several sets of $\beta$ and $\dot{M}$. Based on theoretical predictions typical $\beta$ exponents are between 0.8 and 1.0 for dwarfs and giants and between 0.9 and 1.1 for supergiants \citep{muijres2012}. Based on a study of more than 250 O stars \cite{holgado2018} noted that for supergiants best fitting models have preferences towards $\beta = 1.2$, even though only a lower limit could be determined. Therefore, we only allowed values of $\beta$ to be 0.8, 1.0 and 1.2 for dwarfs and giants and 1.0, 1.2 and 1.5 for supergiants. A more detailed discussion on the effect of varying $\beta$ in the determination of $\dot{M}$ can be found in \cite{markova2005, holgado2018}. 

\subsubsection{Degeneracy of micro-turbulent velocity and He-abundances\label{s:y-mic}}
The micro-turbulent velocity ($\varv_{\rm mic}$) does not only broaden the spectral lines but also modifies the line strength of e.g. He\,{\sc i}-{\sc ii} and N\,{\sc iii}-{\sc iv}-{\sc v} depending on their equivalent widths. As a consequence, not only the derived chemical abundances can be affected when an inaccurate $\varv_{\rm mic}$ is selected but also the effective temperature, if the line is used as a temperature diagnostic. The micro-turbulent velocity can be accurately constrained if the spectra are of high S/N and the number of available spectral lines of the same ion is large enough to achieve a consistent spectroscopic fit to all spectral lines. 
However, in cases where the spectrum has a low S/N ratio the degeneracy is more difficult to resolve. A large micro-turbulent velocity is favoured because of the low S/N, which leads to an underestimation of the derived He-abundance. Therefore, we only allowed typical O stars $\varv_{\rm mic}$ of 5 and 10\,km/s, even though higher velocities are possible as well, in particular for supergiants. 

\subsection{CMFGEN analysis\label{s:cmfgen}}
The three core WNh stars R136a1,\,a2 and a3 plus three supergiants that were also modelled with FASTWIND, R136b, R136a5 and H36, were analysed with the stellar atmosphere and radiative transfer code CMFGEN \citep{hillier1998} using the method described in \cite{bestenlehner2014}. Initial estimates of the stellar parameters were derived with the grid from \cite{bestenlehner2014} with either half solar $\epsilon_{\rm N} = 7.44$ or enriched $8.5$ nitrogen abundances. We computed extra grids of stellar atmospheres around the preferred stellar parameter space of the initial estimates with an extended atomic model and varying $T_{\rm eff}$, $\dot{M}$, $\beta$-type velocity law and helium abundances. Effective temperatures of WR stars are usually defined at $\tau = 10$ or $20$ ($T_{\star}$). In the case of the three WNh stars the differences between $T_{\rm eff}$ and $T_{\star}$ are rather small ($\lesssim 1$\%) and largely depend on the velocity law (Sect.\,\ref{s:winds}). The gravity was fixed for R136a1, a2 and a3 to 4.0 as $\log g$ cannot be derived from emission lines of the optically thick WNh star winds, but varied for R136a5, R136b and H36. Based on the electron scattering wings the wind volume filling factor ($f_{\mathrm{v}}$) was set to 0.1 and the terminal velocities were taken from Paper\,I. The grid of stellar atmospheres contains the following element ions: H\,{\sc i}, He\,{\sc i-ii}, C\,{\sc iii-iv}, N\,{\sc iii-v}, O\,{\sc iii-vi}, Ne\,{\sc iii-vi}, Si\,{\sc iv}, P\,{\sc iv-v}, S\,{\sc iv-vi}, Fe\,{\sc iv-vii} and Ni\,{\sc iv-vi}. $\varv_{\rm mic}$ was set to 10\,km/s.

We used the same line diagnostics as described in Sect.\,\ref{s:fw} for the spectroscopic analysis with FASTWIND (Fig.\,\ref{f:h35}). 
Stellar parameters are given in Table\,\ref{t:sp}. 

\subsection{Luminosity and reddening\label{s:lum_red}}
To derive the bolometric luminosity ($L_{\rm bol}$) and estimate the interstellar extinction towards our targets we match the model spectral energy distribution (SED) in the  optical with $B$ (F438W), $V$ (F555W) from \cite{deMarchi2011} or $B$ (F438W), $V$ (F555W) from \cite{hunter1995} and near-IR $K_{\rm s}$ from \cite{khorrami2017} (top panel of Fig.\,\ref{f:h35}). We extracted intrinsic $U$, $B$, $V$ and $K_{\rm s}$ colours from the modelled SED by applying approximated filter functions for each filter and calculated the extinctions $E(B-V)$ and $E(V-K_{\rm s})$. 

In principle, one should use $R_{5495}$ and $E(4405-5495)$ to define the amount and type of extinction, respectively, instead of $R_V$ and $E(B-V)$ which, in general, depend on both and on the input SED. However, for the case where we are analysing hot stars with low extinction, as it is the case here, there is little difference between $R_{5495}$ and $R_V$ or between $E(4405-5495)$ and $E(B-V)$ \citep[Fig.\,3 of][]{ma-ap2013}. The reddening parameter $R_{V}$ is derived using the following relation inferred from the reddening law by \cite{ma-ap2014}:

\begin{equation}\label{e:r_v1}
	R_V = 1.12 \times E(V-K_{\rm s})/E(B-V) - 0.18.
\end{equation}

In cases where only $U$ and $V$ optical magnitudes are available \citep{hunter1995} we obtained $E(B-V)$ and $R_{V}$ by fitting the $U$, $V$, and $K_{\rm s}$-bands with the model SED and reddening law by \cite{ma-ap2014}. The derived luminosities are anchored on the $K_{\rm s}$-band flux as the extinction near-IR $A_{K_{\rm s}}$ is much smaller than optical $A_{V}$. In this way we are able to determine reliable $L_{\rm bol}$ adopting a distance modulus of 18.48\,mag \citep{pietrzynski2019}.

In some cases we inferred an unusually high $R_V > 5.0$ as a result of crowding. The $B$ and $V$ from \cite{deMarchi2011} showed an inconsistency between crowded regions and stars in relative isolation. $R_V$ defines the overall shape of the reddening law and connects the optical with the near-IR. By using the $K_{\mathrm s}$-band flux to derive the luminosity, the influence of $R_V$ is rather small. There were also a few targets with an unusually low $R_V < 2.5$, which is an indication of a near-IR excess. To tackle the issue we applied sigma clipping to our $R_V$ values and derived an average $R_V = 4.18\pm0.38$ and $A_{K_{\mathrm s}} = 0.21\pm0.03$. The values are similar to what \cite{doran2013} had obtained within the R136 region ($R_V = 4.2$ and $A_{K_{\mathrm s}} = 0.17$). In cases where $R_V > 5.0$ we set $A_{K_{\mathrm s}} = 0.21$ to avoid overestimating the luminosity of the star. We still propagated the potential larger $A_{K_{\mathrm s}}$ value into the upper luminosity error. If $R_V < 2.3$ (2-sigma below the standard $R_V=3.1 - 2\times0.4$) 
we anchored the luminosity on the $V$-band and estimated $A_V$ on the basis of $E(B-V)$ and the average $R_V = 4.18 \pm 0.38$ (H86, H108, H129). No $A_{K}$ is list in Table\,\ref{t:sp} for those stars.

Luminosities, absolute magnitudes and extinction for each star are listed in Table\,\ref{t:sp}.

\subsection{Systematics between FASTWIND and CMFGEN analysis methods\label{s:FW_CMFGEN}}
For our analysis we used two different approaches. The three WN5h stars (Sect.\, \ref{s:cmfgen}) were analysed with the method described in \citet{bestenlehner2014} using a grid of synthetic spectra computed with CMFGEN \citep{hillier1998} and N-abundances of $\epsilon_{\rm N} = 7.44$ and $8.5$. The O stars were analysed with IACOB-GBAT \citep{simon-diaz2011} based on a grid of synthetic spectra computed with FASTWIND \citep{santolaya1997, puls2005, rivero2012a} and N-abundances of $\epsilon_{\rm N} = 7.01$, $8.2$ and $8.5$. 
The grid, to explore the O star parameter space, can be computed faster with FASTWIND than with CMFGEN, but FASTWIND was not designed to analyse stars with strong and optically thick winds such as the three WN5h stars in the core of R136. 

\cite{massey2013} compared both stellar atmosphere codes for the physical properties of SMC and LMC O type stars. The systematic difference is small compared to our error margins. However, systematic differences between the codes might be larger at the transition from optically thin to optically thick winds at the edge of the FASTWIND comfort zone. Two diverse analysis methods were used as well which could add to the systematics (Sect.\,\ref{s:fw} and \ref{s:cmfgen}). 

We compared the results for three objects, H36, R136a5 and R136b, at the transition from optically thin to optically thick winds to identify potential systematics between these two analysis approaches. 
Stellar parameters are given in Table\,\ref{t:sp} and spectral fits are shown in Fig.\,\ref{f:r136a5} to \ref{f:h36}. The results for R136a5 are comparable between the methods. The inferred temperature and surface gravity for R136b are lower for the CMFGEN analysis method. \cite{holgado2018} found a similar systematic toward lower $\log g$ and $T_{\rm eff}$ for CMFGEN, which results from deeper predicted line profiles of He\,{\sc ii} $\lambda$4200/4542. 

A large temperature difference occurs for H36. The FASTWIND analysis method results in an effective temperature of 52\,000K while the CMFGEN one leads to 48\,000K. Model comparison showed that CMFGEN and FASTWIND are very consistent around 50\,000\,K. A test calculation with $T_{\rm eff} = 52$\,000K using CMFGEN showed, that the N\,{\sc v} at $\lambda$4604 and 4620 and N\,{\sc iv} at $\lambda$4058 can be simultaneously fitted with the lower nitrogen abundance of $\epsilon_{\rm N} = 8.2$ from the FASTWIND model. Thus, the different results on $T_{\rm eff}$ are presumable a consequence of the different assumptions on the nitrogen abundance. With the lack of wavelength coverage beyond $\lambda$4850 to observe N\,{\sc v} at $\lambda$4945 we are not able to rule out one of the two possible temperatures or nitrogen abundances. To fit $\mathrm{H}{\alpha}$ and He\,{\sc ii} $\lambda$4686 at the same time an unphysically low helium abundance is required, $Y\sim 20\%$ in mass fraction (Fig.\ref{f:h36}). 
This may point to an excess in the $\mathrm{H}\alpha$ emission. Though we are not sure about the nature of such an excess, it might be due to differential effects of clumping in the $\mathrm{H}\alpha$ and He\,{\sc ii} $\lambda$4686 line forming region or 
could be an indication of binarity. Save for H36, the stellar parameters barely effect the results on the stellar mass and age. The lower temperature based on the CMFGEN fit of H36 would lead to an older age and lower stellar mass (Table\,\ref{t:sp}).

\section{Results}
\label{s:results}

\subsection{Hertzsprung-Russell diagram of R136\label{s:hrd}}
\begin{figure}
\begin{center}
\resizebox{\hsize}{!}{\includegraphics{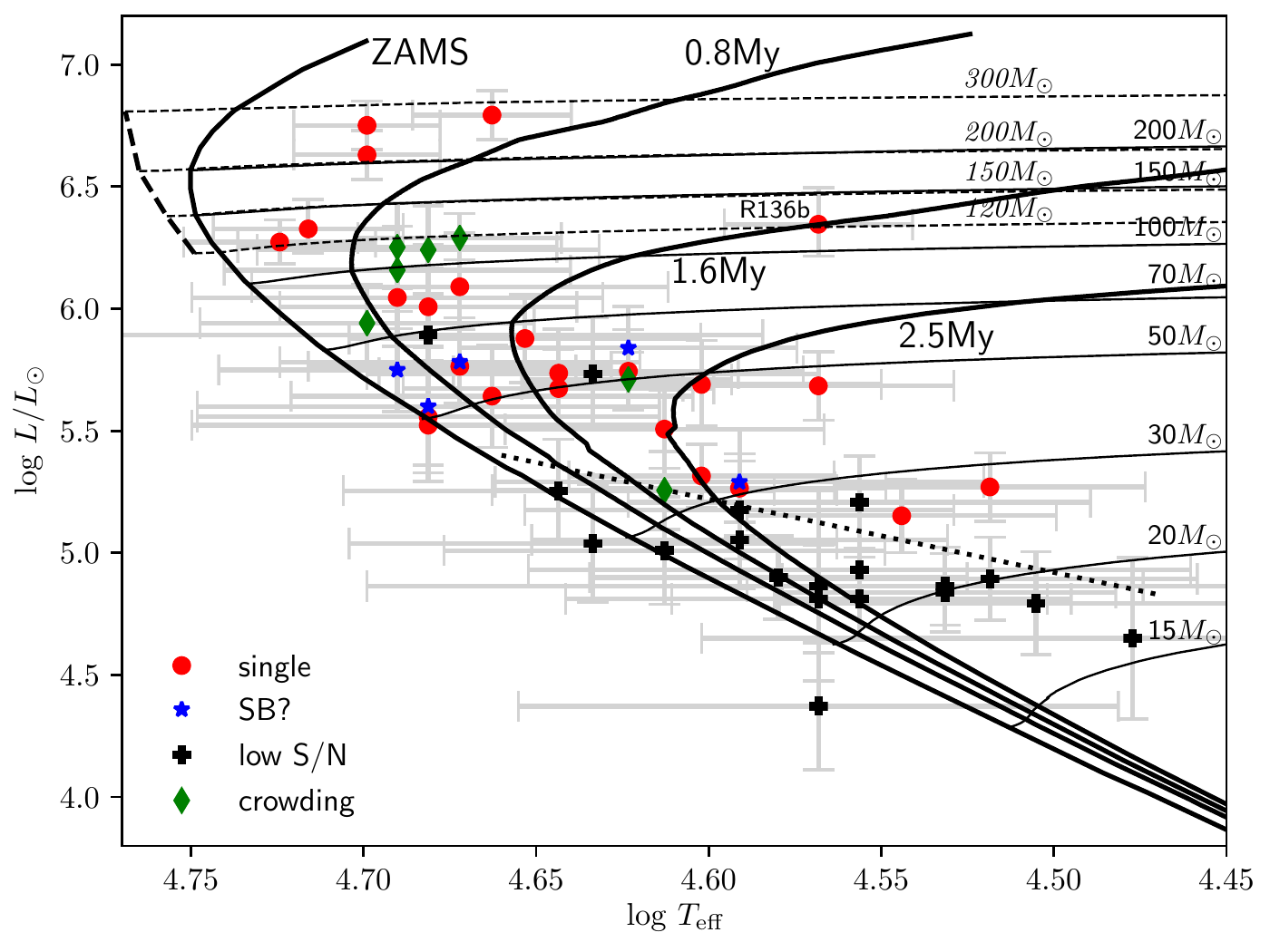}}
\end{center}
\caption{HRD of our analysed stars indicating single stars (red dots), probable spectroscopic binaries (blue stars), stars with low S/N spectra (black pluses) and contaminated objects by nearby stars (green diamonds). Evolutionary tracks are from \citet{brott2011} and \citet{koehler2015} (solid black lines) and \citet{yusof2013} (dashed black lines). Zero-age-main sequence and 0.8, 1.6, 2.5\,My isochrones are shown as well with an initial rotation rate of 180\,km/s. Black dotted line indicates our nominal S/N limit.} 
\label{f:hrd}
\end{figure}
In Fig.\,\ref{f:hrd} we show the Hertzspung-Russell diagram (HRD) for R136. Stars appearing to be single are plotted as red dots (22 stars), probable spectroscopic binaries are blue stars (7 stars), targets with low S/N spectra are black pluses (19 stars) while contaminated spectra by nearby stars as a result of crowding are shown as green diamonds (7 stars). SB2s (H42 and H77) and heavily contaminated/blended (H118) stars are not included in Fig.\,\ref{f:hrd} and were excluded from the analysis as sensible stellar parameters could not be derived. 
Stars below our nominal S/N limit ($\lesssim7$) roughly follow a diagonal line, which is indicated by a black dotted line in Fig.\,\ref{f:hrd}. 
The zero-age main-sequence (ZAMS), 0.8, 1.6 and 2.5 Myr isochrones with $\varv_{\rm rot} =180$\,km/s are visualised in Fig.\,\ref{f:hrd} as well \citep[hereafter {\sc bonn}]{brott2011, koehler2015}. The 
1.6 Myr isochrones correspond to the median age of R136 (Sect.\,\ref{s:r136_age}).
Based on their position in the HRD, stars with luminosities $\log L/L_{\odot} \gtrapprox 5.5$ are all younger than 2.5 Myr. Below this threshold there are stars that are potentially older than 2.5 Myr. However, the isochrones are closer to each other and the position of these stars overlap with 
\onecolumn
\begin{spacing}{1.2}
\begin{landscape}
\begin{center}
\begin{longtable}{@{}l@{~}l@{~}c@{~}c@{~}c@{~}c@{~}c@{}c@{}c@{}c@{}c@{}c@{}c@{}c@{}c@{}c@{}c@{}c@{}c@{}c@{}}
	\caption{Stellar parameters}\label{t:sp}\\
	\hline
    \multicolumn{1}{@{}l@{}}{ID} &  
    \multicolumn{1}{@{}l@{}}{SpT} & 
    \multicolumn{1}{@{}c@{}}{$\log L/L_{\odot}$}  &  
    \multicolumn{1}{@{}c@{}}{$T_{\rm eff}$}    &    
    \multicolumn{1}{@{}c@{}}{$\log g$}   &   
    \multicolumn{1}{@{}c@{}}{$\log\dot{M}/\sqrt{f_{\rm V}}$}   &    
    \multicolumn{1}{@{}c@{}}{$Y$}   & 
    \multicolumn{1}{@{}c@{}}{$\varv \sin i_{\rm max}$} &  
    \multicolumn{1}{@{}c@{}}{$\varv_{\infty}$} & 
    \multicolumn{1}{@{}c@{}}{$R_{\rm eff}$} &
    \multicolumn{1}{@{}c@{}}{$Q_0$} & 
    \multicolumn{1}{@{}c@{}}{$M_{\rm V}$} & 
    \multicolumn{1}{@{}c@{}}{$M_{\rm K}$} & 
    \multicolumn{1}{@{}c@{}}{$A_{\rm K}$} & 
    \multicolumn{1}{@{}c@{}}{$M_{\rm sp}$} & 
    \multicolumn{1}{@{}c@{}}{$ M_{\rm evo}$} &
    \multicolumn{1}{@{}c@{}}{$ M_{\rm evo,ini}$} & 
    \multicolumn{1}{@{}c@{}}{Age} & 
    \multicolumn{1}{@{}c@{}}{comments} & 
    \multicolumn{1}{@{}c@{}}{code} \\
    \multicolumn{1}{@{}l@{}}{} &  
    \multicolumn{1}{@{}c@{}}{} & 
    \multicolumn{1}{@{}c@{}}{}  &  
    \multicolumn{1}{@{}c@{}}{[K]}    &    
    \multicolumn{1}{@{}c@{}}{[cm\,s$^{-2}$]}   &   
    \multicolumn{1}{@{}c@{}}{[$M_{\odot}$yr$^{-1}$]}   &    
    \multicolumn{1}{@{}c@{}}{}   & 
    \multicolumn{1}{@{}c@{}}{[km\,s$^{-1}$]} &  
    \multicolumn{1}{@{}c@{}}{[km\,s$^{-1}$]} & 
    \multicolumn{1}{@{}c@{}}{[$R_{\odot}$]}   &
    \multicolumn{1}{@{}c@{}}{[ph\,s$^{-1}$]} & 
    \multicolumn{1}{@{}c@{}}{[mag]} & 
    \multicolumn{1}{@{}c@{}}{[mag]} & 
    \multicolumn{1}{@{}c@{}}{[mag]} & 
    \multicolumn{1}{@{}c@{}}{[$M_{\odot}$]} & 
    \multicolumn{1}{@{}c@{}}{[$M_{\odot}$]} & 
    \multicolumn{1}{@{}c@{}}{[$M_{\odot}$]} &
    \multicolumn{1}{@{}c@{}}{[$\mathrm{Myr}$]} & 
    \multicolumn{1}{@{}c@{}}{$^{(1)}$} & 
    \multicolumn{1}{@{}c@{}}{} \\
	\hline
	\endfirsthead
	\multicolumn{19}{c}
	{\tablename\ \thetable\ -- \textit{Continued}} \\
		\hline
    \multicolumn{1}{@{}l@{}}{ID} &  
    \multicolumn{1}{@{}l@{}}{SpT} & 
    \multicolumn{1}{@{}c@{}}{$\log L/L_{\odot}$}  &  
    \multicolumn{1}{@{}c@{}}{$T_{\rm eff}$}    &    
    \multicolumn{1}{@{}c@{}}{$\log g$}   &   
    \multicolumn{1}{@{}c@{}}{$\log\dot{M}/\sqrt{f_{\rm V}}$}   &    
    \multicolumn{1}{@{}c@{}}{$Y$}   & 
    \multicolumn{1}{@{}c@{}}{$\varv_{\rm broad}$} &  
    \multicolumn{1}{@{}c@{}}{$\varv_{\infty}$} &
    \multicolumn{1}{@{}c@{}}{$R_{\rm eff}$} & 
    \multicolumn{1}{@{}c@{}}{$Q_0$} & 
    \multicolumn{1}{@{}c@{}}{$M_{\rm V}$} & 
    \multicolumn{1}{@{}c@{}}{$M_{\rm K}$} & 
    \multicolumn{1}{@{}c@{}}{$A_{\rm K}$} & 
    \multicolumn{1}{@{}c@{}}{$M_{\rm sp}$} & 
    \multicolumn{1}{@{}c@{}}{$ M_{\rm evo}$} & 
    \multicolumn{1}{@{}c@{}}{$ M_{\rm evo,ini}$} &
    \multicolumn{1}{@{}c@{}}{Age} & 
    \multicolumn{1}{@{}c@{}}{comments} & 
    \multicolumn{1}{@{}c@{}}{code} \\
    \multicolumn{1}{@{}l@{}}{} &  
    \multicolumn{1}{@{}c@{}}{} & 
    \multicolumn{1}{@{}c@{}}{}  &  
    \multicolumn{1}{@{}c@{}}{[K]}    &    
    \multicolumn{1}{@{}c@{}}{[cm\,s$^{-2}$]}   &   
    \multicolumn{1}{@{}c@{}}{[$M_{\odot}$yr$^{-1}$]}   &    
    \multicolumn{1}{@{}c@{}}{}   & 
    \multicolumn{1}{@{}c@{}}{[km\,s$^{-1}$]} &  
    \multicolumn{1}{@{}c@{}}{[km\,s$^{-1}$]} & 
    \multicolumn{1}{@{}c@{}}{[$R_{\odot}$]}   &
    \multicolumn{1}{@{}c@{}}{[ph\,s$^{-1}$]} & 
    \multicolumn{1}{@{}c@{}}{[mag]} & 
    \multicolumn{1}{@{}c@{}}{[mag]} & 
    \multicolumn{1}{@{}c@{}}{[mag]} & 
    \multicolumn{1}{@{}c@{}}{[$M_{\odot}$]} & 
    \multicolumn{1}{@{}c@{}}{[$M_{\odot}$]} &
    \multicolumn{1}{@{}c@{}}{[$M_{\odot}$]} & 
    \multicolumn{1}{@{}c@{}}{[$\mathrm{Myr}$]} & 
    \multicolumn{1}{@{}c@{}}{$^{(1)}$} & 
    \multicolumn{1}{@{}c@{}}{} \\
	\hline
	\endhead
	\hline \multicolumn{20}{r}{\textit{Continued on next page}} \\
	\endfoot
	\hline 
	\multicolumn{20}{l}{Spectral types are from Paper\,III, (${CD98}$) and ($MH98$) are optical spectral types from \cite{crowther1998} and \cite{massey1998}, respectively,}\\ 
	\multicolumn{20}{l}{(${C16}$) are spectral types from Paper\,I based on ultraviolet spectra. $^{(1)}$apparent single star (s), potential spectroscopic binary (SB?), low S/N (s/n),}\\
	\multicolumn{20}{l}{cross-contamination as a result of crowding (c), only blue-optical wavelength range was used for the spectroscopic analysis (bo),}\\
	\multicolumn{20}{l}{nitrogen abundances: N$7.0 \equiv \epsilon_{\rm N} = 7.01$, N$8.2 \equiv \epsilon_{\rm N} = 8.2$ and N$8.5 \equiv \epsilon_{\rm N} = 8.5$. Uncertainties are $1\sigma$ confidence ranges.}
	\endlastfoot
R136a1&WN5h$^{CD98}$       &$         6.79\pm0.10$&$46000\pm2500 $& --          &$         -3.80\pm0.20$&$0.50\pm0.05$&$ 190$&$2600$&$39.2$&$50.59$&$-8.18$&$-7.68$&$ 0.26$& --                &$215^{+ 45}_{- 31}$&$251^{+ 48}_{- 35}$&$1.0^{+0.2}_{-0.2}$&s,N8.5              &CMFGEN\\
R136a2&WN5h$^{CD98}$       &$         6.75\pm0.10$&$50000\pm2500 $& --          &$         -3.84\pm0.20$&$0.55\pm0.05$&$ 150$&$2425$&$31.6$&$50.59$&$-7.80$&$-7.44$&$ 0.27$& --                &$187^{+ 23}_{- 33}$&$211^{+ 31}_{- 32}$&$1.2^{+0.2}_{-0.2}$&s,N8.5              &CMFGEN\\
R136a3&WN5h$^{CD98}$       &$         6.63\pm0.10$&$50000\pm2500 $& --          &$         -3.83\pm0.20$&$0.55\pm0.05$&$ 150$&$2400$&$27.5$&$50.47$&$-7.52$&$-7.31$&$ 0.26$& --                &$154^{+ 28}_{- 23}$&$181^{+ 29}_{- 31}$&$1.3^{+0.2}_{-0.2}$&s,N8.5              &CMFGEN\\
R136a4&O3\,V((f*))(n)     &$         6.24\pm0.18$&$48000\pm5800 $&$4.1\pm0.2$&$         -5.69\pm0.21$&$0.26^{+0.16}_{-0.06}$&$ 180$&$2475$&$19.1$&$50.06$&$-6.68$&$-5.66$&$ 0.26$&$167^{+ 98}_{- 62}$&$ 86^{+ 27}_{- 20}$&$ 89^{+ 28}_{- 20}$&$0.7^{+0.4}_{-0.6}$&c,N8.2             &FASTWIND\\
\medskip
R136a5&O2\,I(n)f*         &$6.29^{+0.10}_{-0.09}$&$47000\pm3300 $&$4.00\pm0.15$&$-4.52^{+0.19}_{-0.17}$&$0.34\pm0.10$&$ 100$&$3045$&$21.1$&$50.13$&$-6.83$&$-5.97$&$ 0.21$&$162^{+ 61}_{- 45}$&$105^{+ 18}_{- 15}$&$111^{+ 18}_{- 15}$&$1.0^{+0.3}_{-0.3}$&c,N8.5              &FASTWIND\\
R136a5&O2\,I(n)f*         &$6.28^{+0.11}_{-0.10}$&$46000\pm2500 $&$4.00\pm0.25$&$-4.59^{+0.22}_{-0.20}$&$0.30\pm0.05$&$ 100$&$3045$&$21.6$&$50.07$&$-6.86$&$-5.97$&$ 0.21$&$171^{+133}_{- 75}$&$ 96^{+ 19}_{- 13}$&$104^{+ 18}_{- 15}$&$1.2^{+0.3}_{-0.3}$&c,N8.5              &CMFGEN\\
R136a6&O2\,I(n)f*p        &$         6.27\pm0.09$&$53000\pm3500 $&$4.1\pm0.3$&$         -5.15\pm0.17$&$0.26^{+0.12}_{-0.06}$&$ 160$&$2650$&$16.2$&$50.15$&$-6.46$&$-5.46$&$ 0.16$&$121^{+120}_{- 60}$&$112^{+ 17}_{- 15}$&$115^{+ 17}_{- 15}$&$0.4^{+0.3}_{-0.4}$&s,N8.2              &FASTWIND\\
R136a7&O3\,III(f*)$^{MH98}$&$6.25^{+0.18}_{-0.17}$&$49000\pm5500 $&$4.2\pm0.5$&$-5.45^{+0.24}_{-0.20}$&$0.30^{+0.25}_{-0.10}$&$ 250$&$2710$&$18.5$&$50.07$&$-6.59$&$-5.65$&$ 0.21$&$199^{+430}_{-136}$&$ 88^{+ 29}_{- 19}$&$ 93^{+ 28}_{- 21}$&$0.8^{+0.5}_{-0.7}$&c,N8.2              &FASTWIND\\
R136b &O4\,If             &$6.35^{+0.15}_{-0.13}$&$37000\pm2400 $&$3.40\pm0.25$&$-4.61^{+0.23}_{-0.20}$&$0.30^{+0.11}_{-0.10}$&$  85$&$1400$&$36.2$&$50.15$&$-7.75$&$-7.04$&$ 0.21$&$120^{+ 99}_{- 54}$&$ 93^{+ 26}_{- 19}$&$104^{+ 31}_{- 21}$&$1.6^{+0.3}_{-0.3}$&s,N8.5              &FASTWIND\\
\medskip
R136b &O4\,If             &$6.34^{+0.12}_{-0.10}$&$35000\pm2500 $&$3.30\pm0.25$&$-4.55^{+0.22}_{-0.20}$&$0.30\pm0.05$&$  85$&$1400$&$40.0$&$50.10$&$-7.70$&$-7.04$&$ 0.21$&$117^{+ 91}_{- 51}$&$ 93^{+ 24}_{- 13}$&$107^{+ 25}_{- 17}$&$1.7^{+0.2}_{-0.2}$&s,N8.5                       &CMFGEN\\
H30   &O6.5\,Vz           &$         5.68\pm0.14$&$37000\pm3500 $&$3.90\pm0.35$&$         -6.06\pm0.19$&$0.24^{+0.09}_{-0.04}$&$ 170$&$2490$&$16.9$&$49.22$&$-6.06$&$-5.07$&$ 0.23$&$ 83^{+103}_{- 46}$&$ 40^{+  7}_{-  5}$&$ 41^{+  8}_{-  6}$&$2.8^{+0.6}_{-0.6}$&s,\,bo              &FASTWIND\\
H31   &O2\,V((f*))        &$         6.01\pm0.16$&$48000\pm5000 $&$4.00\pm0.25$&$         -5.78\pm0.20$&$0.26^{+0.19}_{-0.06}$&$ 130$&$2815$&$14.6$&$49.84$&$-6.11$&$-5.10$&$ 0.23$&$ 78^{+ 67}_{- 36}$&$ 67^{+ 17}_{- 13}$&$ 69^{+ 18}_{- 13}$&$1.1^{+0.5}_{-0.8}$&s,N8.2              &FASTWIND\\
H35   &O3\,V              &$         5.74\pm0.18$&$44000\pm5600 $&$4.0\pm0.4$&$         -5.88\pm0.21$&$0.24^{+0.10}_{-0.04}$&$ 180$&$2770$&$12.7$&$49.51$&$-5.67$&$-4.68$&$ 0.19$&$ 59^{+ 82}_{- 34}$&$ 47^{+ 11}_{-  9}$&$ 48^{+ 11}_{-  9}$&$1.7^{+0.7}_{-1.1}$&s,N7.0              &FASTWIND\\
H36   &O2\,If*            &$6.33^{+0.12}_{-0.10}$&$52000\pm3400 $&$4.10\pm0.35$&$-4.74^{+0.19}_{-0.17}$&$0.20^{+0.07}_{-0.00}$&$ 125$&$3500$&$18.0$&$50.21$&$-6.62$&$-5.74$&$ 0.21$&$148^{+176}_{- 80}$&$118^{+ 24}_{- 17}$&$122^{+ 23}_{- 18}$&$0.4^{+0.3}_{-0.4}$&s,N8.2              &FASTWIND\\
\medskip
H36   &O2\,If*            &$6.29^{+0.13}_{-0.10}$&$48000\pm2500 $&$4.00\pm0.25$&$-4.78^{+0.22}_{-0.20}$&$0.25^{+0.00}_{-0.05}$&$ 125$&$3500$&$20.2$&$50.12$&$-6.71$&$-5.74$&$ 0.21$&$149^{+116}_{- 65}$&$103^{+ 21}_{- 14}$&$109^{+ 22}_{- 16}$&$1.0^{+0.3}_{-0.3}$&s,N8.5              &CMFGEN\\
H40   &O3\,V              &$         5.88\pm0.18$&$45000\pm5600 $&$3.9\pm0.4$&$         -6.08\pm0.21$&$0.26^{+0.28}_{-0.06}$&$ 150$&$2750$&$14.3$&$49.68$&$-5.98$&$-4.98$&$ 0.21$&$ 59^{+ 86}_{- 35}$&$ 54^{+ 13}_{- 12}$&$ 56^{+ 14}_{- 12}$&$1.6^{+0.7}_{-1.0}$&s,N7.0              &FASTWIND\\
H45   &O4:\,Vz            &$5.84^{+0.17}_{-0.16}$&$42000\pm5000 $&$4.00\pm0.45$&$-6.58^{+0.24}_{-0.20}$&$0.24^{+0.31}_{-0.04}$&$ 170$&$2620$&$15.7$&$49.55$&$-6.09$&$-5.07$&$ 0.21$&$ 90^{+152}_{- 56}$&$ 50^{+ 12}_{-  9}$&$ 52^{+ 12}_{- 10}$&$1.9^{+0.7}_{-1.0}$&SB?                 &FASTWIND\\
H46   &O2-3\,III(f*)      &$6.16^{+0.18}_{-0.17}$&$49000\pm6000 $&$4.20\pm0.35$&$-5.16^{+0.24}_{-0.20}$&$0.24^{+0.13}_{-0.04}$&$ 155$&$3440$&$16.6$&$49.99$&$-6.38$&$-5.41$&$ 0.21$&$160^{+198}_{- 89}$&$ 80^{+ 24}_{- 16}$&$ 83^{+ 24}_{- 18}$&$0.6^{+0.5}_{-0.6}$&c,N8.5              &FASTWIND\\
H47   &O2\,V((f*))        &$6.09^{+0.22}_{-0.21}$&$47000\pm7000 $&$4.0\pm0.4$&$-5.24^{+0.28}_{-0.22}$&$0.24^{+0.22}_{-0.04}$&$ 165$&$3045$&$16.7$&$49.92$&$-6.37$&$-5.38$&$ 0.21$&$102^{+154}_{- 61}$&$ 65^{+ 25}_{- 15}$&$ 68^{+ 25}_{- 17}$&$1.1^{+0.6}_{-0.9}$&s,N8.2              &FASTWIND\\
\medskip
H48   &O2-3\,III(f*)      &$6.05^{+0.21}_{-0.20}$&$49000\pm7200 $&$4.10\pm0.35$&$-5.33^{+0.27}_{-0.22}$&$0.24^{+0.22}_{-0.04}$&$ 150$&$3045$&$14.6$&$49.88$&$-6.13$&$-5.12$&$ 0.21$&$ 98^{+122}_{- 54}$&$ 66^{+ 22}_{- 15}$&$ 68^{+ 23}_{- 16}$&$0.8^{+0.6}_{-0.8}$&s,N8.2              &FASTWIND\\
H49   &O3\,V$^{MH98}$      &$         5.89\pm0.37$&$48000\pm12000$&$4.2\pm1.0$&$         -5.63\pm0.32$&$0.24^{+0.32}_{-0.04}$&$ 155$&$2980$&$12.8$&$49.70$&$-5.80$&$-4.78$&$ 0.27$&$ 94^{+848}_{- 85}$&$ 38^{+ 22}_{- 13}$&$ 39^{+ 23}_{- 14}$&$1.0^{+1.3}_{-1.1}$&s/n                 &FASTWIND\\
H50   &O3-4\,V((f*))      &$         5.71\pm0.11$&$42000\pm3000 $&$3.8\pm0.4$&$         -6.17\pm0.18$&$0.24^{+0.14}_{-0.04}$&$ 200$&$2620$&$13.5$&$49.47$&$-5.79$&$-4.79$&$ 0.26$&$ 42^{+ 59}_{- 25}$&$ 47^{+  6}_{-  6}$&$ 48^{+  7}_{-  6}$&$2.2^{+0.5}_{-0.7}$&c,N7.0              &FASTWIND\\
H52   &O3-4\,Vz           &$         5.67\pm0.16$&$44000\pm4800 $&$4.00\pm0.25$&$         -5.92\pm0.20$&$0.24^{+0.08}_{-0.04}$&$ 180$&$2820$&$11.8$&$49.44$&$-5.52$&$-4.52$&$ 0.23$&$ 51^{+ 42}_{- 23}$&$ 45^{+  9}_{-  8}$&$ 46^{+  9}_{-  8}$&$1.7^{+0.7}_{-1.1}$&s                   &FASTWIND\\
H55   &O2\,V((f*))z       &$         5.76\pm0.15$&$47000\pm5000 $&$3.9\pm0.3$&$         -5.92\pm0.19$&$0.24^{+0.13}_{-0.04}$&$ 130$&$2880$&$11.5$&$49.61$&$-5.59$&$-4.58$&$ 0.24$&$ 38^{+ 36}_{- 19}$&$ 52^{+ 10}_{-  9}$&$ 53^{+ 11}_{-  9}$&$1.5^{+0.6}_{-1.0}$&s,N7.0              &FASTWIND\\
\medskip
H58   &O2-3\,V:           &$         5.94\pm0.16$&$50000\pm5900 $&$4.1\pm0.4$&$         -6.62\pm0.20$&$0.26^{+0.31}_{-0.06}$&$ 150$&$2980$&$12.5$&$49.78$&$-5.84$&$-4.79$&$ 0.21$&$ 71^{+103}_{- 42}$&$ 63^{+ 17}_{- 12}$&$ 66^{+ 16}_{- 13}$&$0.8^{+0.6}_{-0.8}$&c,\,bo,N7.0         &FASTWIND\\
H62   &O2-3\,V            &$         5.75\pm0.17$&$49000\pm6200 $&$4.00\pm0.45$&$         -5.81\pm0.20$&$0.26^{+0.29}_{-0.06}$&$ 170$&$2770$&$10.4$&$49.59$&$-5.41$&$-4.38$&$ 0.24$&$ 39^{+ 72}_{- 25}$&$ 50^{+ 13}_{- 10}$&$ 52^{+ 12}_{- 10}$&$1.1^{+0.7}_{-1.1}$&SB?,N8.2            &FASTWIND\\
H64   &O4-5\,V:           &$5.69^{+0.18}_{-0.17}$&$40000\pm5100 $&$3.9\pm0.3$&$-6.38^{+0.25}_{-0.21}$&$0.24^{+0.34}_{-0.04}$&$ 180$&$1770$&$14.6$&$49.37$&$-5.86$&$-4.86$&$ 0.21$&$ 61^{+ 67}_{- 32}$&$ 41^{+ 10}_{-  7}$&$ 43^{+ 10}_{-  8}$&$2.3^{+0.8}_{-1.0}$&s                   &FASTWIND\\
H65   &O4\,V$^{C16}$      &$5.74^{+0.17}_{-0.16}$&$42000\pm5200 $&$3.90\pm0.55$&$-6.17^{+0.24}_{-0.20}$&$0.24^{+0.35}_{-0.04}$&$ 160$&$2540$&$14.1$&$49.48$&$-5.84$&$-4.85$&$ 0.21$&$ 57^{+141}_{- 41}$&$ 45^{+ 11}_{-  8}$&$ 47^{+ 11}_{-  8}$&$2.1^{+0.8}_{-1.1}$&s/n                 &FASTWIND\\
H66   &O2\,V-III(f*)      &$         5.64\pm0.21$&$46000\pm6600 $&$4.10\pm0.55$&$         -5.65\pm0.22$&$0.24^{+0.21}_{-0.04}$&$ 115$&$2590$&$10.4$&$49.44$&$-5.30$&$-4.31$&$ 0.24$&$ 50^{+135}_{- 36}$&$ 42^{+ 12}_{-  9}$&$ 42^{+ 12}_{-  9}$&$1.3^{+0.8}_{-1.3}$&s,\,bo,N8.2         &FASTWIND\\
\medskip
H68   &O4-5\,Vz           &$5.73^{+0.23}_{-0.22}$&$43000\pm7000 $&$4.0\pm0.4$&$-6.89^{+0.29}_{-0.23}$&$0.24^{+0.28}_{-0.04}$&$ 210$&$1910$&$13.3$&$49.47$&$-5.76$&$-4.74$&$ 0.21$&$ 64^{+101}_{- 39}$&$ 42^{+ 13}_{-  9}$&$ 44^{+ 14}_{- 10}$&$1.8^{+0.9}_{-1.3}$&s/n,\,bo            &FASTWIND\\
H69   &O4-5\,Vz           &$         5.51\pm0.16$&$41000\pm4600 $&$4.1\pm0.3$&$         -6.29\pm0.20$&$0.24^{+0.09}_{-0.04}$&$ 130$&$2580$&$11.2$&$49.18$&$-5.32$&$-4.31$&$ 0.21$&$ 58^{+ 60}_{- 30}$&$ 37^{+  7}_{-  6}$&$ 37^{+  7}_{-  6}$&$2.1^{+0.8}_{-1.3}$&s                   &FASTWIND\\
H70   &O5\,Vz             &$         5.78\pm0.18$&$47000\pm6000 $&$4.20\pm0.35$&$         -5.96\pm0.21$&$0.24^{+0.16}_{-0.04}$&$ 165$&$2670$&$11.7$&$49.57$&$-5.59$&$-4.56$&$ 0.27$&$ 80^{+ 94}_{- 43}$&$ 51^{+ 13}_{- 10}$&$ 52^{+ 13}_{- 10}$&$0.9^{+0.7}_{-0.9}$&SB?,N7.0            &FASTWIND\\
H71   &O2-3\,V((f*))      &$         5.56\pm0.23$&$48000\pm8000 $&$3.9\pm0.7$&$         -7.00\pm0.24$&$0.26^{+0.35}_{-0.06}$&$ 140$&$2475$&$ 8.7$&$49.41$&$-5.03$&$-4.01$&$ 0.22$&$ 22^{+ 88}_{- 17}$&$ 38^{+ 11}_{-  9}$&$ 38^{+ 12}_{-  9}$&$1.2^{+1.1}_{-1.2}$&s,N7.0              &FASTWIND\\
H73   &O9.7-B0\,V         &$         5.27\pm0.14$&$33000\pm3600 $&$4.3\pm0.4$&              --       &$0.24^{+0.10}_{-0.04}$&$ 125$&   -- &$13.2$&$48.31$&$-5.30$&$-4.32$&$ 0.21$&$127^{+184}_{- 75}$&$ 26^{+  4}_{-  3}$&$ 27^{+  4}_{-  3}$&$4.0^{+1.2}_{-1.4}$&s                   &FASTWIND\\
\medskip
H75   &O6\,V              &$         5.29\pm0.22$&$39000\pm6900 $&$4.3\pm0.5$&$         -6.41\pm0.23$&$0.24^{+0.28}_{-0.04}$&$ 145$&$2550$&$ 9.7$&$48.84$&$-4.92$&$-3.90$&$ 0.19$&$ 68^{+142}_{- 46}$&$ 28^{+  7}_{-  5}$&$ 28^{+  7}_{-  6}$&$1.9^{+1.2}_{-1.9}$&SB?                 &FASTWIND\\
H78   &O4:\,V             &$         5.60\pm0.24$&$48000\pm8000 $&$4.20\pm0.45$&$         -6.00\pm0.24$&$0.24^{+0.33}_{-0.04}$&$ 105$&$2375$&$ 9.1$&$49.41$&$-5.07$&$-4.05$&$ 0.24$&$ 48^{+ 87}_{- 31}$&$ 39^{+ 13}_{-  9}$&$ 40^{+ 13}_{- 10}$&$0.7^{+1.4}_{-0.7}$&SB?,N8.2            &FASTWIND\\
H80   &O8\,V              &$         5.15\pm0.15$&$35000\pm3800 $&$3.8\pm0.4$&$         -7.16\pm0.20$&$0.24^{+0.13}_{-0.04}$&$ 155$&$1655$&$10.2$&$48.58$&$-4.91$&$-3.92$&$ 0.20$&$ 24^{+ 39}_{- 15}$&$ 25^{+  4}_{-  3}$&$ 25^{+  4}_{-  3}$&$4.2^{+1.4}_{-1.7}$&s                   &FASTWIND\\
H86   &O5:\,V             &$         5.26\pm0.16$&$41000\pm5000 $&$3.8\pm0.4$&$         -6.02\pm0.20$&$0.24^{+0.07}_{-0.04}$&$ 175$&$2475$&$ 8.4$&$49.01$&$-4.69$&$-3.74$&--    &$ 16^{+ 23}_{-  9}$&$ 29^{+  5}_{-  5}$&$ 30^{+  5}_{-  5}$&$2.7^{+1.2}_{-1.8}$&c                   &FASTWIND\\
H90   &O4:\,V:            &$         5.32\pm0.13$&$40000\pm3700 $&$4.1\pm0.5$&$         -6.15\pm0.19$&$0.24^{+0.19}_{-0.04}$&$ 130$&$2475$&$ 9.5$&$48.96$&$-4.91$&$-3.91$&$ 0.25$&$ 41^{+ 83}_{- 27}$&$ 31^{+  5}_{-  4}$&$ 32^{+  5}_{-  4}$&$2.5^{+1.0}_{-1.6}$&s                   &FASTWIND\\
\medskip
H92   &O6\,Vz             &$         5.26\pm0.14$&$39000\pm4000 $&$4.00\pm0.45$&$         -7.06\pm0.19$&$0.24^{+0.23}_{-0.04}$&$ 150$&$2080$&$ 9.4$&$48.87$&$-4.87$&$-3.86$&$ 0.24$&$ 32^{+ 56}_{- 21}$&$ 30^{+  4}_{-  4}$&$ 30^{+  5}_{-  4}$&$2.8^{+1.1}_{-1.7}$&s                   &FASTWIND\\
H94   &O4-5\,Vz           &$         5.52\pm0.23$&$48000\pm8200 $&$4.20\pm0.45$&$         -6.50\pm0.24$&$0.24^{+0.33}_{-0.04}$&$ 170$&$2490$&$ 8.3$&$49.32$&$-4.90$&$-3.85$&$ 0.21$&$ 40^{+ 71}_{- 26}$&$ 37^{+ 11}_{-  9}$&$ 37^{+ 11}_{-  8}$&$0.6^{+1.6}_{-0.6}$&s,N7.0              &FASTWIND\\
H108  &O\,Vn              &$         5.04\pm0.24$&$43000\pm7600 $&$4.20\pm0.55$&$         -7.31\pm0.26$&$0.24^{+0.31}_{-0.04}$&$ 260$&$1040$&$ 6.0$&$48.75$&$-4.00$&$-2.98$&--    &$ 21^{+ 49}_{- 14}$&$ 23^{+  6}_{-  5}$&$ 23^{+  6}_{-  5}$&$0.0^{+3.5}_{-0.0}$&s/n                 &FASTWIND\\
H112  &O7-9\,Vz           &$         5.21\pm0.19$&$36000\pm6000 $&$4.3\pm0.7$&              --       &$0.26^{+0.35}_{-0.06}$&$ 160$&   -- &$10.3$&$48.57$&$-4.93$&$-3.92$&$ 0.21$&$ 77^{+311}_{- 62}$&$ 25^{+  6}_{-  4}$&$ 26^{+  6}_{-  5}$&$3.3^{+1.5}_{-2.4}$&s/n,\,bo            &FASTWIND\\
H114  &O5-6\,V            &$         5.25\pm0.21$&$44000\pm6800 $&$4.2\pm0.5$&$         -7.33\pm0.23$&$0.26^{+0.35}_{-0.06}$&$ 100$&$1770$&$ 7.3$&$48.98$&$-4.48$&$-3.44$&$ 0.21$&$ 31^{+ 67}_{- 21}$&$ 29^{+  7}_{-  6}$&$ 29^{+  7}_{-  6}$&$0.8^{+2.0}_{-0.8}$&s/n                 &FASTWIND\\
\medskip
H116  &O7\,V              &$         4.84\pm0.16$&$34000\pm4100 $&$3.70\pm0.55$&$         -7.71\pm0.24$&$0.26^{+0.35}_{-0.06}$&$ 185$&$ 960$&$ 7.6$&$48.22$&$-4.21$&$-3.23$&$ 0.19$&$ 10^{+ 26}_{-  7}$&$ 19^{+  3}_{-  2}$&$ 19^{+  3}_{-  3}$&$5.0^{+2.1}_{-3.0}$&s/n                 &FASTWIND\\
H120  &--                 &$         4.81\pm0.22$&$37000\pm6800 $&$4.3\pm0.6$&              --       &$0.24^{+0.22}_{-0.04}$&$ 150$&   -- &$ 6.2$&$48.25$&$-3.87$&$-2.86$&$ 0.12$&$ 28^{+ 81}_{- 21}$&$ 19^{+  4}_{-  4}$&$ 19^{+  4}_{-  3}$&$1.1^{+3.7}_{-1.1}$&s/n                 &FASTWIND\\
H121  &O9.5\,V            &$         4.86\pm0.16$&$34000\pm4800 $&$4.20\pm0.65$&              --       &$0.30^{+0.35}_{-0.10}$&$ 150$&   -- &$ 7.8$&$48.05$&$-4.22$&$-3.22$&$ 0.20$&$ 35^{+118}_{- 27}$&$ 20^{+  3}_{-  3}$&$ 20^{+  3}_{-  3}$&$4.1^{+1.9}_{-3.1}$&s/n,\,bo            &FASTWIND\\
H123  &O6\,V              &$         5.01\pm0.22$&$41000\pm6500 $&$4.10\pm0.45$&$         -7.48\pm0.24$&$0.26^{+0.35}_{-0.06}$&$ 120$&$1615$&$ 6.3$&$48.67$&$-4.09$&$-3.07$&$ 0.19$&$ 18^{+ 31}_{- 12}$&$ 23^{+  5}_{-  4}$&$ 23^{+  5}_{-  4}$&$1.5^{+2.1}_{-1.6}$&s/n                 &FASTWIND\\
H129  &--                 &$         4.37\pm0.26$&$37000\pm8200 $&$4.00\pm0.65$&              --       &$0.24^{+0.35}_{-0.04}$&$ 135$&   -- &$ 3.7$&$47.91$&$-2.76$&$-1.80$&--    &$  5^{+ 18}_{-  4}$&$ 13^{+  3}_{-  3}$&$ 13^{+  3}_{-  3}$&$0.1^{+8.6}_{-0.1}$&s/n                 &FASTWIND\\
\medskip
H132  &O7:\,V             &$         5.05\pm0.20$&$39000\pm5800 $&$4.0\pm0.5$&              --       &$0.26^{+0.35}_{-0.06}$&$ 115$&   -- &$ 7.4$&$48.68$&$-4.32$&$-3.34$&$ 0.21$&$ 20^{+ 44}_{- 14}$&$ 23^{+  5}_{-  4}$&$ 24^{+  5}_{-  4}$&$2.6^{+1.5}_{-2.4}$&s/n                 &FASTWIND\\
H134  &O7\,Vz             &$         4.81\pm0.17$&$36000\pm4800 $&$4.00\pm0.35$&$         -7.67\pm0.23$&$0.24^{+0.30}_{-0.04}$&$ 105$&$1170$&$ 6.6$&$48.25$&$-3.96$&$-2.96$&$ 0.18$&$ 16^{+ 20}_{-  9}$&$ 19^{+  3}_{-  3}$&$ 20^{+  3}_{-  3}$&$3.5^{+1.9}_{-2.8}$&s/n                 &FASTWIND\\
H135  &B                  &$         4.89\pm0.17$&$33000\pm4900 $&$4.0\pm0.5$&              --       &$0.26^{+0.35}_{-0.06}$&$ 125$&   -- &$ 8.6$&$48.05$&$-4.37$&$-3.40$&$ 0.23$&$ 27^{+ 54}_{- 18}$&$ 19^{+  3}_{-  3}$&$ 20^{+  3}_{-  3}$&$4.8^{+2.1}_{-3.0}$&s/n                 &FASTWIND\\
H139  &--                 &$         4.90\pm0.17$&$38000\pm5100 $&$4.00\pm0.45$&              --       &$0.26^{+0.35}_{-0.06}$&$  85$&   -- &$ 6.5$&$48.48$&$-4.01$&$-3.04$&$ 0.19$&$ 15^{+ 29}_{- 10}$&$ 21^{+  4}_{-  3}$&$ 21^{+  4}_{-  3}$&$2.6^{+1.7}_{-2.5}$&s/n,\,bo            &FASTWIND\\
H141  &O5-6\,V$^{C16}$    &$         4.79\pm0.21$&$32000\pm6000 $&$3.6\pm0.7$&              --       &$0.24^{+0.28}_{-0.04}$&$ 120$&   -- &$ 8.1$&$48.07$&$-4.24$&$-3.29$&$ 0.20$&$ 10^{+ 38}_{-  8}$&$ 17^{+  4}_{-  3}$&$ 17^{+  4}_{-  3}$&$5.9^{+2.9}_{-3.9}$&s/n                 &FASTWIND\\
\medskip
H143  &O8-9\,V-III        &$         5.18\pm0.20$&$39000\pm6000 $&$4.2\pm0.5$&$         -6.35\pm0.23$&$0.24^{+0.35}_{-0.04}$&$ 150$&$1480$&$ 8.5$&$48.78$&$-4.62$&$-3.65$&$ 0.27$&$ 42^{+ 90}_{- 28}$&$ 26^{+  6}_{-  4}$&$ 26^{+  6}_{-  4}$&$2.2^{+1.3}_{-2.1}$&s/n                 &FASTWIND\\
H159  &--                 &$         4.93\pm0.28$&$36000\pm8900 $&$4.3\pm0.9$&              --       &$0.24^{+0.35}_{-0.04}$&$ 150$&   -- &$ 7.5$&$48.30$&$-4.24$&$-3.23$&$ 0.25$&$ 41^{+277}_{- 36}$&$ 18^{+  6}_{-  4}$&$ 19^{+  6}_{-  4}$&$2.6^{+2.9}_{-2.6}$&s/n,\,bo            &FASTWIND\\
H162  &--                 &$         4.87\pm0.39$&$37000\pm13000$&$4.3\pm1.0$&              --       &$0.24^{+0.35}_{-0.04}$&$ 150$&   -- &$ 6.6$&$48.36$&$-3.98$&$-3.03$&$ 0.20$&$ 32^{+284}_{- 28}$&$ 15^{+  6}_{-  4}$&$ 15^{+  6}_{-  4}$&$1.7^{+5.6}_{-1.7}$&s/n                 &FASTWIND\\
H173  &O9+\,V$^{C16}$   &$         4.65\pm0.33$&$30000\pm10000$&$4.30\pm0.85$&              --       &$0.24^{+0.25}_{-0.04}$&$ 140$&   -- &$ 7.8$&$47.14$&$-3.97$&$-3.03$&$ 0.17$&$ 45^{+271}_{- 38}$&$ 13^{+  4}_{-  3}$&$ 13^{+  4}_{-  3}$&$4.0^{+5.2}_{-4.0}$&s/n                 &FASTWIND\\
\end{longtable}
\end{center}
\end{landscape}
\end{spacing}
\twocolumn 
\noindent both isochrones within their uncertainties. An age determination based on the location in the HRD is inaccurate as the uncertainties of the stellar parameters of those stars are large. We applied the more sophisticated tool such as BONNSAI\footnote{\url{https://www.astro.uni-bonn.de/stars/bonnsai/}} \citep{schneider2014} to quantify the age distribution in R136, which is described in Sect.\,\ref{s:masses_ages}.

At the high mass end we also added the 120, 150, 200 and 300\,$M_{\odot}$ tracks from \citet[hereafter {\sc geneva}]{yusof2013}. {\sc bonn} and {\sc geneva} tracks are comparable, but there is an offset for the location of the ZAMS. The difference between tracks is that the stellar structure models by the {\sc bonn} group allow stellar envelope inflation to occur as a result of their proximity to the Eddington limit \citep{sanyal2015}. Therefore, these tracks have cooler effective temperatures already at the high mass ZAMS ($\log L/L_{\odot} \gtrsim 6.5$). The mass range of our targets are between 10 and 300\,$M_{\odot}$ based on the evolutionary tracks. Most stars populate the region near the ZAMS. R136b appears to be isolated from the rest of our sample. This star has the lowest determined surface gravity of our entire sample. To match the strong nitrogen lines in the spectrum of R136b a high N-abundance similar to the WNh stars was required, which indicates that carbon and oxygen are largely converted into nitrogen as a result of the CNO cycle 
(Fig.\,\ref{f:r136b}).



\subsection{Stellar masses and ages \label{s:masses_ages}}

\begin{figure}
\begin{center}
\resizebox{\hsize}{!}{\includegraphics{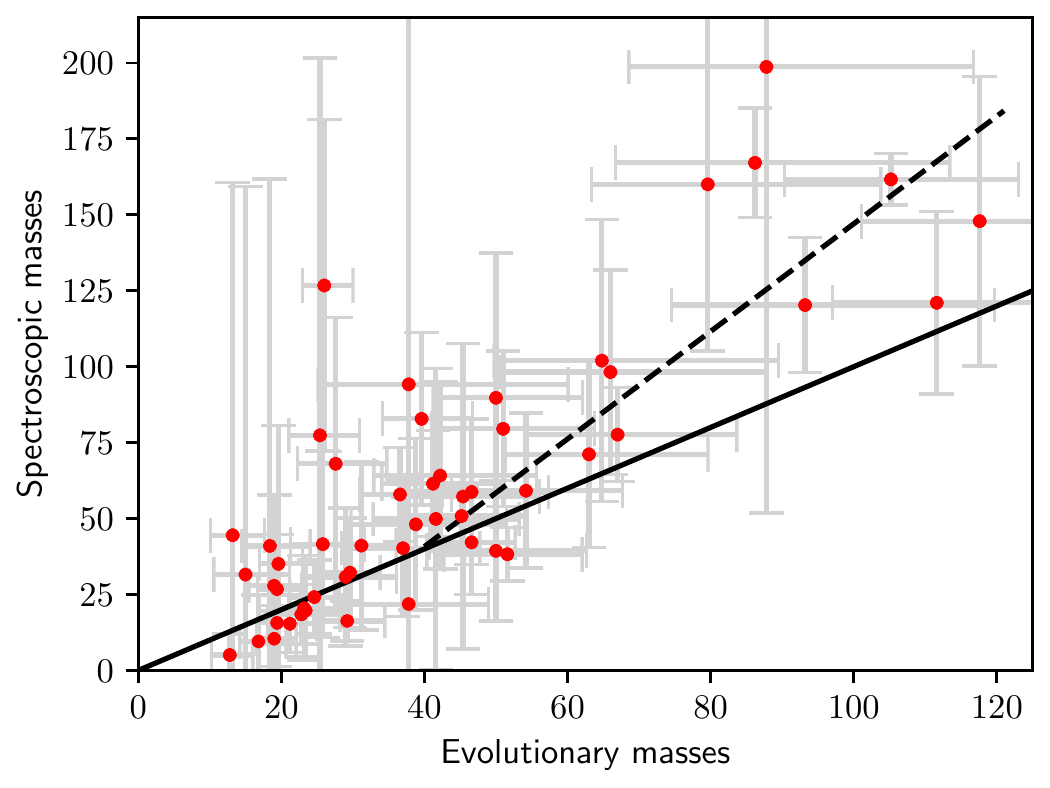}}
\end{center}
\caption{Spectroscopic versus current evolutionary masses: even though both masses mostly agreement within their uncertainties, a systematic offset develops toward higher masses (dashed line).}
\label{f:masses}
\end{figure}
\begin{figure}
\begin{center}
\resizebox{\hsize}{!}{\includegraphics{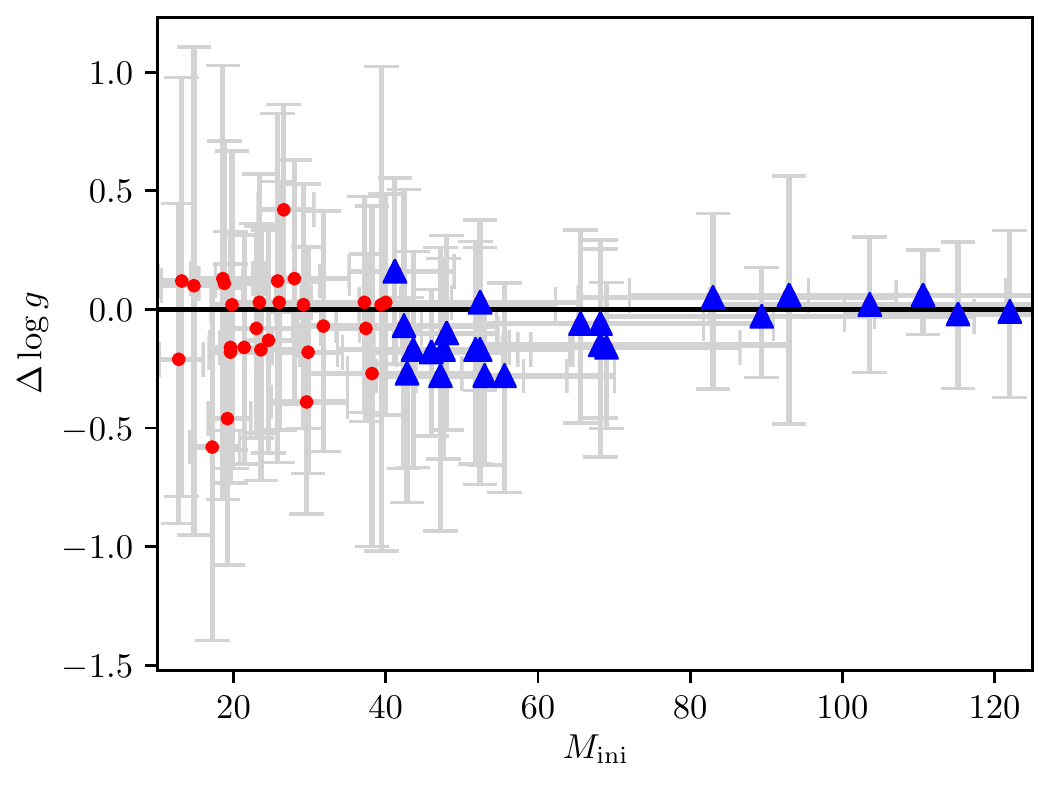}}
\end{center}
\caption{Spectroscopic minus evolutionary gravities against initial stellar mass (red dots). Stars with initial mass $>40 M_{\odot}$ are shown as blue triangles. Up to $\sim 80\,M_{\odot}$ spectroscopic gravities are systematically smaller than evolutionary gravities, which is expected for the negative mass-discrepancy.}
\label{f:logg}
\end{figure}
\begin{figure}
\begin{center}
\resizebox{\hsize}{!}{\includegraphics{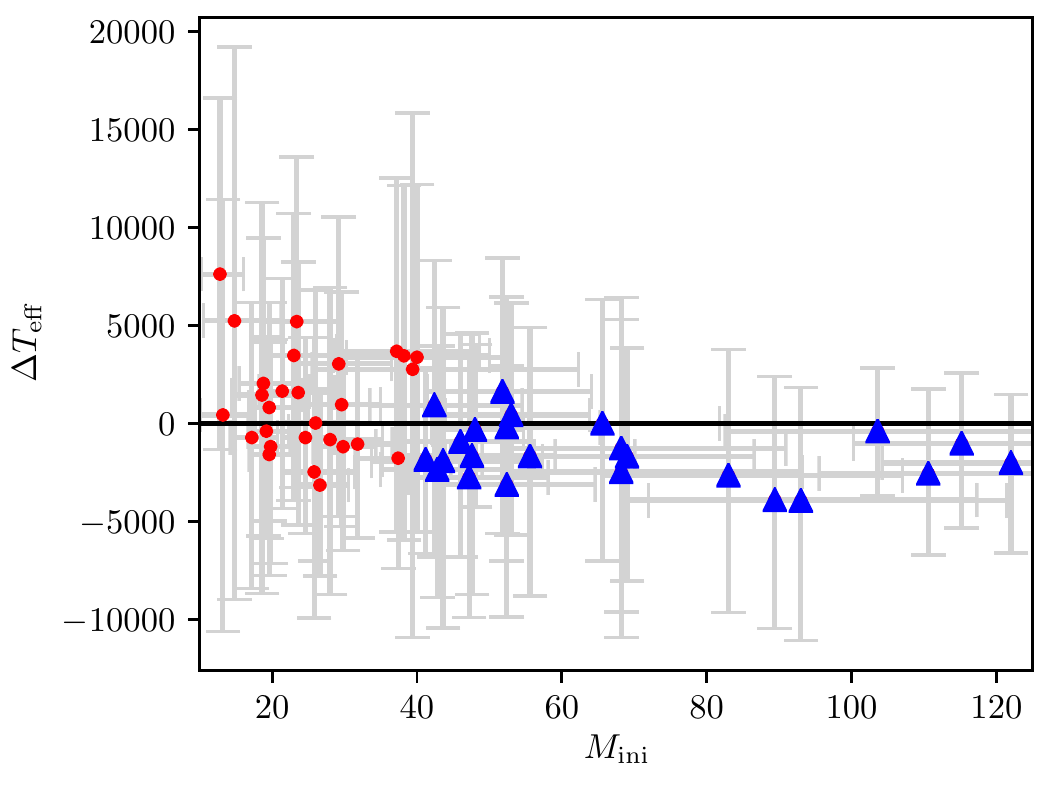}}
\end{center}
\caption{Spectroscopic minus evolutionary temperature against initial stellar mass (red dots). Stars with initial mass $>40 M_{\odot}$ are shown as blue triangles. Evolutionary temperatures are systematically larger for stars more massive than $\sim 40\,M_{\odot}$.}
\label{f:teff}
\end{figure}

Stellar evolutionary masses and ages for our targets are derived with the BONN Stellar Astrophysics Interface \citep[BONNSAI,][]{schneider2014}. BONNSAI is a Bayesian tool to calculate the probability distributions of fundamental stellar parameters for a given set of observed stellar parameters including their uncertainties. It also provides predictions of unobserved quantities and tests stellar evolutionary models. Our input to BONNSAI were luminosity, effective temperature, surface gravity and helium abundances. WNh stars analysed with CMFGEN had too strong stellar winds to constrain $\log g$. For those objects the surface gravity was not an input parameter to BONNSAI. 

In Fig.\,\ref{f:masses} we compare the spectroscopic masses with the evolutionary masses ($M_{\rm evo}$) derived with BONNSAI. The three WNh stars have no $\log g$ determination and are excluded from Fig.\,\ref{f:masses} to \ref{f:teff}. For the evolutionary masses we used the mode of the probability distribution function (PDF). Spectroscopic masses were calculated with the surface gravities and radii given in Table\,\ref{t:sp}. 
With increasing stellar mass we see a systematic trend toward larger spectroscopic than evolutionary masses (positive mass-discrepancy), especially for $M_{\rm evo} \gtrsim 70\,M_{\odot}$. 

To investigate this further we compare the differences of spectroscopic $\log g$ versus the mode of the BONNSAI probability distribution of $\log g$. 
From Fig.\,\ref{f:logg} it seems that up to $80\,M_{\odot}$ the stellar models prefer higher $\log g$ values which would also place the stars closer to the ZAMS. For those objects spectroscopic gravities are lower than evolutionary values, which is the typical case for the negative mass-discrepancy. Already three decades ago \cite{herrero1992} report that evolutionary masses are systematically larger than spectroscopic masses. However, our sample does not allow us to draw any conclusion on the negative mass-discrepancy. Stars with $M_{\rm ini}>80\,M_{\odot}$ the gravities agree and no systematic trend is visible. 

The only other variable quantity, which goes into the calculation of the spectroscopic mass, is the stellar radius, which is defined by the luminosity and temperature ($M \propto gR^2 \propto gL/T^4$). We compared the spectroscopic luminosities with those with the highest probability by BONNSAI (mode of PDF) and found a systematic offset of $-0.08$\,dex (Fig.\,\ref{f:lum}). For the given set of stellar parameters stellar evolution models systematically under-predict the stellar luminosity, which leads to a lower evolutionary mass. However, the systematic occurs over the whole mass and luminosity range. 

The picture is different for the temperature (Fig.\,\ref{f:teff}). The temperatures of stars with $M_{\rm ini} > 40\,M_{\odot}$ are systematically over-predicted by the evolutionary models. This could be a result of the stellar wind of the most massive stars. The outer-boundary in stellar structure calculations is approximated by a plane-parallel grey atmosphere without wind. The effect of wind-blanketing is neglected as well, which alters the temperature and ionisation structure of the stellar atmosphere \citep{hummer1982,kudritzki1989b}. The mass-loss rate is only a parameter, which removes mass from the star. With increase stellar mass and luminosity the mass-loss rates increases as well and the stellar wind becomes more and more optically thick. The photosphere which is defined at an optical depth of $\tau = 2/3$ gradually shifts into the stellar winds and is then also referred as a pseudo photosphere. A comparison with plane-parallel stellar atmosphere models without winds computed with CMFGEN showed that this temperature offset for those stars in our sample is between a few hundred to around 1000\,K, which is well within the temperature uncertainties. This discrepancy largely depends on the $\beta$ exponent of the velocity law rather than the mass-loss rate. The strong dependence on the temperature might explain the discrepancy between spectroscopic and evolutionary masses for stars with ($M_{\rm ini} \gtrsim 80\,M_{\odot}$), where spectroscopic and evolutionary gravities largely agree.

By considering only stars with masses greater than $80\,M_{\odot}$ we investigate how these systematics add up. For those objects the average mass ratio of $M_{\rm spec}/M_{\rm evo} = 1.52$. The average $g_{\rm spec}/g_{\rm evo} = 1.04$, $L_{\rm spec}/L_{\rm evo} = 1.01$ and $(T_{\rm evo}/T_{\rm spec})^4 = 1.20$, which leads to $M_{\rm spec}/M_{\rm evo} = 1.27$. The systematics described above can only partially explain the observed positive mass-discrepancy. We conclude that the high spectroscopic masses cannot be reproduced by current stellar models. Relevant physics might be not included or not well enough understood. However, mixing or binary mass transfer would even increase the mass-discrepancy as it would lead to even lower evolutionary masses. 

To summarize BONNSAI systematically under predicts $L$ over the whole mass range. The temperatures for the most massive stars ($\gtrsim 40 M_{\odot}$) are over-predicted while gravities are over-predicted for stars less massive than $\sim 80\,M_{\odot}$. The shift to higher temperatures and gravities can also implicate younger ages. At luminosities above $\log L/L_{\odot} \sim 5$ the isochrones fan out and changes in the temperature are less critical on the resulting age (Fig.\,\ref{f:hrd}). The systematics could be a result that star occupied different location in the HRD for a different period of time. 
This information is provided to BONNSAI by the evolutionary tracks which are evaluated when determining e.g. stellar masses and ages for given sets of observables. As our uncertainties are rather large the probability where a star is most likely to be located in the HRD becomes more relevant. By looking at the predicted HRD by BONNSAI most stars are placed near the ZAMS, where they also spend most of the time during their MS lifetime (Fig.~\ref{f:hrd_evo}). 

The positive mass-discrepancy between spectroscopic and evolutionary masses was already observed in Galactic O-type stars ($\gtrsim35\,M_{\odot}$) by \cite{markova2018}. \cite{markova2018} compared the spectroscopic with the derived evolutionary masses using the {\sc bonn} \citep{brott2011} and {\sc geneva} \citep{ekstroem2012} evolutionary tracks. The mass discrepancy is more pronounced for masses based on the {\sc geneva} tracks. In our analysis we used the {\sc bonn} tracks with LMC composition and find a clear trend towards larger evolutionary masses $\gtrsim40\,M_{\odot}$. \cite{markova2018} proposed that the positive mass-discrepancy can be explained in terms of overestimated mass-loss rates in evolutionary model calculations on the basis of the \cite{vink2000, vink2001} mass-loss prescriptions. Based on the physical properties of the individual components of spectroscopic-eclipsing binary system HD\,166734 and their N-abundance ratio \citep{mahy2017} \cite{higgins2019} excluded mass-loss rates which lie outside 0.5 to 1.5 times the \cite{vink2000,vink2001} mass-loss prescription. In addition, rotational mixing is necessary and they favoured larger overshooting parameters of the order of $\alpha = 0.5$ compared to the {\sc bonn} $\alpha = 0.335$ and {\sc genva} $\alpha = 0.1$ evolutionary models.

\subsection{Stellar winds\label{s:winds}}
\begin{figure}
\begin{center}
\resizebox{\hsize}{!}{\includegraphics{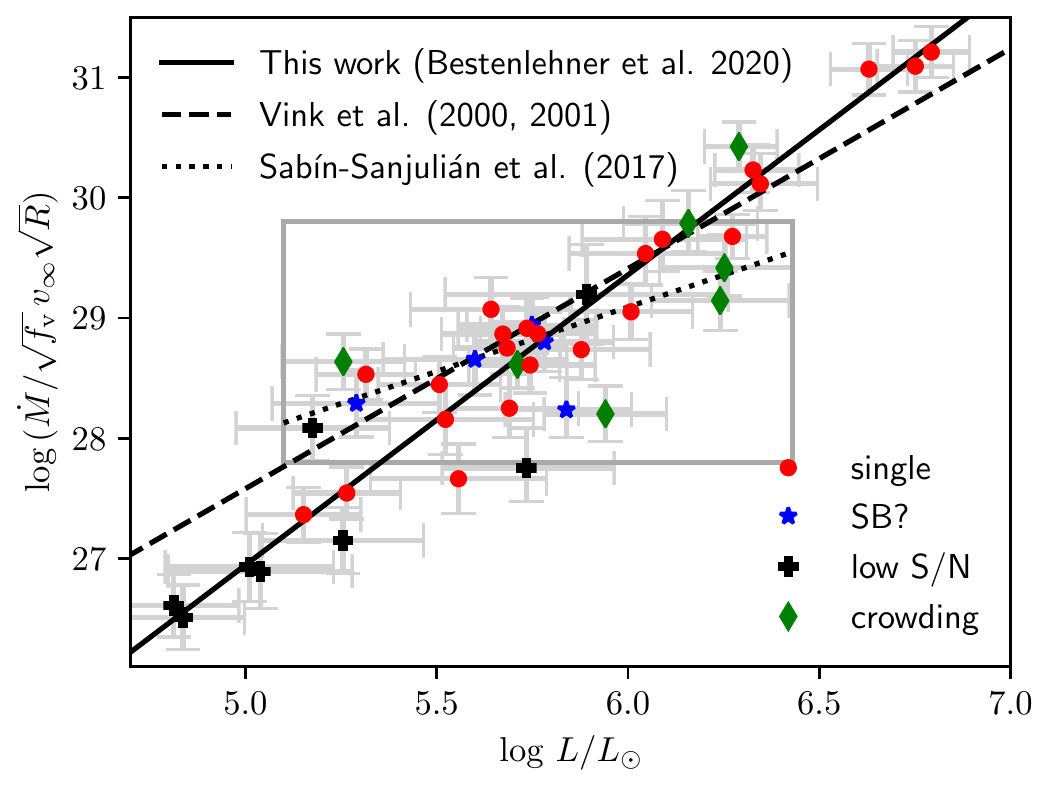}}
\end{center}
\caption{Wind-momentum -- luminosity relation. Black solid line is a linear fit through our sample. The theoretical prediction by \citet{vink2000,  vink2001} is shown as black dashed line. We indicated the empirical found by \citet{sabin2017} as black dotted line while the grey lined box marks the parameter space of their study.}
\label{f:wlr}
\end{figure}
\begin{table*}
	\centering
	\caption{Coefficients for WLR of the form $\log (\dot{M}/\sqrt{f_{\mathrm{v}}} \varv_{\infty}\sqrt{R/R_{\odot}}) = m_0 \log  (L/L_{\odot})+C_0$. Coefficients listed below the horizontal line are not shown in Fig.\,\ref{f:wlr} and are given for reference.}
	\label{t:wlr}
	\begin{tabular}{lccc} 
		\hline
		                & $m_0$         & $C_0$          & Source\\
		\hline
		All stars       & $2.41\pm0.13$ & $14.88\pm0.74$ & Fig.\,\ref{f:wlr}\\
		LMC prediction & $1.83\pm0.04$ & $18.43\pm0.26$ & \citet{vink2000, vink2001}\\
		VFTS O dwarfs   & $1.07\pm0.18$ & $22.67\pm0.99$ & \citet{sabin2017}\\
		\hline
		Apparent single stars       & $2.34\pm0.13$ & $15.37\pm0.75$ & This study\\
		VFTS O giants/supergiants   & $1.78\pm0.14$ & $19.17\pm0.79$ & \citet{ramirez2017} \\
		VFTS O stars ($\log L/L_{\odot}>5.5$)   & $1.45\pm0.16$ & $20.70\pm0.88$ & \citet{bestenlehner2014} \\
		\hline
	\end{tabular}
\end{table*}

The usual wind-momentum rate is given by the product of mass-loss rate and terminal wind velocity ($\dot{M} \varv_{\infty}$). \cite{kudritzki1995} introduced the modified wind-momentum ($\dot{M} \varv_{\infty}\sqrt{R}$). The latter is expected to be nearly independent of stellar mass and to primarily depend on the stellar luminosity for fixed metallicity \citep[for more details]{puls1996}. The modified wind-momentum -- luminosity relation (WLR) allows us to compare the wind properties of population of hot massive stars.

Figure\,\ref{f:wlr} shows the WLR for our sample. We assumed an unclumped, homogeneous wind (i.e. $f_{\rm V} = 1.0$). Mass-loss rates of the three WNh stars were corrected accordingly to their adopted $f_{\rm V} = 0.1$ ($\dot{M}/\sqrt{f_{\rm V}}$). 
We excluded stars which had no $\varv_{\infty}$ measurement in Paper\,I. We derived an observed WLR of the form $\log (\dot{M}/\sqrt{f_{\mathrm{v}}} \varv_{\infty}\sqrt{R/R_{\odot}}) = m_0 \log  (L/L_{\odot})+C_0$ with coefficients $m_0$ and $C_0$ given in Table\,\ref{t:wlr}. Fits through all and apparent single stars marginal diverge. The black solid line is an orthogonal-distance regression fit through our data considering abscissa and ordinate errors\footnote{In many other studies, a more conventional regression is done considering only errors in the modified wind momentum rate, see discussion in \cite{markova2004}.}. The WLR covers a luminosity range of 2\,dex from faint late O stars with weak winds up to the extremely bright WNh stars with optically thick winds. The theoretical prediction by \cite{vink2000, vink2001} is less steep than found empirically. Predicted mass-loss rates are higher at the low luminosity end, while for the three WNh stars they are lower. Taking into account the observed wind inhomogeneity of $f_{\mathrm{v}}\sim0.1$ for WNh stars the mass-loss prediction are in good agreement, even though the mass-loss prescription was based on models with $4.5 \geq \log L/L_{\odot} \geq 6.25$. If we assume a similar clumping factor for the O stars, which is supported by radiation-hydrodynamical models including the line-de-shadowing instability \citep{sundqvist2018}, the predicted mass-loss rates for O stars would be still higher. Another cause for clumping might be the result of sub-surface convection \citep{cantiello2009}.

A volume filling factor $f_{\mathrm{v}}\sim0.1$ is also supported by \cite{bestenlehner2020}. The author derived a mass-loss recipe which predicts how the mass-loss rate scales with metallicity and at which Eddington parameter ($\Gamma_{\rm e}$, considering only the electron scattering opacity) the transition from optically thin O star to optically thick WNh star winds occurs. With the definition of the transition mass-loss rate, introduced by \cite{vink2012}, \cite{bestenlehner2020} was able to calibrate the absolute mass-loss rate-scale for the chemical composition of the Tarantula Nebula and obtained a volume filling factor $f_V = 0.23^{+0.40}_{-0.15}$ for the sample studied here. 

In the context of VFTS \cite{ramirez2017} report for the O giants/supergiants a WLR slope in agreement with the prediction by \cite{vink2000, vink2001} while \cite{bestenlehner2014} notes a less steep slope for the most luminous O-type stars ($\log L/L_{\odot}> 5.5$, Table\,\ref{t:wlr}). \cite{sabin2017} found in their sample of O dwarfs an even shallower WLR 
for stars with $\log L/L_{\odot} > 5.1$ (black dotted line). In the grey box we indicate the parameter space of the stars by \cite{sabin2017}. Interestingly, in this parameter range our WLR is less tight. Our targets inside the grey box seem to follow the WLR by \cite{sabin2017} and we are able to confirm their findings by considering only stars which lie in their parameter space. For the most massive and luminous objects with a high mass-loss rate and optically thick winds, the steeper slope could be the result of the increasing efficiency of multi-line (ML) scattering in dense stellar winds \citep{friend1983, puls1987, lucy1993}, which might increase $\dot{M}$ significantly (factors of up to $\sim 3$ are not unlikely) compared to the O star winds. On the low luminosity side we might begin to see the weak-wind domain \citep[e.g.][]{puls2008}, which gives rise to lower than predicted $\dot{M}$ and also a steeper slope towards standard conditions. On one hand the most luminous stars with ML scattering and on the other hand the less luminous stars in the potential weak-wind domain result in an overall steeper WLR slope than expected and relative to the conditions in the grey rectangle.


Mass-loss rates from objects with weak winds can be uncertain derived from optical spectra. A future study of the stellar wind parameters including ultraviolet diagnostics will be presented in Paper IV.

\subsection{Comparison with previous studies\label{s:cwps}}

\cite{massey1998} and \cite{deKoter1998} used the Faint Object Spectrograph (FOS) and the Goddard High Resolution Spectrograph (GHRS) aboard of HST to obtain UV and optical spectra of individual stars in and around R136. More recently, \cite{crowther2010} combined those archival data with near-infrared VLT/SINFONI spectra plus VLT/MAD K-band photometry to perform a multi-wavelength spectroscopic analysis. These studies derived temperatures and luminosities and estimated initial stellar masses using evolutionary models. A summary of stellar parameters of stars in common with our study is given in Table\,\ref{t:com_pre_work}.

From HST/FOS spectra \cite{massey1998} estimated properties of 11 stars in common using two different spectral type -- temperature scales. The absolute bolometric magnitudes $M_{\rm bol}$ based on the temperature scale by \cite{vacca1996} are in better agreement with our results and only listed in Table\,\ref{t:com_pre_work}. The temperature scale was based on unblanketed stellar atmosphere models, which results in 2\,000 to 8\,000\,K higher temperatures and 0.2 to 0.35\,dex higher luminosities estimates \citep{martins2005}. 
As a result of this luminosities are in agreement with our results within $+0.2$\,dex. The only exceptions are H36 and H46 which are in our spectroscopic analysis around 0.4\,dex more luminous. The reason for this agreement might be that our luminosity scale is anchored on $K$-band photometry leading to systematically higher $L$ while \cite{massey1998} relied on optical WFPC2 photometry which are more affected by extinction, even though they determined extinction parameters. Estimated initial masses agree reasonable well up to $\sim 100\,M_{\odot}$ but are systematically lower at higher masses (Table\,\ref{t:com_pre_work}). \cite{massey1998} used the evolutionary models by \cite{schaerer1993} extending up to $120\,M_{\odot}$ which were extrapolated for more luminous and massive stars.

11 stars are in common with \cite{deKoter1997, deKoter1998}. They used the  ISA-WIND non-LTE model atmosphere code \citep{deKoter1993} to derive temperature and mass-loss rates from HST/FOS and HST/GHRS data. Their temperatures are systematically lower than ours and result in lower luminosities (Table\,\ref{t:com_pre_work}). Evolutionary models from \cite{meynet1994} extending to $120\,M_{\odot}$ were applied to estimate the initial masses. The most massive star in their sample is R136a1, $M_{\rm ini} = 120\,M_{\odot}$, which has according to our analysis $M_{\rm ini} = 250\,M_{\odot}$. With decreasing luminosities differences become small and H55 agrees well with our results (Table\,\ref{t:com_pre_work}).

\cite{crowther2010} spectroscopically analysed the 4 brightest stars in R136 combining HST/GHRS, HST/FOS and VLT/SINFONI data. There is an overlap of three stars with our sample. For the stars in common they derived systematically higher temperatures, in particular for R136a1 where $\Delta T_{\rm eff} \approx 7\,000K$ (Table\,\ref{t:com_pre_work}). In the hotter model the N\,{\sc v} $\lambda 4604$ and $4620$ lines are too strong. The N-abundance can be reduced to match the line intensity, but N\,{\sc iv} $\lambda 4058$ would be then to weak. As R136a1 shows an enriched He composition at the surface a reduce N-abundances would also contradict the findings by \cite{rivero2012b} and \cite{grin2017} that the same process should be responsible to bring up both materials to the surface. Our luminosity for R136a1 is 0.15\,dex lower while R136a2 and R136a3 agree within 0.05\,dex. The initial masses were derived with evolutionary models published later in \cite{yusof2013} extending up to $500\,M_{\odot}$. Initial masses agree within their uncertainties, but with the largest discrepancy for R136a1. \cite{crowther2010} obtained $320^{+100}_{-40}\,M_{\odot}$ and we derive $251^{+50}_{-35}\,M_{\odot}$ which is a result of the large difference of the determined effective temperature.

\section{Discussion}
\label{s:disc}

\subsection{Helium enrichment: mixing or mass-loss?\label{s:he-abund}}

\begin{figure}
\begin{center}
\resizebox{\hsize}{!}{\includegraphics{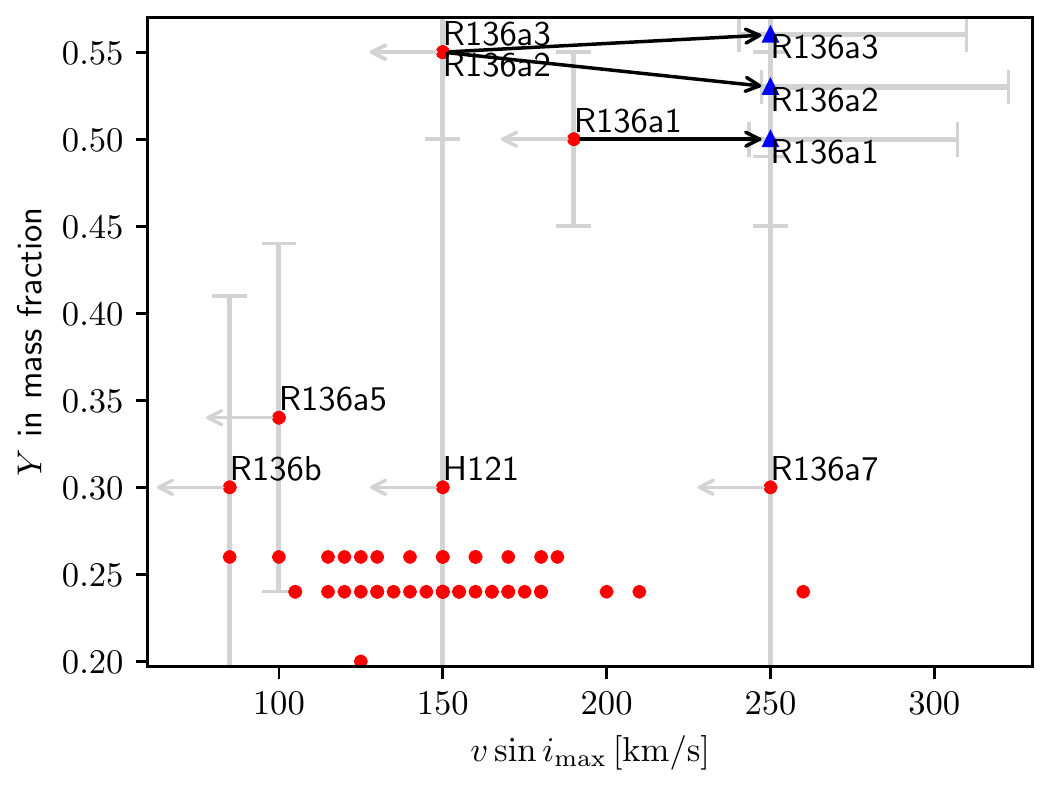}}
            \caption{Surface helium abundances versus upper limits of the projected rotation (red dots and arrows pointing to the left). There is no clear correlation between $\varv \sin i_{\rm max}$ and $Y$. Blue triangles indicated the most probable current day rotation rates for the three WNh stars, which largely excludes $v_{\rm rot} < 250$\,km/s.}
\label{f:vbroad_y}
\end{center}
\end{figure}

\begin{figure}
\begin{center}
\resizebox{\hsize}{!}{\includegraphics{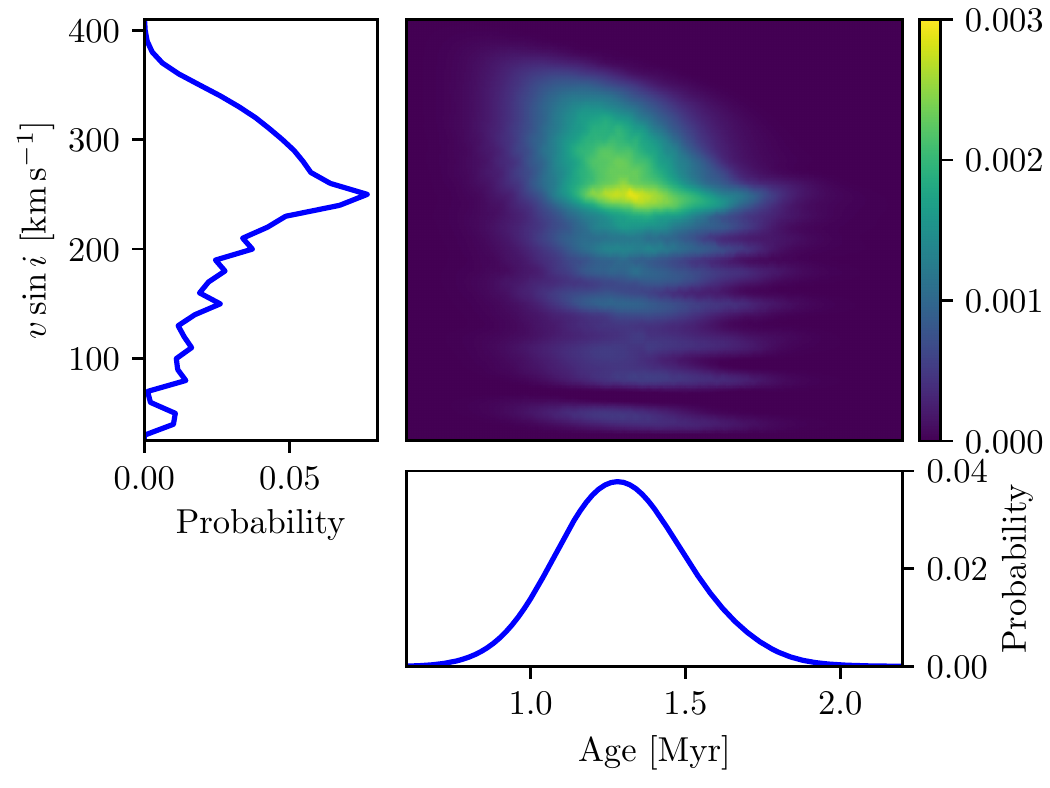}}
\end{center}
\caption{R136a3: probability distributions for stellar age and current projected-rotational velocity ($\varv \sin i$). We assumed a flat prior distribution for $\varv \sin i$. The most probable projected-rotational velocity is $\sim$250km/s. The stellar age is well constrained.}
\label{f:age-vrot}
\end{figure}

\begin{figure}
\begin{center}
\resizebox{\hsize}{!}{\includegraphics{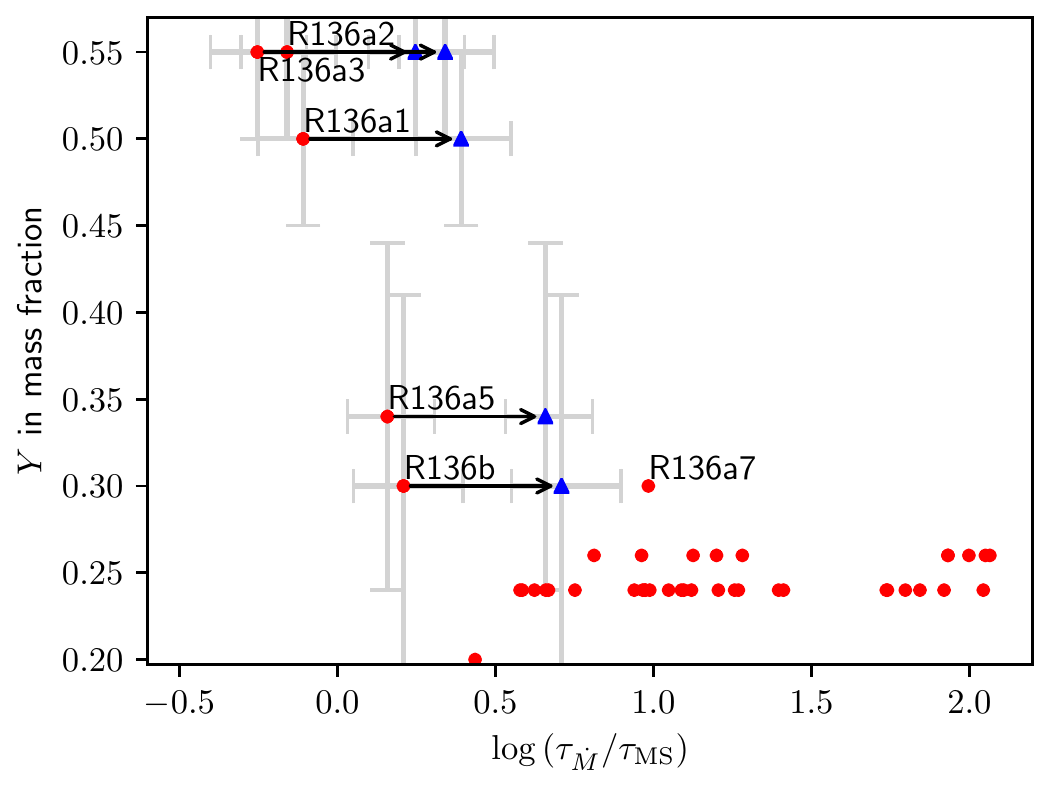}}
\caption{Surface helium abundances versus mass-loss timescale for homogeneous (red dots) and clumped winds (blue triangles) over main-sequence lifetime. Main-sequence lifetimes are estimated based on the most probable initial mass according to the models of \citet{brott2011} and \citet{koehler2015}.}
\label{f:tau_y}\end{center}
\end{figure}


In this section we discuss the surface helium enrichment and HRD position. To simultaneously explain the observed He-abundance and HRD location of the star we require enhanced mixing due to rotational mixing or high enough mass-loss rates to shed away the outer hydrogen layers to uncover the helium rich layers.

Figure\,\ref{f:vbroad_y} shows the helium mass fraction ($Y$) against the line broadening. The quality of our data does not allow us to disentangle projected rotational velocity and macro-turbulent velocity (Sect.\,\ref{s:line_broad}). Therefore, the line projected rotational velocities stated on the abscissa are upper limits of the actual $\varv \sin i$. In agreement with \cite{bestenlehner2014} $Y$ does not correlated with $\varv \sin i_{\rm max}$, which is an indication that rotational mixing might be not the dominant process for the helium enrichment at the stellar surface. 

With the exception of the WNh stars evolutionary models are able to reproduce helium composition and HRD location of the O-type stars without the necessity of high rotation rates. With blue triangles we indicated the predicted current rotation rates by BONNSAI of the WNh stars. These stars have a most probable rotation rate of 250\,km\,s$^{-1}$ excluding essentially lower rotation rates. Through projection effects lower $\varv \sin i$ values can be observed and we cannot exclude that our WNh star actually rotate much faster as the inclination is unknown (Fig.\,\ref{f:age-vrot}). However, the 11 helium enriched Of/WN and WNh stars of \cite{bestenlehner2014} also do not show high $\varv \sin i$ values. The stellar models by \cite{koehler2015}, which fit $L$, $T_{\rm eff}$ and $Y$, evolve chemically homogeneously due to rotational mixing and reproduce those stellar parameters at about the same time ($\sim 1.2$\,Myr, Fig.\,\ref{f:Y_Mini}).

In Fig.\,\ref{f:age-vrot} we visualise the probability distributions of R136a3 with the largest joint $Y$ for stellar age and current projected-rotational velocity provided by BONNSAI assuming a flat $\varv_{\rm rot}$ prior. $\varv \sin i_{\rm max}$ of R136a3 is $\sim 150$km/s and in the 2$\sigma$ confidence range of the most probable $\varv \sin i = 250^{+80}_{-55}$\,km/s from Fig.\,\ref{f:age-vrot}. Its surface abundance $Y$ and age of R136a3 is well determined at around 1.3 Myr. An older age or flat PDF extending to older ages could have provided a probability that the star might have spun down and transported the helium enriched material from the core to the surface on a longer timescale. 
There is one exception which might be chemically enriched due to rotational mixing. R136a7 shows a helium enriched chemical composition and has one of the highest upper limits for $\varv \sin i_{\rm max}$. This could be interpreted as R136a7 being a mass gainer or mass gainer or merger product but the uncertainties on $Y$ are large.

All helium enriched stars ($Y\geq0.3$) have in common that they show emission line features in their spectra indicating high $\dot{M}$. \cite{herrero2004} and \cite{vink2015} proposed that $\dot{M}$ dominates the evolution for stars above $60\,M_{\odot}$. \cite{bestenlehner2014} studied $Y$ at the stellar surface as a function of mass-loss rate over stellar mass ($\dot{M}/M$), which can be interpreted as the inverse mass-loss timescale ($\tau_{\dot{M}}$). They found that for $\log\dot{M}/M \gtrsim -6.5$ there is a well defined correlation between $Y$ and $\dot{M}/M$ when the mass-loss timescale ($\tau_{\dot{M}}\lesssim 3$\,Myr) is comparable to the main-sequence lifetime. In agreement with \cite{bestenlehner2014} we find a similar correlation for $\log\dot{M}/M \gtrsim -6.7$ (Fig.\,\ref{f:mdot-mass_He}), but we take it a step further. Based on the initial masses given by BONNSAI we estimated the main-sequence lifetime ($\tau_{\rm MS}$) according to \cite{brott2011, koehler2015} and examine the helium enrichment as a function of $\tau_{\dot{M}}/\tau_{\rm MS}$.

Figure\,\ref{f:tau_y} shows $Y$ versus $\tau_{\dot{M}}/\tau_{\rm MS}$. Only the three WNh stars have shorter mass-loss timescale than MS lifetime, even though the MS lifetime decreases with increasing stellar mass. We find a correlation of $Y$ with $\tau_{\dot{M}}/\tau_{\rm MS}$ at $\log (\tau_{\dot{M}}/\tau_{\rm MS}) \lesssim 0.2$ (red dots in Fig.\,\ref{f:tau_y}). A star evolves quasi chemical homogeneously when $\tau_{\dot{M}} < \tau_{\rm MS}$ because the MS lifetime corresponds to the nuclear fusion timescale of hydrogen in the core. However, if we account for wind inhomogeneity and correct the mass-loss rates for a volume filling factor $f_{\rm V}=0.1$ derived from the electron scattering wings of the emission line stars, $\tau_{\dot{M}}$ increases by factor of $\sqrt{10}$ or 0.5\,dex (blue triangles in Fig.\,\ref{f:tau_y}). The correlation already occurs at $\log\tau_{\dot{M}}/\tau_{\rm MS} \lesssim 0.7$ near the location of R136b. After considering the wind inhomogeneity no star has $\tau_{\dot{M}} < \tau_{\rm MS}$ and evolves virtually chemical homogeneously.

R136a3 has the smallest $\tau_{\dot{M}}/\tau_{\rm MS}$ ratio 
and has already lost ~25\% of its initial mass. 
The observed $Y=0.55$ might be produced by mass loss only, but the mass-loss rates would need to be significantly increased, which is not really justified (Sect.\,\ref{s:winds}). A much higher mass-loss rate would not only mean a much higher initial mass but also the star would spin down on very short timescale and low $\varv_{\rm rot}$ would be expected. Fig.\,\ref{f:vbroad_y} shows that this is not the case, in particular for R136a1. $\dot{M}$ steeply increases when the star approaches the Eddington limit \citep[e.g.][]{graefener2008, graefener2011, vink2011, bestenlehner2014, bestenlehner2020}. A period of extensive mass loss at the beginning of the evolution of VMS might help to solve the current tension \citep[e.g.][]{bestenlehner2014, schneider2018b}.

\cite{vink2010}, \cite{castro2014} and \cite{mcevoy2015} suggested additional core-overshooting for massive stars to bring the predicted location of the TAMS by \cite{brott2011} in agreement with the observations. With increasing stellar mass the convective core increases as well. The WNh stars in access of $100\,M_{\odot}$ are largely convective and the amount of core-overshooting might be less relevant. \cite{higgins2019} reported that the nitrogen compositions of the binary HD\,166734 consisting of two O supergiants with masses between 30 and 40\,$M_{\odot}$ could not be reproduced by mass loss and core-overshooting alone. At least some amount of rational mixing is necessary to transport the right amount of nitrogen to the surface to match the observed compositions of both stars. \cite{rivero2012b} and \cite{grin2017} found a correlation between He and N enrichment suggesting that the same process should be responsible to dredge-up both elements.

Chemical homogeneously evolving models due to rotational mixing well reproduce the observed $L$, $T_{\rm eff}$ and $Y$ of the WNh stars. All the stars need relatively fast initial rotation in excess of 300\,km/s. Such fast rotation is not found at lower masses in 30 Doradus and appears to be in conflict \citep{schneider2018b}. Stellar models with enhanced mixing predict lower evolutionary masses, which might be the reason for the observed positive mass-discrepancy in Sect.\,\ref{s:masses_ages}. However, only the WNh stars, where we were not able to derive $M_{\rm spec}$, require high initial $\varv_{\rm rot}$ to reproduce the observables. This might be an indication that some physical conditions are not well enough understood or missing. We can conclude that the evolution of the most massive is dominated by mass loss, as seen by the tight correlation in Fig.\,\ref{f:tau_y} and \ref{f:mdot-mass_He}. Therefore, we do not expect a correlation of $Y$ with $\varv \sin i_{\rm max}$ which is supported by Fig.\,\ref{f:vbroad_y}, but an additional mixing process such as rotational mixing or other mixing process appears to be still necessary to reproduce the observables. 

\subsection{Cluster ages of R136\label{s:r136_age}}
\begin{figure}
\begin{center}
\resizebox{\hsize}{!}{\includegraphics{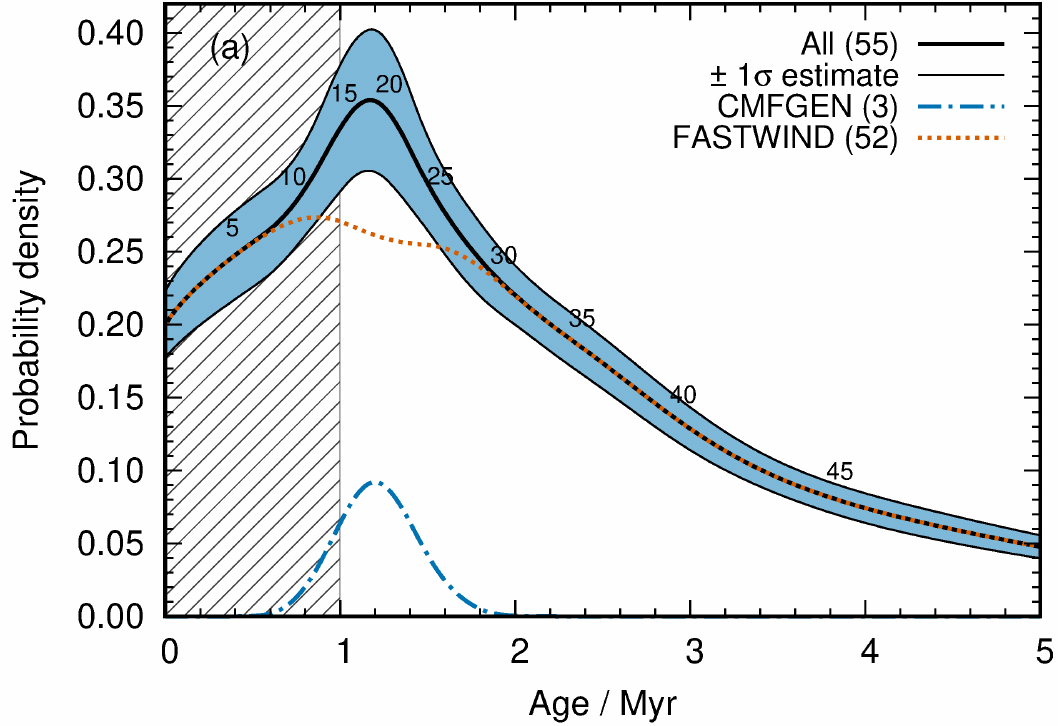}}
\end{center}
\caption{Probability density functions of stellar ages (black solid line and $\pm 1 \sigma$ estimate blue shaded). Blue dotted-dashed line: PDF of stars analysed with CMFGEN. Red dotted line: PDF of stars analysed with FASTWIND. Numbers are cumulative counts. The population of R136 can be roughly divided into two, a younger one with an age $< 2.5$ Myr and an older population with an age $>2.5$ Myr. Hatched area correspond to the minimum age R136 of 1\,Myr.}
\label{f:age_func}
\end{figure}
\begin{figure}
\begin{center}
\resizebox{\hsize}{!}{\includegraphics{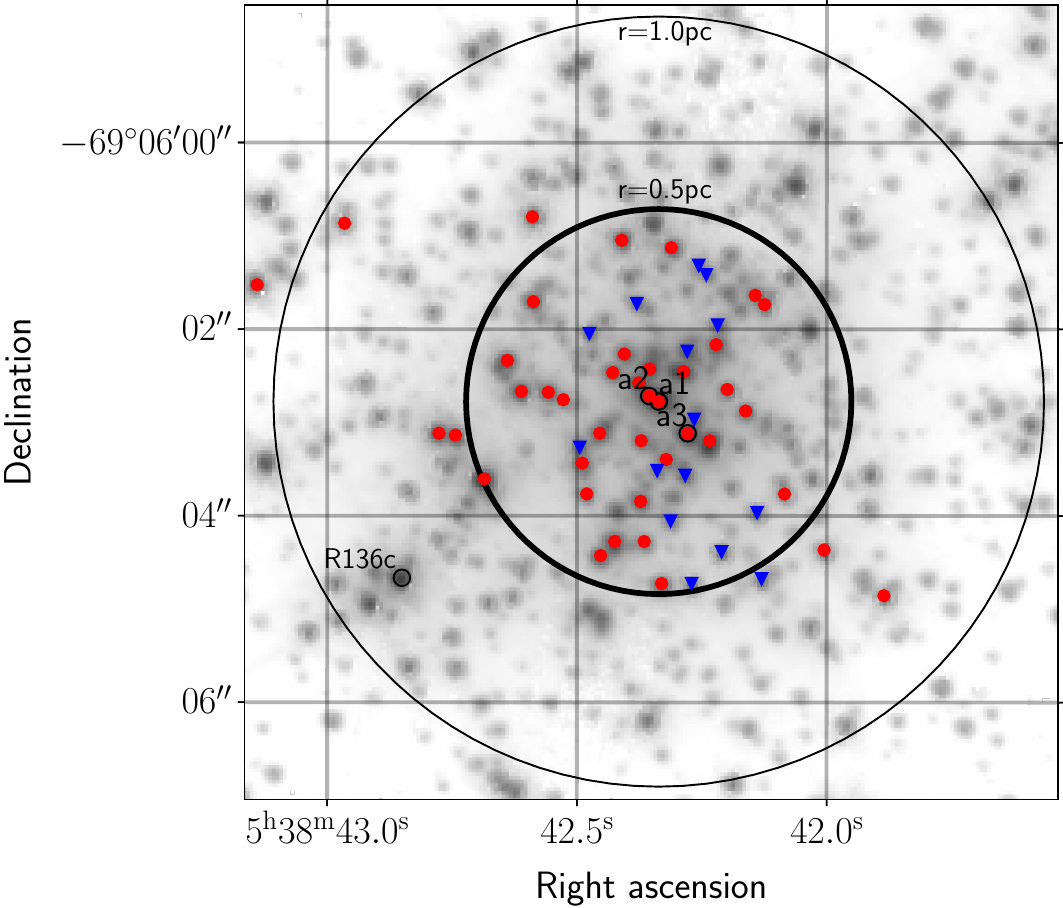}}
\end{center}
\caption{Spatial distribution of our targets in R136. Stars indicated by a red dot likely belong to a younger population while blue triangles to an older ($\gtrsim 2.5$ Myr). The position of older and younger stars are randomly distributed. Black bold solid circle of radius 0.5\,parsec and black solid circle of radius 1.0\,parsec are centred on R136a1. The background image was taken with HST/WFC3 using the F555W filter.}
\label{f:ages}
\end{figure}
\begin{figure}
\begin{center}
\resizebox{\hsize}{!}{\includegraphics{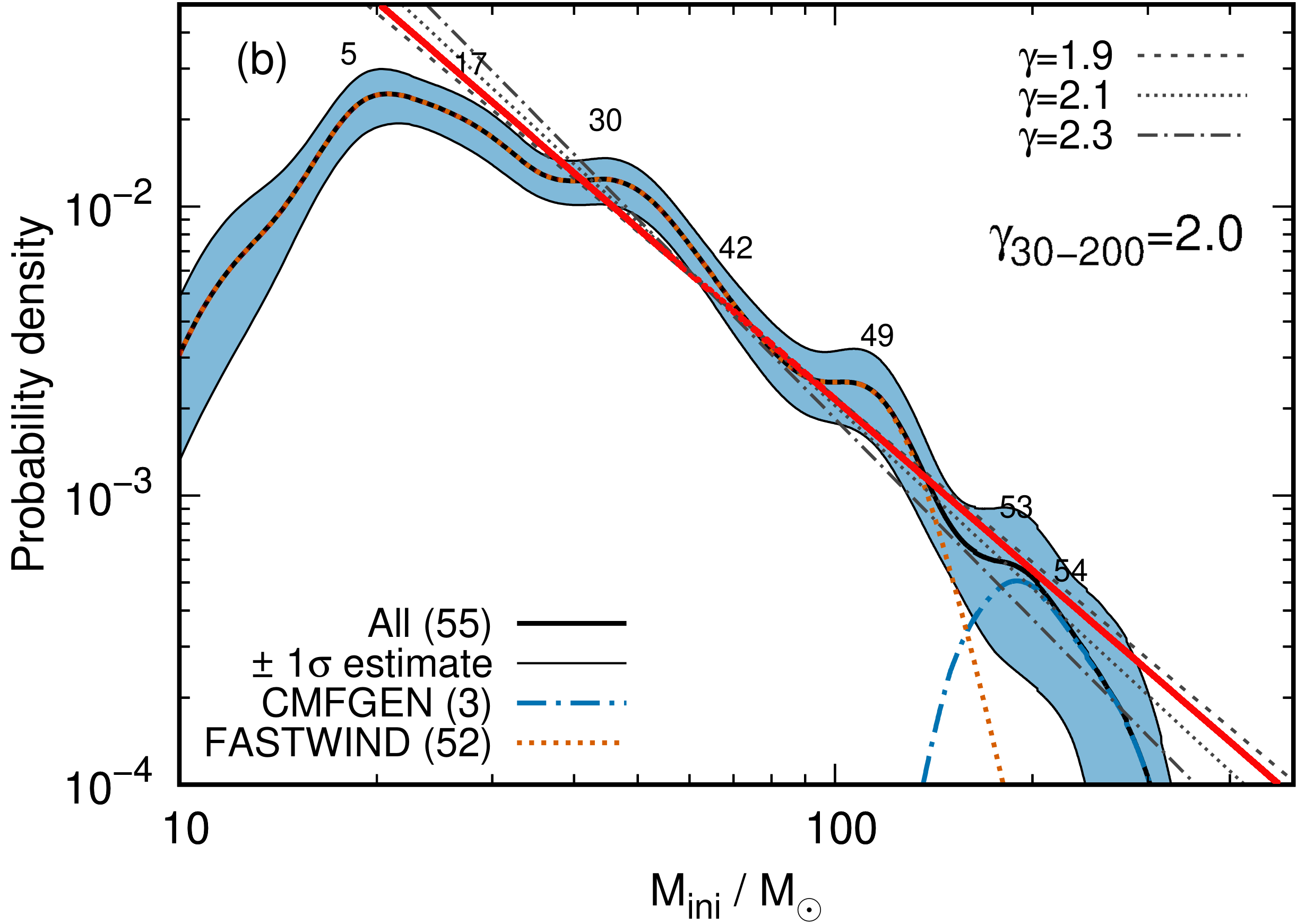}}
\end{center}
\caption{Probability density functions of initial stellar masses. Results are indicated as in Fig.\,\ref{f:ages}. 
Red solid line is the best fit with slope $\gamma \approx 2$ derived over the 30 to 200 mass range. Slopes of $\gamma=1.9$, 2.1 and 2.3 seemed to work similarly well (grey dashed, dotted and dotted-dashed lines). 
Our sample is complete down to $30 - 40\,M_{\odot}$. 7 stars are more massive than $100\,M_{\odot}$.}
\label{f:imf}
\end{figure}

In Fig.\,\ref{f:age_func} we show the probability density functions of ages. 
This shows a young stellar population up to 2.5 Myr (38 stars) and an older population extending beyond 2.5 Myr (17 stars). Most but not all of the older objects have a low S/N spectrum. We found a median age around 1.6 Myr of R136, similar to what had been found in Paper\,I from ultraviolet calibrations. Based on pre-MS stars and their associated tracks \cite{cignoni2015} established that the star-formation rate in R136 peaked between 1 and 2 Myr. \cite{sabbi2012} identified a slightly older group located $\sim 5.4$\,pc to the north-east potentially merging into R136. Their analysis suggests that the  majority of stars in the north-east clump were formed between 2 and 5 Myr ago while R136 is not older than $\sim$2 Myr. 

Figure\,\ref{f:ages} indicates that the older population ($>2.5$\,Myr) representing 1/3 of the stars are spatially well distributed within 0.5\,pc of R136. R136 is located in an extended H{\sc ii} region NGC\,2070, which contributes to the projected stellar population of R136. \cite{henault2012} find that this contribution is $\lesssim 5$\% in the inner 1.25\,pc corresponding to only $\lesssim 3$ stars. If this older population is part of a more diffuse surrounding population, it should become more dominant when moving out to larger radii. 

If those stars are descended from the north-east clump, their number is still rather large and a noticeable over-density of older stars should be found in the proximity to R136. \cite{castro2018} observed the surrounding region centred on R136 with Multi Unit Spectroscopic Explorer (MUSE) on the Very Large Telescope. The four field mosaic covers a box of $\sim 30 \times 30$\,pc and the spectroscopic analysis of Castro et al. (submitted) might confirm the older foreground population or the north east clump merging into R136. Currently, we are not able to establish if the apparently older stars originated from the north-east clump, an older foreground population due to the 3 dimensional nature of 30 Doradus, or that R136 consists of a multiple age population.  

The age probability distribution function, black solid line in Fig.\,\ref{f:age_func}, culminates at $\sim$\,1.2\,Myr suggesting that the star-formation rate in R136 peaked at around this time. The prominent peak is mainly caused by the 3 most massive WNh stars (blue dashed-dotted line) while the O star distribution (orange dotted line) is flatter suggesting a more continuous star formation rate up to 2\,Myr or larger age errors. 

Considering the surface helium mass fraction of the 3 WNh stars R136a1, a2 and a3 we can estimate a lower age boundary for R136. The helium shown at the surface has to be produced in the core first due to nuclear fusion. As probably more helium has been produced than visible at the surface this boundary is a lower limit. Under these assumptions R136 must be older than 1.0\,Myr which is indicated as the hatched area in Fig.\,\ref{f:age_func}. \cite{lennon2018} report that the proper motion of VFTS\,016 is consistent with an ejection from R136. If VFTS\,016 is ejected from R136 during or shortly after the cluster was formed, it would set a lower age limit of 1.3\,Myr based on its current distance to R136 and proper motion.

Even though our sample contains stars with ages up to $\lesssim 6$\,Myr, the majority of stars in R136 has an age between 1 and 2\,Myr. Based on the minimum age of 1.0\,Myr using $Y$, the lower age limit of 1.3\,Myr by dynamical ejection of VFTS\,016 and the distribution of the probability distribution function we can assume a cluster age of R136 between 1.0 to 2.0\,Myr.


\subsection{Initial mass function and upper mass limit\label{s:r136_imf}}

In Figure\,\ref{f:imf} we show the initial mass function of R136. Our sample is complete down to 30 to 40\,$M_{\odot}$, which largely represents the stars younger than 2.5 Myr (Fig.\,\ref{f:age_func}). Similar to \cite{schneider2018} we assumed a power-law function of the form $\xi(M) \propto M^{-\gamma}$ with the stellar mass ($M$) and exponent $\gamma$ to fit the slope of the initial mass function (IMF). To accurately determine the slope of the initial mass function (IMF) it is crucial to know down to which stellar mass the sample is complete, or alternatively the completeness of a given mass bin.  
We fitted power-laws over the mass range 30 to 200\,$M_{\odot}$ to the distribution of initial masses. The best fit is indicated as a red line in Fig.\,\ref{f:imf} and has an exponent $\gamma=2.0\pm0.3$. Slopes of $\gamma=1.9$, 2.1 and 2.3 seemed to work similarly well (grey dashed, dotted and dotted-dashed lines). All IMFs show a clear change in slope at $M\lesssim 30\,M_{\odot}$. On the one hand this may reflect incompleteness in this mass regime. We removed two SB2s (H42 and H77) from our sample which would add 4 stars to the mass range between 30 and 40\,$M_{\odot}$ based on the estimated properties from Paper\,I assuming equal mass binaries and similar stellar parameters suggested from their optical spectral types. On the other hand it may point to R136 being a composite of stellar population. 


\cite{schneider2018} derived $\gamma \approx 1.90^{+0.37}_{-0.26}$ for a stellar sample in the wider 30 Doradus region which is complete down to 15\,$M_{\odot}$. Our slope is in line with theirs, but the uncertainties are significantly larger. For the solar neighbourhood \cite{salpeter1955} obtained a slope of $\gamma \approx 2.35$ from stellar populations with masses up to 17\,$M_{\odot}$ (B0V star).  The most common IMFs to simulate and interpret clusters and galaxies are \cite{kroupa2001} and \cite{chabrier2003}. Both studies suggest a $\gamma \approx 2.3$. A shallower slope at the high mass end would predict more massive stars. 
However, the uncertainties in our analysis are too large to firmly suggest this. 


Seven stars have initial masses above 100\,$M_{\odot}$. Three of them are more massive than 150\,$M_{\odot}$ with the most massive two exceeding 200\,$M_{\odot}$. \cite{figer2005} proposed a canonical upper mass limit of $\sim$\,150\,$M_{\odot}$, which is challenged by these findings. If we include R136c \citep[alias VFTS\,1025,][]{bestenlehner2014, schneider2018} as a cluster member, we find that three stars in R136 exceed $200\,M_{\odot}$. The upper mass limit might still be valid if those very massive stars are stellar merger products \citep{banerjee2012b}, although all would need to be have merged within $1-2$\,Myr after formation. \cite{banerjee2012b} simulated a handful of clusters with identical initial conditions. Even though none of their simulations is able to predict these numbers, we cannot exclude the merger scenario, as results considerably varied between their simulations. 

VFTS\,682 is a very massive star in apparent isolation with a current day mass $\sim150\,M_{\odot}$ \citep{bestenlehner2011}. It is a candidate runaway star from R136 \citep{renzo2019}. This supports the existence of very massive stars $> 150\,M_{\odot}$ in general and in particular in the core of R136. \cite{tehrani2019} discovered that Mk34 is likely the most massive binary system known today, and is located just to the east of R136 with a projected distance of $\sim$\,3\,pc. Mk34 consists of two WN5h stars showing a similar spectrum to VFTS\,682. The combined mass of the system exceeds 250\,$M_{\odot}$. Even though most stars are found in binaries or higher-order systems \citep{sana2012Sci, sana2013, ma-ap2019}, it is still possible that in some rare cases a system like Mk34 merged during the formation process or on the MS and formed a single very massive star exceeding 200\,$M_{\odot}$. 

Even though the uncertainties are large, there is no clear evidence from the IMF of Fig.\,\ref{f:imf} that the most massive stars are stellar mergers. Based on the most massive stars in R136, NGC\,3603 and the Arches Cluster \cite{crowther2010} revised the upper mass limit. \cite{bestenlehner2020} finds that the mass-loss rates of the most massive stars might be underestimated by a factor of $\sim$2 in the {\sc bonn} models. This could mean that the actual initial masses of those stars are even larger suggesting a higher upper mass limit. It has been suggested that the first stars in the universe had masses in excess of 1000\,$M_{\odot}$ \citep[e.g.][]{bromm1999}. Based on Monte Carlo radiative transfer models \cite{vink2018} proposed a metallicity-dependent upper mass limit with higher stellar masses in metal poorer environments. With the current number and properties of known VMS in spatially-resolved clusters in the Milky Way and Magellanic Clouds it is difficult to find an indisputable answer to the question of the upper mass limit of stars.


\subsection{Ionising fluxes and mechanical feedback\label{s:io_mech}}

In this section we compare our integrated ionising fluxes and mechanical feedback with \cite{doran2013}. The ionising flux ($Q_0$) is measured in photons per second (ph/s) while the mechanical feedback is given by the stellar wind luminosity ($L_{\rm SW} = \frac{1}{2} \dot{M} \varv_{\infty}^2$) in erg/s. \cite{doran2013} applied a template method to estimate the stellar parameters and used theoretical mass-loss predictions by \cite{vink2001} to evaluate $\dot{M}$. They assigned typical $\varv_{\infty}$ values based on averaged values by \cite{prinja1990}. \cite{doran2013} estimated the ionising and mechanical output for the entire Tarantula Nebula within a radius of 150\,pc around R136a1 and emphasised that the few most massive and luminous stars dominate the overall ionising and mechanical budget of 30 Doradus. In this work we calculated the mechanical feedback with the values given in Table\,\ref{t:sp} where we explicitly list the ionising fluxes. 

Figure\,\ref{f:ages} shows, that our sample is likely complete in terms of stars contributing to the cumulative $Q_0$ and $L_{\rm sw}$ within a radius of 0.5\,pc around R136a1. We derived an integrated $\log Q_0\,\mathrm{[ph/s]} = 51.44$ and $L_{\rm SW}\,\mathrm{[erg/s]} = 39.07$. \cite{doran2013} obtained an integrated $\log Q_0 = 51.36$ and $\log L_{\rm SW} = 38.58$ using their Table\,D.2. Both results are similar, but we find a 0.5\,dex higher stellar wind luminosity. Accounting for a volume filling factor ($f_{\rm v} = 0.1$) our result would be 0.5\,dex lower and in agreement ($\log L_{\rm SW} = 38.57$). Extending the sampled region to 1\,pc we find $\log Q_0 = 51.46$ and $\log L_{\rm SW} = 38.57$ while \cite{doran2013} find $51.48$ and $38.68$, respectively. 
The main contributor in this range is the WN5h star R136c \citep{bestenlehner2014}, which essentially accounts for the increase in $Q_0$ and $L_{\rm SW}$ in \cite{doran2013} and is not in our sample.

In our sample we have 7 VMS with masses greater than 100\,$M_{\odot}$. The VMS account for $\sim$\,57\% of the ionising flux and $\sim$\,90\% of the stellar wind luminosity relative to all 55 stars.  R136 contributes $\sim$\,27\% of the overall ionising flux and $\sim$\,19\% of the overall mechanical feedback to the Tarantula Nebula. We conclude that the cluster R136 is the major contributor to the stellar feedback in the Tarantula Nebula. We confirm that the ionising and mechanical feedback is dominated by the most massive stars at the top of the IMF.

\subsection{The stellar population of R136 and the Tarantula Nebula\label{s:vfts}}
\begin{figure}
\begin{center}
\resizebox{\hsize}{!}{\includegraphics{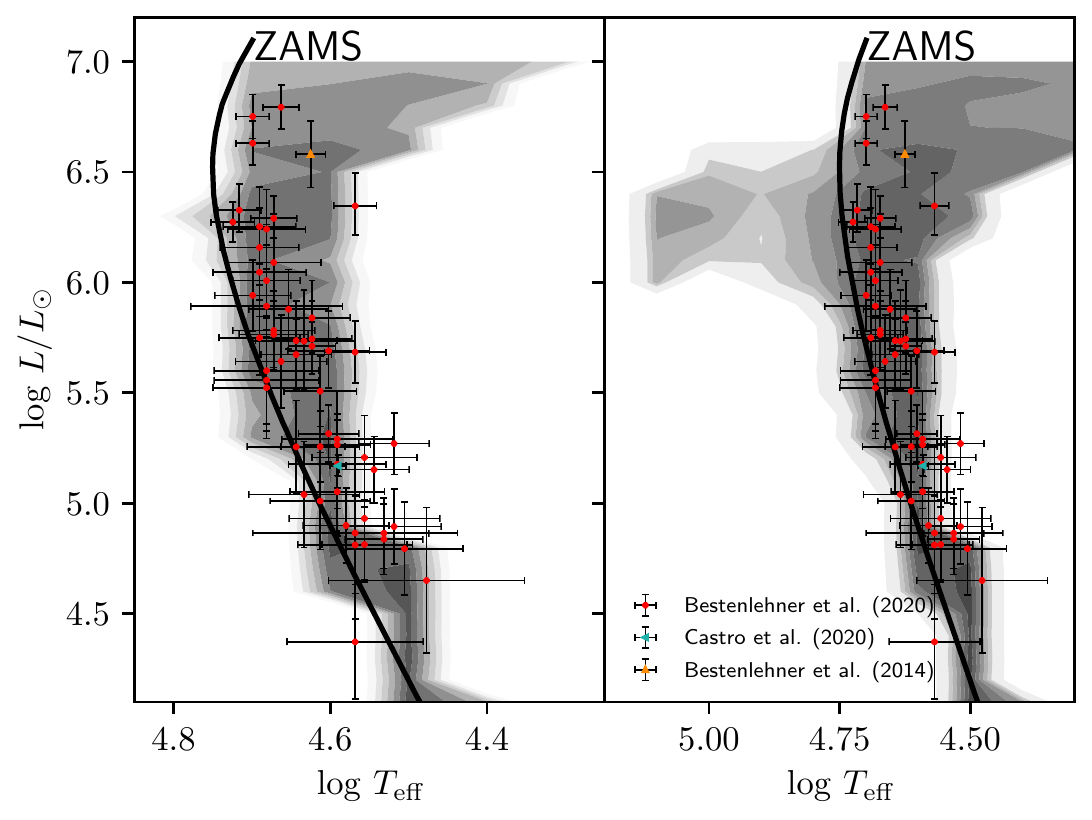}}
\end{center}
\caption{HRD of R136 and overlaid binary population synthesis. The stellar population of R136 within in a radius of 1\,pc from R136a1 compared to the predicted binary stellar population of an age between 1.0 to 2.0 (left panel) and 1.0 to 2.3\,Myr (right panel) from BPASS \citep[grey shaded contours]{eldridge2017, stevance2020}. Each contour represents an order of magnitude difference in stellar number density.
}
\label{f:r136_hrd}
\end{figure}
\begin{figure*}
\begin{center}
\resizebox{\hsize}{!}{\includegraphics{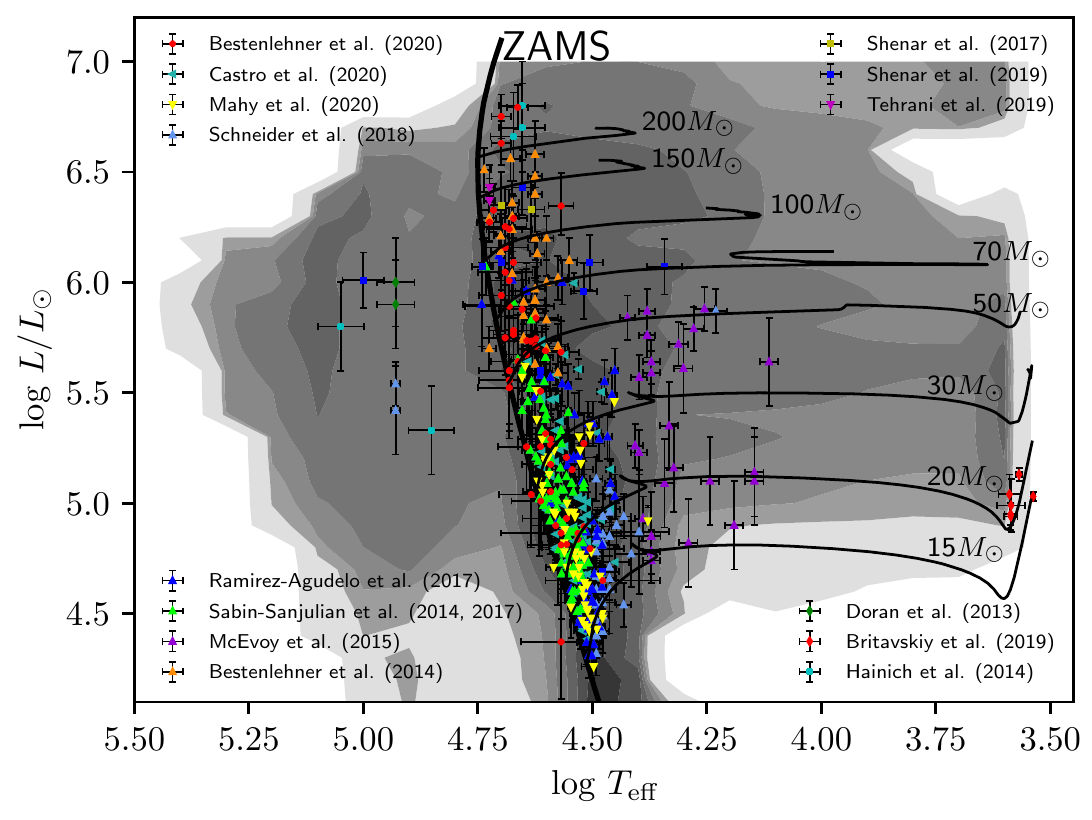}}
\end{center}
\caption{HRD of the Tarantula Nebula and overlaid binary population synthesis. Census of massive stars ($>15\,M_{\odot}$) in 30 Doradus in the LMC within a radius of 10 arcmin (150\,pc) from R136a1. The age of the observed stellar population is 1 to 12 Myr \citep{schneider2018}. Black solid lines are {\sc bonn} evolutionary tracks and ZAMS \citep{brott2011, koehler2015}. The grey shaded contours visualise the predicted binary stellar population from BPASS \citep{eldridge2017, stevance2020} for the observed age range. Each contour represents an order of magnitude difference in stellar number density. The figure includes single and binaries stars from this work (Bestenlehner et al. 2020), Castro et al. (submitted), \citet{mahy2020}, \citet{schneider2018}, \citet{ramirez2017}, \citet{sabin2014, sabin2017}, \citet{mcevoy2015}, \citet{bestenlehner2014}, \citet{doran2013}, \citet{britavskiy2019}, \citet{hainich2014}, \citet{shenar2017}, \citet{shenar2019} and \citet{tehrani2019}.
}
\label{f:ngc2070_hrd}
\end{figure*}

The integrated light of star-forming galaxies is dominated by massive stars \citep[Sect.\,\ref{s:io_mech} or e.g.][]{crowther2019}. The interpretation is based on population synthesis models like STARBURST99 \citep{leitherer1999} or BPASS \citep{eldridge2017}. In this section we compare the stellar population of R136 (Fig.\,\ref{f:r136_hrd}) within a radius of 1\,pc and the Tarantula Nebula (Fig.\,\ref{f:ngc2070_hrd}) within 150\,pc (10 arcmin) from R136a1 \citep{walborn1991} to the population synthesis prediction from BPASS \citep[v.2.2.1,][]{stanway2018}. We downloaded the commonly used and publicly available BPASS output\footnote{\url{https://drive.google.com/drive/folders/1BS2w9hpdaJeul6-YtZum--F4gxWIPYXl}} and visualised it with the python package Hoki which has been designed to interface with the BPASS models and their outputs \citep{stevance2020}. The publicly available BPASS models for binary population synthesis used the binary, period and mass distribution according to \cite{sana2012Sci, moe2017}. Both single and binary models employ in the mass range of our sample a standard \cite{salpeter1955} IMF with an exponent $\gamma = 2.35$ and an upper mass limit of 300\,$M_{\odot}$. For more details on the nature of BPASS models we refer the reader to \cite{eldridge2017, stanway2018}.

In the left panel of Fig.\,\ref{f:r136_hrd} we show an HR diagram of R136 in which we have overlaid the BPASS population synthesis prediction assuming a single starburst from 1 to 2\,Myr. 
The population synthesis contours are truncated at $\log L/L_{\odot} < 7.0$ corresponding to the upper mass limit of 300\,$M_{\odot}$. Our sample well populate the region near the ZAMS over the whole luminosity range as predicted by BPASS. If we extend the starburst to 2.3\,Myr the first {\it classical} WR stars become visible to the hot side of the ZAMS (Fig.\ref{f:r136_hrd}, right panel). Their absence gives us an upper age limit of 2.2\,Myr for R136 which confirms our findings from Sect.\,\ref{s:r136_age}.

Now we consider R136 in the wider context of the Tarantula Nebula. We compiled an HR diagram of 460 stars using stellar parameters of stars more massive than 15\,$M_{\odot}$ with ages up to $\sim 12$\,Myr from the literature (Fig.\,\ref{f:ngc2070_hrd}, caption for references). The sample includes apparent single as well as binary stars within 10\,arcmin (150\,pc) from R136a1 including stars from NGC\,2070 and NGC\,2060. We overlaid the contours of the BPASS binary population synthesis prediction of a 1 to 12\,Myr old stellar population using the star-formation history (SFH) of the Tarantula Nebula by \cite{schneider2018}. As this SFH is not implemented into BPASS we divided the age range into 0.02\,Myr age bins. The age bins were weighted according to the SFH and stacked. Each contour represents an order of magnitude difference in stellar number density.

A significant number of stars densely populate the region near the ZAMS ($\sim 400$), where the BPASS models predicts the highest number densities. The number of around 50 blue supergiants (BSG) is roughly expect based on the BPASS contours. However,  only one of those supergiants which is part of a binary system is more luminous than $\log L/L_{\odot} \gtrsim 5.8$, even though a population of more luminous BSG are predicted. No yellow/red supergiants (Y/RSG) are observed in the Hertzsprung gap between 5000 and 12\,000\,K. Because only very few are expected due to our sample size, we are not able to quantify an actual disagreement between observation and BPASS prediction. All 6 RSG are less luminous than $\log L/L_{\odot} = 5.3$, even though at least a similar number should be observed above this threshold.

Most stars in Fig.\,\ref{f:ngc2070_hrd} have been observed in the context of VFTS \citep{evans2011}. The selection criterion was a magnitude cut which includes cool stars as well. Any RSG more luminous than $\log L/L_{\odot} = 5.5$ should have been picked up. However, the observations are in line with the empirical RSG upper luminosity limit of $\log L/L_{\odot} \approx 5.5$ in the LMC \citep{davies2018}, but there total number might be on the lower side.

Turning to the hot side of the ZAMS there are several {\it classical} WR stars in the Tarantula Nebula around R136 (Fig.\,\ref{f:ngc2070_hrd}). Comparing the number densities of WR stars relative to the one near the ZAMS the count of 7 WR stars is rather low. At least a factor of 2 or 3 more WR stars could be expected. 
Single star population synthesis models predict that WR stars should be all more luminous than $\log L/L_{\odot} \gtrsim 5.5$ (Fig.\,\ref{f:ngc2070_hrd_sin}). In contrast the binary synthesis models predicts also stars below $\log L/L_{\odot} < 4.5$, but the shown sample is incomplete below $\log L/L_{\odot} \sim 5$ on the hot side of the ZAMS. 
The binary evolution channel seems to be important to form less luminous WR stars or helium stars. A general discussion on the formation of WR stars via binary evolution and the transition between WR and He stars can be found in \cite{shenar2020}. 

Based on binary population synthesis models we find an upper age limit for R136 of 2.2\,Myr. Most stars populate the region near and to the cooler side of the ZAMS covering the entire luminosity rage. In the wider context of the Tarantula Nebula the number of {\it classical} WR stars is lower than expected based on the BPASS models. We observe a discrepancy between the predicted stellar number densities by BPASS of luminous blue ($\log L/L_{\odot} \gtrsim 5.8$) and red supergiants ($\log L/L_{\odot} \gtrsim 5.3$). A potential top-heavy IMF in comparison to the standard Salpeter IMF would increase the discrepancies between number of WR stars and more luminous B/RSG, which is suggested by \cite{schneider2018} and this study. This has not only an impact on the predicted radiative and mechanical output of the Tarantula Nebula but also on the analysis and interpretation of unresolved stellar populations in star-forming galaxies. 

\section{Conclusion}
\label{s:con}

In this study we have spectroscopically analysed 55 stars in R136, the central cluster in the Tarantula Nebula in the Large Magellanic Cloud. The sample is complete down to about 40\,$M_{\odot}$, including seven very massive stars over 100 solar masses. The slope of the wind-luminosity relation is $2.41\pm0.13$ which is steeper than the usually observed value of $\sim 1.8$ \citep[e.g.][]{mokiem2007, ramirez2017} and predicted value of 1.83 \citep{vink2000, vink2001} in the LMC.

The most luminous stars ($\log L/L_{\odot} > 6.3$) are helium enriched at the stellar surface. Luminosities, temperatures and He-abundances of the three WNh stars are well reproduce by chemical homogeneously evolving stellar models due to rotational mixing. We find a tight correlation of helium surface composition with the ratio of the mass-loss over main-sequence timescale indicating the importance of mass loss during their evolution. We conclude that mass loss dominates the evolution of the most massive stars, but rotational mixing or other mixing processes might be still necessary. 

There is an indication that the initial mass function of massive stars in R136 might be top heavy with a power-law exponent $\gamma \sim 2.0\pm 0.3$ by comparison to the standard Salpeter exponent, although slopes of 1.9, 2.1 and 2.3 work similarly well due to the large uncertainties. Based on the chemical composition of the most massive stars we derived a lower age limit of 1.0\,Myr for R136. Because there are no $\it classical$ WR stars in our sample of R136 we estimate an upper age limit of 2.2\,Myr. We conclude that the age of R136 is between 1 and 2\,Myr.

Based on evolutionary models the most massive star R136a1 had an initial mass of $250^{+50}_{-35}\,M_{\odot}$ and a current day mass of $215^{+45}_{-30}\,M_{\odot}$. Stars more massive than 40\,$M_{\odot}$ exhibit larger spectroscopic masses than evolutionary masses. This positive mass-discrepancy problem was already observed for Milky Way stars at a similar stellar mass \citep[$\gtrsim35\,M_{\odot}$][]{markova2018}.

The ionising ($\log Q_0\,[{\rm ph/s}] = 51.4$) and mechanical ($\log L_{\rm SW}\,[{\rm erg/s}] = 38.6$) output of R136 is dominated by the most massive stars. The seven most massive stars account for $\sim$\,57\% of the ionising flux and $\sim$\,90\% of the stellar wind luminosity of R136. R136 as a whole contributes around 1/4th of the ionising flux and around 1/5th of the mechanical feedback to the overall budget of the Tarantula Nebula. 

BPASS population synthesis predictions of R136 are in good agreement, which might be the result of the relative young age of R136. 
In the wider context of the Tarantula Nebula binary evolution is required on the basis of BPASS models to match the least luminous WR stars. In addition, BPASS predicts larger stellar number densities for WR stars and luminous blue ($\log L/L_{\odot} \gtrsim 5.8$) and red supergiants ($\log L/L_{\odot} \gtrsim 5.3$), which would considerably contribute to the radiative and mechanical output of the Tarantula Nebula. A potential top-heavy IMF would amplify the discrepancy between observation and prediction and has implications for the analysis and interpretation of unresolved stellar populations in star-forming galaxies.

\section*{Acknowledgements}
We thank the referee, Tomer Shenar, providing helpful comments and suggestions which improved the clarity and content of the manuscript.
We thank 
Chris Evans and Hugues Sana for comments to this manuscript. We would like to thank Nolan Walborn for the spectral classification, which will be published in Paper\,III. Financial support to Azalee Bostroem from Nolan Walborn's $HST$ grant NAS 5-26555 and thanks to Azalee Bostroem for the work on the data reduction. JMB acknowledges financial support from the University of Sheffield. JMA acknowledges support from the Spanish Government Ministerio de Ciencia through grant PGC2018-095\,049-B-C22. SS-D and AHD acknowledge funding from the Spanish Government Ministerio de Ciencia e Innovaci\'on through grants PGC-2018-091\,3741-B-C22, SEV 2015-0548 and CEX2019-000920-S, and from the Canarian Agency for Research, Innovation and Information Society (ACIISI), of the Canary Islands Government, and the European Regional Development Fund (ERDF), under grant with reference ProID2017010115.

\section*{Data availability}

The data underlying this article are available in the article and in its online supplementary material. Reduced spectroscopic data will be published in Paper\,III.




\bibliographystyle{mnras}
\bibliography{reference} 

\begin{thebibliography}{}
\makeatletter
\relax
\def\mn@urlcharsother{\let\do\@makeother \do\$\do\&\do\#\do\^\do\_\do\%\do\~}
\def\mn@doi{\begingroup\mn@urlcharsother \@ifnextchar [ {\mn@doi@}
  {\mn@doi@[]}}
\def\mn@doi@[#1]#2{\def\@tempa{#1}\ifx\@tempa\@empty \href
  {http://dx.doi.org/#2} {doi:#2}\else \href {http://dx.doi.org/#2} {#1}\fi
  \endgroup}
\def\mn@eprint#1#2{\mn@eprint@#1:#2::\@nil}
\def\mn@eprint@arXiv#1{\href {http://arxiv.org/abs/#1} {{\tt arXiv:#1}}}
\def\mn@eprint@dblp#1{\href {http://dblp.uni-trier.de/rec/bibtex/#1.xml}
  {dblp:#1}}
\def\mn@eprint@#1:#2:#3:#4\@nil{\def\@tempa {#1}\def\@tempb {#2}\def\@tempc
  {#3}\ifx \@tempc \@empty \let \@tempc \@tempb \let \@tempb \@tempa \fi \ifx
  \@tempb \@empty \def\@tempb {arXiv}\fi \@ifundefined
  {mn@eprint@\@tempb}{\@tempb:\@tempc}{\expandafter \expandafter \csname
  mn@eprint@\@tempb\endcsname \expandafter{\@tempc}}}

\bibitem[\protect\citeauthoryear{{Asplund}, {Grevesse}  \& {Sauval}}{{Asplund}
  et~al.}{2005}]{asplund2005}
{Asplund} M.,  {Grevesse} N.,   {Sauval} A.~J.,  2005, in {T.~G.~Barnes III \&
  F.~N.~Bash} ed.,  Astronomical Society of the Pacific Conference Series Vol.
  336, Cosmic Abundances as Records of Stellar Evolution and Nucleosynthesis.
  p.~25

\bibitem[\protect\citeauthoryear{{Banerjee}, {Kroupa}  \& {Oh}}{{Banerjee}
  et~al.}{2012}]{banerjee2012b}
{Banerjee} S.,  {Kroupa} P.,   {Oh} S.,  2012, \mn@doi [\mnras]
  {10.1111/j.1365-2966.2012.21672.x}, \href
  {http://adsabs.harvard.edu/abs/2012MNRAS.426.1416B} {426, 1416}

\bibitem[\protect\citeauthoryear{{Bestenlehner}}{{Bestenlehner}}{2020}]{bestenlehner2020}
{Bestenlehner} J.~M.,  2020, \mn@doi [\mnras] {10.1093/mnras/staa474}, \href
  {https://ui.adsabs.harvard.edu/abs/2020MNRAS.493.3938B} {493, 3938}

\bibitem[\protect\citeauthoryear{{Bestenlehner} et~al.,}{{Bestenlehner}
  et~al.}{2011}]{bestenlehner2011}
{Bestenlehner} J.~M.,  et~al., 2011, \mn@doi [\aap]
  {10.1051/0004-6361/201117043}, \href
  {http://adsabs.harvard.edu/abs/2011A%26A...530L..14B} {530, L14}

\bibitem[\protect\citeauthoryear{{Bestenlehner} et~al.,}{{Bestenlehner}
  et~al.}{2014}]{bestenlehner2014}
{Bestenlehner} J.~M.,  et~al., 2014, \mn@doi [\aap]
  {10.1051/0004-6361/201423643}, \href
  {http://adsabs.harvard.edu/abs/2014A%26A...570A..38B} {570, A38}

\bibitem[\protect\citeauthoryear{{Bressert} et~al.,}{{Bressert}
  et~al.}{2012}]{bressert2012}
{Bressert} E.,  et~al., 2012, \mn@doi [\aap] {10.1051/0004-6361/201117247},
  \href {http://adsabs.harvard.edu/abs/2012A%26A...542A..49B} {542, A49}

\bibitem[\protect\citeauthoryear{{Britavskiy} et~al.,}{{Britavskiy}
  et~al.}{2019}]{britavskiy2019}
{Britavskiy} N.,  et~al., 2019, \mn@doi [\aap] {10.1051/0004-6361/201834564},
  \href {https://ui.adsabs.harvard.edu/abs/2019A&A...624A.128B} {624, A128}

\bibitem[\protect\citeauthoryear{{Bromm}, {Coppi}  \& {Larson}}{{Bromm}
  et~al.}{1999}]{bromm1999}
{Bromm} V.,  {Coppi} P.~S.,   {Larson} R.~B.,  1999, \mn@doi [\apjl]
  {10.1086/312385}, \href {http://adsabs.harvard.edu/a7L...5B} {527, L5}

\bibitem[\protect\citeauthoryear{{Brott} et~al.,}{{Brott}
  et~al.}{2011}]{brott2011}
{Brott} I.,  et~al., 2011, \mn@doi [\aap] {10.1051/0004-6361/201016113}, \href
  {http://adsabs.harvard.edu/abs/2011A%26A...530A.115B} {530, A115}

\bibitem[\protect\citeauthoryear{{Campbell}, {Evans}, {Mackey}, {Gieles},
  {Alves}, {Ascenso}, {Bastian}  \& {Longmore}}{{Campbell}
  et~al.}{2010}]{campbell2010}
{Campbell} M.~A.,  {Evans} C.~J.,  {Mackey} A.~D.,  {Gieles} M.,  {Alves} J.,
  {Ascenso} J.,  {Bastian} N.,   {Longmore} A.~J.,  2010, \mn@doi [\mnras]
  {10.1111/j.1365-2966.2010.16447.x}, \href
  {http://adsabs.harvard.edu/abs/2010MNRAS.tmp..416C} {pp 416--+}

\bibitem[\protect\citeauthoryear{{Cantiello} et~al.,}{{Cantiello}
  et~al.}{2009}]{cantiello2009}
{Cantiello} M.,  et~al., 2009, \mn@doi [\aap] {10.1051/0004-6361/200911643},
  \href {https://ui.adsabs.harvard.edu/abs/2009A&A...499..279C} {499, 279}

\bibitem[\protect\citeauthoryear{{Castro}, {Fossati}, {Langer},
  {Sim{\'o}n-D{\'{\i}}az}, {Schneider}  \& {Izzard}}{{Castro}
  et~al.}{2014}]{castro2014}
{Castro} N.,  {Fossati} L.,  {Langer} N.,  {Sim{\'o}n-D{\'{\i}}az} S.,
  {Schneider} F.~R.~N.,   {Izzard} R.~G.,  2014, \mn@doi [\aap]
  {10.1051/0004-6361/201425028}, \href
  {http://adsabs.harvard.edu/abs/2014A%26A...570L..13C} {570, L13}

\bibitem[\protect\citeauthoryear{{Castro}, {Crowther}, {Evans}, {Mackey},
  {Castro-Rodriguez}, {Vink}, {Melnick}  \& {Selman}}{{Castro}
  et~al.}{2018}]{castro2018}
{Castro} N.,  {Crowther} P.~A.,  {Evans} C.~J.,  {Mackey} J.,
  {Castro-Rodriguez} N.,  {Vink} J.~S.,  {Melnick} J.,   {Selman} F.,  2018,
  \mn@doi [\aap] {10.1051/0004-6361/201732084}, \href
  {https://ui.adsabs.harvard.edu/abs/2018A&A...614A.147C} {614, A147}

\bibitem[\protect\citeauthoryear{{Chabrier}}{{Chabrier}}{2003}]{chabrier2003}
{Chabrier} G.,  2003, \mn@doi [\pasp] {10.1086/376392}, \href
  {https://ui.adsabs.harvard.edu/abs/2003PASP..115..763C} {115, 763}

\bibitem[\protect\citeauthoryear{{Cignoni} et~al.,}{{Cignoni}
  et~al.}{2015}]{cignoni2015}
{Cignoni} M.,  et~al., 2015, \mn@doi [\apj] {10.1088/0004-637X/811/2/76}, \href
  {http://adsabs.harvard.edu/abs/2015ApJ...811...76C} {811, 76}

\bibitem[\protect\citeauthoryear{{Crowther}}{{Crowther}}{2019}]{crowther2019}
{Crowther} P.~A.,  2019, \mn@doi [Galaxies] {10.3390/galaxies7040088}, \href
  {https://ui.adsabs.harvard.edu/abs/2019Galax...7...88C} {7, 88}

\bibitem[\protect\citeauthoryear{{Crowther} \& {Dessart}}{{Crowther} \&
  {Dessart}}{1998}]{crowther1998}
{Crowther} P.~A.,  {Dessart} L.,  1998, \mn@doi [\mnras]
  {10.1046/j.1365-8711.1998.01400.x}, \href
  {http://adsabs.harvard.edu/abs/1998MNRAS.296..622C} {296, 622}

\bibitem[\protect\citeauthoryear{{Crowther}, {Schnurr}, {Hirschi}, {Yusof},
  {Parker}, {Goodwin}  \& {Kassim}}{{Crowther} et~al.}{2010}]{crowther2010}
{Crowther} P.~A.,  {Schnurr} O.,  {Hirschi} R.,  {Yusof} N.,  {Parker} R.~J.,
  {Goodwin} S.~P.,   {Kassim} H.~A.,  2010, \mn@doi [\mnras]
  {10.1111/j.1365-2966.2010.17167.x}, \href
  {http://adsabs.harvard.edu/abs/2010MNRAS.408..731C} {408, 731}

\bibitem[\protect\citeauthoryear{{Crowther} et~al.,}{{Crowther}
  et~al.}{2016}]{crowther2016}
{Crowther} P.~A.,  et~al., 2016, \mn@doi [\mnras] {10.1093/mnras/stw273}, \href
  {http://adsabs.harvard.edu/abs/2016MNRAS.458..624C} {458, 624}

\bibitem[\protect\citeauthoryear{{Davies}, {Crowther}  \& {Beasor}}{{Davies}
  et~al.}{2018}]{davies2018}
{Davies} B.,  {Crowther} P.~A.,   {Beasor} E.~R.,  2018, \mn@doi [\mnras]
  {10.1093/mnras/sty1302}, \href
  {https://ui.adsabs.harvard.edu/abs/2018MNRAS.478.3138D} {478, 3138}

\bibitem[\protect\citeauthoryear{{De Marchi} et~al.,}{{De Marchi}
  et~al.}{2011}]{deMarchi2011}
{De Marchi} G.,  et~al., 2011, \mn@doi [\apj] {10.1088/0004-637X/739/1/27},
  \href {http://adsabs.harvard.edu/abs/2011ApJ...739...27D} {739, 27}

\bibitem[\protect\citeauthoryear{{Doran} et~al.,}{{Doran}
  et~al.}{2013}]{doran2013}
{Doran} E.~I.,  et~al., 2013, \mn@doi [\aap] {10.1051/0004-6361/201321824},
  \href {http://adsabs.harvard.edu/abs/2013A%26A...558A.134D} {558, A134}

\bibitem[\protect\citeauthoryear{{Ekstr{\"o}m} et~al.,}{{Ekstr{\"o}m}
  et~al.}{2012}]{ekstroem2012}
{Ekstr{\"o}m} S.,  et~al., 2012, \mn@doi [\aap] {10.1051/0004-6361/201117751},
  \href {http://adsabs.harvard.edu/abs/2012A%26A...537A.146E} {537, A146}

\bibitem[\protect\citeauthoryear{{Eldridge}, {Stanway}, {Xiao}, {McClelland},
  {Taylor}, {Ng}, {Greis}  \& {Bray}}{{Eldridge} et~al.}{2017}]{eldridge2017}
{Eldridge} J.~J.,  {Stanway} E.~R.,  {Xiao} L.,  {McClelland} L.~A.~S.,
  {Taylor} G.,  {Ng} M.,  {Greis} S.~M.~L.,   {Bray} J.~C.,  2017, \mn@doi
  [\pasa] {10.1017/pasa.2017.51}, \href
  {http://adsabs.harvard.edu/abs/2017PASA...34...58E} {34, e058}

\bibitem[\protect\citeauthoryear{{Evans} et~al.,}{{Evans}
  et~al.}{2011}]{evans2011}
{Evans} C.~J.,  et~al., 2011, \mn@doi [\aap] {10.1051/0004-6361/201116782},
  \href {http://adsabs.harvard.edu/abs/2011A%26A...530A.108E} {530, A108}

\bibitem[\protect\citeauthoryear{{Figer}}{{Figer}}{2005}]{figer2005}
{Figer} D.~F.,  2005, \mn@doi [\nat] {10.1038/nature03293}, \href
  {http://adsabs.harvard.edu/abs/2005Natur.434..192F} {434, 192}

\bibitem[\protect\citeauthoryear{{Friend} \& {Castor}}{{Friend} \&
  {Castor}}{1983}]{friend1983}
{Friend} D.~B.,  {Castor} J.~I.,  1983, \mn@doi [\apj] {10.1086/161289}, \href
  {https://ui.adsabs.harvard.edu/abs/1983ApJ...272..259F} {272, 259}

\bibitem[\protect\citeauthoryear{{Gr{\"a}fener} \& {Hamann}}{{Gr{\"a}fener} \&
  {Hamann}}{2008}]{graefener2008}
{Gr{\"a}fener} G.,  {Hamann} W.,  2008, \mn@doi [\aap]
  {10.1051/0004-6361:20066176}, \href
  {http://adsabs.harvard.edu/abs/2008A%26A...482..945G} {482, 945}

\bibitem[\protect\citeauthoryear{{Gr{\"a}fener}, {Koesterke}  \&
  {Hamann}}{{Gr{\"a}fener} et~al.}{2002}]{graefener2002}
{Gr{\"a}fener} G.,  {Koesterke} L.,   {Hamann} W.~R.,  2002, \mn@doi [\aap]
  {10.1051/0004-6361:20020269}, \href
  {https://ui.adsabs.harvard.edu/abs/2002A&A...387..244G} {387, 244}

\bibitem[\protect\citeauthoryear{{Gr{\"a}fener}, {Vink}, {de Koter}  \&
  {Langer}}{{Gr{\"a}fener} et~al.}{2011}]{graefener2011}
{Gr{\"a}fener} G.,  {Vink} J.~S.,  {de Koter} A.,   {Langer} N.,  2011, \mn@doi
  [\aap] {10.1051/0004-6361/201116701}, \href
  {http://adsabs.harvard.edu/abs/2011A%26A...535A..56G} {535, A56}

\bibitem[\protect\citeauthoryear{{Grin} et~al.,}{{Grin}
  et~al.}{2017}]{grin2017}
{Grin} N.~J.,  et~al., 2017, \mn@doi [\aap] {10.1051/0004-6361/201629225},
  \href {http://adsabs.harvard.edu/abs/2017A%26A...600A..82G} {600, A82}

\bibitem[\protect\citeauthoryear{{Hainich} et~al.,}{{Hainich}
  et~al.}{2014}]{hainich2014}
{Hainich} R.,  et~al., 2014, \mn@doi [\aap] {10.1051/0004-6361/201322696},
  \href {http://adsabs.harvard.edu/abs/2014A%26A...565A..27H} {565, A27}

\bibitem[\protect\citeauthoryear{{Hamann} \& {Gr{\"a}fener}}{{Hamann} \&
  {Gr{\"a}fener}}{2004}]{hamann2004}
{Hamann} W.-R.,  {Gr{\"a}fener} G.,  2004, \mn@doi [\aap]
  {10.1051/0004-6361:20040506}, \href
  {http://adsabs.harvard.edu/abs/2004A%26A...427..697H} {427, 697}

\bibitem[\protect\citeauthoryear{{Heger} \& {Woosley}}{{Heger} \&
  {Woosley}}{2002}]{heger2002}
{Heger} A.,  {Woosley} S.~E.,  2002, \mn@doi [\apj] {10.1086/338487}, \href
  {http://adsabs.harvard.edu/abs/2002ApJ...567..532H} {567, 532}

\bibitem[\protect\citeauthoryear{{H{\'e}nault-Brunet}
  et~al.,}{{H{\'e}nault-Brunet} et~al.}{2012}]{henault2012}
{H{\'e}nault-Brunet} V.,  et~al., 2012, \mn@doi [\aap]
  {10.1051/0004-6361/201219471}, \href
  {http://adsabs.harvard.edu/abs/2012A%26A...546A..73H} {546, A73}

\bibitem[\protect\citeauthoryear{{Herrero} \& {Lennon}}{{Herrero} \&
  {Lennon}}{2004}]{herrero2004}
{Herrero} A.,  {Lennon} D.~J.,  2004, in {Maeder} A.,  {Eenens} P.,  eds,  IAU
  Symposium Vol. 215, Stellar Rotation. p.~209

\bibitem[\protect\citeauthoryear{{Herrero}, {Kudritzki}, {Vilchez}, {Kunze},
  {Butler}  \& {Haser}}{{Herrero} et~al.}{1992}]{herrero1992}
{Herrero} A.,  {Kudritzki} R.~P.,  {Vilchez} J.~M.,  {Kunze} D.,  {Butler} K.,
   {Haser} S.,  1992, \aap, \href
  {https://ui.adsabs.harvard.edu/abs/1992A&A...261..209H} {261, 209}

\bibitem[\protect\citeauthoryear{{Higgins} \& {Vink}}{{Higgins} \&
  {Vink}}{2019}]{higgins2019}
{Higgins} E.~R.,  {Vink} J.~S.,  2019, \mn@doi [\aap]
  {10.1051/0004-6361/201834123}, \href
  {https://ui.adsabs.harvard.edu/abs/2019A&A...622A..50H} {622, A50}

\bibitem[\protect\citeauthoryear{{Hillier} \& {Miller}}{{Hillier} \&
  {Miller}}{1998}]{hillier1998}
{Hillier} D.~J.,  {Miller} D.~L.,  1998, \mn@doi [\apj] {10.1086/305350}, \href
  {http://adsabs.harvard.edu/abs/1998ApJ...496..407H} {496, 407}

\bibitem[\protect\citeauthoryear{{Holgado} et~al.,}{{Holgado}
  et~al.}{2018}]{holgado2018}
{Holgado} G.,  et~al., 2018, \mn@doi [\aap] {10.1051/0004-6361/201731543},
  \href {https://ui.adsabs.harvard.edu/abs/2018A&A...613A..65H} {613, A65}

\bibitem[\protect\citeauthoryear{{Hummer}}{{Hummer}}{1982}]{hummer1982}
{Hummer} D.~G.,  1982, \mn@doi [\apj] {10.1086/160027}, \href
  {https://ui.adsabs.harvard.edu/abs/1982ApJ...257..724H} {257, 724}

\bibitem[\protect\citeauthoryear{{Hunter}, {Shaya}, {Holtzman}, {Light},
  {O'Neil}  \& {Lynds}}{{Hunter} et~al.}{1995}]{hunter1995}
{Hunter} D.~A.,  {Shaya} E.~J.,  {Holtzman} J.~A.,  {Light} R.~M.,  {O'Neil}
  Jr. E.~J.,   {Lynds} R.,  1995, \mn@doi [\apj] {10.1086/175950}, \href
  {http://adsabs.harvard.edu/abs/1995ApJ...448..179H} {448, 179}

\bibitem[\protect\citeauthoryear{{Hunter} et~al.,}{{Hunter}
  et~al.}{2007}]{hunter2007}
{Hunter} I.,  et~al., 2007, \mn@doi [\aap] {10.1051/0004-6361:20066148}, \href
  {https://ui.adsabs.harvard.edu/abs/2007A&A...466..277H} {466, 277}

\bibitem[\protect\citeauthoryear{{Khorrami} et~al.,}{{Khorrami}
  et~al.}{2017}]{khorrami2017}
{Khorrami} Z.,  et~al., 2017, \mn@doi [\aap] {10.1051/0004-6361/201629279},
  \href {http://adsabs.harvard.edu/abs/2017A%26A...602A..56K} {602, A56}

\bibitem[\protect\citeauthoryear{{K{\"o}hler} et~al.,}{{K{\"o}hler}
  et~al.}{2015}]{koehler2015}
{K{\"o}hler} K.,  et~al., 2015, \mn@doi [\aap] {10.1051/0004-6361/201424356},
  \href {http://adsabs.harvard.edu/abs/2015A%26A...573A..71K} {573, A71}

\bibitem[\protect\citeauthoryear{{Korn}, {Keller}, {Kaufer}, {Langer},
  {Przybilla}, {Stahl}  \& {Wolf}}{{Korn} et~al.}{2002}]{korn2002}
{Korn} A.~J.,  {Keller} S.~C.,  {Kaufer} A.,  {Langer} N.,  {Przybilla} N.,
  {Stahl} O.,   {Wolf} B.,  2002, \mn@doi [\aap] {10.1051/0004-6361:20020116},
  \href {http://adsabs.harvard.edu/abs/2002A%26A...385..143K} {385, 143}

\bibitem[\protect\citeauthoryear{{Kozyreva}, {Yoon}  \& {Langer}}{{Kozyreva}
  et~al.}{2014}]{kozyreva2014}
{Kozyreva} A.,  {Yoon} S.-C.,   {Langer} N.,  2014, \mn@doi [\aap]
  {10.1051/0004-6361/201423641}, \href
  {http://adsabs.harvard.edu/abs/2014A%26A...566A.146K} {566, A146}

\bibitem[\protect\citeauthoryear{{Kroupa}}{{Kroupa}}{2001}]{kroupa2001}
{Kroupa} P.,  2001, \mn@doi [\mnras] {10.1046/j.1365-8711.2001.04022.x}, \href
  {https://ui.adsabs.harvard.edu/abs/2001MNRAS.322..231K} {322, 231}

\bibitem[\protect\citeauthoryear{{Krumholz}}{{Krumholz}}{2015}]{krumholz2015}
{Krumholz} M.~R.,  2015, {The Formation of Very Massive Stars}.
p.~43, \mn@doi{10.1007/978-3-319-09596-7_3}

\bibitem[\protect\citeauthoryear{{Kudritzki}, {Cabanne}, {Husfeld}, {Niemela},
  {Groth}, {Puls}  \& {Herrero}}{{Kudritzki} et~al.}{1989}]{kudritzki1989b}
{Kudritzki} R.~P.,  {Cabanne} M.~L.,  {Husfeld} D.,  {Niemela} V.~S.,  {Groth}
  H.~G.,  {Puls} J.,   {Herrero} A.,  1989, \aap, \href
  {https://ui.adsabs.harvard.edu/abs/1989A&A...226..235K} {226, 235}

\bibitem[\protect\citeauthoryear{{Kudritzki}, {Lennon}  \& {Puls}}{{Kudritzki}
  et~al.}{1995}]{kudritzki1995}
{Kudritzki} R.~P.,  {Lennon} D.~J.,   {Puls} J.,  1995, in {Walsh} J.~R.,
  {Danziger} I.~J.,  eds, Science with the VLT. p.~246

\bibitem[\protect\citeauthoryear{{Langer}}{{Langer}}{2009}]{langer2009}
{Langer} N.,  2009, \mn@doi [\nat] {10.1038/462579a}, \href
  {http://adsabs.harvard.edu/abs/2009Natur.462..579L} {462, 579}

\bibitem[\protect\citeauthoryear{{Langer}}{{Langer}}{2012}]{langer2012}
{Langer} N.,  2012, \mn@doi [\araa] {10.1146/annurev-astro-081811-125534},
  \href {http://adsabs.harvard.edu/abs/2012ARA%26A..50..107L} {50, 107}

\bibitem[\protect\citeauthoryear{{Langer}, {Norman}, {de Koter}, {Vink},
  {Cantiello}  \& {Yoon}}{{Langer} et~al.}{2007}]{langer2007}
{Langer} N.,  {Norman} C.~A.,  {de Koter} A.,  {Vink} J.~S.,  {Cantiello} M.,
  {Yoon} S.-C.,  2007, \mn@doi [\aap] {10.1051/0004-6361:20078482}, \href
  {http://adsabs.harvard.edu/abs/2007A%26A...475L..19L} {475, L19}

\bibitem[\protect\citeauthoryear{{Leitherer} et~al.,}{{Leitherer}
  et~al.}{1999}]{leitherer1999}
{Leitherer} C.,  et~al., 1999, \mn@doi [\apjs] {10.1086/313233}, \href
  {https://ui.adsabs.harvard.edu/abs/1999ApJS..123....3L} {123, 3}

\bibitem[\protect\citeauthoryear{{Lennon} et~al.,}{{Lennon}
  et~al.}{2018}]{lennon2018}
{Lennon} D.~J.,  et~al., 2018, \mn@doi [\aap] {10.1051/0004-6361/201833465},
  \href {https://ui.adsabs.harvard.edu/abs/2018A&A...619A..78L} {619, A78}

\bibitem[\protect\citeauthoryear{{Lucy} \& {Abbott}}{{Lucy} \&
  {Abbott}}{1993}]{lucy1993}
{Lucy} L.~B.,  {Abbott} D.~C.,  1993, \mn@doi [\apj] {10.1086/172402}, \href
  {https://ui.adsabs.harvard.edu/abs/1993ApJ...405..738L} {405, 738}

\bibitem[\protect\citeauthoryear{{Mahy}, {Damerdji}, {Gosset}, {Nitschelm},
  {Eenens}, {Sana}  \& {Klotz}}{{Mahy} et~al.}{2017}]{mahy2017}
{Mahy} L.,  {Damerdji} Y.,  {Gosset} E.,  {Nitschelm} C.,  {Eenens} P.,  {Sana}
  H.,   {Klotz} A.,  2017, \mn@doi [\aap] {10.1051/0004-6361/201730674}, \href
  {https://ui.adsabs.harvard.edu/abs/2017A&A...607A..96M} {607, A96}

\bibitem[\protect\citeauthoryear{{Mahy} et~al.,}{{Mahy}
  et~al.}{2020}]{mahy2020}
{Mahy} L.,  et~al., 2020, \mn@doi [\aap] {10.1051/0004-6361/201936151}, \href
  {https://ui.adsabs.harvard.edu/abs/2020A&A...634A.118M} {634, A118}

\bibitem[\protect\citeauthoryear{{Ma{\'\i}z Apell{\'a}niz}}{{Ma{\'\i}z
  Apell{\'a}niz}}{2013}]{ma-ap2013}
{Ma{\'\i}z Apell{\'a}niz} J.,  2013, in {Guirado} J.~C.,  {Lara} L.~M.,
  {Quilis} V.,   {Gorgas} J.,  eds, Highlights of Spanish Astrophysics VII. pp
  583--589 (\mn@eprint {arXiv} {1209.2560})

\bibitem[\protect\citeauthoryear{{Ma{\'{\i}}z Apell{\'a}niz}
  et~al.,}{{Ma{\'{\i}}z Apell{\'a}niz} et~al.}{2014}]{ma-ap2014}
{Ma{\'{\i}}z Apell{\'a}niz} J.,  et~al., 2014, \mn@doi [\aap]
  {10.1051/0004-6361/201423439}, \href
  {http://adsabs.harvard.edu/abs/2014A%26A...564A..63M} {564, A63}

\bibitem[\protect\citeauthoryear{{Ma{\'\i}z Apell{\'a}niz} et~al.,}{{Ma{\'\i}z
  Apell{\'a}niz} et~al.}{2019}]{ma-ap2019}
{Ma{\'\i}z Apell{\'a}niz} J.,  et~al., 2019, \mn@doi [\aap]
  {10.1051/0004-6361/201935359}, \href
  {https://ui.adsabs.harvard.edu/abs/2019A&A...626A..20M} {626, A20}

\bibitem[\protect\citeauthoryear{{Markova}, {Puls}, {Repolust}  \&
  {Markov}}{{Markova} et~al.}{2004}]{markova2004}
{Markova} N.,  {Puls} J.,  {Repolust} T.,   {Markov} H.,  2004, \mn@doi [\aap]
  {10.1051/0004-6361:20031463}, \href
  {https://ui.adsabs.harvard.edu/abs/2004A&A...413..693M} {413, 693}

\bibitem[\protect\citeauthoryear{{Markova}, {Puls}, {Scuderi}  \&
  {Markov}}{{Markova} et~al.}{2005}]{markova2005}
{Markova} N.,  {Puls} J.,  {Scuderi} S.,   {Markov} H.,  2005, \mn@doi [\aap]
  {10.1051/0004-6361:20041774}, \href
  {https://ui.adsabs.harvard.edu/abs/2005A&A...440.1133M} {440, 1133}

\bibitem[\protect\citeauthoryear{{Markova}, {Puls}  \& {Langer}}{{Markova}
  et~al.}{2018}]{markova2018}
{Markova} N.,  {Puls} J.,   {Langer} N.,  2018, \mn@doi [\aap]
  {10.1051/0004-6361/201731361}, \href
  {https://ui.adsabs.harvard.edu/abs/2018A&A...613A..12M} {613, A12}

\bibitem[\protect\citeauthoryear{{Martins} \& {Palacios}}{{Martins} \&
  {Palacios}}{2013}]{martins2013}
{Martins} F.,  {Palacios} A.,  2013, \mn@doi [\aap]
  {10.1051/0004-6361/201322480}, \href
  {http://adsabs.harvard.edu/abs/2013A%26A...560A..16M} {560, A16}

\bibitem[\protect\citeauthoryear{{Martins}, {Schaerer}  \& {Hillier}}{{Martins}
  et~al.}{2005}]{martins2005}
{Martins} F.,  {Schaerer} D.,   {Hillier} D.~J.,  2005, \mn@doi [\aap]
  {10.1051/0004-6361:20042386}, \href
  {http://adsabs.harvard.edu/abs/2005A%26A...436.1049M} {436, 1049}

\bibitem[\protect\citeauthoryear{{Massey} \& {Hunter}}{{Massey} \&
  {Hunter}}{1998}]{massey1998}
{Massey} P.,  {Hunter} D.~A.,  1998, \mn@doi [\apj] {10.1086/305126}, \href
  {http://adsabs.harvard.edu/abs/1998ApJ...493..180M} {493, 180}

\bibitem[\protect\citeauthoryear{{Massey}, {Neugent}, {Hillier}  \&
  {Puls}}{{Massey} et~al.}{2013}]{massey2013}
{Massey} P.,  {Neugent} K.~F.,  {Hillier} D.~J.,   {Puls} J.,  2013, \mn@doi
  [\apj] {10.1088/0004-637X/768/1/6}, \href
  {http://adsabs.harvard.edu/abs/2013ApJ...768....6M} {768, 6}

\bibitem[\protect\citeauthoryear{{McEvoy} et~al.,}{{McEvoy}
  et~al.}{2015}]{mcevoy2015}
{McEvoy} C.~M.,  et~al., 2015, \mn@doi [\aap] {10.1051/0004-6361/201425202},
  \href {https://ui.adsabs.harvard.edu/abs/2015A%26A...575A..70M} {575, A70}

\bibitem[\protect\citeauthoryear{{Meynet}, {Maeder}, {Schaller}, {Schaerer}  \&
  {Charbonnel}}{{Meynet} et~al.}{1994}]{meynet1994}
{Meynet} G.,  {Maeder} A.,  {Schaller} G.,  {Schaerer} D.,   {Charbonnel} C.,
  1994, \aaps, \href {https://ui.adsabs.harvard.edu/abs/1994A&AS..103...97M}
  {103, 97}

\bibitem[\protect\citeauthoryear{{Moe} \& {Di Stefano}}{{Moe} \& {Di
  Stefano}}{2017}]{moe2017}
{Moe} M.,  {Di Stefano} R.,  2017, \mn@doi [\apjs] {10.3847/1538-4365/aa6fb6},
  \href {https://ui.adsabs.harvard.edu/abs/2017ApJS..230...15M} {230, 15}

\bibitem[\protect\citeauthoryear{{Mokiem} et~al.,}{{Mokiem}
  et~al.}{2007}]{mokiem2007}
{Mokiem} M.~R.,  et~al., 2007, \mn@doi [\aap] {10.1051/0004-6361:20066489},
  \href {http://adsabs.harvard.edu/abs/2007A%26A...465.1003M} {465, 1003}

\bibitem[\protect\citeauthoryear{{Muijres}, {Vink}, {de Koter}, {M{\"u}ller}
  \& {Langer}}{{Muijres} et~al.}{2012}]{muijres2012}
{Muijres} L.~E.,  {Vink} J.~S.,  {de Koter} A.,  {M{\"u}ller} P.~E.,   {Langer}
  N.,  2012, \mn@doi [\aap] {10.1051/0004-6361/201015818}, \href
  {https://ui.adsabs.harvard.edu/abs/2012A&A...537A..37M} {537, A37}

\bibitem[\protect\citeauthoryear{{Pellegrini}, {Baldwin}  \&
  {Ferland}}{{Pellegrini} et~al.}{2011}]{pellegrini2011}
{Pellegrini} E.~W.,  {Baldwin} J.~A.,   {Ferland} G.~J.,  2011, \mn@doi [\apj]
  {10.1088/0004-637X/738/1/34}, \href
  {https://ui.adsabs.harvard.edu/abs/2011ApJ...738...34P} {738, 34}

\bibitem[\protect\citeauthoryear{{Pietrzy{\'n}ski} et~al.,}{{Pietrzy{\'n}ski}
  et~al.}{2019}]{pietrzynski2019}
{Pietrzy{\'n}ski} G.,  et~al., 2019, \mn@doi [\nat]
  {10.1038/s41586-019-0999-4}, \href
  {https://ui.adsabs.harvard.edu/abs/2019Natur.567..200P} {567, 200}

\bibitem[\protect\citeauthoryear{{Prinja}, {Barlow}  \& {Howarth}}{{Prinja}
  et~al.}{1990}]{prinja1990}
{Prinja} R.~K.,  {Barlow} M.~J.,   {Howarth} I.~D.,  1990, \mn@doi [\apj]
  {10.1086/169224}, \href {http://adsabs.harvard.edu/abs/1990ApJ...361..607P}
  {361, 607}

\bibitem[\protect\citeauthoryear{{Puls}}{{Puls}}{1987}]{puls1987}
{Puls} J.,  1987, \aap, \href
  {https://ui.adsabs.harvard.edu/abs/1987A&A...184..227P} {184, 227}

\bibitem[\protect\citeauthoryear{{Puls} et~al.,}{{Puls}
  et~al.}{1996}]{puls1996}
{Puls} J.,  et~al., 1996, \aap, \href
  {http://adsabs.harvard.edu/abs/1996A%26A...305..171P} {305, 171}

\bibitem[\protect\citeauthoryear{{Puls}, {Urbaneja}, {Venero}, {Repolust},
  {Springmann}, {Jokuthy}  \& {Mokiem}}{{Puls} et~al.}{2005}]{puls2005}
{Puls} J.,  {Urbaneja} M.~A.,  {Venero} R.,  {Repolust} T.,  {Springmann} U.,
  {Jokuthy} A.,   {Mokiem} M.~R.,  2005, \mn@doi [\aap]
  {10.1051/0004-6361:20042365}, \href
  {http://adsabs.harvard.edu/abs/2005A%26A...435..669P} {435, 669}

\bibitem[\protect\citeauthoryear{{Puls}, {Vink}  \& {Najarro}}{{Puls}
  et~al.}{2008}]{puls2008}
{Puls} J.,  {Vink} J.~S.,   {Najarro} F.,  2008, \mn@doi [\aapr]
  {10.1007/s00159-008-0015-8}, \href
  {https://ui.adsabs.harvard.edu/abs/2008A&ARv..16..209P} {16, 209}

\bibitem[\protect\citeauthoryear{{Ramachandran} et~al.,}{{Ramachandran}
  et~al.}{2019}]{ramachandran2019}
{Ramachandran} V.,  et~al., 2019, \mn@doi [\aap] {10.1051/0004-6361/201935365},
  \href {https://ui.adsabs.harvard.edu/abs/2019A&A...625A.104R} {625, A104}

\bibitem[\protect\citeauthoryear{{Ram{\'{\i}}rez-Agudelo}
  et~al.,}{{Ram{\'{\i}}rez-Agudelo} et~al.}{2017}]{ramirez2017}
{Ram{\'{\i}}rez-Agudelo} O.~H.,  et~al., 2017, \mn@doi [\aap]
  {10.1051/0004-6361/201628914}, \href
  {http://adsabs.harvard.edu/abs/2017A%26A...600A..81R} {600, A81}

\bibitem[\protect\citeauthoryear{{Renzo} et~al.,}{{Renzo}
  et~al.}{2019}]{renzo2019}
{Renzo} M.,  et~al., 2019, \mn@doi [\mnras] {10.1093/mnrasl/sly194}, \href
  {https://ui.adsabs.harvard.edu/abs/2019MNRAS.482L.102R} {482, L102}

\bibitem[\protect\citeauthoryear{{Rivero Gonz{\'a}lez}, {Puls}  \&
  {Najarro}}{{Rivero Gonz{\'a}lez} et~al.}{2011}]{rivero2011}
{Rivero Gonz{\'a}lez} J.~G.,  {Puls} J.,   {Najarro} F.,  2011, \mn@doi [\aap]
  {10.1051/0004-6361/201117101}, \href
  {http://adsabs.harvard.edu/abs/2011A%26A...536A..58R} {536, A58}

\bibitem[\protect\citeauthoryear{{Rivero Gonz{\'a}lez}, {Puls}, {Najarro}  \&
  {Brott}}{{Rivero Gonz{\'a}lez} et~al.}{2012a}]{rivero2012a}
{Rivero Gonz{\'a}lez} J.~G.,  {Puls} J.,  {Najarro} F.,   {Brott} I.,  2012a,
  \mn@doi [\aap] {10.1051/0004-6361/201117790}, \href
  {http://adsabs.harvard.edu/abs/2012A%26A...537A..79R} {537, A79}

\bibitem[\protect\citeauthoryear{{Rivero Gonz{\'a}lez}, {Puls}, {Massey}  \&
  {Najarro}}{{Rivero Gonz{\'a}lez} et~al.}{2012b}]{rivero2012b}
{Rivero Gonz{\'a}lez} J.~G.,  {Puls} J.,  {Massey} P.,   {Najarro} F.,  2012b,
  \mn@doi [\aap] {10.1051/0004-6361/201218955}, \href
  {http://adsabs.harvard.edu/abs/2012A%26A...543A..95R} {543, A95}

\bibitem[\protect\citeauthoryear{{Sabbi} et~al.,}{{Sabbi}
  et~al.}{2012}]{sabbi2012}
{Sabbi} E.,  et~al., 2012, \mn@doi [\apjl] {10.1088/2041-8205/754/2/L37}, \href
  {http://adsabs.harvard.edu/abs/2012ApJ...754L..37S} {754, L37}

\bibitem[\protect\citeauthoryear{{Sab{\'{\i}}n-Sanjuli{\'a}n}
  et~al.,}{{Sab{\'{\i}}n-Sanjuli{\'a}n} et~al.}{2014}]{sabin2014}
{Sab{\'{\i}}n-Sanjuli{\'a}n} C.,  et~al., 2014, \mn@doi [\aap]
  {10.1051/0004-6361/201322798}, \href
  {http://adsabs.harvard.edu/abs/2014A%26A...564A..39S} {564, A39}

\bibitem[\protect\citeauthoryear{{Sab{\'{\i}}n-Sanjuli{\'a}n}
  et~al.,}{{Sab{\'{\i}}n-Sanjuli{\'a}n} et~al.}{2017}]{sabin2017}
{Sab{\'{\i}}n-Sanjuli{\'a}n} C.,  et~al., 2017, \mn@doi [\aap]
  {10.1051/0004-6361/201629210}, \href
  {http://adsabs.harvard.edu/abs/2017A%26A...601A..79S} {601, A79}

\bibitem[\protect\citeauthoryear{{Salpeter}}{{Salpeter}}{1955}]{salpeter1955}
{Salpeter} E.~E.,  1955, \mn@doi [\apj] {10.1086/145971}, \href
  {http://adsabs.harvard.edu/abs/1955ApJ...121..161S} {121, 161}

\bibitem[\protect\citeauthoryear{{Sana} et~al.,}{{Sana}
  et~al.}{2012}]{sana2012Sci}
{Sana} H.,  et~al., 2012, \mn@doi [Science] {10.1126/science.1223344}, \href
  {http://ads.ari.uni-heidelberg.de/abs/2012Sci...337..444S} {337, 444}

\bibitem[\protect\citeauthoryear{{Sana} et~al.,}{{Sana}
  et~al.}{2013}]{sana2013}
{Sana} H.,  et~al., 2013, \mn@doi [\aap] {10.1051/0004-6361/201219621}, \href
  {http://adsabs.harvard.edu/abs/2013A%26A...550A.107S} {550, A107}

\bibitem[\protect\citeauthoryear{{Santolaya-Rey}, {Puls}  \&
  {Herrero}}{{Santolaya-Rey} et~al.}{1997}]{santolaya1997}
{Santolaya-Rey} A.~E.,  {Puls} J.,   {Herrero} A.,  1997, \aap, \href
  {http://adsabs.harvard.edu/abs/1997A%26A...323..488S} {323, 488}

\bibitem[\protect\citeauthoryear{{Sanyal}, {Grassitelli}, {Langer}  \&
  {Bestenlehner}}{{Sanyal} et~al.}{2015}]{sanyal2015}
{Sanyal} D.,  {Grassitelli} L.,  {Langer} N.,   {Bestenlehner} J.~M.,  2015,
  \mn@doi [\aap] {10.1051/0004-6361/201525945}, \href
  {http://adsabs.harvard.edu/abs/2015A%26A...580A..20S} {580, A20}

\bibitem[\protect\citeauthoryear{{Schaerer}, {Meynet}, {Maeder}  \&
  {Schaller}}{{Schaerer} et~al.}{1993}]{schaerer1993}
{Schaerer} D.,  {Meynet} G.,  {Maeder} A.,   {Schaller} G.,  1993, \aaps, \href
  {https://ui.adsabs.harvard.edu/abs/1993A&AS...98..523S} {98, 523}

\bibitem[\protect\citeauthoryear{{Schmutz}, {Hamann}  \&
  {Wessolowski}}{{Schmutz} et~al.}{1989}]{schmutz1989}
{Schmutz} W.,  {Hamann} W.,   {Wessolowski} U.,  1989, \aap, \href
  {http://adsabs.harvard.edu/abs/1989A%26A...210..236S} {210, 236}

\bibitem[\protect\citeauthoryear{{Schneider}, {Langer}, {de Koter}, {Brott},
  {Izzard}  \& {Lau}}{{Schneider} et~al.}{2014a}]{schneider2014}
{Schneider} F.~R.~N.,  {Langer} N.,  {de Koter} A.,  {Brott} I.,  {Izzard}
  R.~G.,   {Lau} H.~H.~B.,  2014a, \mn@doi [\aap]
  {10.1051/0004-6361/201424286}, \href
  {http://adsabs.harvard.edu/abs/2014A%26A...570A..66S} {570, A66}

\bibitem[\protect\citeauthoryear{{Schneider} et~al.,}{{Schneider}
  et~al.}{2014b}]{schneider2014b}
{Schneider} F.~R.~N.,  et~al., 2014b, \mn@doi [\apj]
  {10.1088/0004-637X/780/2/117}, \href
  {http://adsabs.harvard.edu/abs/2014ApJ...780..117S} {780, 117}

\bibitem[\protect\citeauthoryear{{Schneider} et~al.,}{{Schneider}
  et~al.}{2018a}]{schneider2018}
{Schneider} F.~R.~N.,  et~al., 2018a, \mn@doi [Science]
  {10.1126/science.aan0106}, \href
  {http://adsabs.harvard.edu/abs/2018Sci...359...69S} {359, 69}

\bibitem[\protect\citeauthoryear{{Schneider} et~al.,}{{Schneider}
  et~al.}{2018b}]{schneider2018b}
{Schneider} F.~R.~N.,  et~al., 2018b, \mn@doi [\aap]
  {10.1051/0004-6361/201833433}, \href
  {https://ui.adsabs.harvard.edu/abs/2018A%26A...618A..73S} {618, A73}

\bibitem[\protect\citeauthoryear{{Shenar} et~al.,}{{Shenar}
  et~al.}{2017}]{shenar2017}
{Shenar} T.,  et~al., 2017, \mn@doi [\aap] {10.1051/0004-6361/201629621}, \href
  {https://ui.adsabs.harvard.edu/abs/2017A&A...598A..85S} {598, A85}

\bibitem[\protect\citeauthoryear{{Shenar} et~al.,}{{Shenar}
  et~al.}{2019}]{shenar2019}
{Shenar} T.,  et~al., 2019, \mn@doi [\aap] {10.1051/0004-6361/201935684}, \href
  {https://ui.adsabs.harvard.edu/abs/2019A&A...627A.151S} {627, A151}

\bibitem[\protect\citeauthoryear{{Shenar}, {Gilkis}, {Vink}, {Sana}  \& {Sand
  er}}{{Shenar} et~al.}{2020}]{shenar2020}
{Shenar} T.,  {Gilkis} A.,  {Vink} J.~S.,  {Sana} H.,   {Sand er} A.~A.~C.,
  2020, \mn@doi [\aap] {10.1051/0004-6361/201936948}, \href
  {https://ui.adsabs.harvard.edu/abs/2020A&A...634A..79S} {634, A79}

\bibitem[\protect\citeauthoryear{{Sim{\'o}n-D{\'{\i}}az} \&
  {Herrero}}{{Sim{\'o}n-D{\'{\i}}az} \& {Herrero}}{2014}]{simon-diaz2014}
{Sim{\'o}n-D{\'{\i}}az} S.,  {Herrero} A.,  2014, \mn@doi [\aap]
  {10.1051/0004-6361/201322758}, \href
  {http://adsabs.harvard.edu/abs/2014A%26A...562A.135S} {562, A135}

\bibitem[\protect\citeauthoryear{{Sim{\'o}n-D{\'{\i}}az}, {Castro}, {Herrero},
  {Puls}, {Garcia}  \& {Sab{\'{\i}}n-Sanjuli{\'a}n}}{{Sim{\'o}n-D{\'{\i}}az}
  et~al.}{2011}]{simon-diaz2011}
{Sim{\'o}n-D{\'{\i}}az} S.,  {Castro} N.,  {Herrero} A.,  {Puls} J.,  {Garcia}
  M.,   {Sab{\'{\i}}n-Sanjuli{\'a}n} C.,  2011, in Journal of Physics
  Conference Series. p. 012021 (\mn@eprint {arXiv} {1111.1341}),
  \mn@doi{10.1088/1742-6596/328/1/012021}

\bibitem[\protect\citeauthoryear{{Stanway} \& {Eldridge}}{{Stanway} \&
  {Eldridge}}{2018}]{stanway2018}
{Stanway} E.~R.,  {Eldridge} J.~J.,  2018, \mn@doi [\mnras]
  {10.1093/mnras/sty1353}, \href
  {https://ui.adsabs.harvard.edu/abs/2018MNRAS.479...75S} {479, 75}

\bibitem[\protect\citeauthoryear{{Stevance}, {Eldridge}  \&
  {Stanway}}{{Stevance} et~al.}{2020}]{stevance2020}
{Stevance} H.,  {Eldridge} J.,   {Stanway} E.,  2020, \mn@doi [The Journal of
  Open Source Software] {10.21105/joss.01987}, \href
  {https://ui.adsabs.harvard.edu/abs/2020JOSS....5.1987S} {5, 1987}

\bibitem[\protect\citeauthoryear{{Sundqvist} \& {Puls}}{{Sundqvist} \&
  {Puls}}{2018}]{sundqvist2018}
{Sundqvist} J.~O.,  {Puls} J.,  2018, \mn@doi [\aap]
  {10.1051/0004-6361/201832993}, \href
  {https://ui.adsabs.harvard.edu/abs/2018A&A...619A..59S} {619, A59}

\bibitem[\protect\citeauthoryear{{Tehrani}, {Crowther}, {Bestenlehner},
  {Littlefair}, {Pollock}, {Parker}  \& {Schnurr}}{{Tehrani}
  et~al.}{2019}]{tehrani2019}
{Tehrani} K.~A.,  {Crowther} P.~A.,  {Bestenlehner} J.~M.,  {Littlefair} S.~P.,
   {Pollock} A.~M.~T.,  {Parker} R.~J.,   {Schnurr} O.,  2019, \mn@doi [\mnras]
  {10.1093/mnras/stz147}, \href
  {https://ui.adsabs.harvard.edu/abs/2019MNRAS.484.2692T} {484, 2692}

\bibitem[\protect\citeauthoryear{{Vacca}, {Garmany}  \& {Shull}}{{Vacca}
  et~al.}{1996}]{vacca1996}
{Vacca} W.~D.,  {Garmany} C.~D.,   {Shull} J.~M.,  1996, \mn@doi [\apj]
  {10.1086/177020}, \href
  {https://ui.adsabs.harvard.edu/abs/1996ApJ...460..914V} {460, 914}

\bibitem[\protect\citeauthoryear{{Vink}}{{Vink}}{2015}]{vink2015}
{Vink} J.~S.,  2015, {Very Massive Stars in the Local Universe}.
 Vol. 412, \mn@doi{10.1007/978-3-319-09596-7, }

\bibitem[\protect\citeauthoryear{{Vink}}{{Vink}}{2018}]{vink2018}
{Vink} J.~S.,  2018, \mn@doi [\aap] {10.1051/0004-6361/201832773}, \href
  {https://ui.adsabs.harvard.edu/abs/2018A&A...615A.119V} {615, A119}

\bibitem[\protect\citeauthoryear{{Vink} \& {Gr{\"a}fener}}{{Vink} \&
  {Gr{\"a}fener}}{2012}]{vink2012}
{Vink} J.~S.,  {Gr{\"a}fener} G.,  2012, \mn@doi [\apjl]
  {10.1088/2041-8205/751/2/L34}, \href
  {http://adsabs.harvard.edu/abs/2012ApJ...751L..34V} {751, L34}

\bibitem[\protect\citeauthoryear{{Vink}, {de Koter}  \& {Lamers}}{{Vink}
  et~al.}{2000}]{vink2000}
{Vink} J.~S.,  {de Koter} A.,   {Lamers} H.~J.~G.~L.~M.,  2000, \aap, \href
  {http://adsabs.harvard.edu/abs/2000A%26A...362..295V} {362, 295}

\bibitem[\protect\citeauthoryear{{Vink}, {de Koter}  \& {Lamers}}{{Vink}
  et~al.}{2001}]{vink2001}
{Vink} J.~S.,  {de Koter} A.,   {Lamers} H.~J.~G.~L.~M.,  2001, \mn@doi [\aap]
  {10.1051/0004-6361:20010127}, \href
  {http://adsabs.harvard.edu/abs/2001A%26A...369..574V} {369, 574}

\bibitem[\protect\citeauthoryear{{Vink}, {Brott}, {Gr{\"a}fener}, {Langer}, {de
  Koter}  \& {Lennon}}{{Vink} et~al.}{2010}]{vink2010}
{Vink} J.~S.,  {Brott} I.,  {Gr{\"a}fener} G.,  {Langer} N.,  {de Koter} A.,
  {Lennon} D.~J.,  2010, \mn@doi [\aap] {10.1051/0004-6361/201014205}, \href
  {https://ui.adsabs.harvard.edu/abs/2010A&A...512L...7V} {512, L7}

\bibitem[\protect\citeauthoryear{{Vink}, {Muijres}, {Anthonisse}, {de Koter},
  {Gr{\"a}fener}  \& {Langer}}{{Vink} et~al.}{2011}]{vink2011}
{Vink} J.~S.,  {Muijres} L.~E.,  {Anthonisse} B.,  {de Koter} A.,
  {Gr{\"a}fener} G.,   {Langer} N.,  2011, \mn@doi [\aap]
  {10.1051/0004-6361/201116614}, \href
  {http://adsabs.harvard.edu/abs/2011A%26A...531A.132V} {531, A132}

\bibitem[\protect\citeauthoryear{{Vink} et~al.,}{{Vink}
  et~al.}{2015}]{vink2015:IAU}
{Vink} J.~S.,  et~al., 2015, \mn@doi [Highlights of Astronomy]
  {10.1017/S1743921314004657}, \href
  {https://ui.adsabs.harvard.edu/abs/2015HiA....16...51V} {16, 51}

\bibitem[\protect\citeauthoryear{{Walborn}}{{Walborn}}{1991}]{walborn1991}
{Walborn} N.~R.,  1991, in {Haynes} R.,  {Milne} D.,  eds,  IAU Symposium Vol.
  148, The Magellanic Clouds. p.~145

\bibitem[\protect\citeauthoryear{{Yusof} et~al.,}{{Yusof}
  et~al.}{2013}]{yusof2013}
{Yusof} N.,  et~al., 2013, \mn@doi [\mnras] {10.1093/mnras/stt794}, \href
  {http://adsabs.harvard.edu/abs/2013MNRAS.433.1114Y} {433, 1114}

\bibitem[\protect\citeauthoryear{{de Koter}, {Schmutz}  \& {Lamers}}{{de Koter}
  et~al.}{1993}]{deKoter1993}
{de Koter} A.,  {Schmutz} W.,   {Lamers} J.~G.~L.~M.,  1993, \aap, \href
  {https://ui.adsabs.harvard.edu/abs/1993A&A...277..561D} {277, 561}

\bibitem[\protect\citeauthoryear{{de Koter}, {Heap}  \& {Hubeny}}{{de Koter}
  et~al.}{1997}]{deKoter1997}
{de Koter} A.,  {Heap} S.~R.,   {Hubeny} I.,  1997, \mn@doi [\apj]
  {10.1086/303736}, \href {http://adsabs.harvard.edu/abs/1997ApJ...477..792D}
  {477, 792}

\bibitem[\protect\citeauthoryear{{de Koter}, {Heap}  \& {Hubeny}}{{de Koter}
  et~al.}{1998}]{deKoter1998}
{de Koter} A.,  {Heap} S.~R.,   {Hubeny} I.,  1998, \mn@doi [\apj]
  {10.1086/306503}, \href
  {https://ui.adsabs.harvard.edu/abs/1998ApJ...509..879D} {509, 879}

\makeatother
\end{thebibliography}


\bsp	
\label{lastpage}


\clearpage
\newpage
\onecolumn
\appendix
\setcounter{section}{19}
%
\section*{Supplementary material: Additional Tables and Figures}
%
\twocolumn

\begin{table}
	\centering
	\caption{Comparison of stellar parameters to previous studies using spectra of spatially resolved stars. Used spectra and wavelength ranges: \citet{massey1998}: optical (HST/FOS), \citet{deKoter1998}: UV (HST/GHRS) + optical (HST/FOS), \citet{crowther2010}: UV (HST/GHRS), optical (HST/FOS) + near-IR (VLT/SINFONI), this study: optical (HST/STIS).}
	\label{t:com_pre_work}
	\begin{tabular}{lcccl} 
		\hline
		ID     & $\log L/L_{\odot}$ & $T_{\rm eff}$ & $M_{\rm ini}$ & Source\\
		\hline
		R136a1 & 6.79              & 46000         & 251           & This study\\
		       & 6.94              & 52700         & 320           & \citet{crowther2010}\\
		       & 6.33              & 44000         & 120           & \citet{deKoter1997}\\
        	\medskip
		       & 6.66              & --            & 155           & \citet{massey1998}\\
		R136a2 & 6.75              & 50000         & 211           & This study\\
		       & 6.78              & 52650         & 240           & \citet{crowther2010}\\
		\medskip
		       & 6.54              & --            & 140           & \citet{massey1998}\\
		R136a3 & 6.63              & 50000         & 181           & This study\\
		       & 6.58              & 52700         & 165           & \citet{crowther2010}\\
		       & 6.25              & 42000         & 110           & \citet{deKoter1997}\\
		\medskip
		       & 6.54              & --            & 137           & \citet{massey1998}\\
		R136a4 & 6.24              & 48000         &  89           & This study\\ 
		\medskip
		       & 6.06              & --            &  90           & \citet{massey1998}\\ 
		R136a5 & 6.29              & 46000         & 111           & This study\\
		       & 6.03              & 43200         &  76           & \citet{deKoter1998}\\
		\medskip
		       & 6.10              & --            &  93           & \citet{massey1998}\\
		R136a6 & 6.27              & 53000         & 115           & This study\\
		\medskip
		       & 6.10              & --            &  95           & \citet{massey1998}\\
		R136a7 & 6.25              & 49000         &  93           & This study\\
		       & 5.86              & 40700         &  59           & \citet{deKoter1998}\\
		\medskip
		       & 6.06              & --            &  88           & \citet{massey1998}\\
		R136b  & 6.35              & 37000         & 104           & This study\\
		\medskip
		       & 6.38              & --            & 121           & \citet{massey1998}\\
		H31    & 6.01              & 48000         &  69           & This study\\
		\medskip
		       & 5.90              & --            &  76           & \citet{massey1998}\\
		H36    & 6.33              & 52000         & 122           & This study\\
		       & 5.80              & 43500         &  57           & \citet{deKoter1998}\\
		\medskip
		       & 5.90              & --            &  77           & \citet{massey1998}\\
		H40	   & 5.88              & 45000         &  56           & This study\\
		\medskip
		       & 5.62              & 45000         &  49           & \citet{deKoter1998}\\
		H46    & 6.16              & 49000         &  83           & This study\\
		       & 5.82              & 43500         &  58           & \citet{deKoter1998}\\
		\medskip
		       & 5.78              & --            &  61           & \citet{massey1998}\\
		H47    & 6.09              & 47000         &  68           & This study\\
		\medskip
		       & 5.72              & 44000         &  53           & \citet{deKoter1998}\\
		H50	   & 5.71              & 42000         &  48           & This study\\
		\medskip
		       & 5.58              & 42200         &  44           & \citet{deKoter1998}\\
		H55    & 5.76              & 47000         &  53           & This study\\
		\medskip
		       & 5.66              & 45400         &  51           & \citet{deKoter1998}\\
		H58	   & 5.94              & 50000         &  66           & This study\\
		\medskip
		       & 5.50              & 42300         &  41           & \citet{deKoter1998}\\
		H70	   & 5.78              & 48000         &  52           & This study\\
		       & 5.44              & 40400         &  37           & \citet{deKoter1998}\\
		\hline
	\end{tabular}
\end{table}

\begin{figure}
\begin{center}
\resizebox{\hsize}{!}{\includegraphics{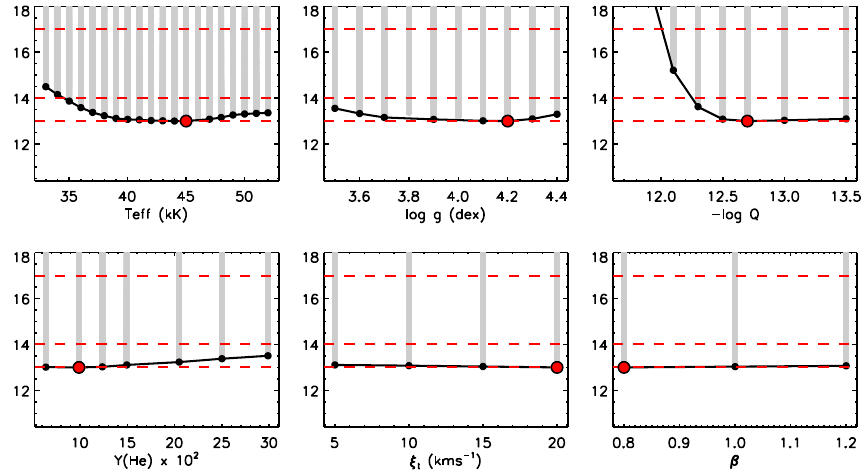}}
\end{center}
\caption{Example $\chi^2$ distributions of stellar parameters for a low S/N star. All stellar parameters seemed to be degenerated.}
\label{f:chi_square}
\end{figure}

\begin{figure}
\begin{center}
\resizebox{\hsize}{!}{\includegraphics{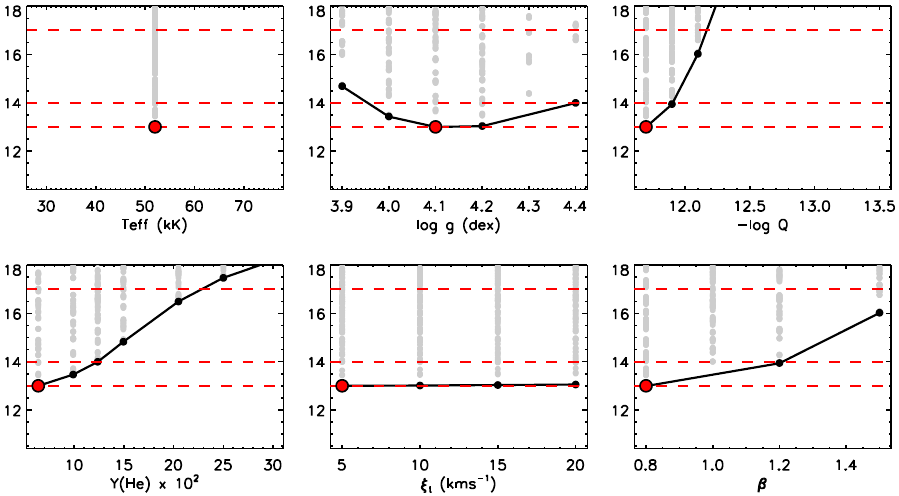}}
\end{center}
\caption{Example $\chi^2$ distributions of stellar parameters for a star hotter than 45\,000\,K. The temperature has been adjusted based on the N\,{\sc iii-iv-v} ionisation balance.}
\label{f:chi_square_hot}
\end{figure}

\begin{figure}
\begin{center}
\resizebox{\hsize}{!}{\includegraphics{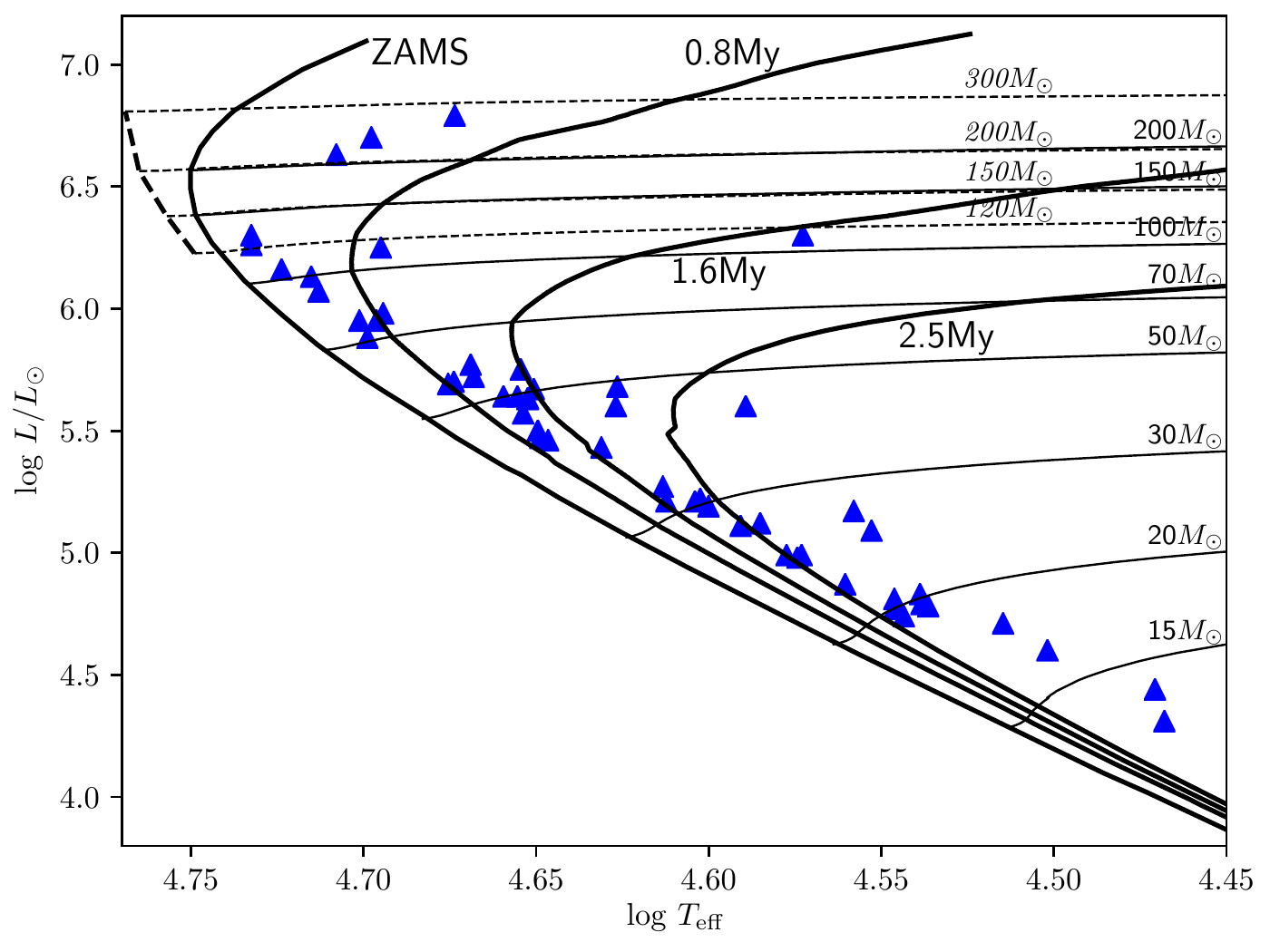}}
\end{center}
\caption{Evolutionary HRD based on the output of BONNSAI. Stars are preferentially located near the ZAMS.}
\label{f:hrd_evo}
\end{figure}

\begin{figure*}
\begin{center}
\resizebox{0.9\hsize}{!}{\includegraphics{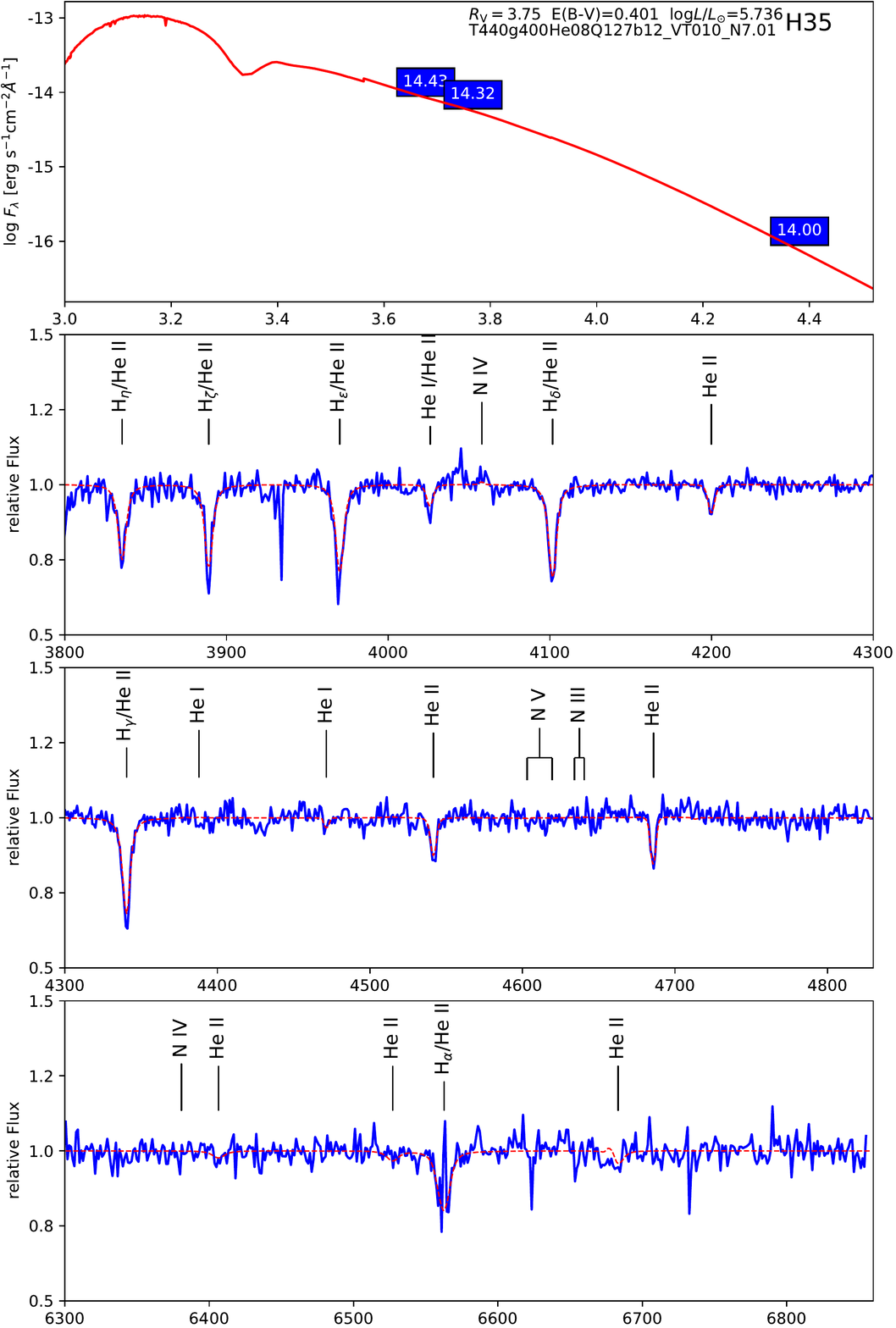}}
\end{center}
\caption{Example spectroscopic fit of H35. Top panel: red solid line is the model spectral energy distribution. Blue boxes are the  optical with $B$ (F438W), $V$ (F555W) from \citet{deMarchi2011} and near-IR $K_{\rm s}$ from \citet{khorrami2017}. 
Panel 2 to 4: blue solid is the observed HST spectrum while the red dashed line is the fitted synthetic spectrum.}
\label{f:h35}
\end{figure*}

\begin{figure*}
\begin{center}
\resizebox{0.8\hsize}{!}{\includegraphics{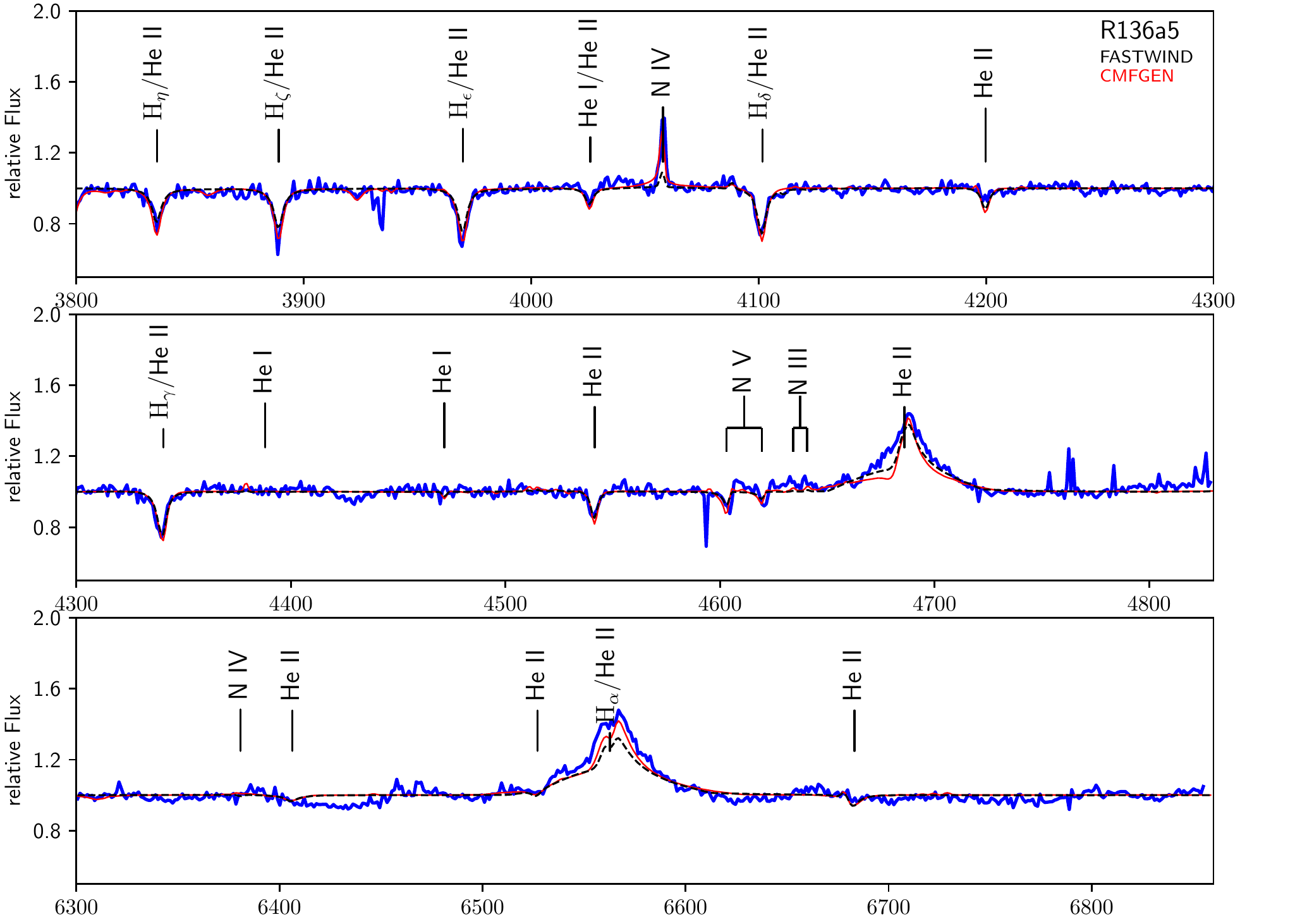}}
\end{center}
\caption{Spectroscopic fit to the data of R136a5. Blue solid line is the observed HST/STIS spectrum. Black dashed line is the synthetic spectrum computed with FASTWIND. Red solid line is the synthetic spectrum computed with CMFGEN. Stellar parameters are given in Table\,1.}
\label{f:r136a5}
\end{figure*}
\begin{figure*}
\begin{center}
\resizebox{0.8\hsize}{!}{\includegraphics{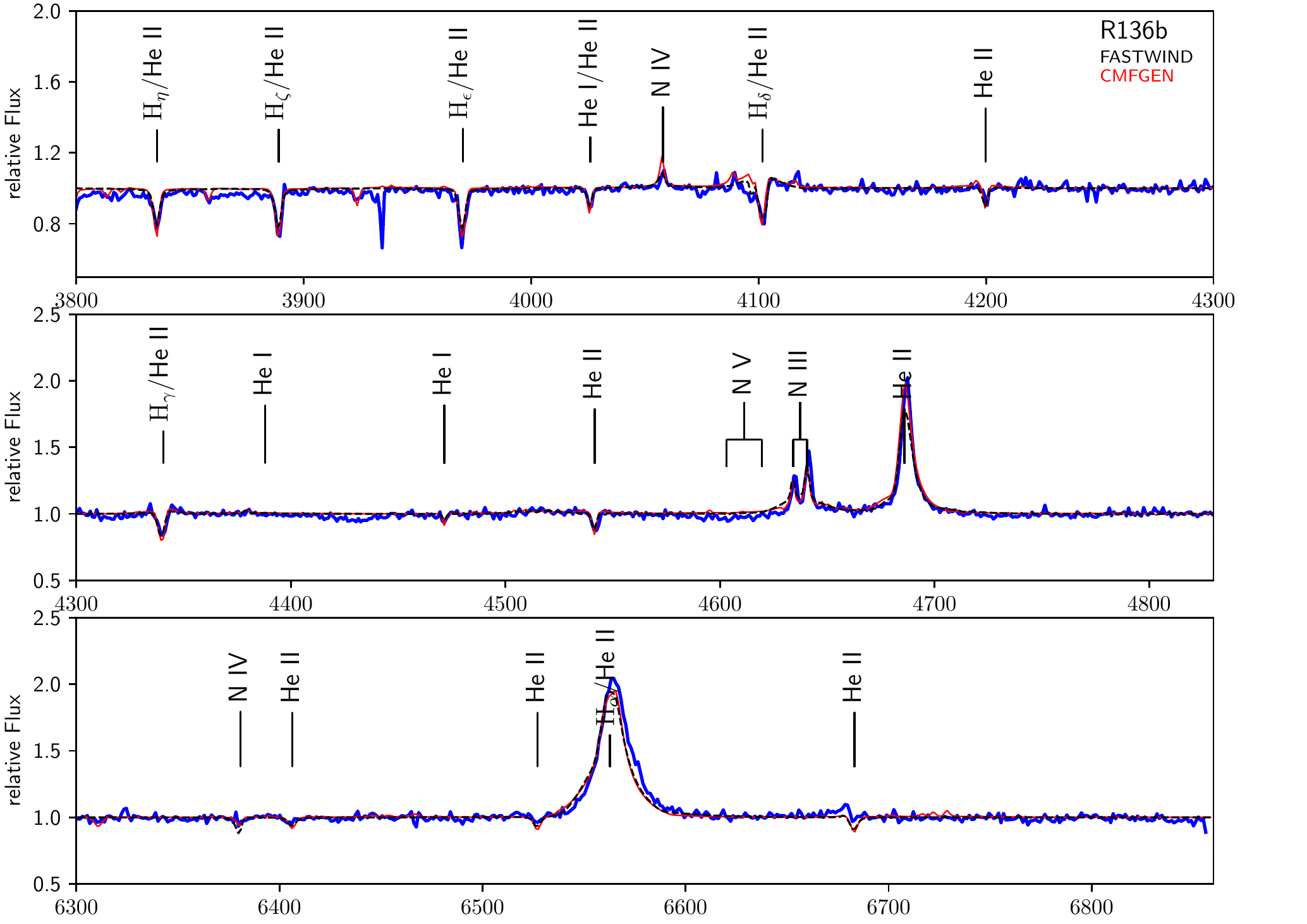}}
\end{center}
\caption{Spectroscopic fit to the data of R136b. Blue solid line is the observed HST/STIS spectrum. Black dashed line is the synthetic spectrum computed with FASTWIND. Red solid line is the synthetic spectrum computed with CMFGEN. Stellar parameters are given in Table\,1.}
\label{f:r136b}
\end{figure*}
\begin{figure*}
\begin{center}
\resizebox{0.8\hsize}{!}{\includegraphics{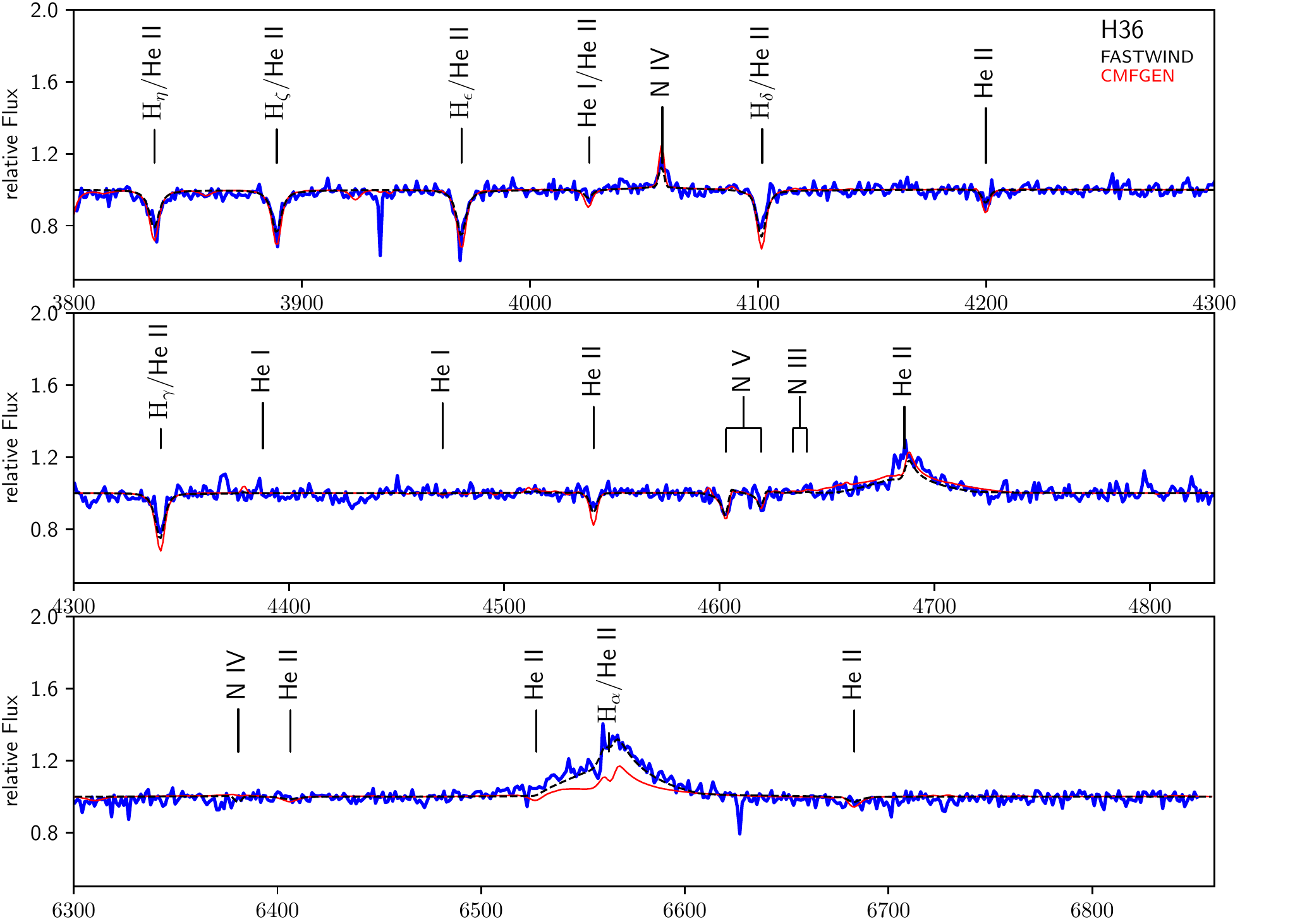}}
\end{center}
\caption{Spectroscopic fit to the data of H36. Blue solid line is the observed HST/STIS spectrum. Black dashed line is the synthetic spectrum computed with FASTWIND. Red solid line is the synthetic spectrum computed with CMFGEN. Stellar parameters are given in Table\,1.}
\label{f:h36}
\end{figure*}

\begin{figure}
\begin{center}
\resizebox{\hsize}{!}{\includegraphics{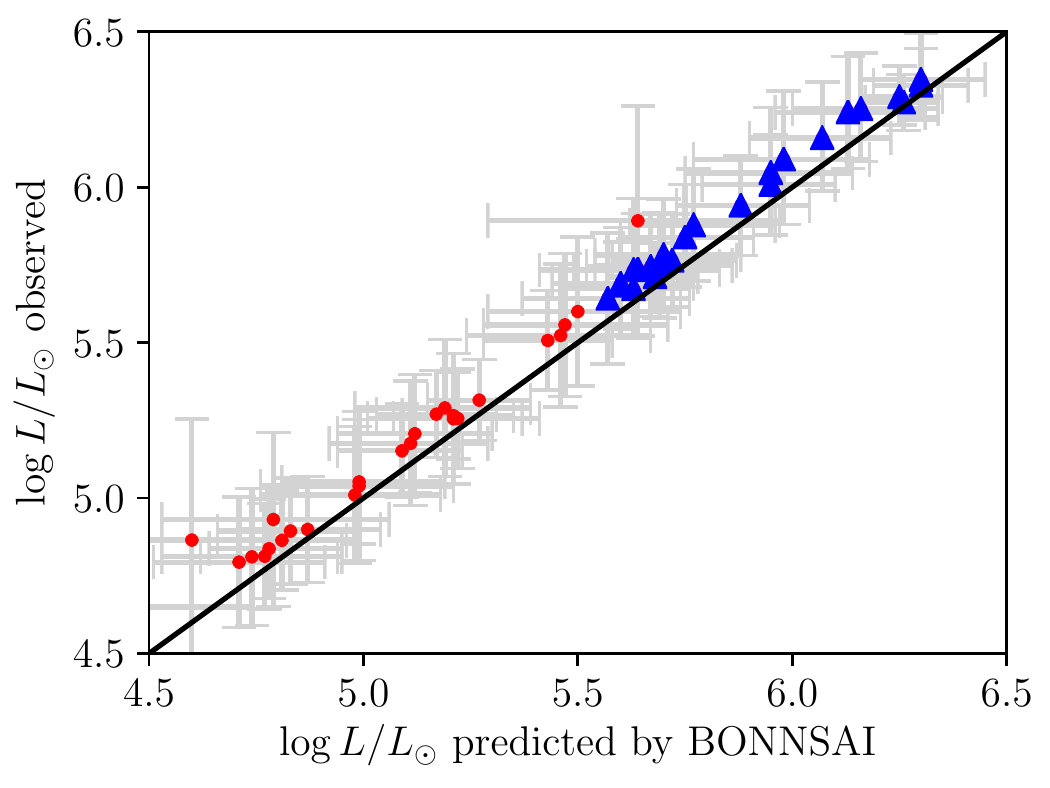}}
\end{center}
\caption{Observed $\log L$ against predicted by BONNSAI: there is a good agreement between both $\log L$s, but BONNSAI tends to systematically under-predict the observed luminosities. Stars with initial mass $> 40\,M_{\odot}$ are shown as blue triangles.}
\label{f:lum}
\end{figure}

\begin{figure}
\begin{center}
\resizebox{\hsize}{!}{\includegraphics{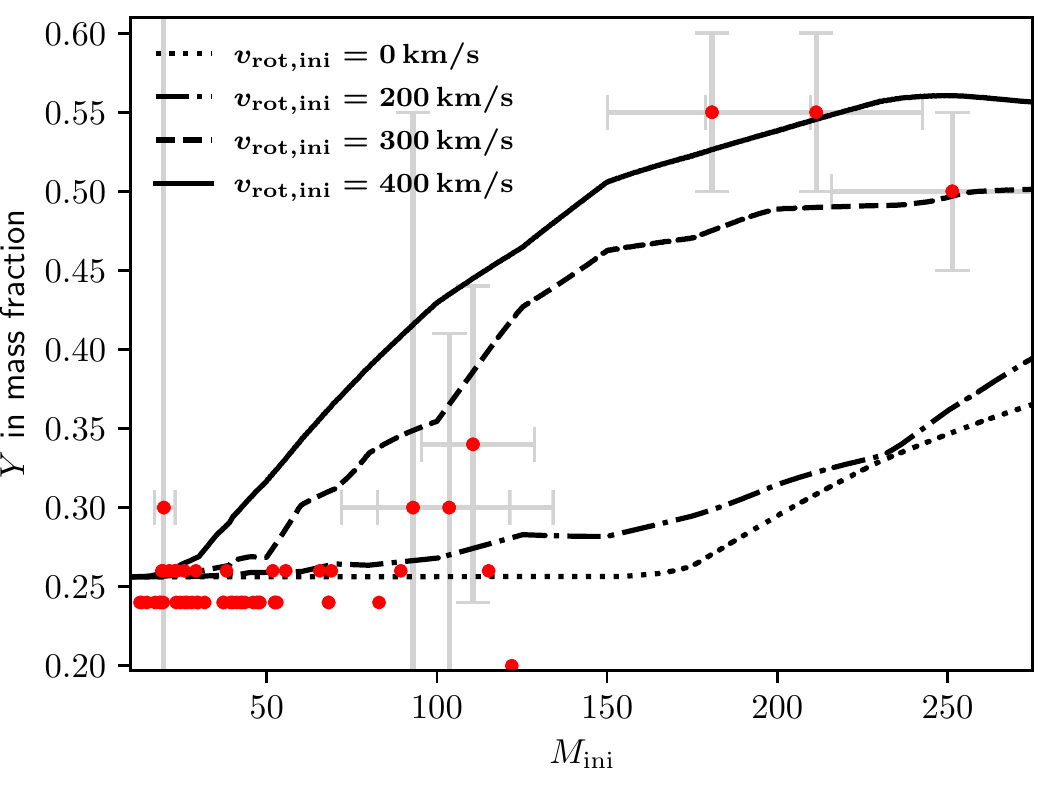}}
\end{center}
\caption{Helium abundances versus initial mass. 1.2\,Myr isochrones are overlaid with initial rotation velocity ($\varv_{\rm rot,ini}$) of 0, 200, 300 and 400\,km/s. The He composition of the WNh stars can be reproduced at a similar by varying only $\varv_{\rm rot,ini}$. }
\label{f:Y_Mini}
\end{figure}

\begin{figure*}
\centering
\begin{minipage}{0.475\linewidth}
\resizebox{\hsize}{!}{\includegraphics{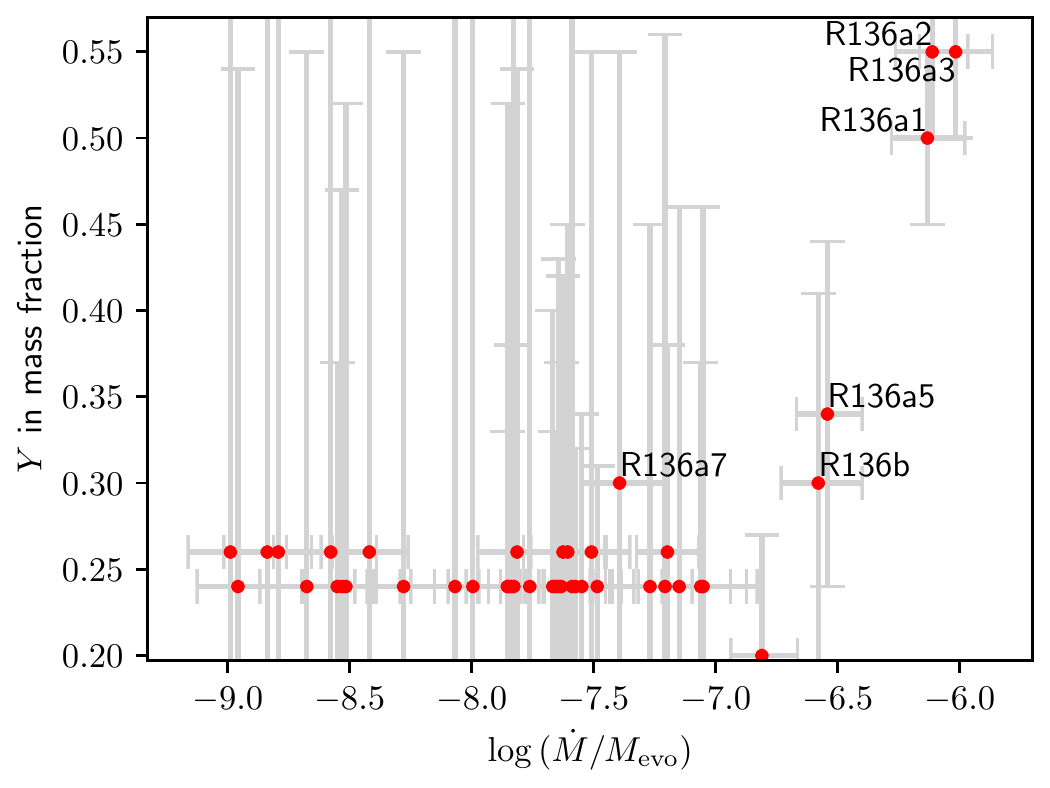}}
\caption{Mass-loss rate over mass versus helium abundance.
{\color{white}{aa  aaa  aaa  aaa  aa aaa}}
}
\label{f:mdot-mass_He}
\end{minipage}
\hspace{0.04\linewidth}
\begin{minipage}{0.475\linewidth}
\resizebox{\hsize}{!}{\includegraphics{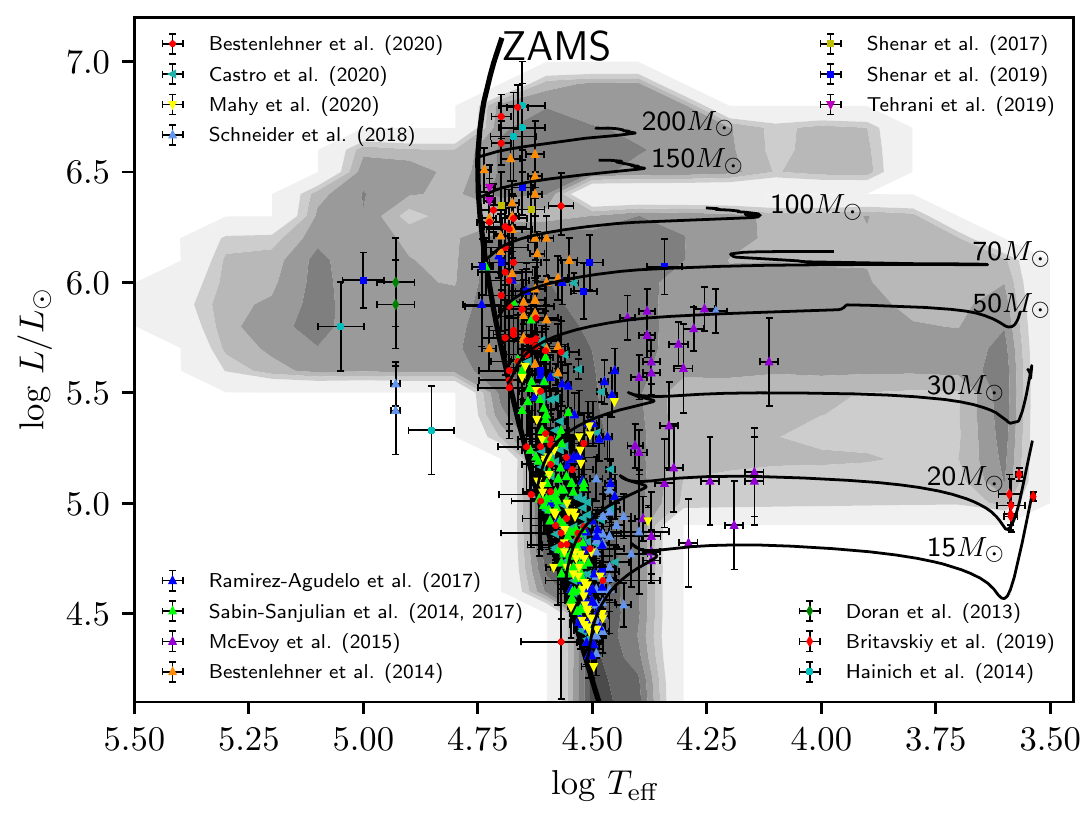}}
\caption{Same as Fig.\,14 but with an overlaid single star population synthesis.
}
\label{f:ngc2070_hrd_sin}
\end{minipage}
\end{figure*}

%


\begin{figure*}
\begin{center}
\resizebox{0.825\hsize}{!}{\includegraphics{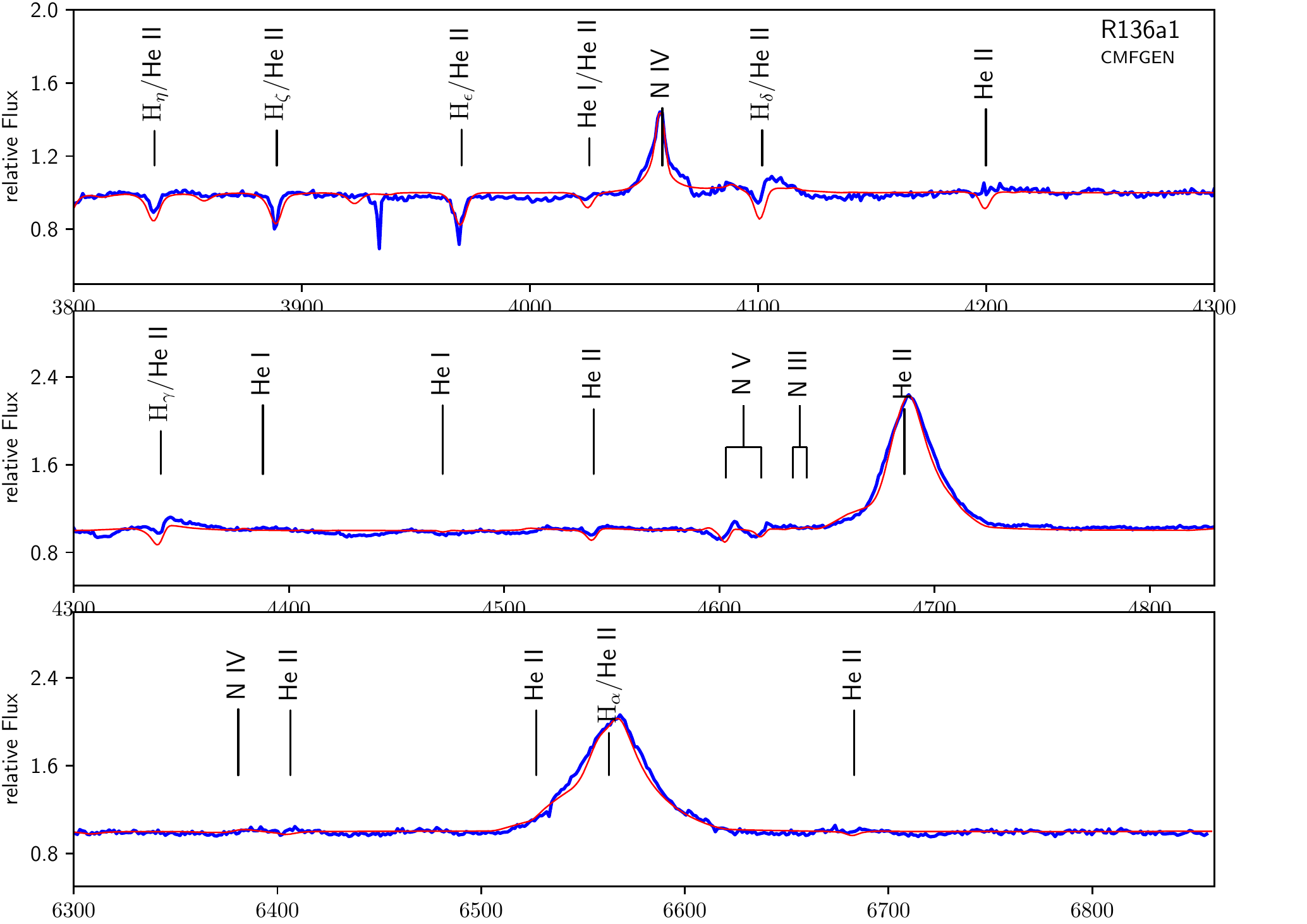}}
\end{center}
\caption{Spectroscopic fit to the data of R136a1. Blue solid line is the observed HST/STIS spectrum. Red solid line is the synthetic spectrum computed with CMFGEN. Stellar parameters are given in Table\,1.}
\label{f:r136a1}
\end{figure*}
\clearpage
\begin{figure*}
\begin{center}
\resizebox{0.825\hsize}{!}{\includegraphics{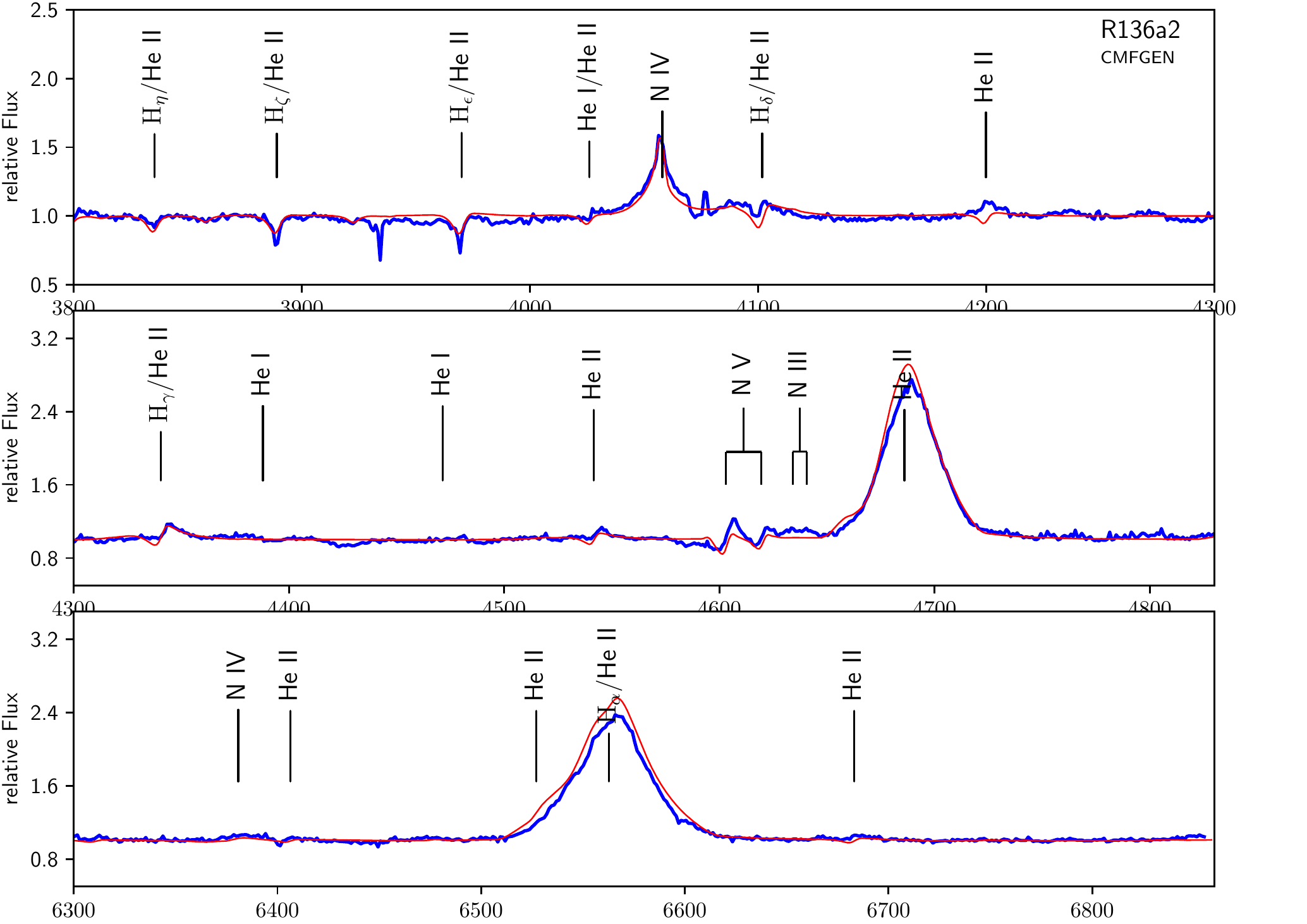}}
\end{center}
\caption{Spectroscopic fit to the data of R136a2. Blue solid line is the observed HST/STIS spectrum. Red solid line is the synthetic spectrum computed with CMFGEN. Stellar parameters are given in Table\,1.}
\end{figure*}

\begin{figure*}
\begin{center}
\resizebox{0.825\hsize}{!}{\includegraphics{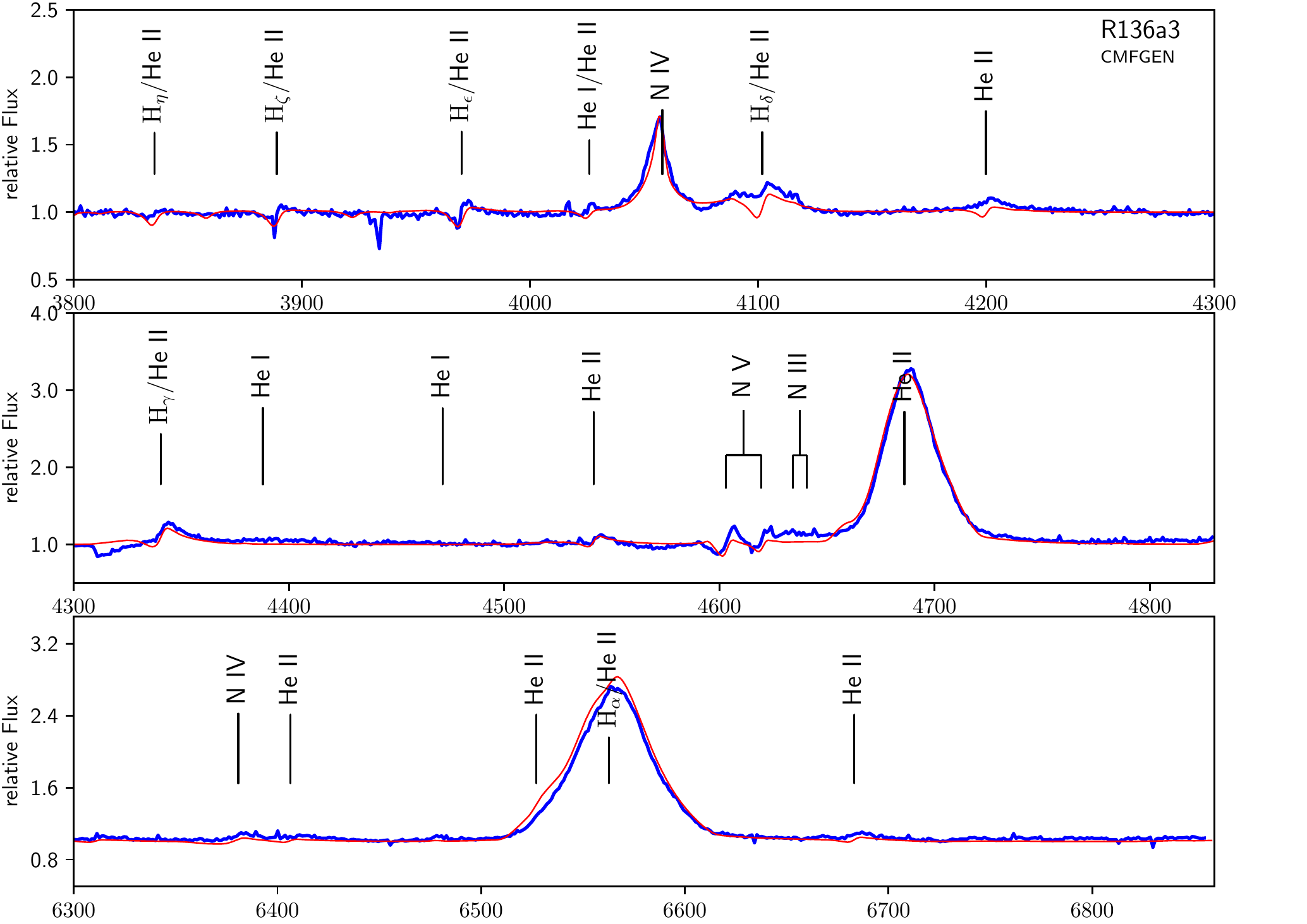}}
\end{center}
\caption{Spectroscopic fit to the data of R136a3. Blue solid line is the observed HST/STIS spectrum. Red solid line is the synthetic spectrum computed with CMFGEN. Stellar parameters are given in Table\,1.}
\label{f:r136a3}
\end{figure*}
\begin{figure*}
\begin{center}
\resizebox{0.825\hsize}{!}{\includegraphics{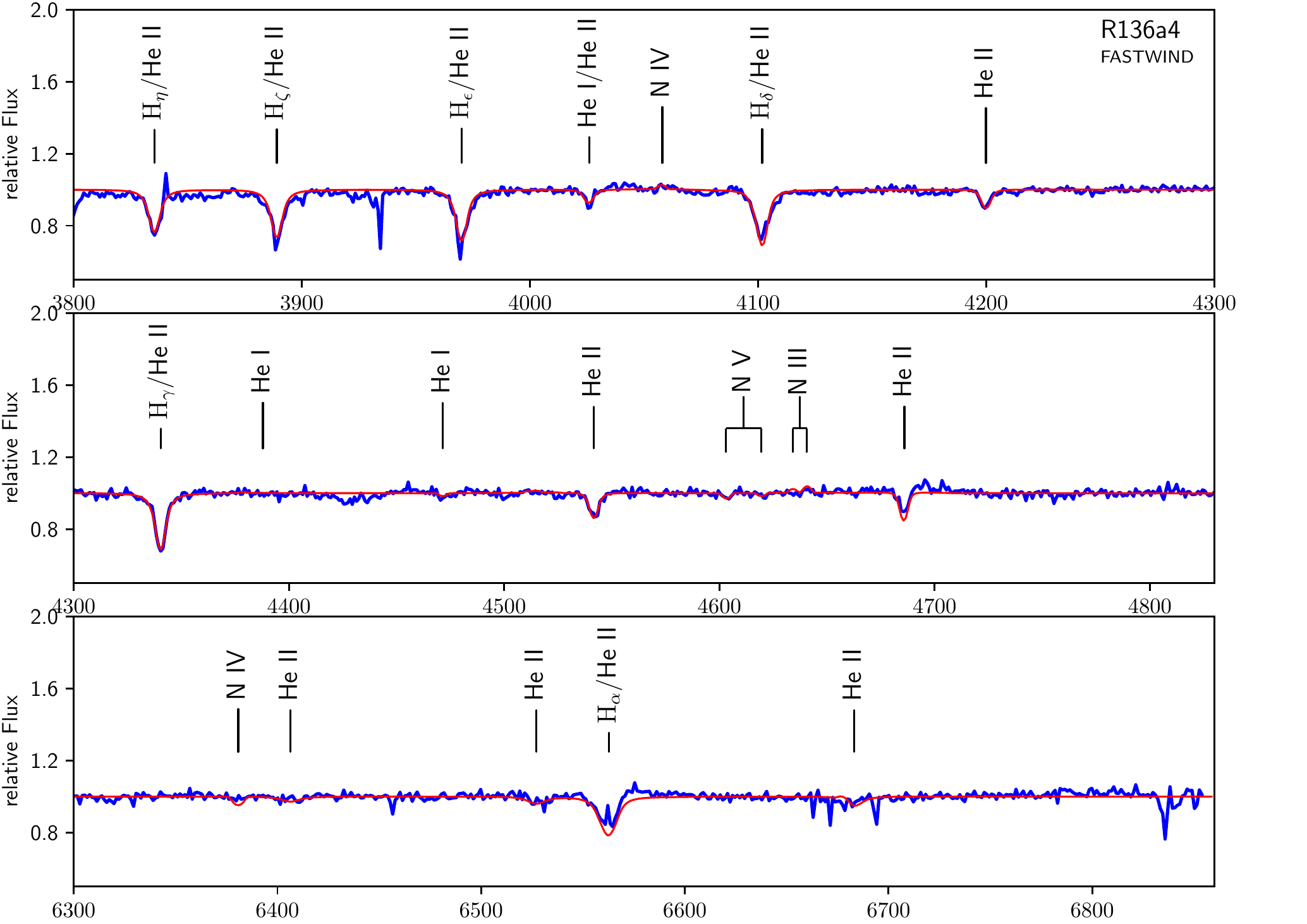}}
\end{center}
\caption{Spectroscopic fit to the data of R136a4. Blue solid line is the observed HST/STIS spectrum. Red solid line is the synthetic spectrum computed with FASTWIND. Stellar parameters are given in Table\,1.}
\end{figure*}

\begin{figure*}
\begin{center}
\resizebox{0.825\hsize}{!}{\includegraphics{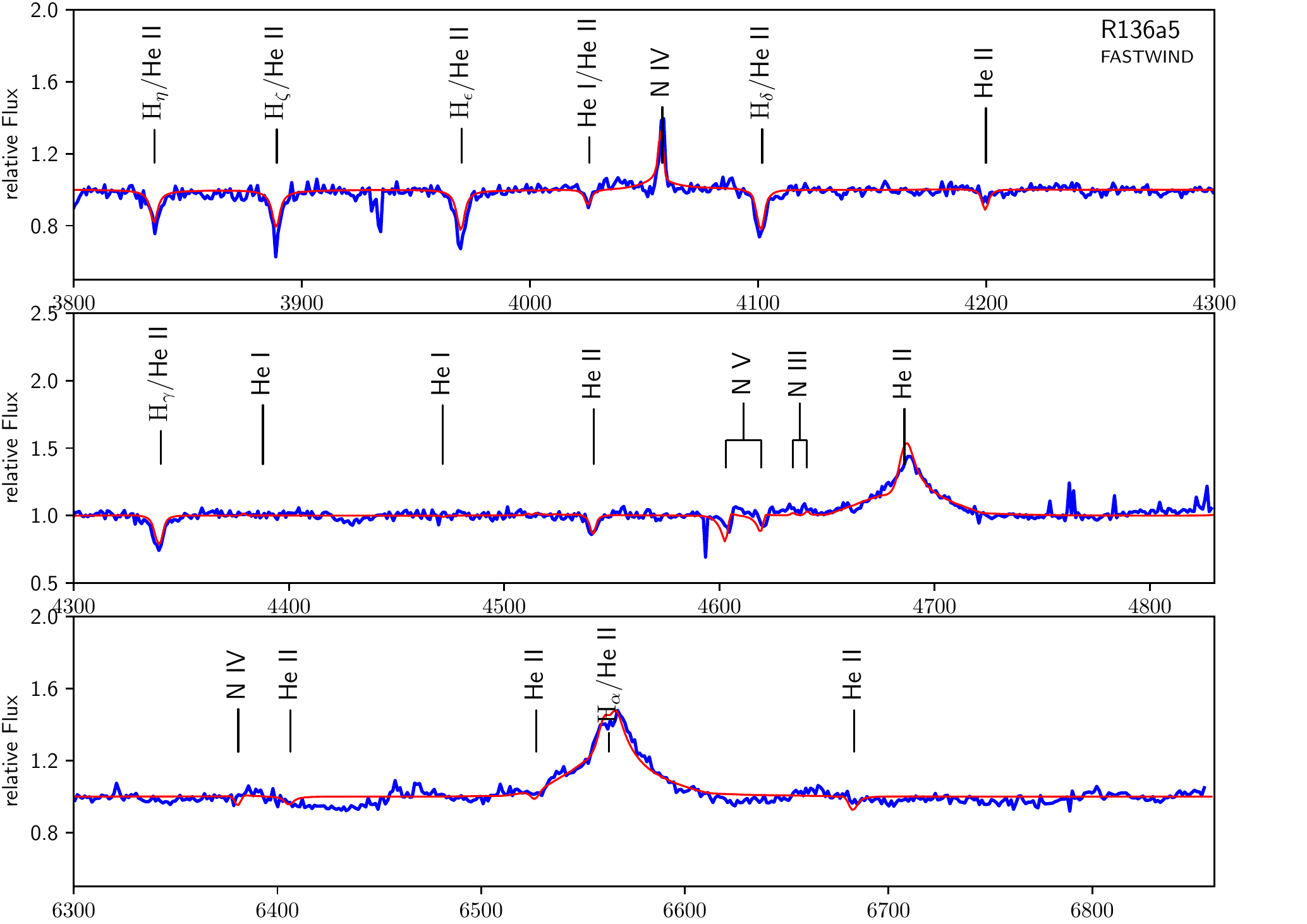}}
\end{center}
\caption{Spectroscopic fit to the data of R136a5. Blue solid line is the observed HST/STIS spectrum. Red solid line is the synthetic spectrum computed with FASTWIND. Stellar parameters are given in Table\,1.}
\end{figure*}
\begin{figure*}
\begin{center}
\resizebox{0.825\hsize}{!}{\includegraphics{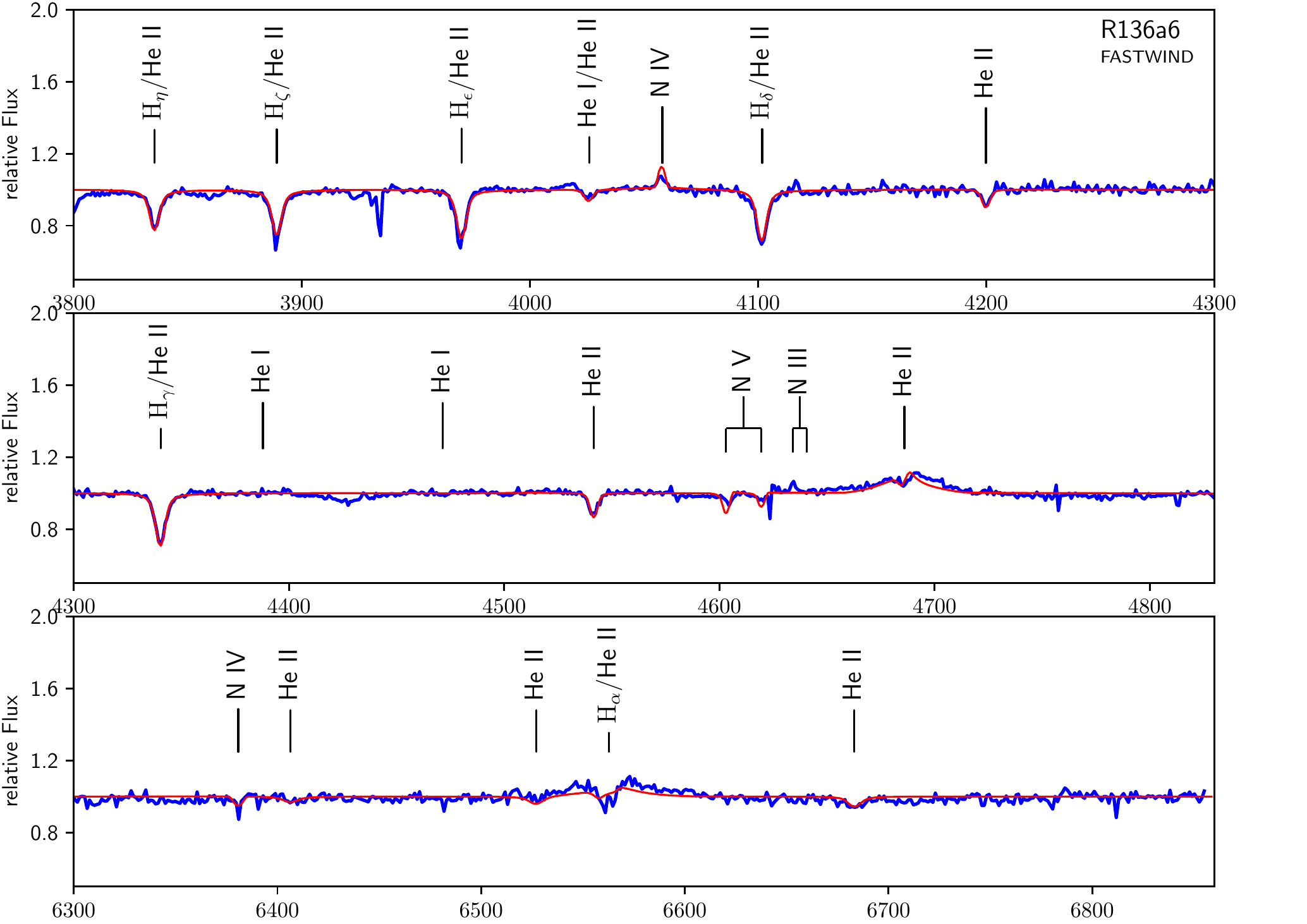}}
\end{center}
\caption{Spectroscopic fit to the data of R136a6. Blue solid line is the observed HST/STIS spectrum. Red solid line is the synthetic spectrum computed with FASTWIND. Stellar parameters are given in Table\,1.}
\end{figure*}

\clearpage
\begin{figure*}
\begin{center}
\resizebox{0.825\hsize}{!}{\includegraphics{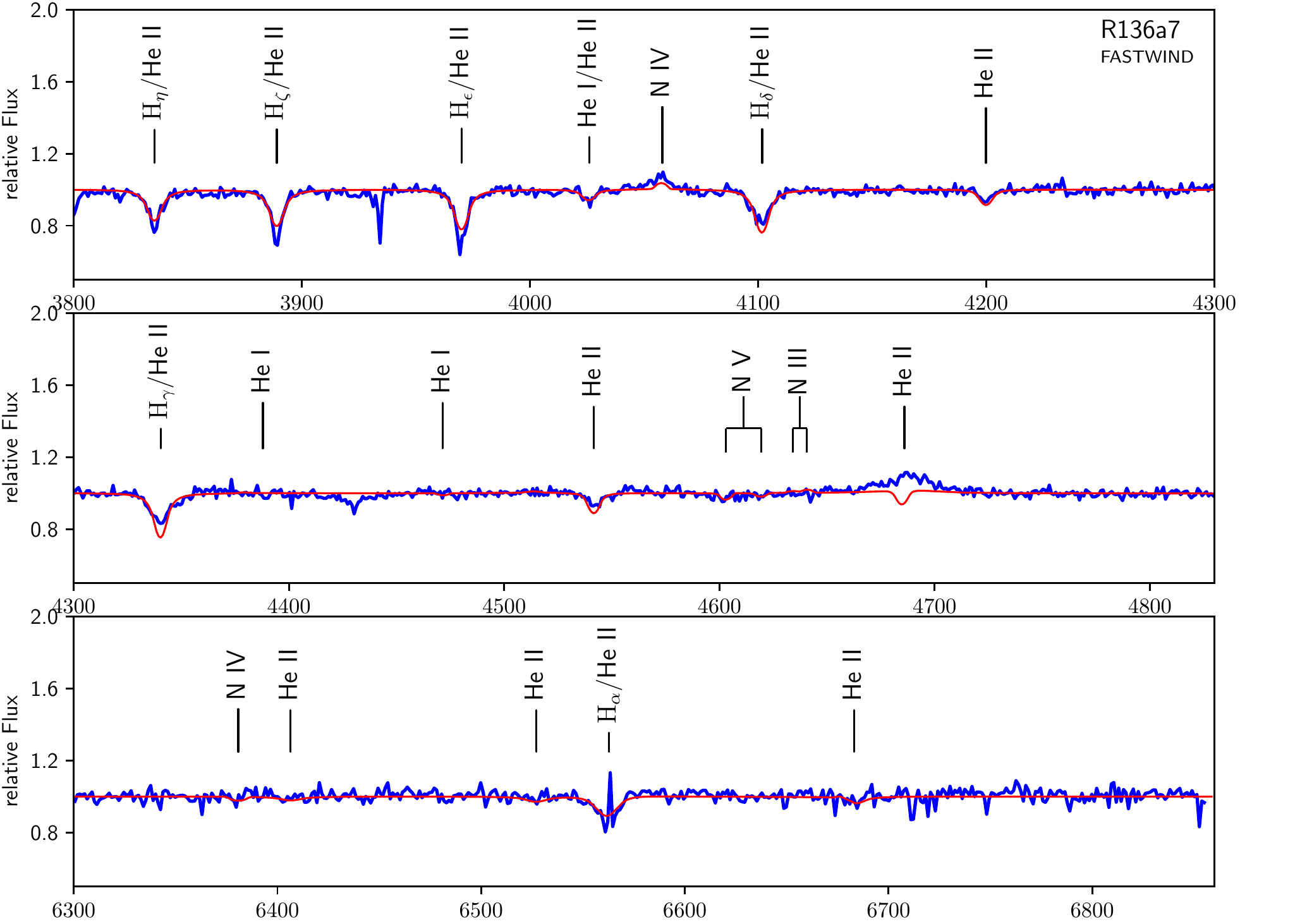}}
\end{center}
\caption{Spectroscopic fit to the data of R136a7. Blue solid line is the observed HST/STIS spectrum. Red solid line is the synthetic spectrum computed with FASTWIND. Stellar parameters are given in Table\,1.}
\end{figure*}
\begin{figure*}
\begin{center}
\resizebox{0.825\hsize}{!}{\includegraphics{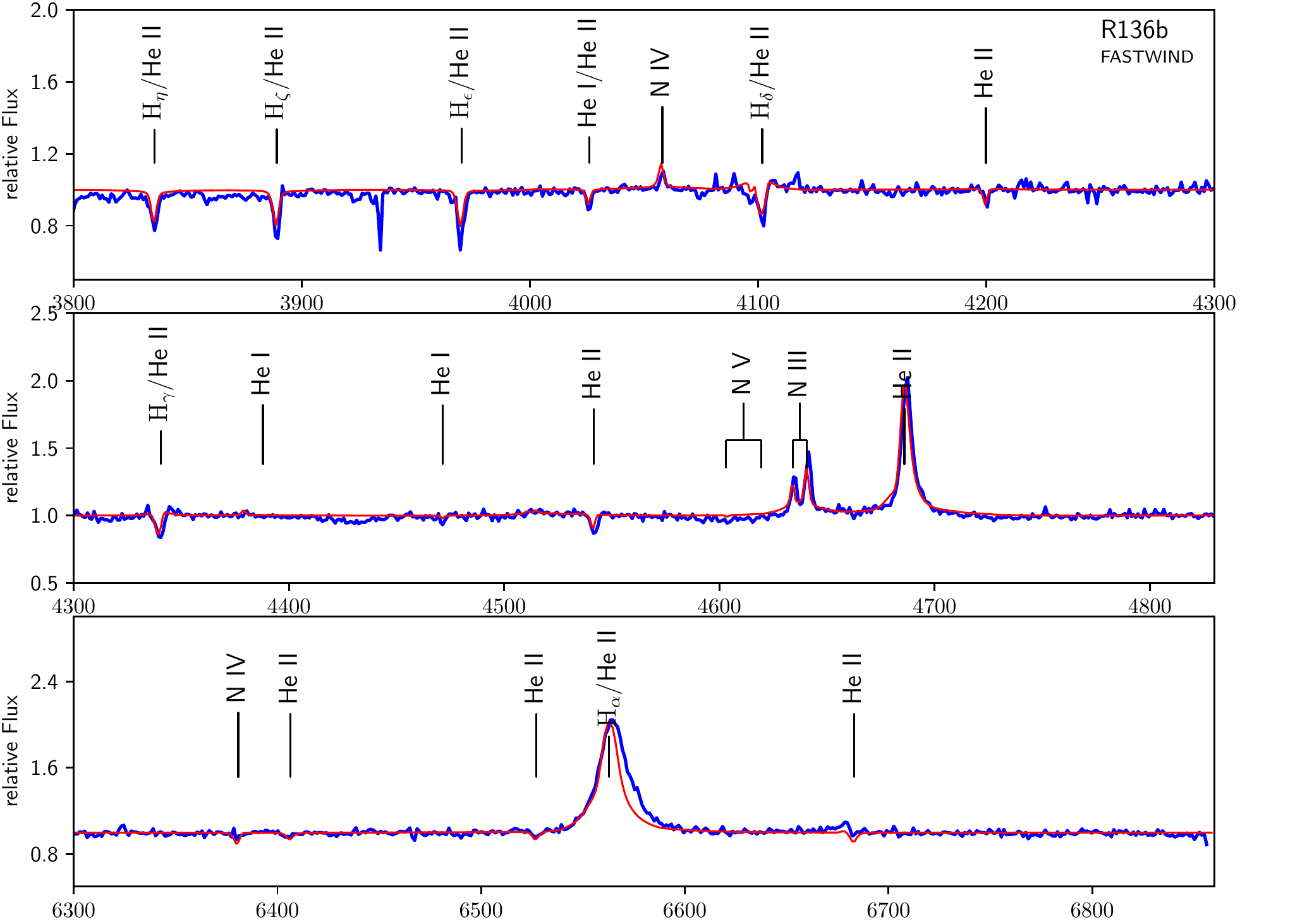}}
\end{center}
\caption{Spectroscopic fit to the data of R136b. Blue solid line is the observed HST/STIS spectrum. Red solid line is the synthetic spectrum computed with FASTWIND. Stellar parameters are given in Table\,1.}
\end{figure*}

\begin{figure*}
\begin{center}
\resizebox{0.825\hsize}{!}{\includegraphics{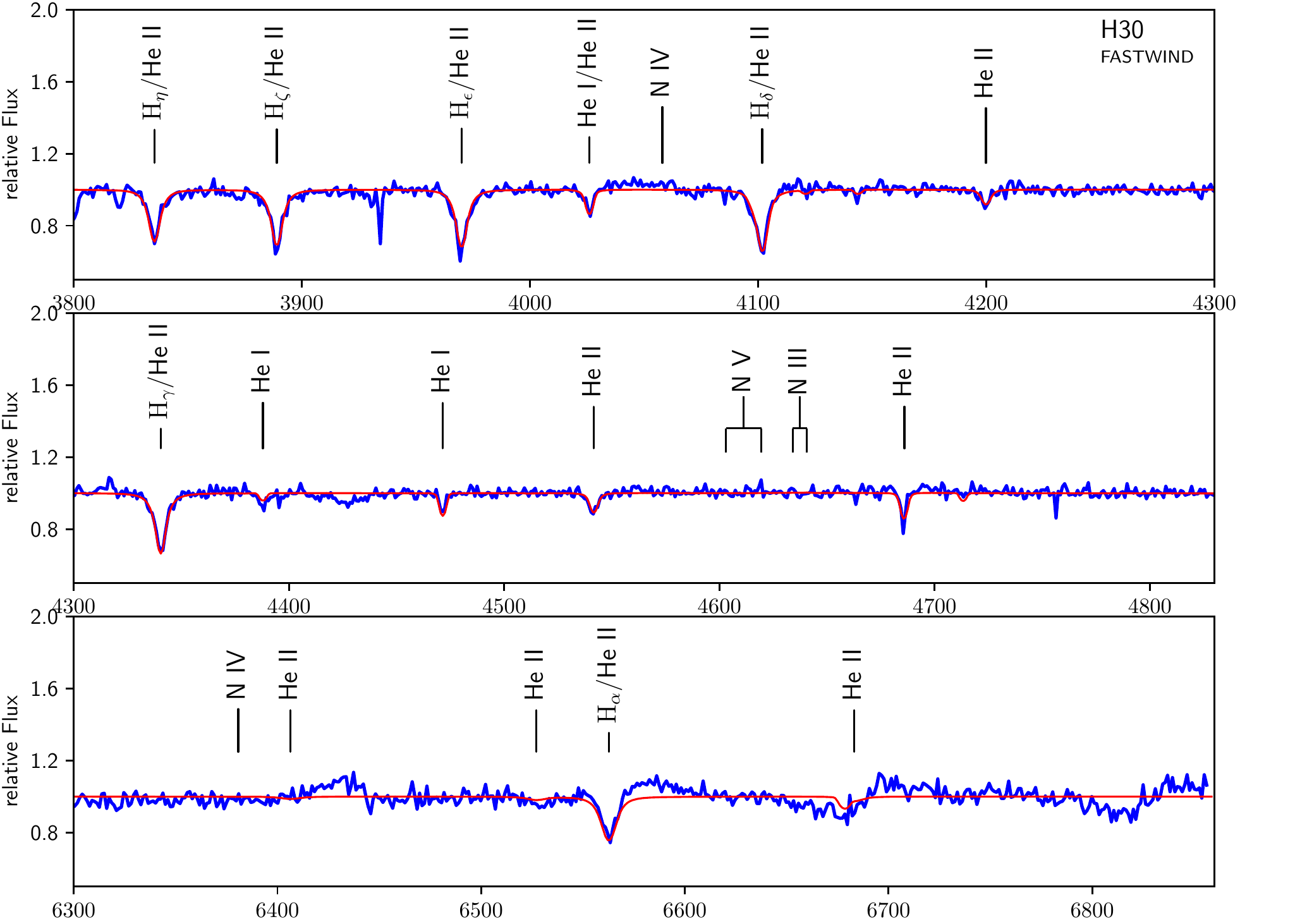}}
\end{center}
\caption{Spectroscopic fit to the data of H30. Blue solid line is the observed HST/STIS spectrum. Red solid line is the synthetic spectrum computed with FASTWIND. Stellar parameters are given in Table\,1.}
\end{figure*}
\begin{figure*}
\begin{center}
\resizebox{0.825\hsize}{!}{\includegraphics{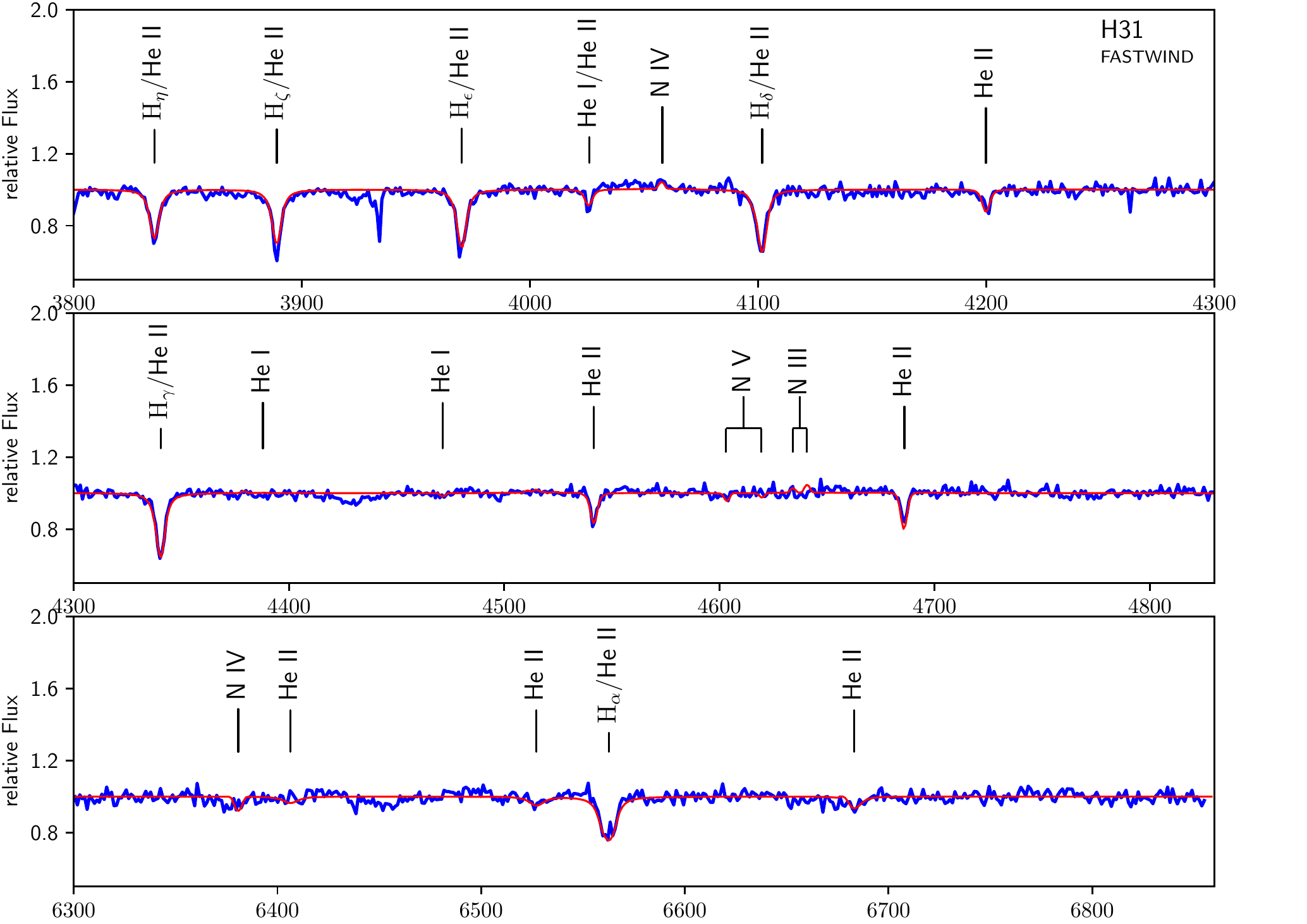}}
\end{center}
\caption{Spectroscopic fit to the data of H31. Blue solid line is the observed HST/STIS spectrum. Red solid line is the synthetic spectrum computed with FASTWIND. Stellar parameters are given in Table\,1.}
\end{figure*}

\clearpage
\begin{figure*}
\begin{center}
\resizebox{0.825\hsize}{!}{\includegraphics{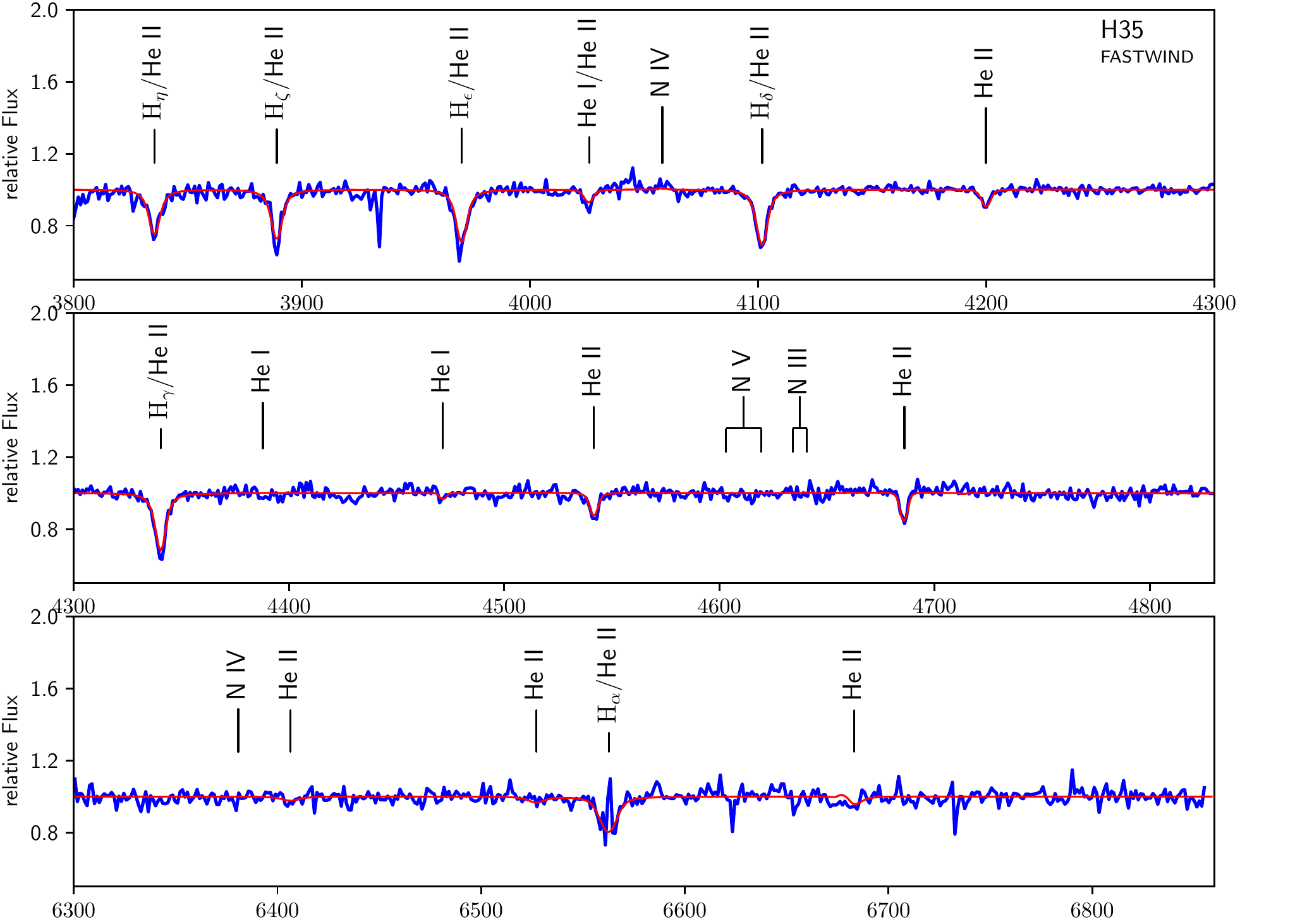}}
\end{center}
\caption{Spectroscopic fit to the data of H35. Blue solid line is the observed HST/STIS spectrum. Red solid line is the synthetic spectrum computed with FASTWIND. Stellar parameters are given in Table\,1.}
\end{figure*}
\begin{figure*}
\begin{center}
\resizebox{0.825\hsize}{!}{\includegraphics{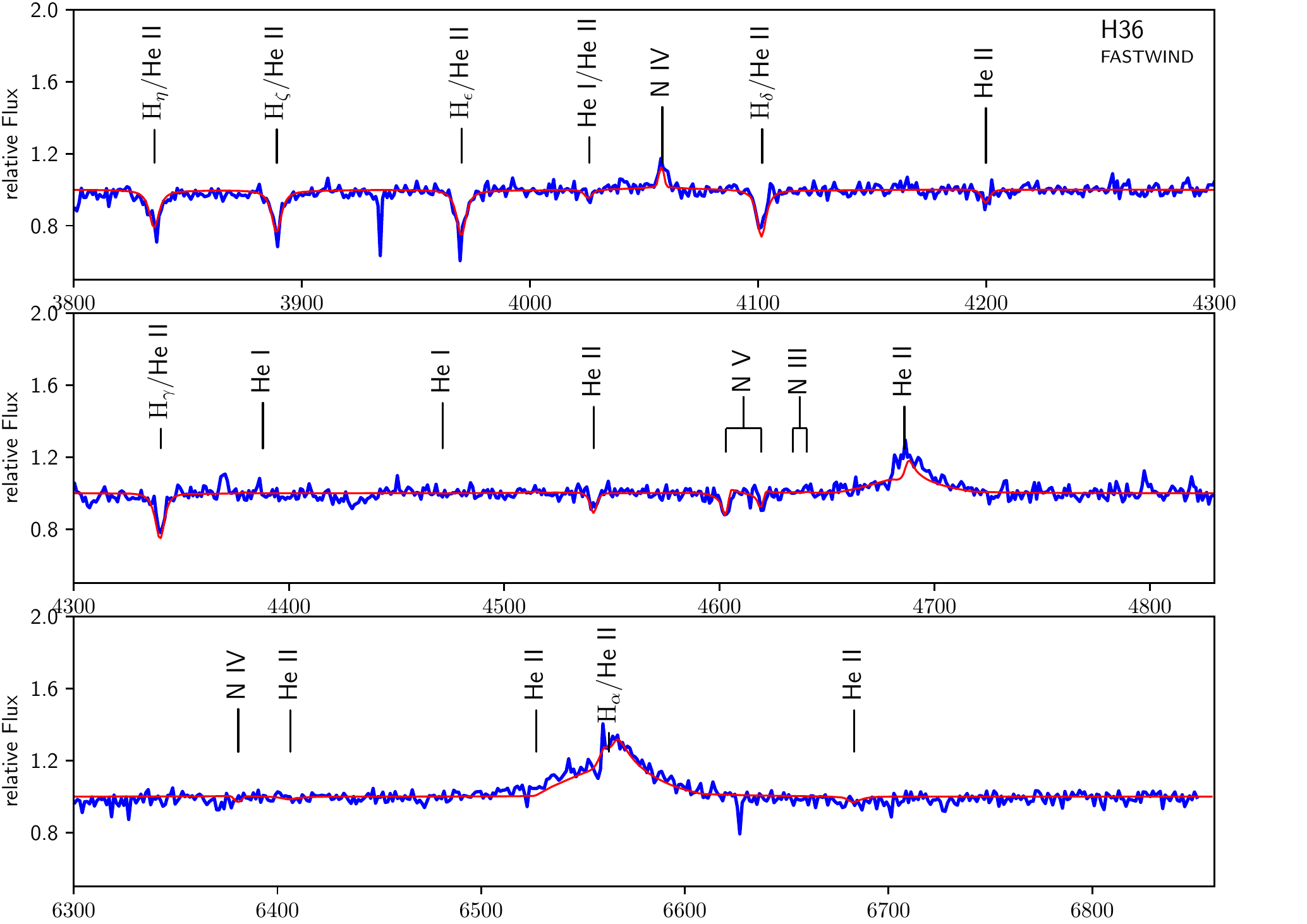}}
\end{center}
\caption{Spectroscopic fit to the data of H36. Blue solid line is the observed HST/STIS spectrum. Red solid line is the synthetic spectrum computed with FASTWIND. Stellar parameters are given in Table\,1.}
\end{figure*}

\begin{figure*}
\begin{center}
\resizebox{0.825\hsize}{!}{\includegraphics{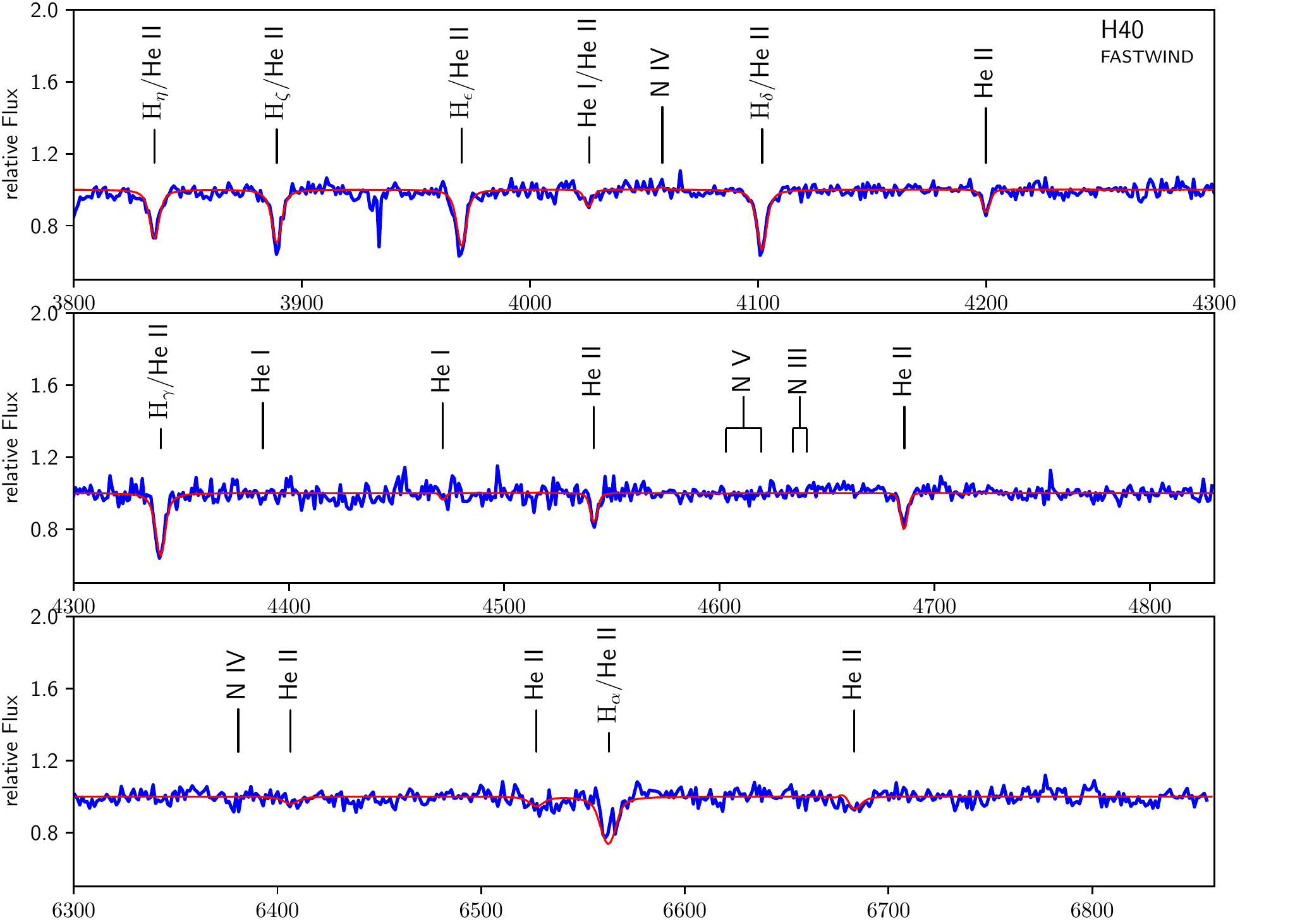}}
\end{center}
\caption{Spectroscopic fit to the data of H40. Blue solid line is the observed HST/STIS spectrum. Red solid line is the synthetic spectrum computed with FASTWIND. Stellar parameters are given in Table\,1.}
\end{figure*}
\begin{figure*}
\begin{center}
\resizebox{0.825\hsize}{!}{\includegraphics{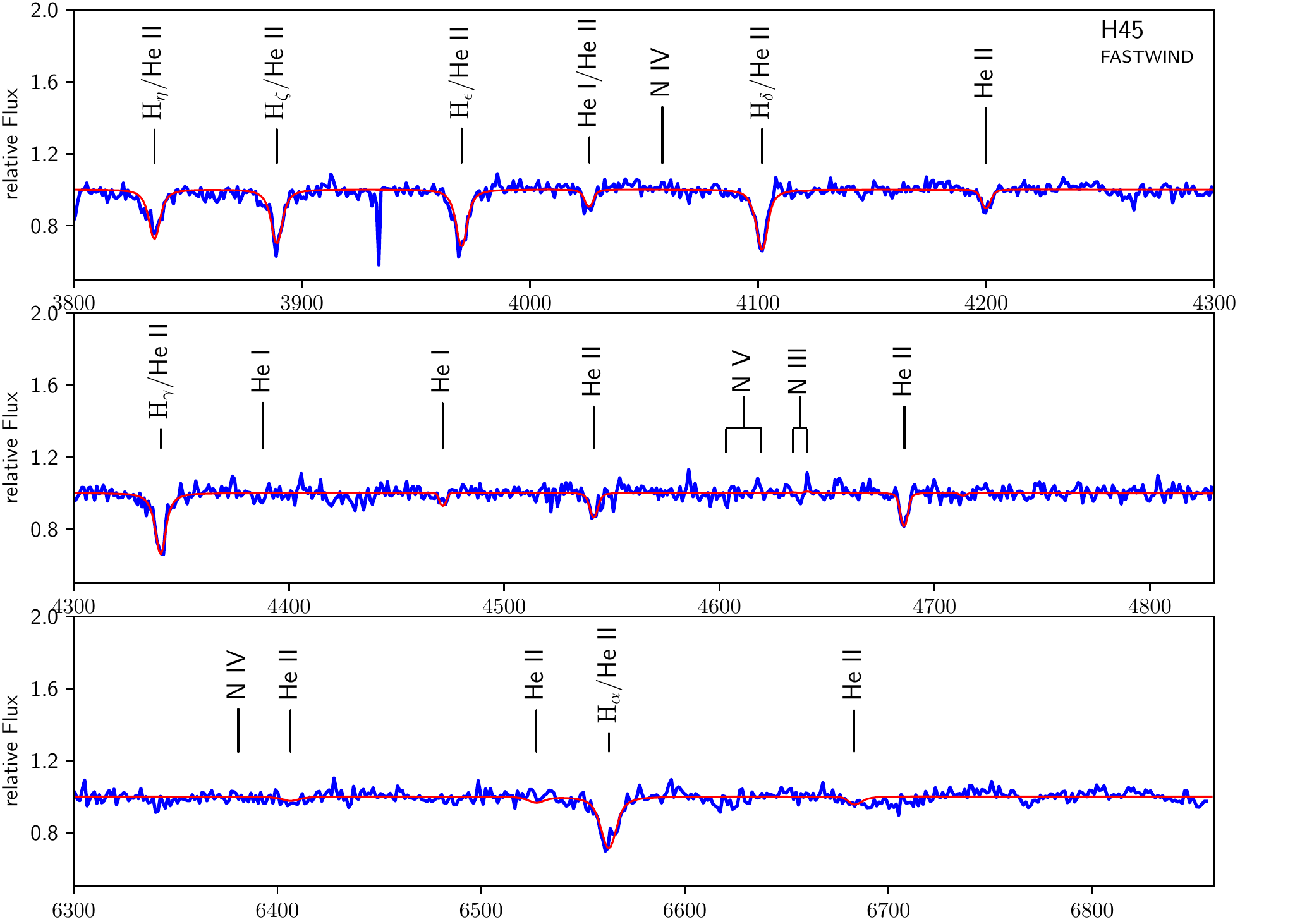}}
\end{center}
\caption{Spectroscopic fit to the data of H45. Blue solid line is the observed HST/STIS spectrum. Red solid line is the synthetic spectrum computed with FASTWIND. Stellar parameters are given in Table\,1.}
\end{figure*}

\begin{figure*}
\begin{center}
\resizebox{0.825\hsize}{!}{\includegraphics{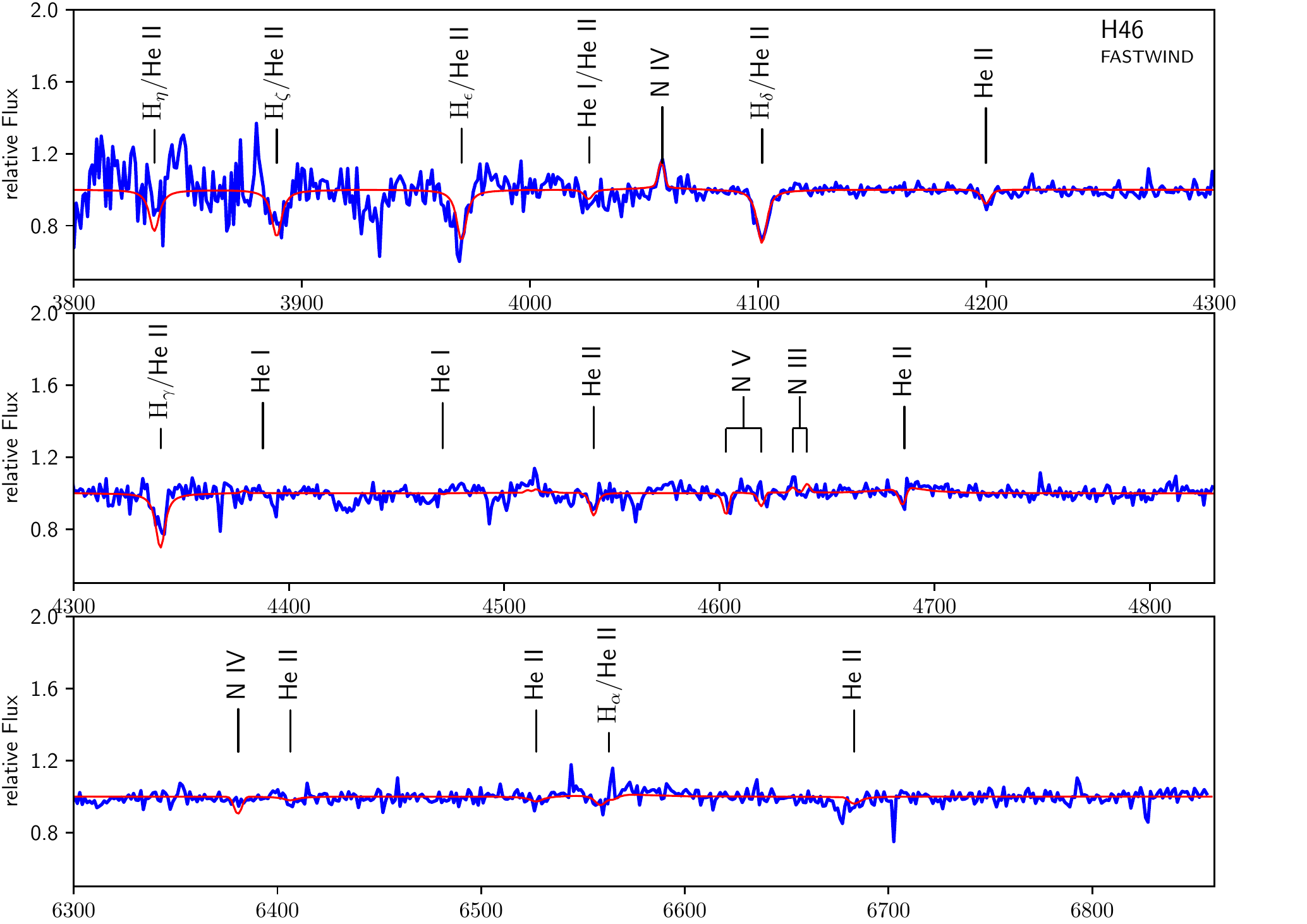}}
\end{center}
\caption{Spectroscopic fit to the data of H46. Blue solid line is the observed HST/STIS spectrum. Red solid line is the synthetic spectrum computed with FASTWIND. Stellar parameters are given in Table\,1.}
\end{figure*}
\begin{figure*}
\begin{center}
\resizebox{0.825\hsize}{!}{\includegraphics{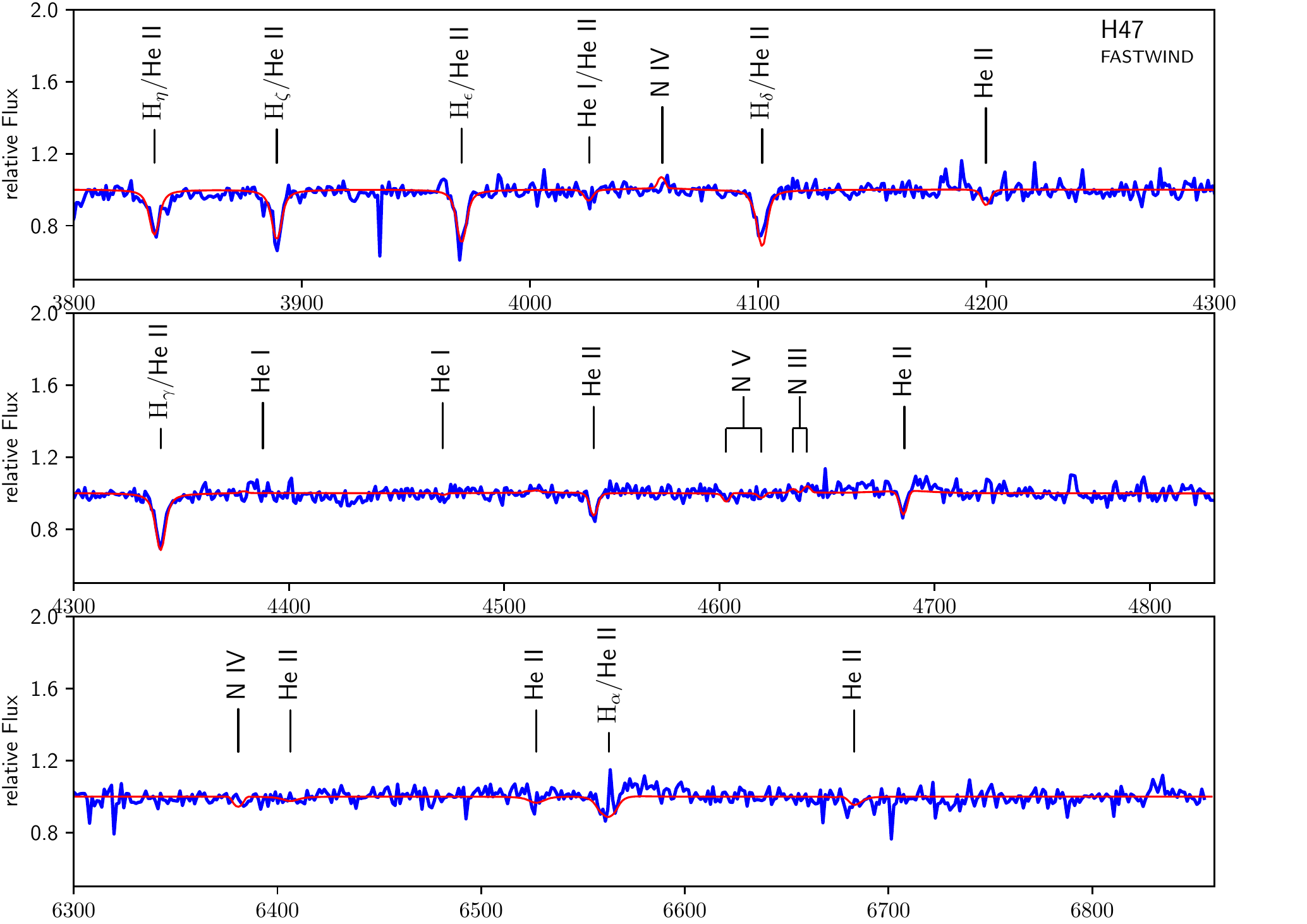}}
\end{center}
\caption{Spectroscopic fit to the data of H47. Blue solid line is the observed HST/STIS spectrum. Red solid line is the synthetic spectrum computed with FASTWIND. Stellar parameters are given in Table\,1.}
\end{figure*}

\begin{figure*}
\begin{center}
\resizebox{0.825\hsize}{!}{\includegraphics{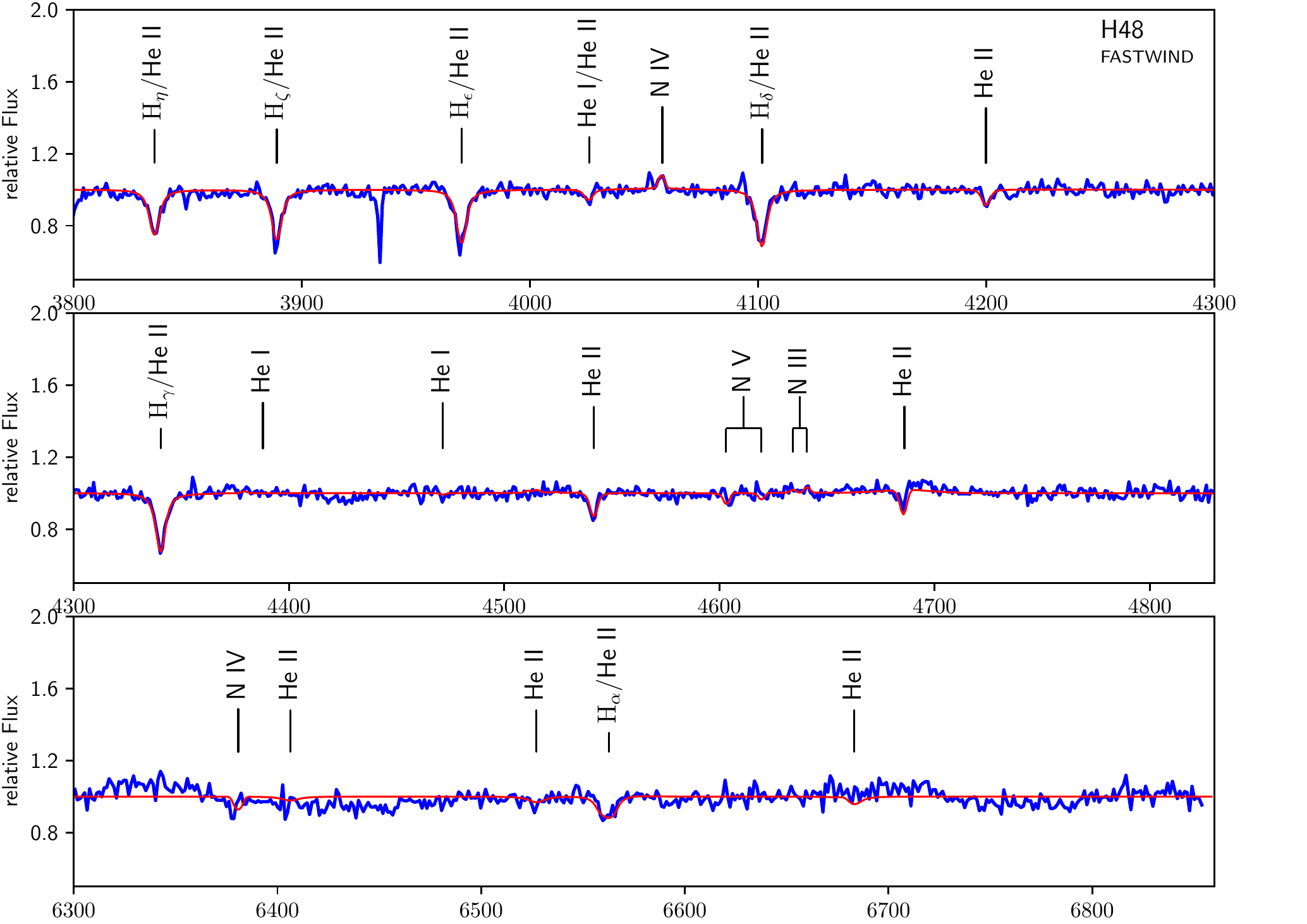}}
\end{center}
\caption{Spectroscopic fit to the data of H48. Blue solid line is the observed HST/STIS spectrum. Red solid line is the synthetic spectrum computed with FASTWIND. Stellar parameters are given in Table\,1.}
\end{figure*}
\begin{figure*}
\begin{center}
\resizebox{0.825\hsize}{!}{\includegraphics{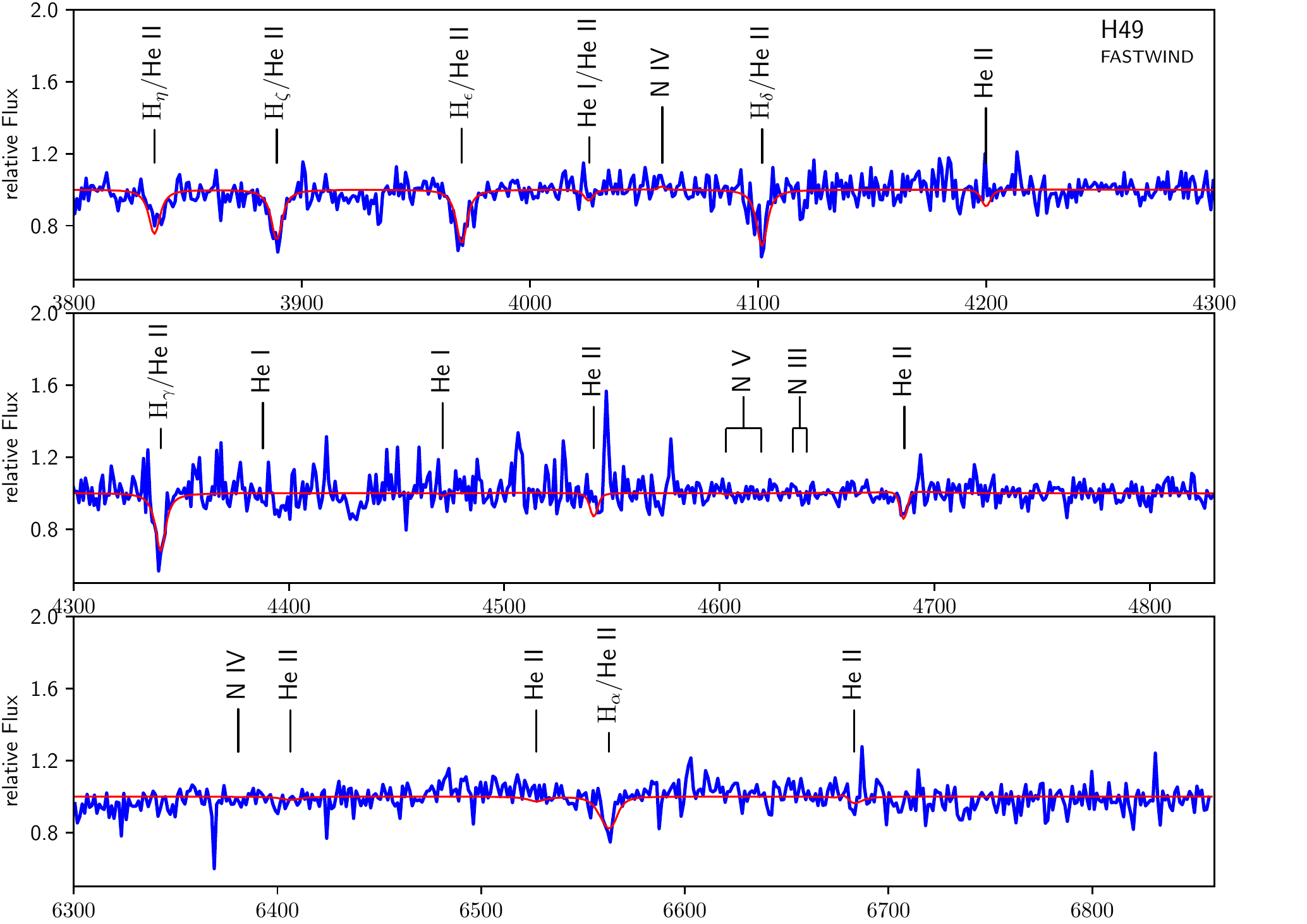}}
\end{center}
\caption{Spectroscopic fit to the data of H49. Blue solid line is the observed HST/STIS spectrum. Red solid line is the synthetic spectrum computed with FASTWIND. Stellar parameters are given in Table\,1.}
\end{figure*}

\clearpage
\begin{figure*}
\begin{center}
\resizebox{0.825\hsize}{!}{\includegraphics{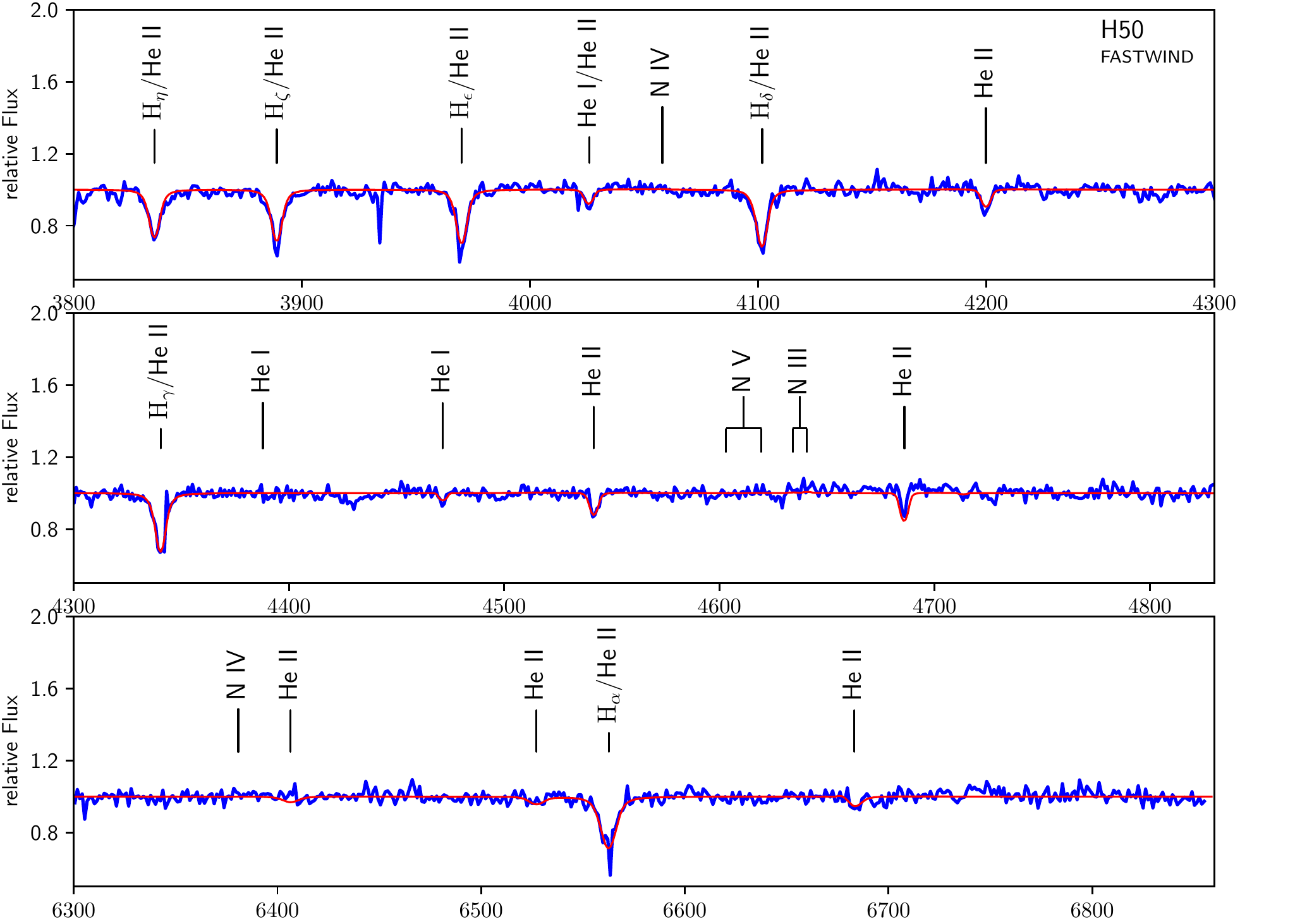}}
\end{center}
\caption{Spectroscopic fit to the data of H50. Blue solid line is the observed HST/STIS spectrum. Red solid line is the synthetic spectrum computed with FASTWIND. Stellar parameters are given in Table\,1.}
\end{figure*}
\begin{figure*}
\begin{center}
\resizebox{0.825\hsize}{!}{\includegraphics{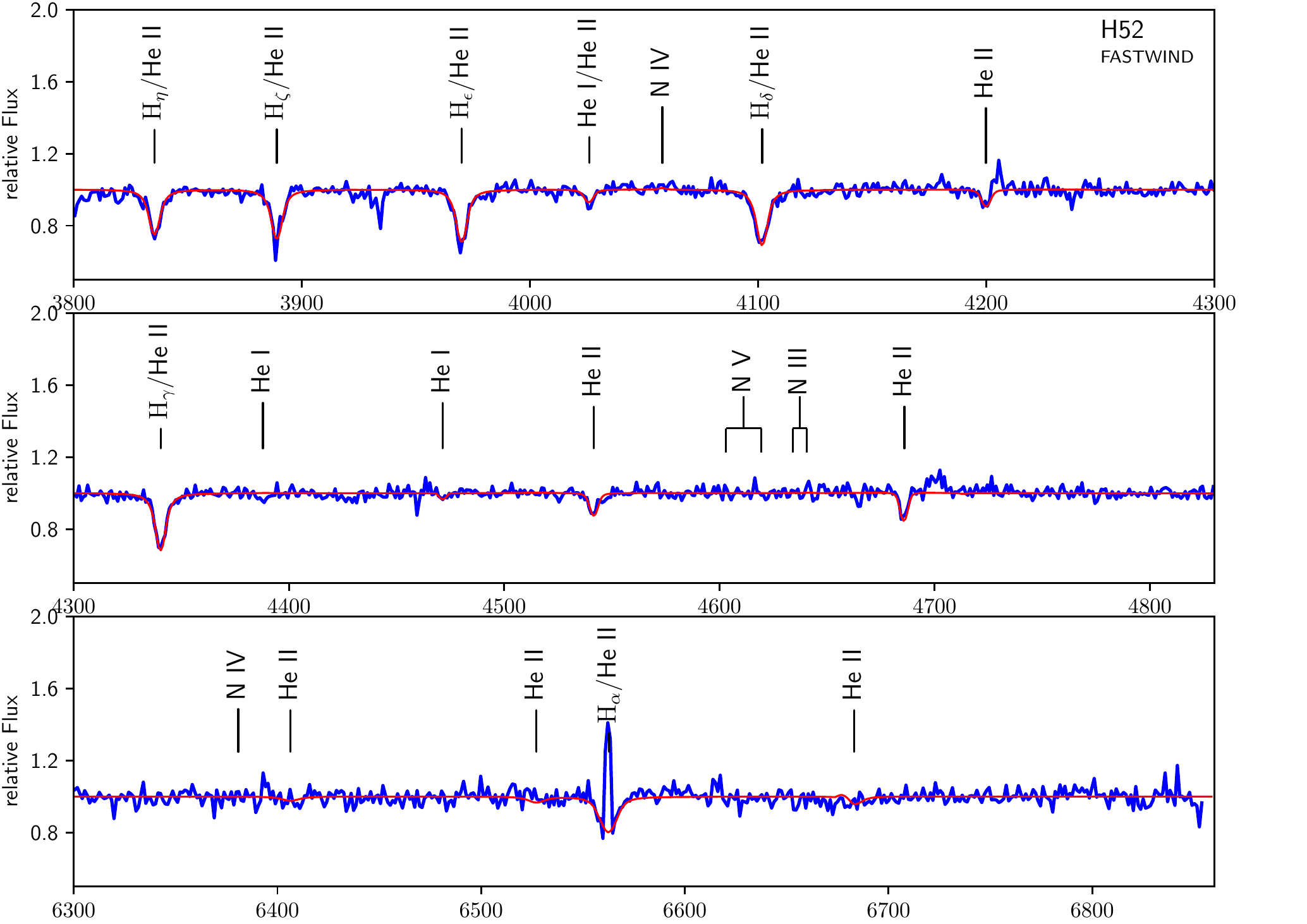}}
\end{center}
\caption{Spectroscopic fit to the data of H52. Blue solid line is the observed HST/STIS spectrum. Red solid line is the synthetic spectrum computed with FASTWIND. Stellar parameters are given in Table\,1.}
\end{figure*}

\begin{figure*}
\begin{center}
\resizebox{0.825\hsize}{!}{\includegraphics{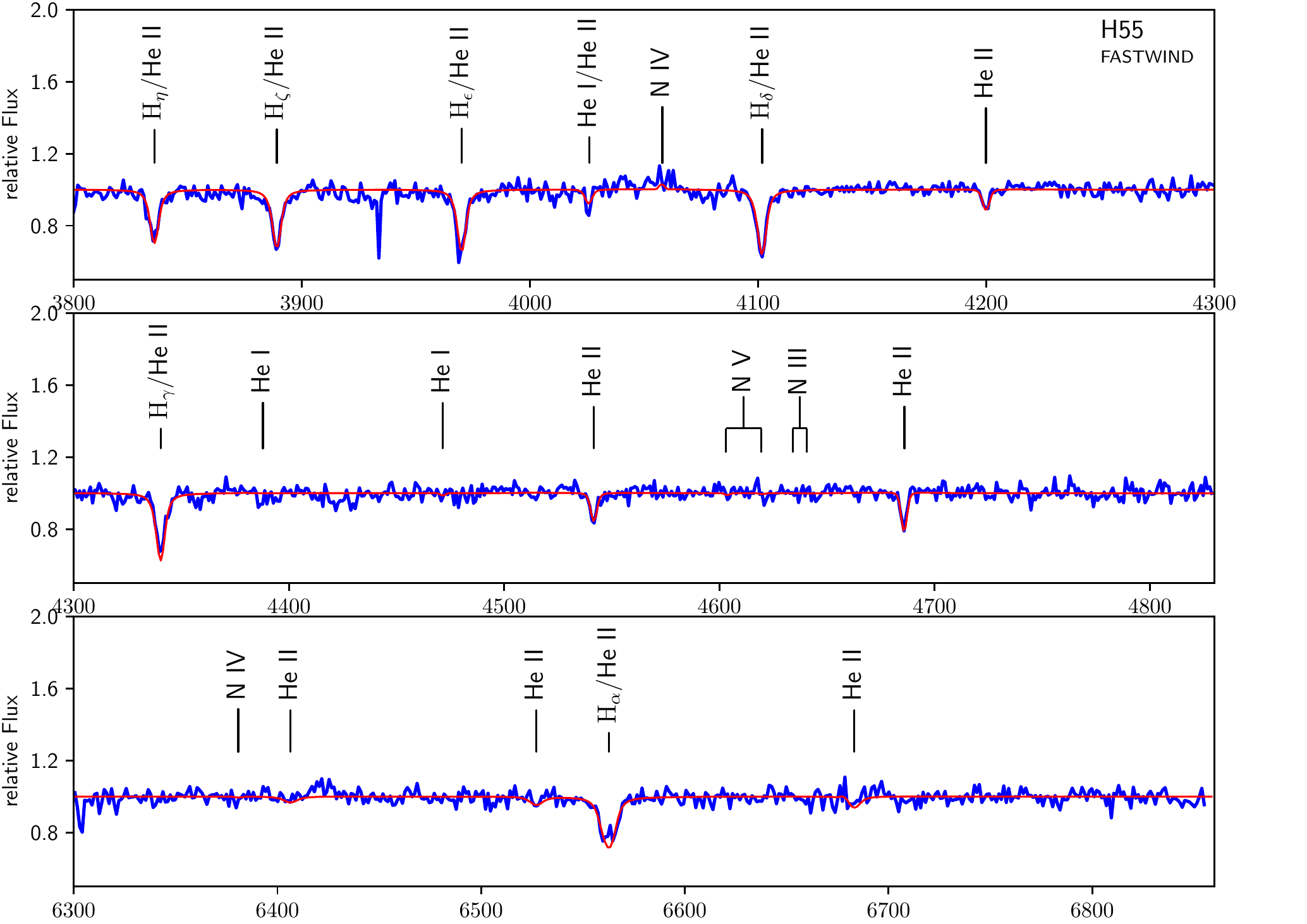}}
\end{center}
\caption{Spectroscopic fit to the data of H55. Blue solid line is the observed HST/STIS spectrum. Red solid line is the synthetic spectrum computed with FASTWIND. Stellar parameters are given in Table\,1.}
\end{figure*}
\begin{figure*}
\begin{center}
\resizebox{0.825\hsize}{!}{\includegraphics{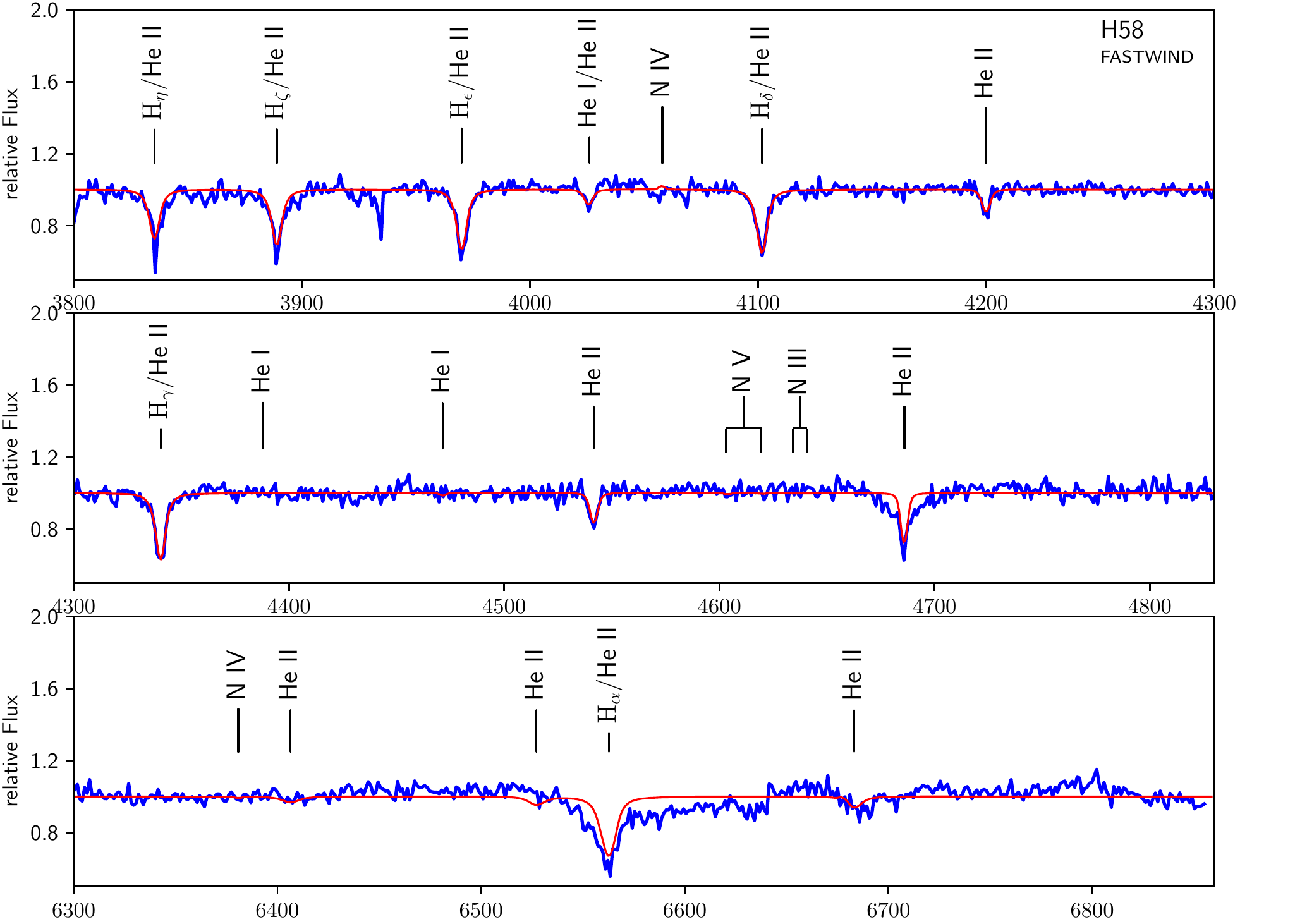}}
\end{center}
\caption{Spectroscopic fit to the data of H58. Blue solid line is the observed HST/STIS spectrum. Red solid line is the synthetic spectrum computed with FASTWIND. Stellar parameters are given in Table\,1.}
\end{figure*}

\begin{figure*}
\begin{center}
\resizebox{0.825\hsize}{!}{\includegraphics{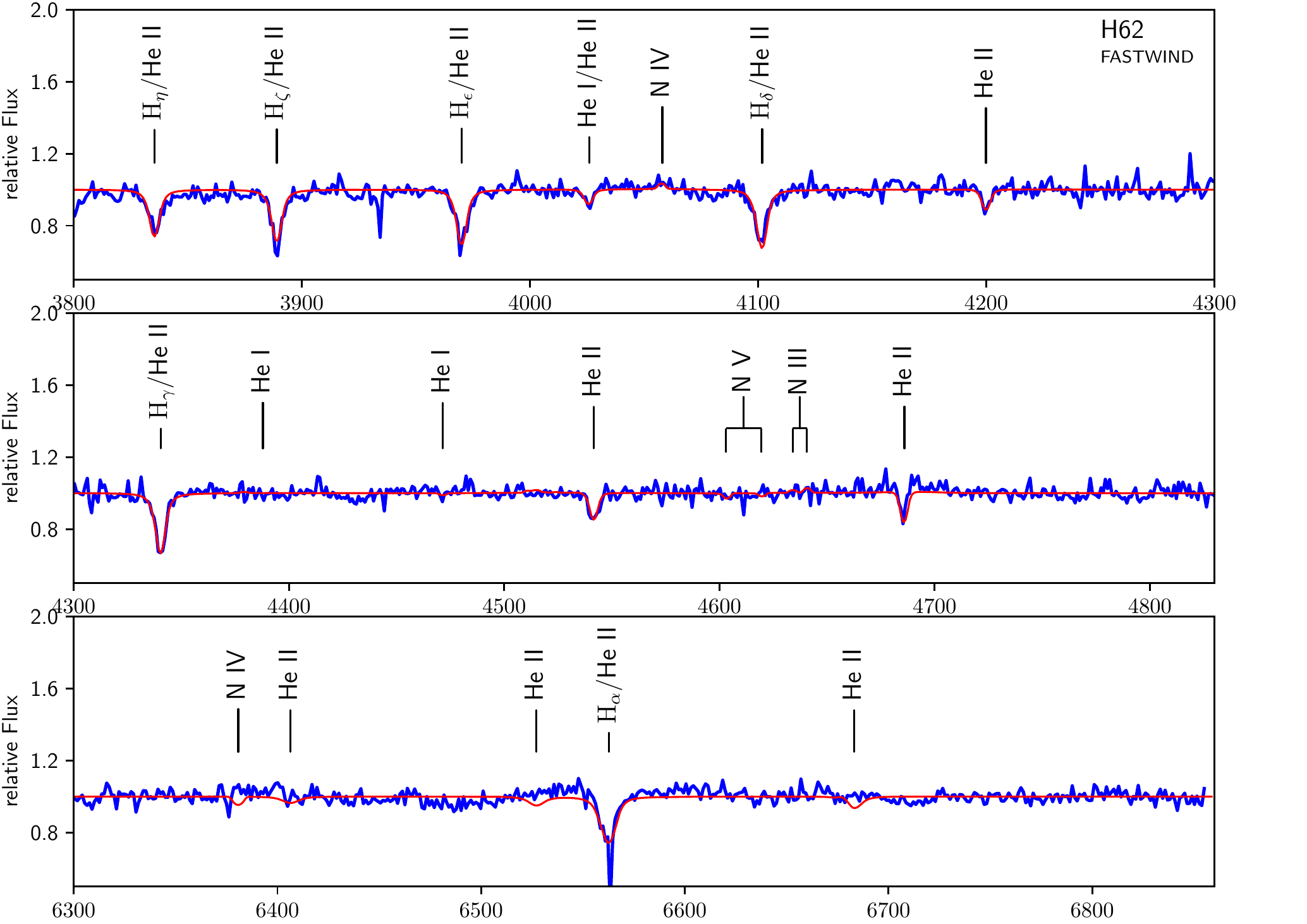}}
\end{center}
\caption{Spectroscopic fit to the data of H62. Blue solid line is the observed HST/STIS spectrum. Red solid line is the synthetic spectrum computed with FASTWIND. Stellar parameters are given in Table\,1.}
\end{figure*}
\begin{figure*}
\begin{center}
\resizebox{0.825\hsize}{!}{\includegraphics{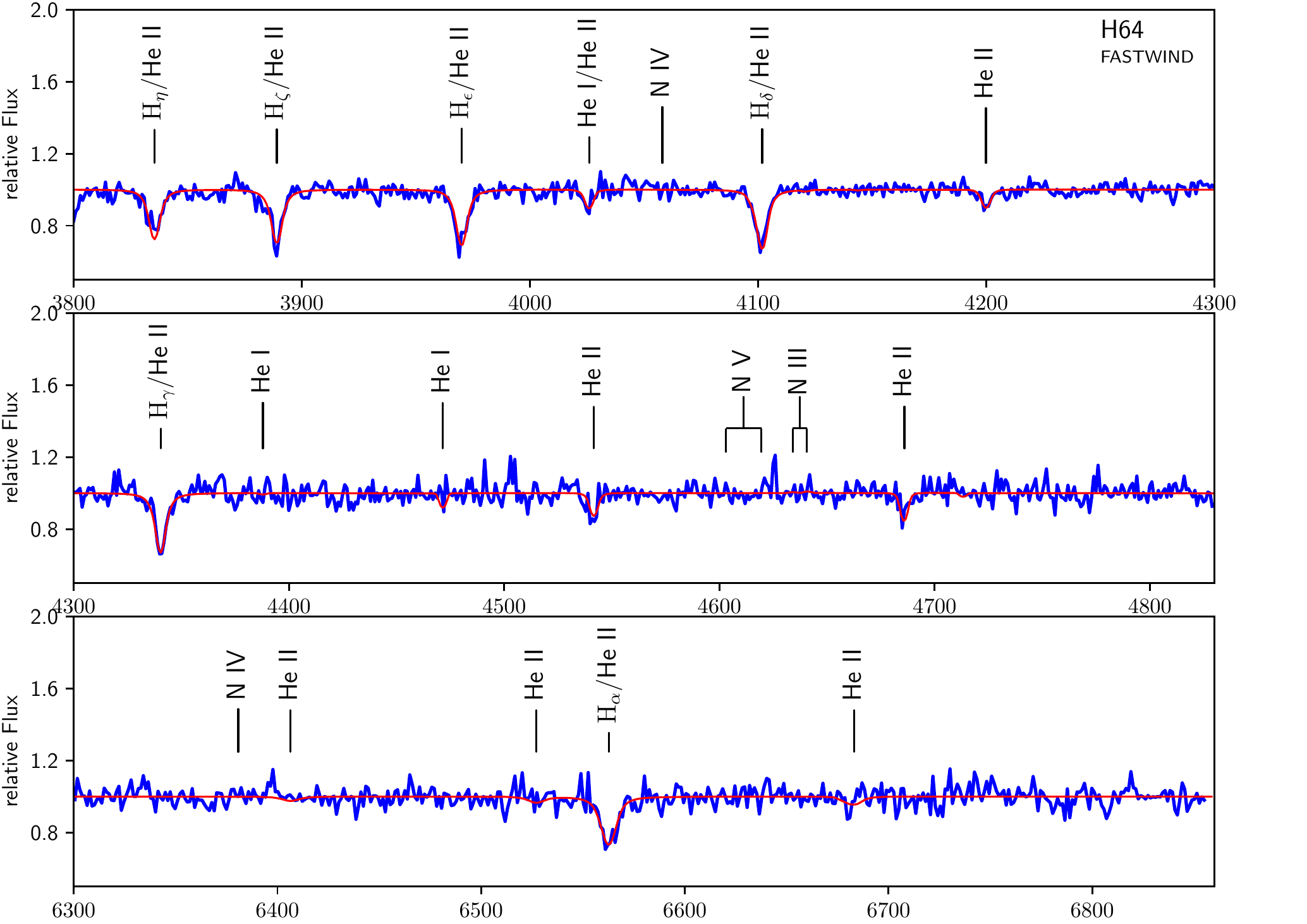}}
\end{center}
\caption{Spectroscopic fit to the data of H64. Blue solid line is the observed HST/STIS spectrum. Red solid line is the synthetic spectrum computed with FASTWIND. Stellar parameters are given in Table\,1.}
\end{figure*}

\begin{figure*}
\begin{center}
\resizebox{0.825\hsize}{!}{\includegraphics{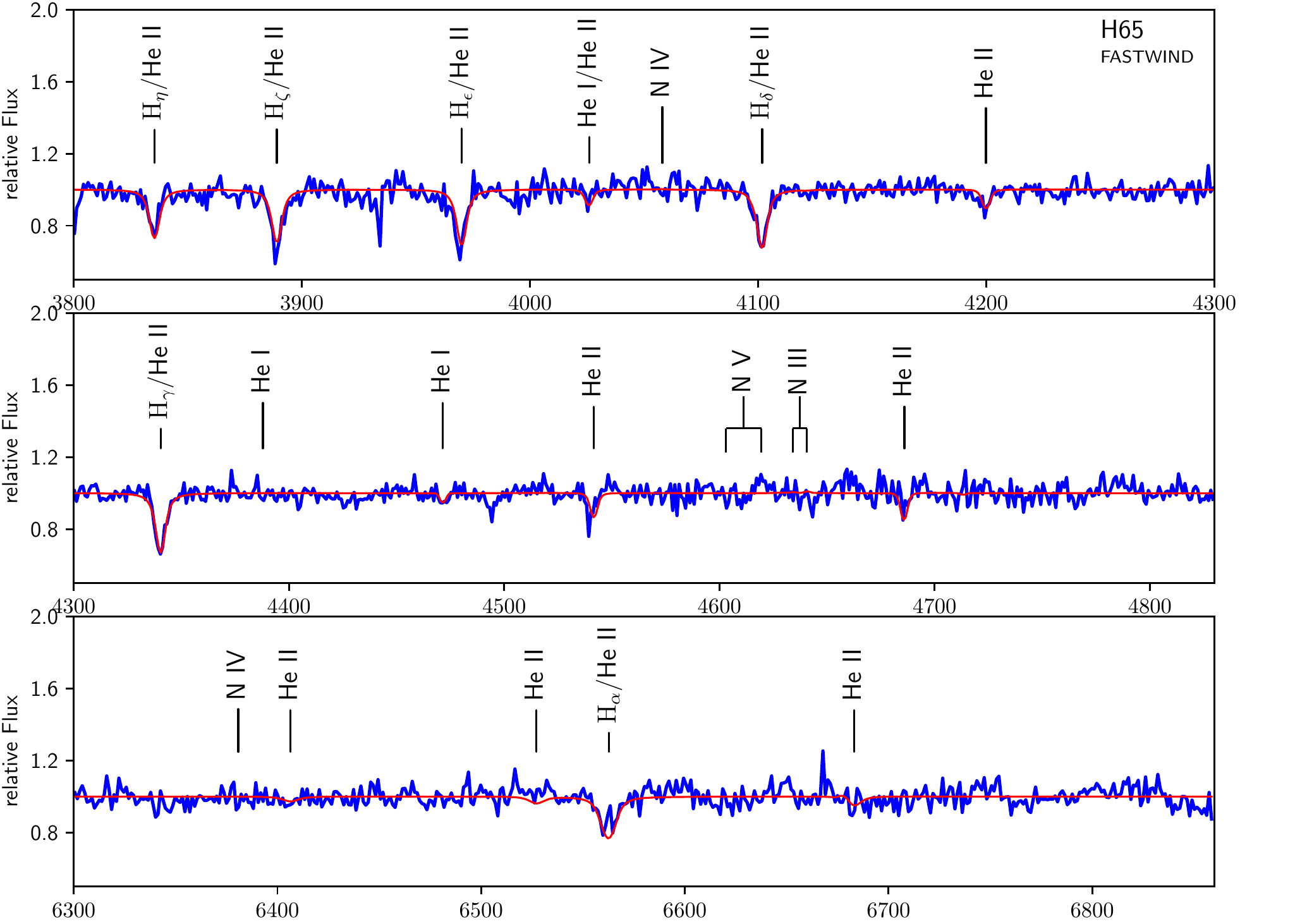}}
\end{center}
\caption{Spectroscopic fit to the data of H65. Blue solid line is the observed HST/STIS spectrum. Red solid line is the synthetic spectrum computed with FASTWIND. Stellar parameters are given in Table\,1.}
\end{figure*}
\begin{figure*}
\begin{center}
\resizebox{0.825\hsize}{!}{\includegraphics{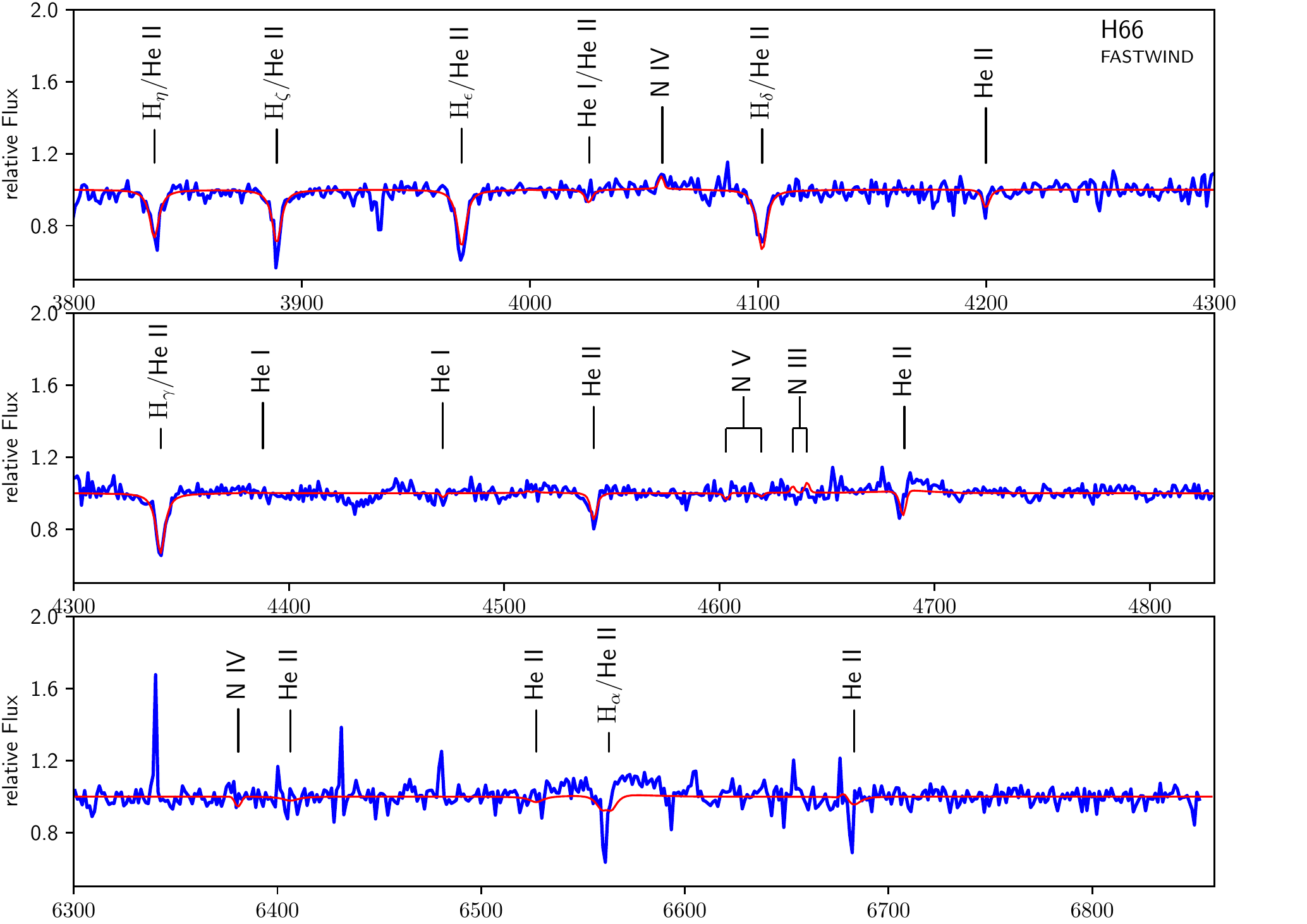}}
\end{center}
\caption{Spectroscopic fit to the data of H66. Blue solid line is the observed HST/STIS spectrum. Red solid line is the synthetic spectrum computed with FASTWIND. Stellar parameters are given in Table\,1.}
\end{figure*}

\begin{figure*}
\begin{center}
\resizebox{0.825\hsize}{!}{\includegraphics{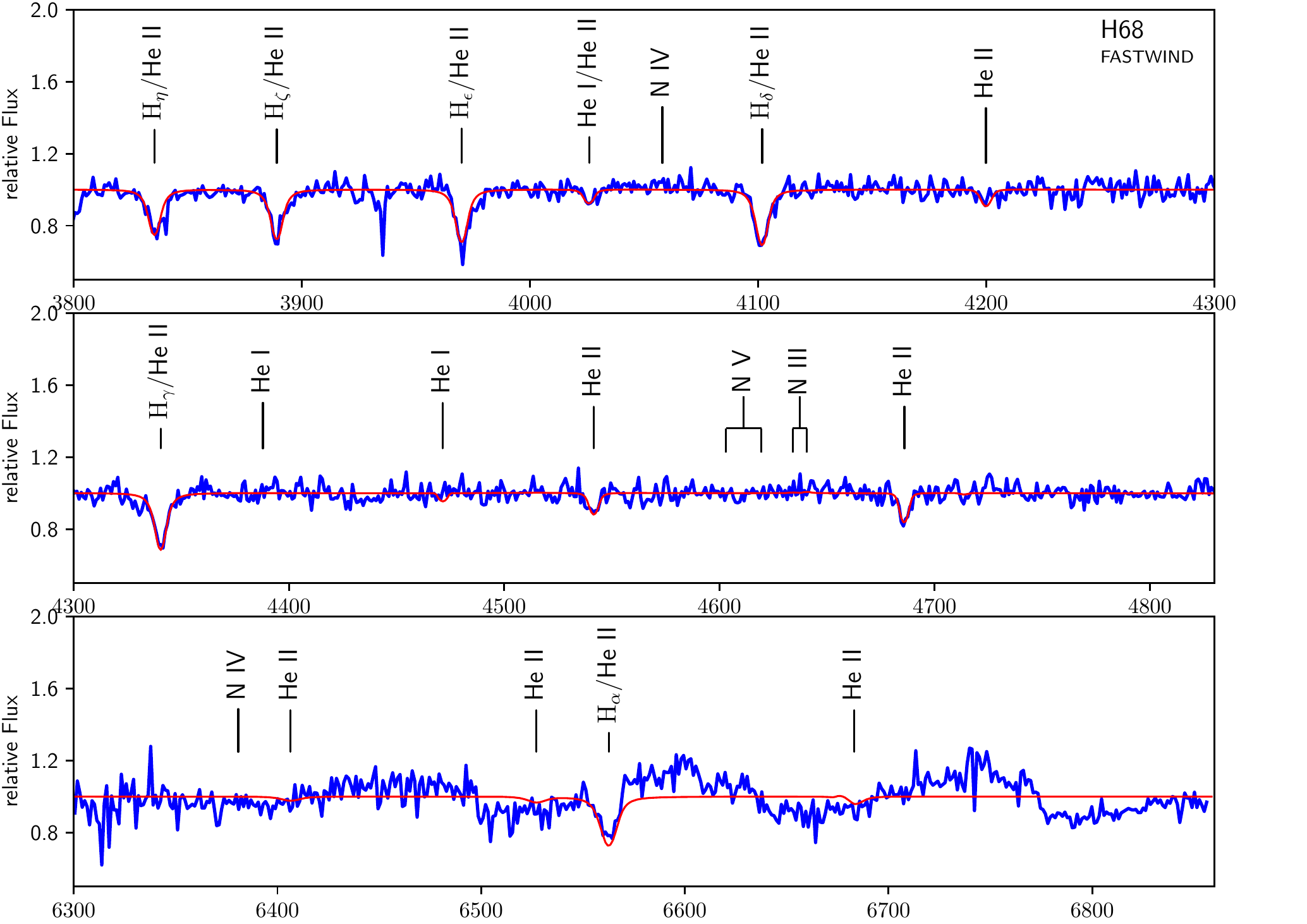}}
\end{center}
\caption{Spectroscopic fit to the data of H68. Blue solid line is the observed HST/STIS spectrum. Red solid line is the synthetic spectrum computed with FASTWIND. Stellar parameters are given in Table\,1.}
\end{figure*}
\begin{figure*}
\begin{center}
\resizebox{0.825\hsize}{!}{\includegraphics{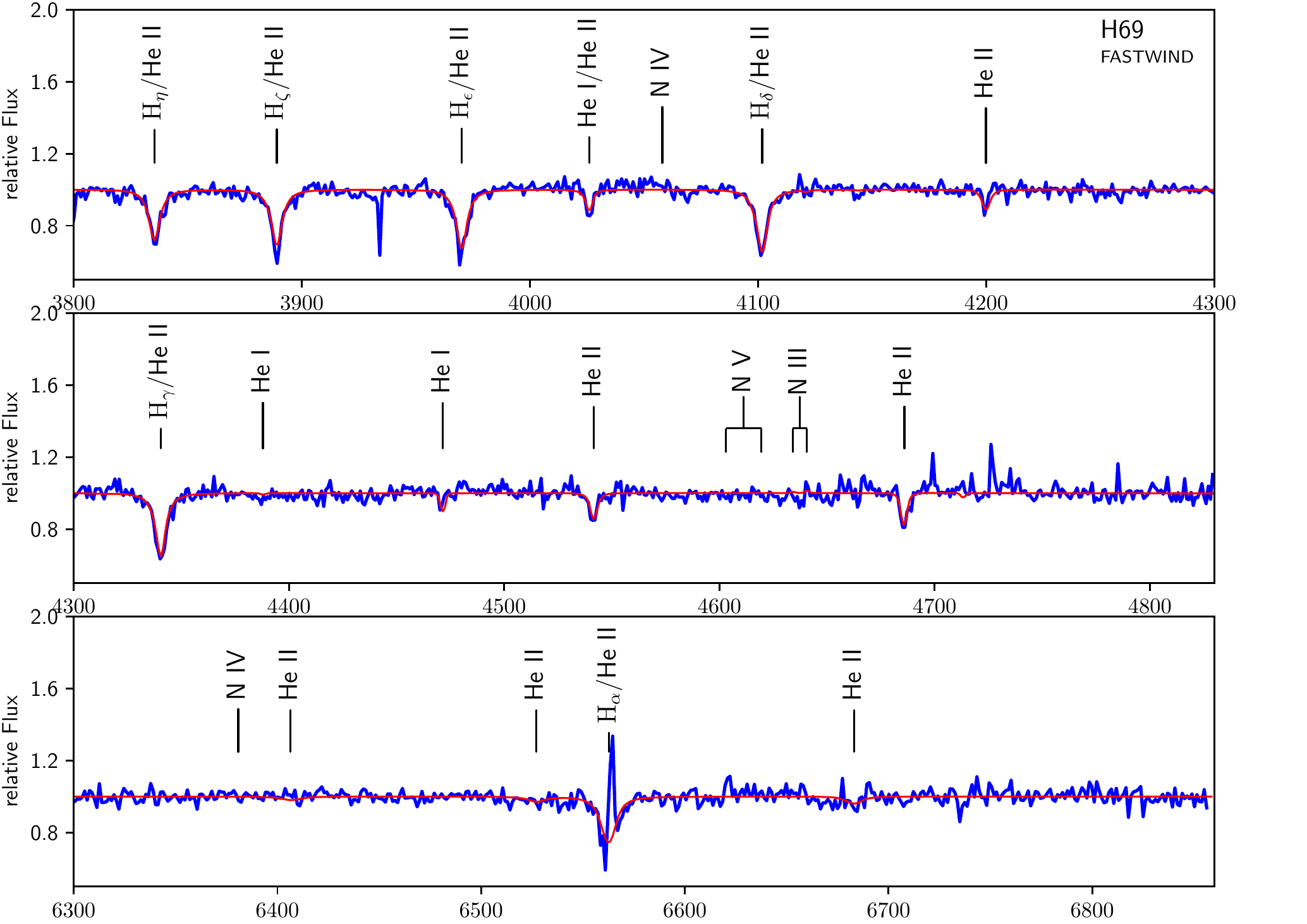}}
\end{center}
\caption{Spectroscopic fit to the data of H69. Blue solid line is the observed HST/STIS spectrum. Red solid line is the synthetic spectrum computed with FASTWIND. Stellar parameters are given in Table\,1.}
\end{figure*}

\begin{figure*}
\begin{center}
\resizebox{0.825\hsize}{!}{\includegraphics{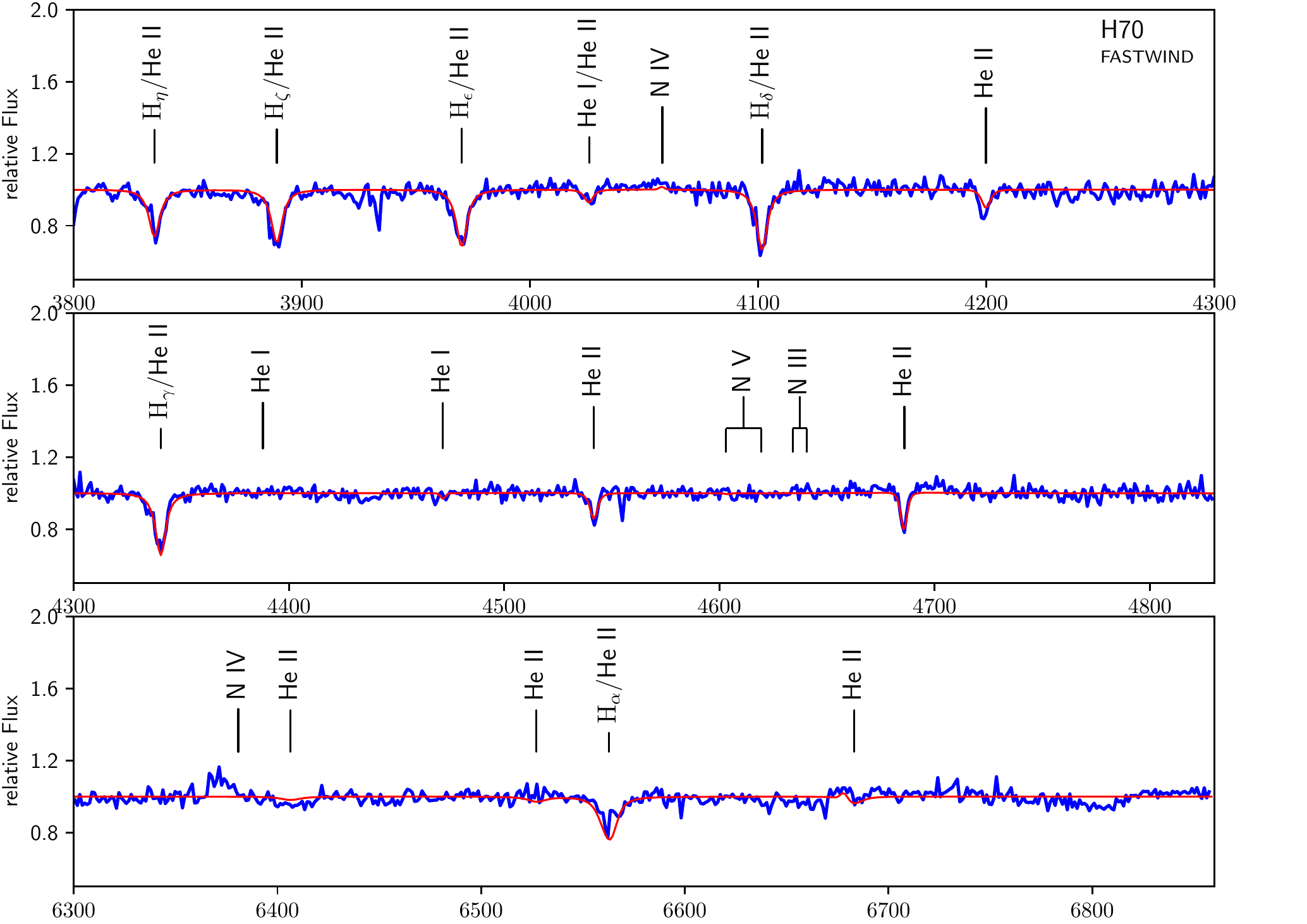}}
\end{center}
\caption{Spectroscopic fit to the data of H70. Blue solid line is the observed HST/STIS spectrum. Red solid line is the synthetic spectrum computed with FASTWIND. Stellar parameters are given in Table\,1.}
\end{figure*}
\begin{figure*}
\begin{center}
\resizebox{0.825\hsize}{!}{\includegraphics{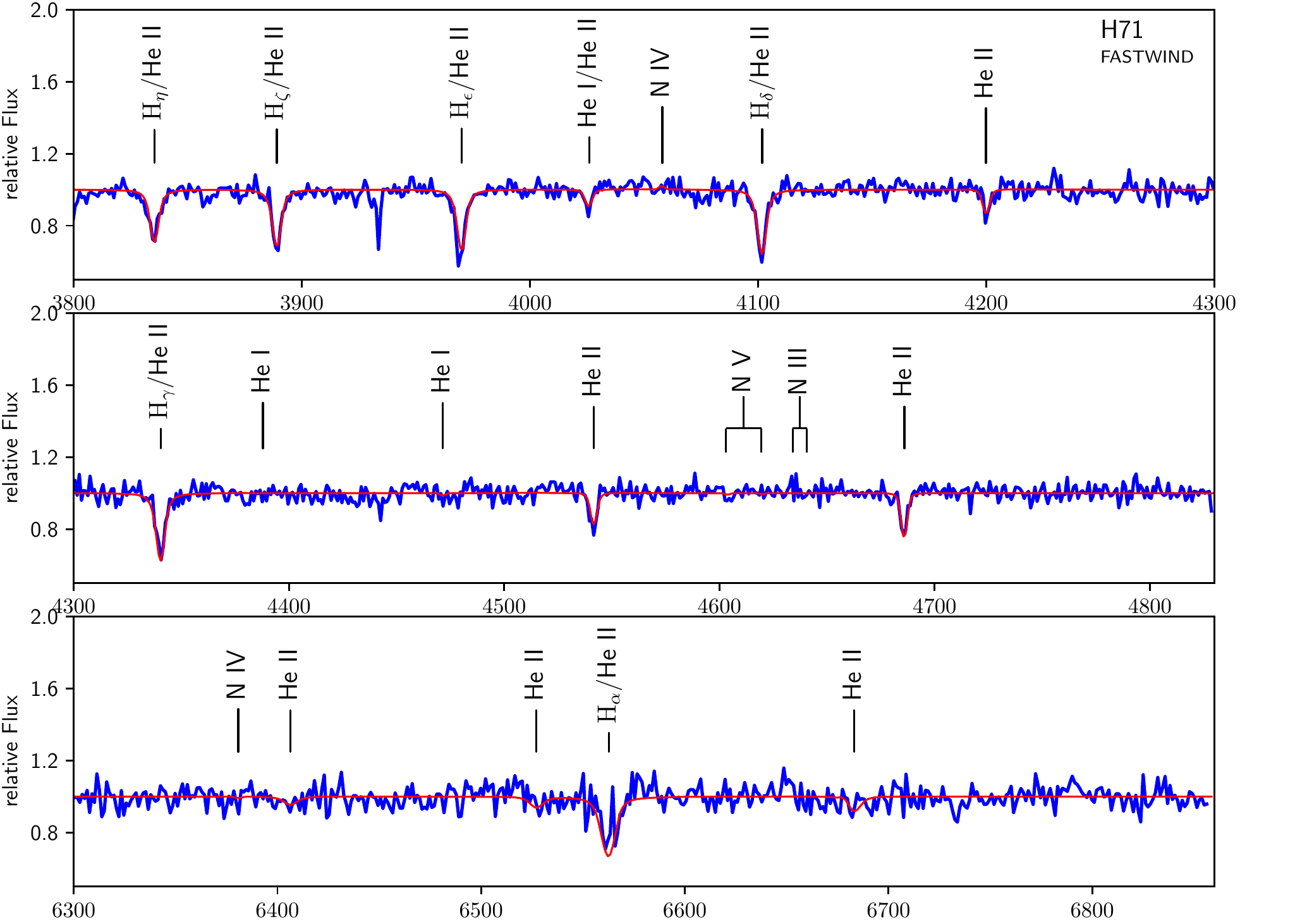}}
\end{center}
\caption{Spectroscopic fit to the data of H71. Blue solid line is the observed HST/STIS spectrum. Red solid line is the synthetic spectrum computed with FASTWIND. Stellar parameters are given in Table\,1.}
\end{figure*}

\clearpage
\begin{figure*}
\begin{center}
\resizebox{0.825\hsize}{!}{\includegraphics{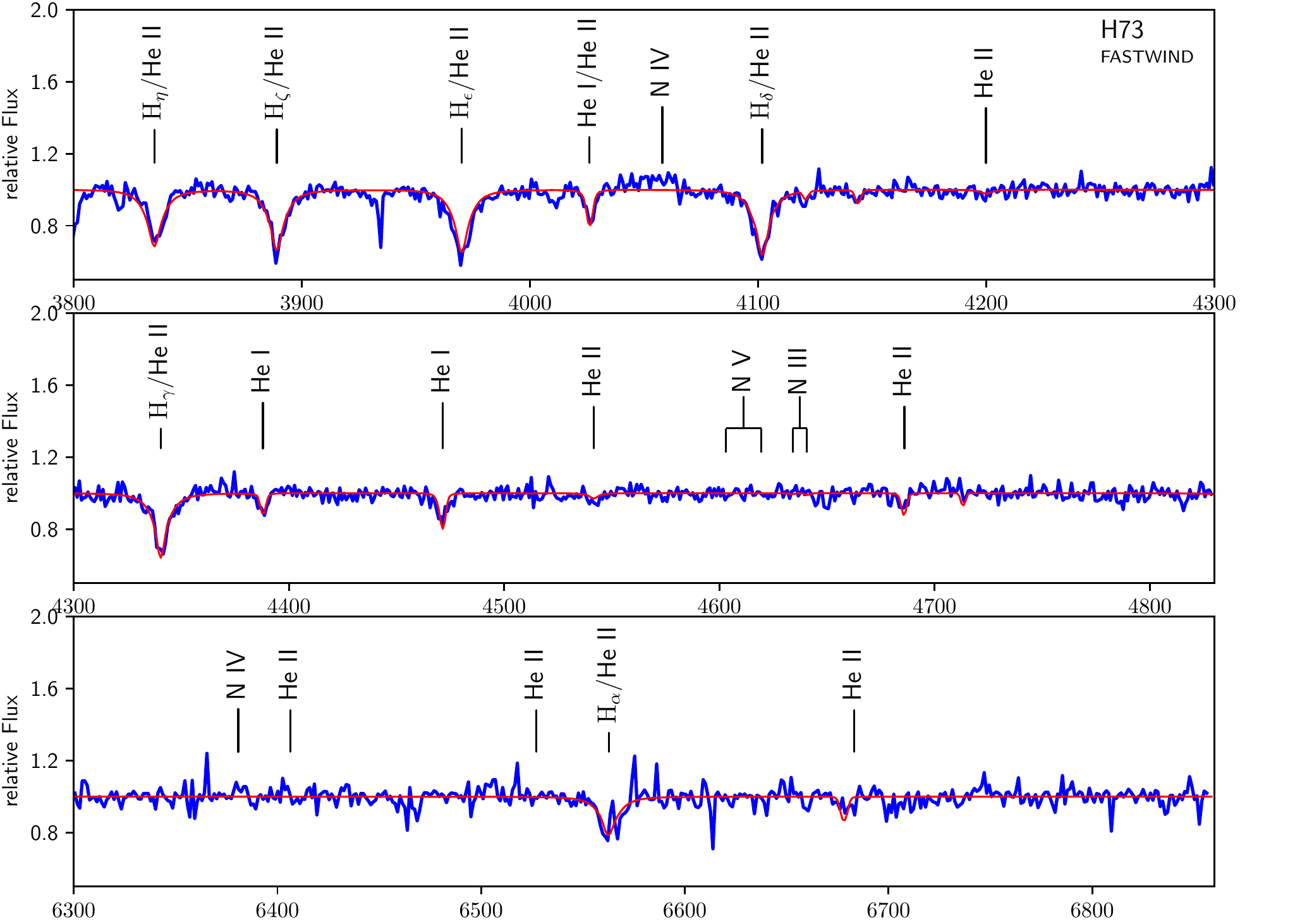}}
\end{center}
\caption{Spectroscopic fit to the data of H73. Blue solid line is the observed HST/STIS spectrum. Red solid line is the synthetic spectrum computed with FASTWIND. Stellar parameters are given in Table\,1.}
\end{figure*}
\begin{figure*}
\begin{center}
\resizebox{0.825\hsize}{!}{\includegraphics{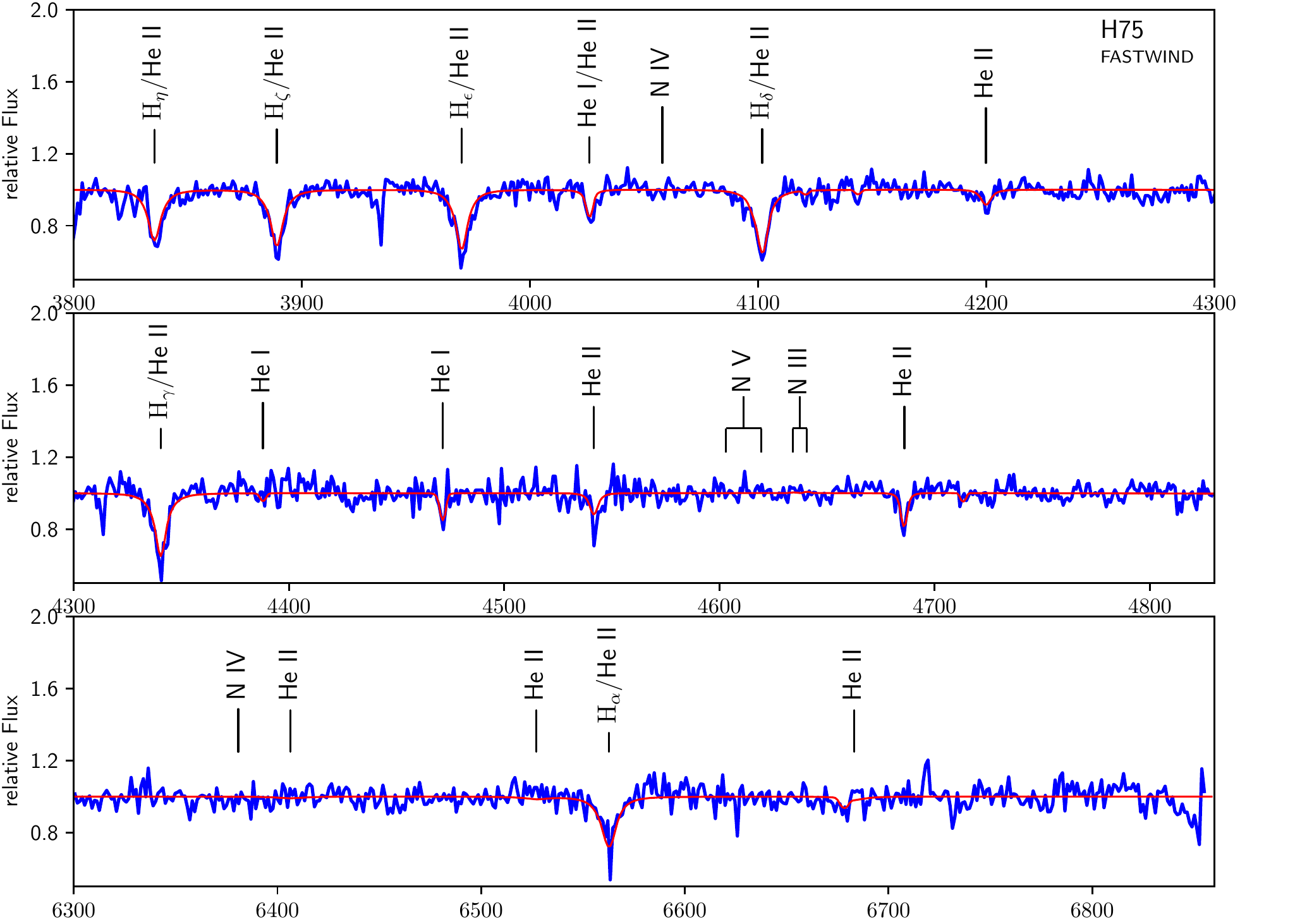}}
\end{center}
\caption{Spectroscopic fit to the data of H75. Blue solid line is the observed HST/STIS spectrum. Red solid line is the synthetic spectrum computed with FASTWIND. Stellar parameters are given in Table\,1.}
\end{figure*}

\begin{figure*}
\begin{center}
\resizebox{0.825\hsize}{!}{\includegraphics{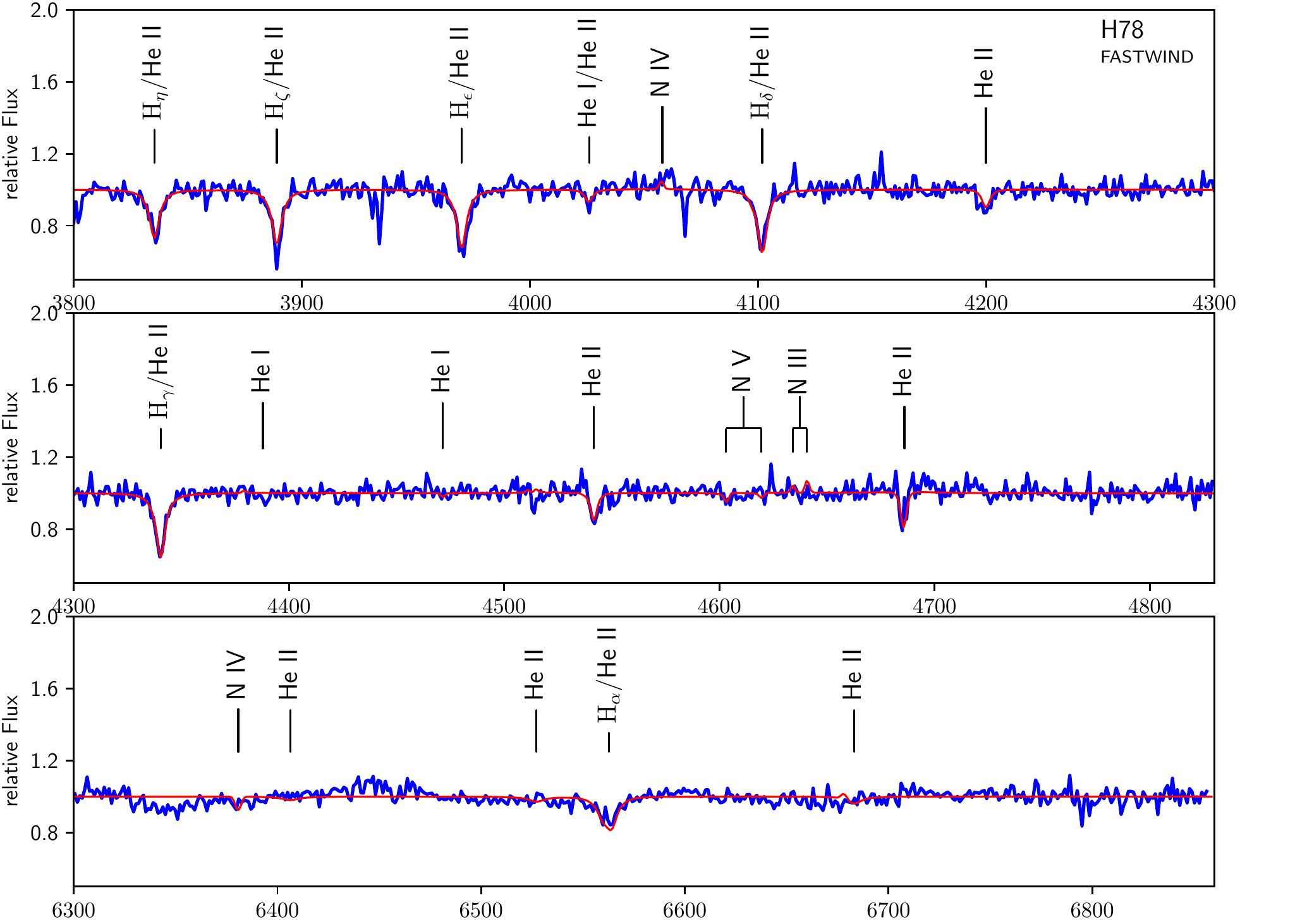}}
\end{center}
\caption{Spectroscopic fit to the data of H78. Blue solid line is the observed HST/STIS spectrum. Red solid line is the synthetic spectrum computed with FASTWIND. Stellar parameters are given in Table\,1.}
\end{figure*}
\begin{figure*}
\begin{center}
\resizebox{0.825\hsize}{!}{\includegraphics{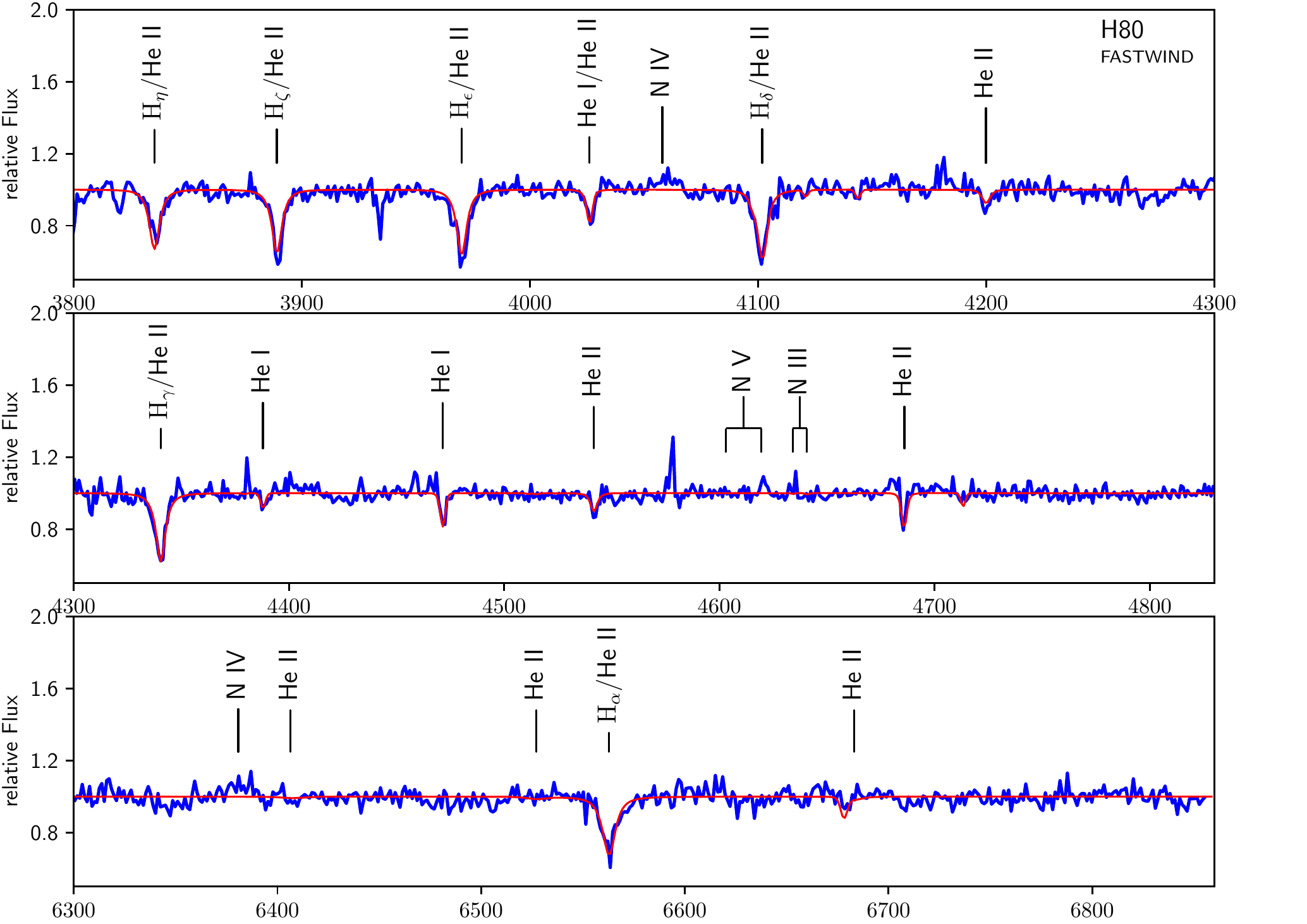}}
\end{center}
\caption{Spectroscopic fit to the data of H80. Blue solid line is the observed HST/STIS spectrum. Red solid line is the synthetic spectrum computed with FASTWIND. Stellar parameters are given in Table\,1.}
\end{figure*}

\begin{figure*}
\begin{center}
\resizebox{0.825\hsize}{!}{\includegraphics{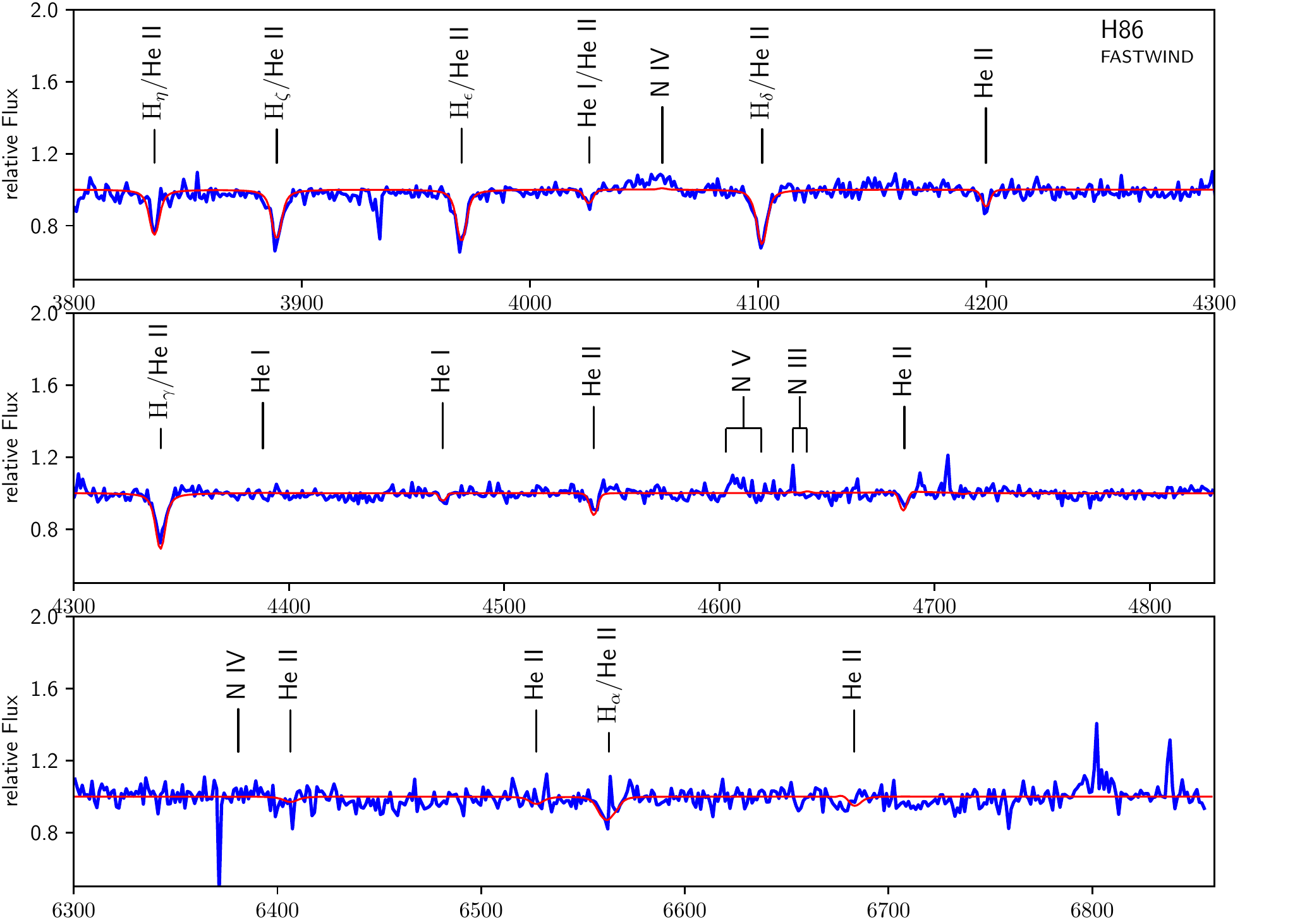}}
\end{center}
\caption{Spectroscopic fit to the data of H86. Blue solid line is the observed HST/STIS spectrum. Red solid line is the synthetic spectrum computed with FASTWIND. Stellar parameters are given in Table\,1.}
\end{figure*}
\begin{figure*}
\begin{center}
\resizebox{0.825\hsize}{!}{\includegraphics{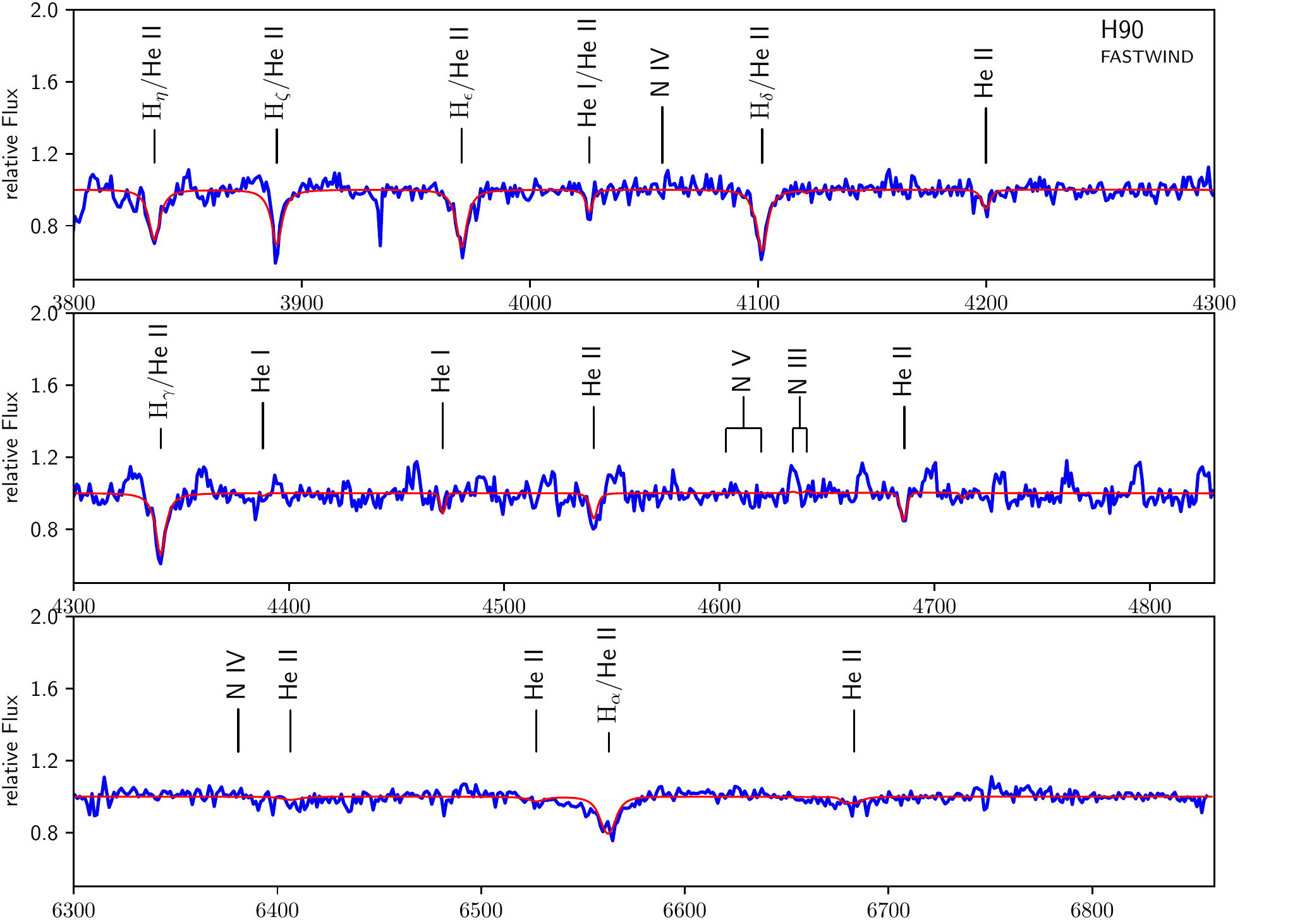}}
\end{center}
\caption{Spectroscopic fit to the data of H90. Blue solid line is the observed HST/STIS spectrum. Red solid line is the synthetic spectrum computed with FASTWIND. Stellar parameters are given in Table\,1.}
\end{figure*}

\clearpage

\begin{figure*}
\begin{center}
\resizebox{0.825\hsize}{!}{\includegraphics{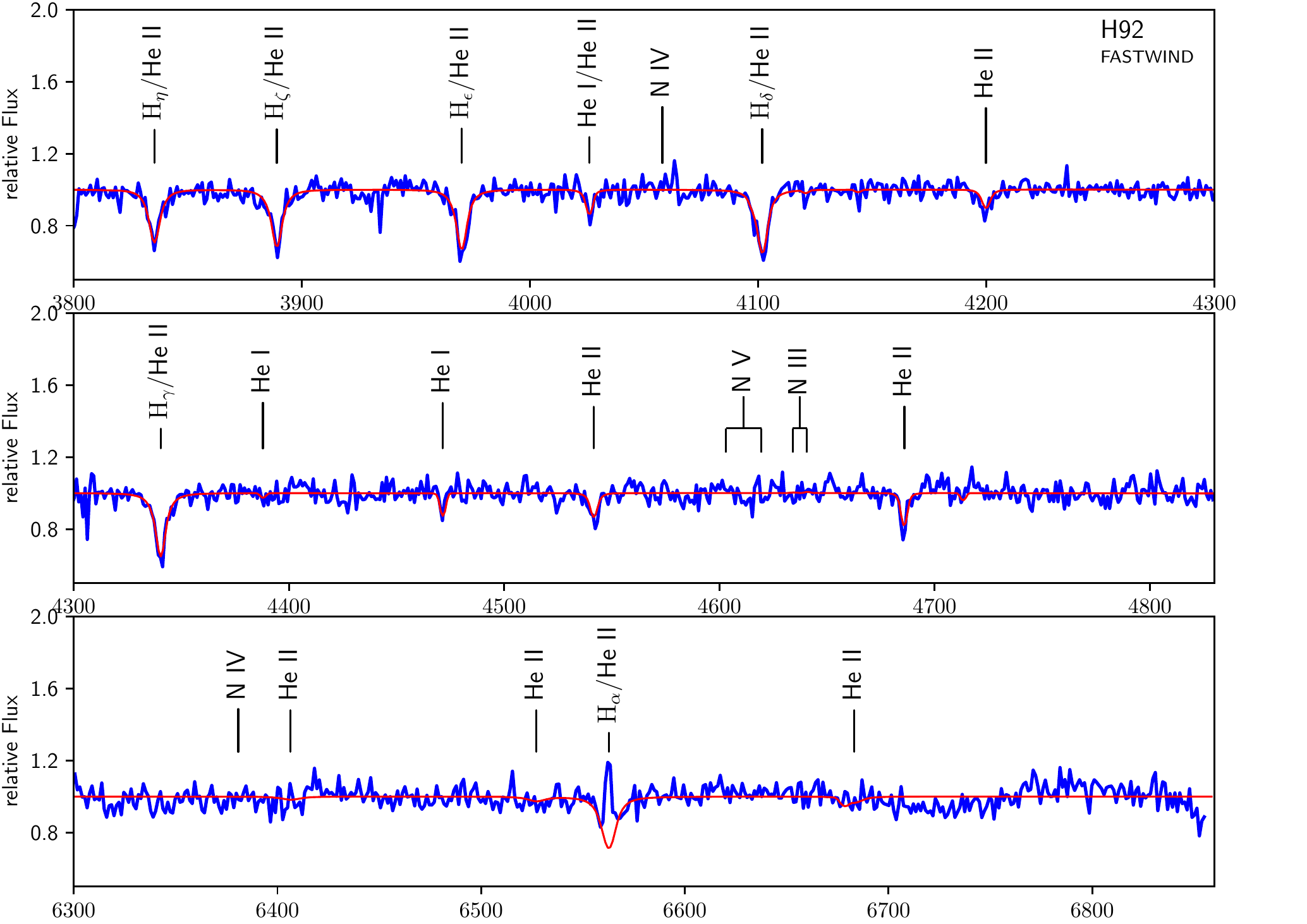}}
\end{center}
\caption{Spectroscopic fit to the data of H92. Blue solid line is the observed HST/STIS spectrum. Red solid line is the synthetic spectrum computed with FASTWIND. Stellar parameters are given in Table\,1.}
\end{figure*}
\begin{figure*}
\begin{center}
\resizebox{0.825\hsize}{!}{\includegraphics{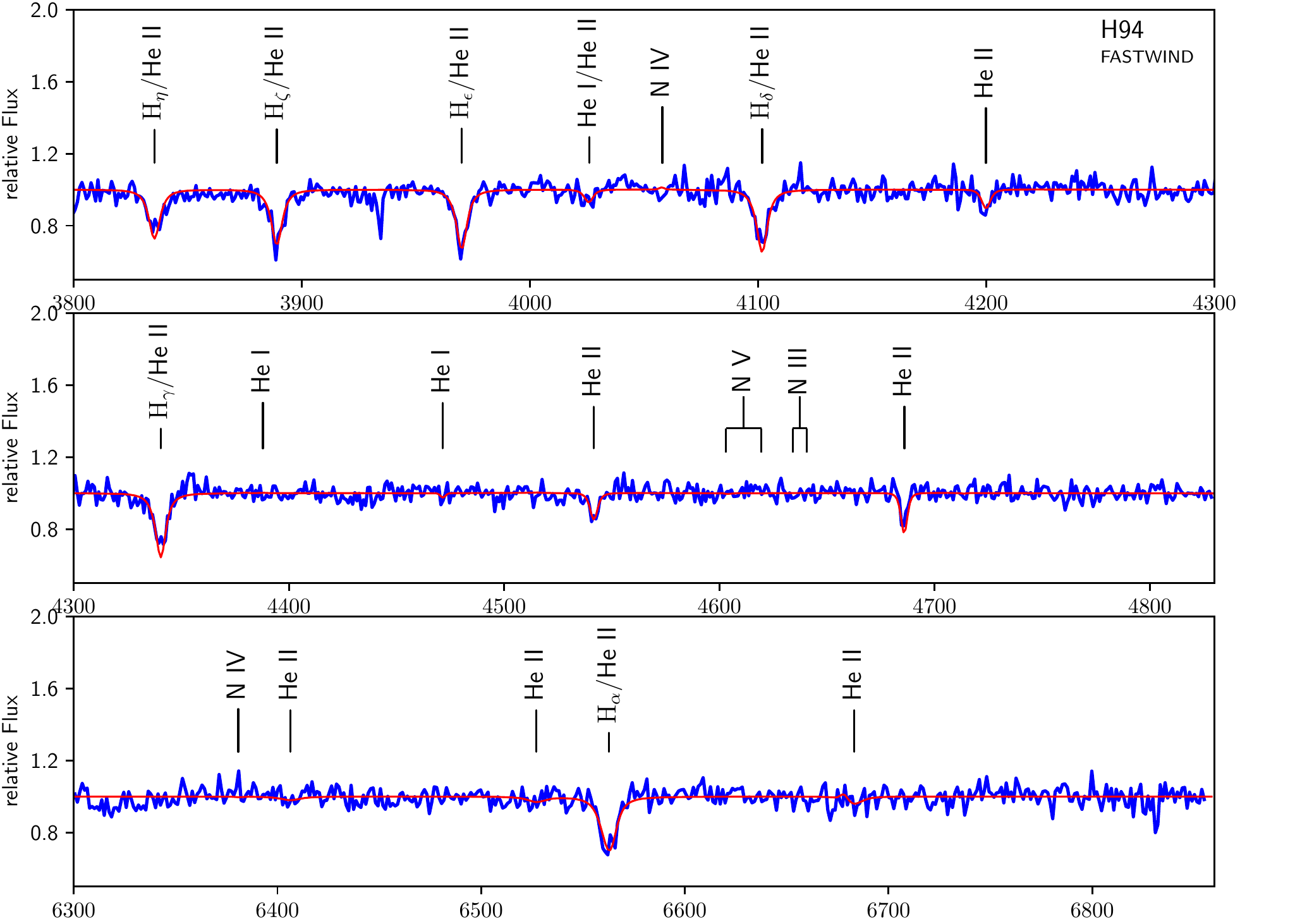}}
\end{center}
\caption{Spectroscopic fit to the data of H94. Blue solid line is the observed HST/STIS spectrum. Red solid line is the synthetic spectrum computed with FASTWIND. Stellar parameters are given in Table\,1.}
\end{figure*}

\begin{figure*}
\begin{center}
\resizebox{0.825\hsize}{!}{\includegraphics{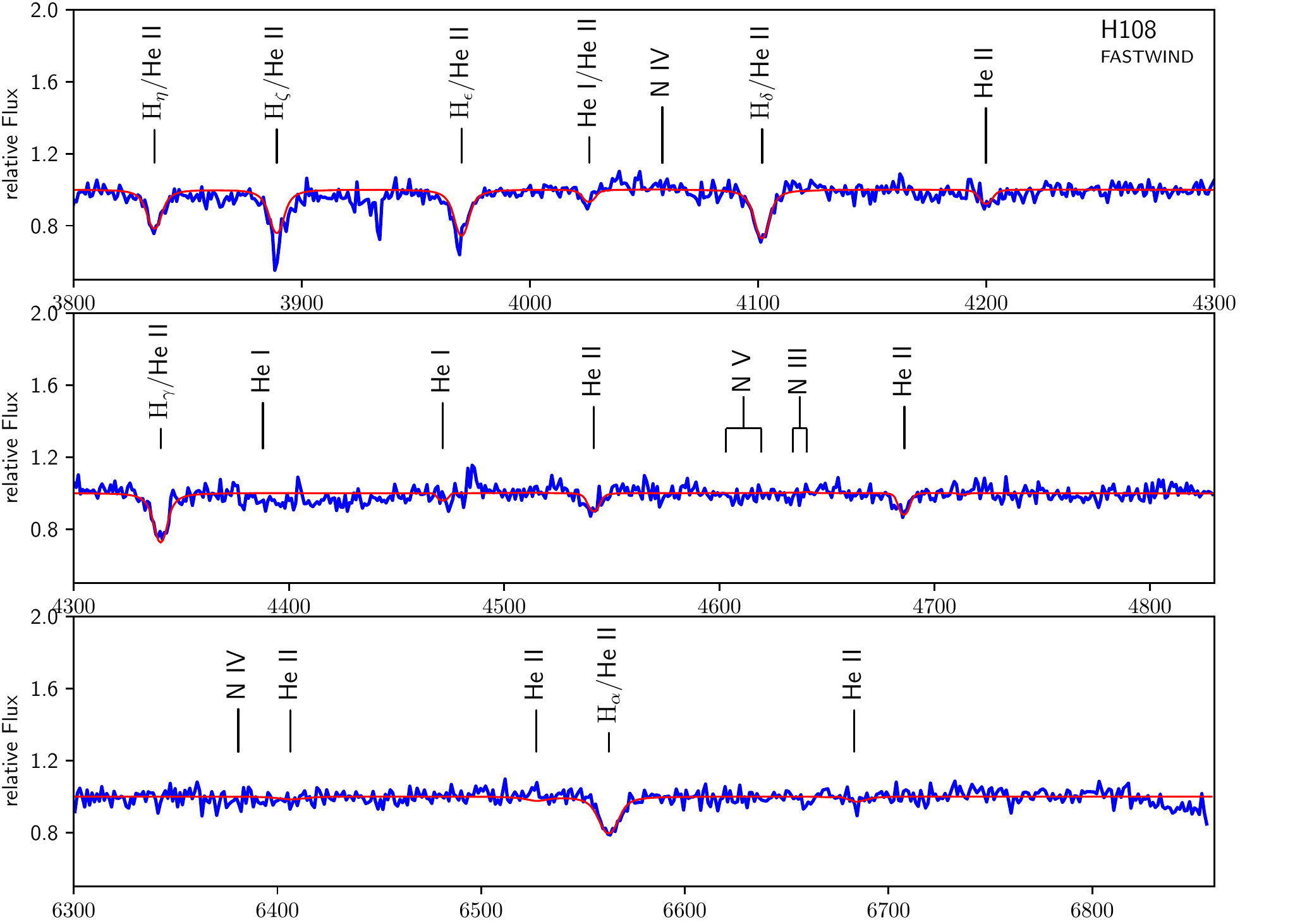}}
\end{center}
\caption{Spectroscopic fit to the data of H108. Blue solid line is the observed HST/STIS spectrum. Red solid line is the synthetic spectrum computed with FASTWIND. Stellar parameters are given in Table\,1.}
\end{figure*}
\begin{figure*}
\begin{center}
\resizebox{0.825\hsize}{!}{\includegraphics{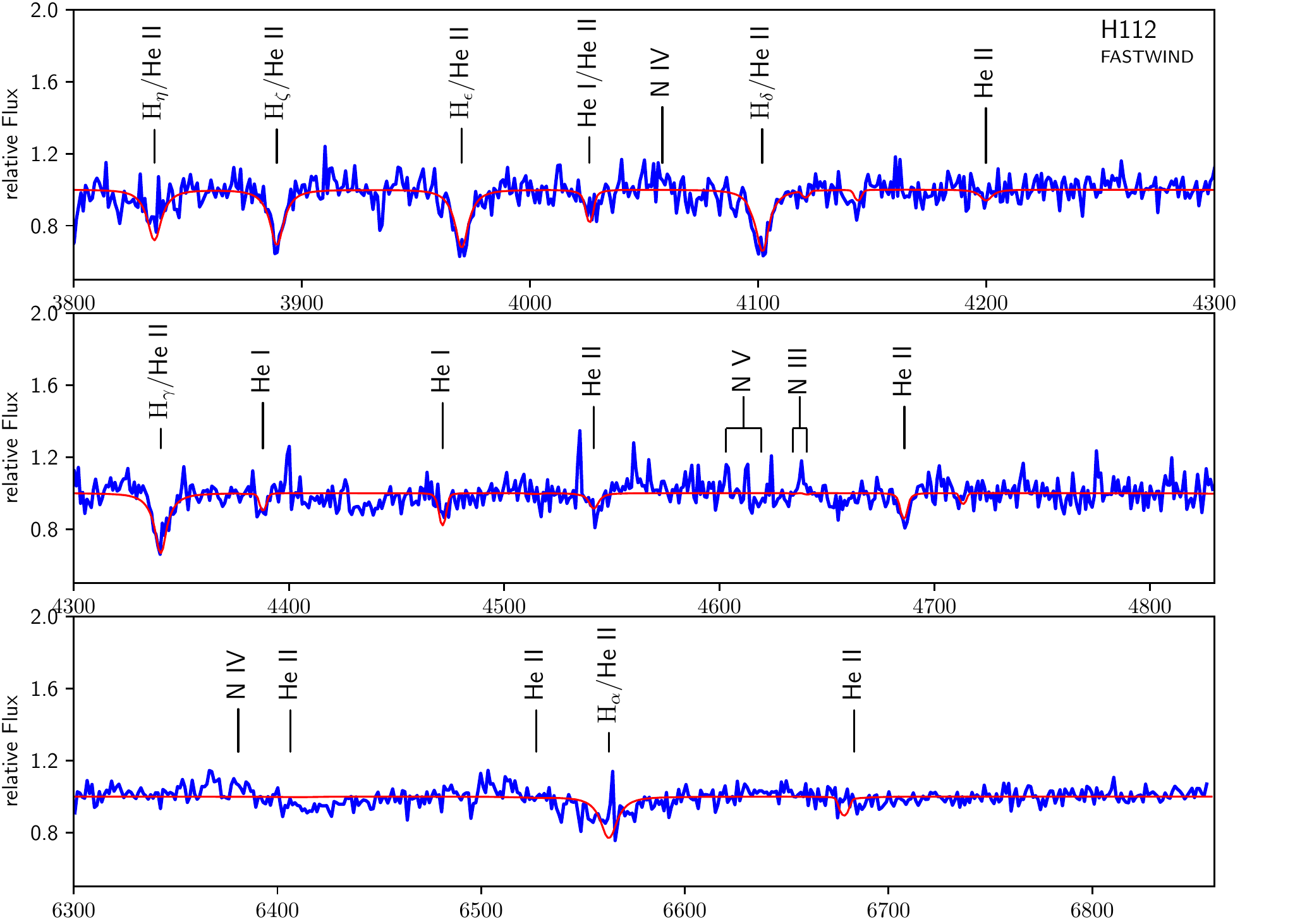}}
\end{center}
\caption{Spectroscopic fit to the data of H112. Blue solid line is the observed HST/STIS spectrum. Red solid line is the synthetic spectrum computed with FASTWIND. Stellar parameters are given in Table\,1.}
\end{figure*}

\begin{figure*}
\begin{center}
\resizebox{0.825\hsize}{!}{\includegraphics{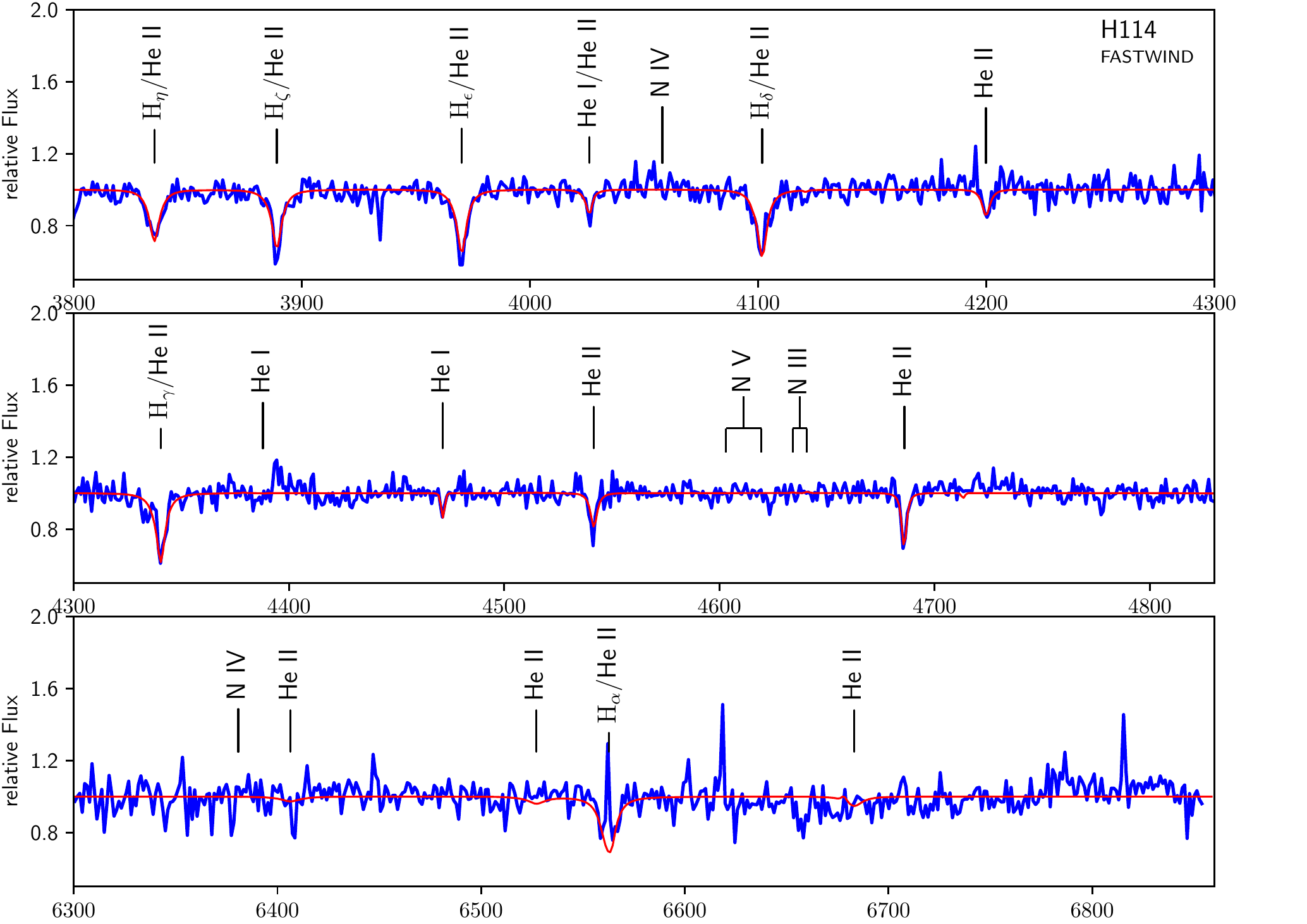}}
\end{center}
\caption{Spectroscopic fit to the data of H114. Blue solid line is the observed HST/STIS spectrum. Red solid line is the synthetic spectrum computed with FASTWIND. Stellar parameters are given in Table\,1.}
\end{figure*}
\begin{figure*}
\begin{center}
\resizebox{0.825\hsize}{!}{\includegraphics{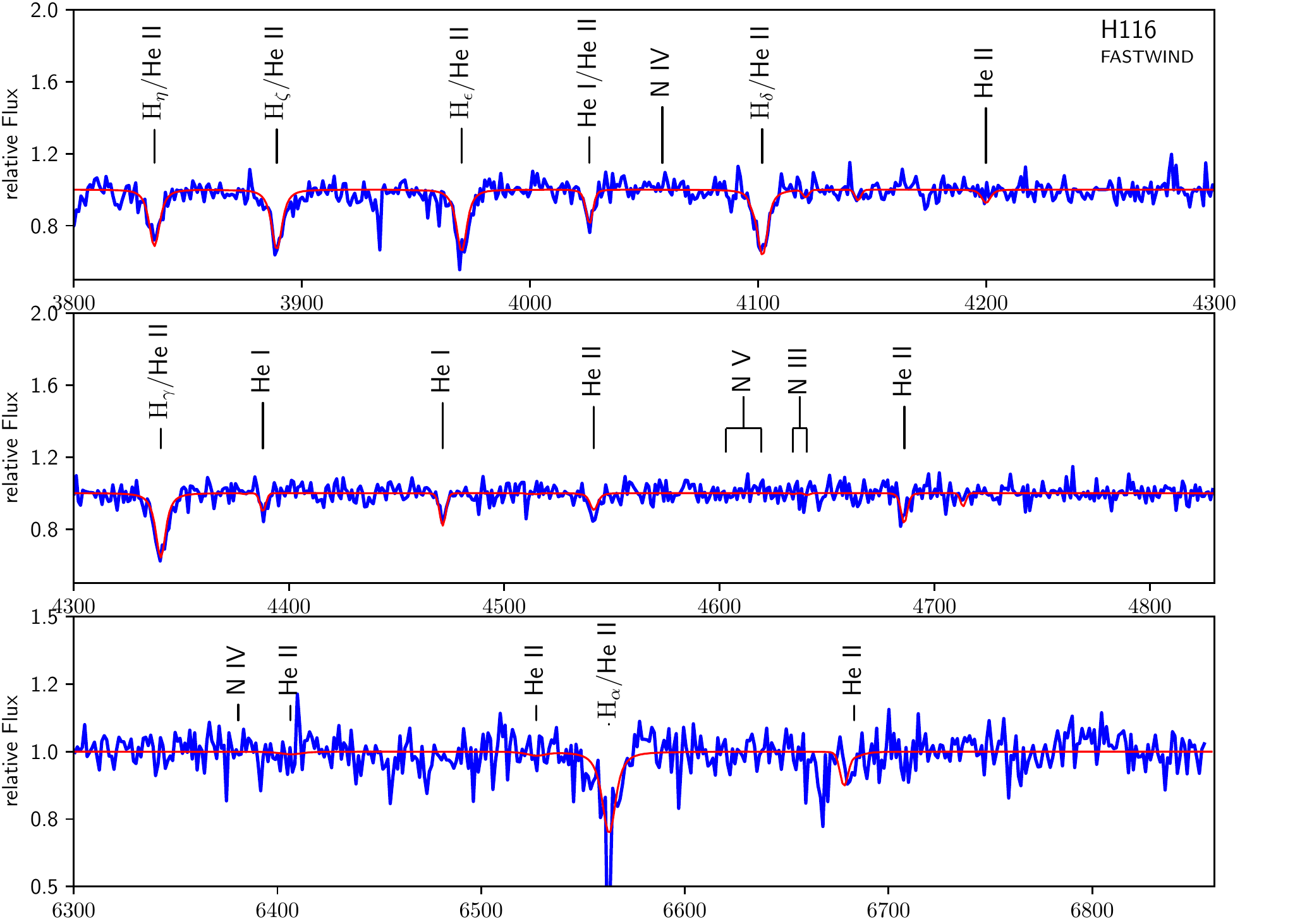}}
\end{center}
\caption{Spectroscopic fit to the data of H116. Blue solid line is the observed HST/STIS spectrum. Red solid line is the synthetic spectrum computed with FASTWIND. Stellar parameters are given in Table\,1.}
\end{figure*}

\clearpage
\begin{figure*}
\begin{center}
\resizebox{0.825\hsize}{!}{\includegraphics{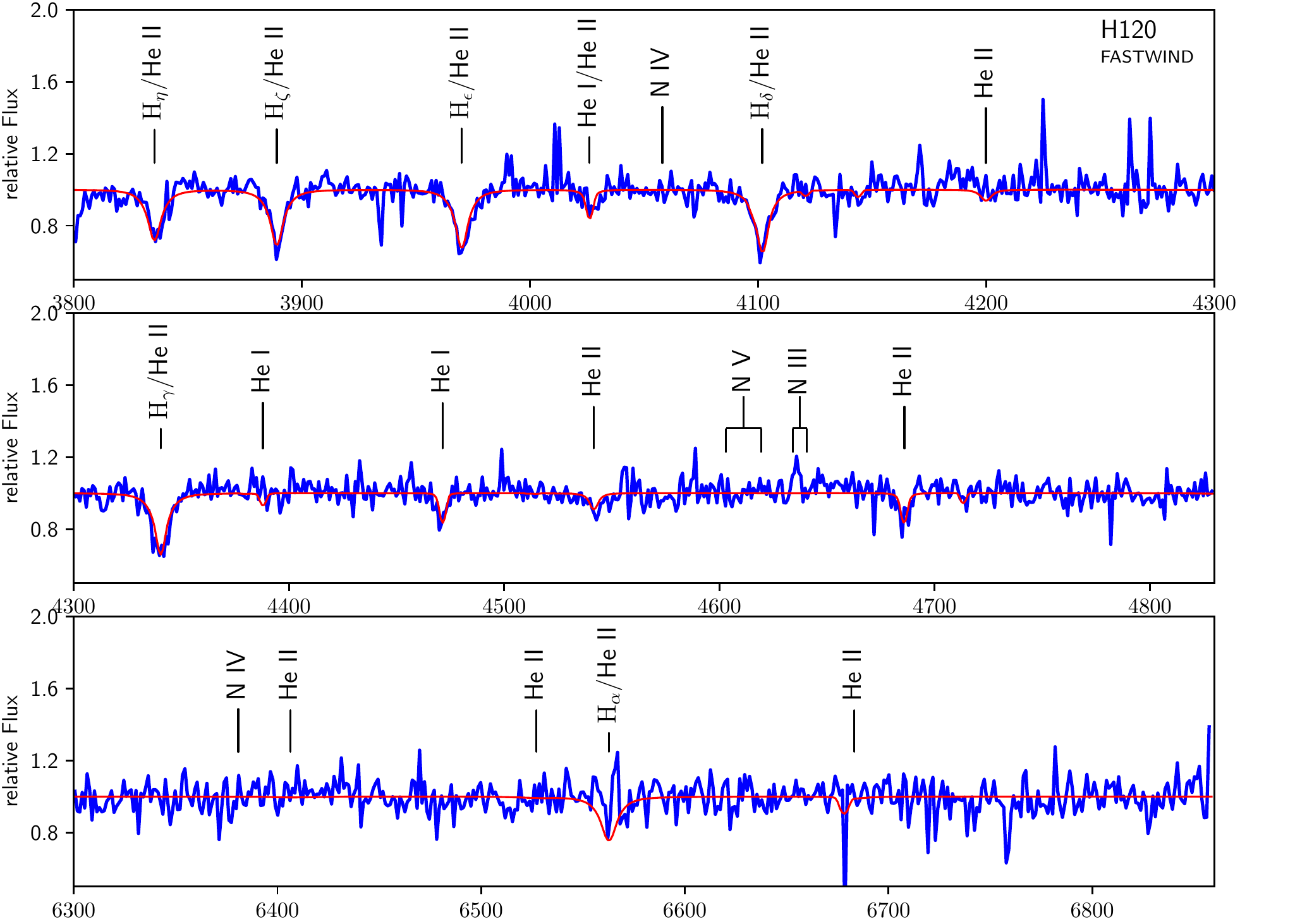}}
\end{center}
\caption{Spectroscopic fit to the data of H120. Blue solid line is the observed HST/STIS spectrum. Red solid line is the synthetic spectrum computed with FASTWIND. Stellar parameters are given in Table\,1.}
\end{figure*}
\begin{figure*}
\begin{center}
\resizebox{0.825\hsize}{!}{\includegraphics{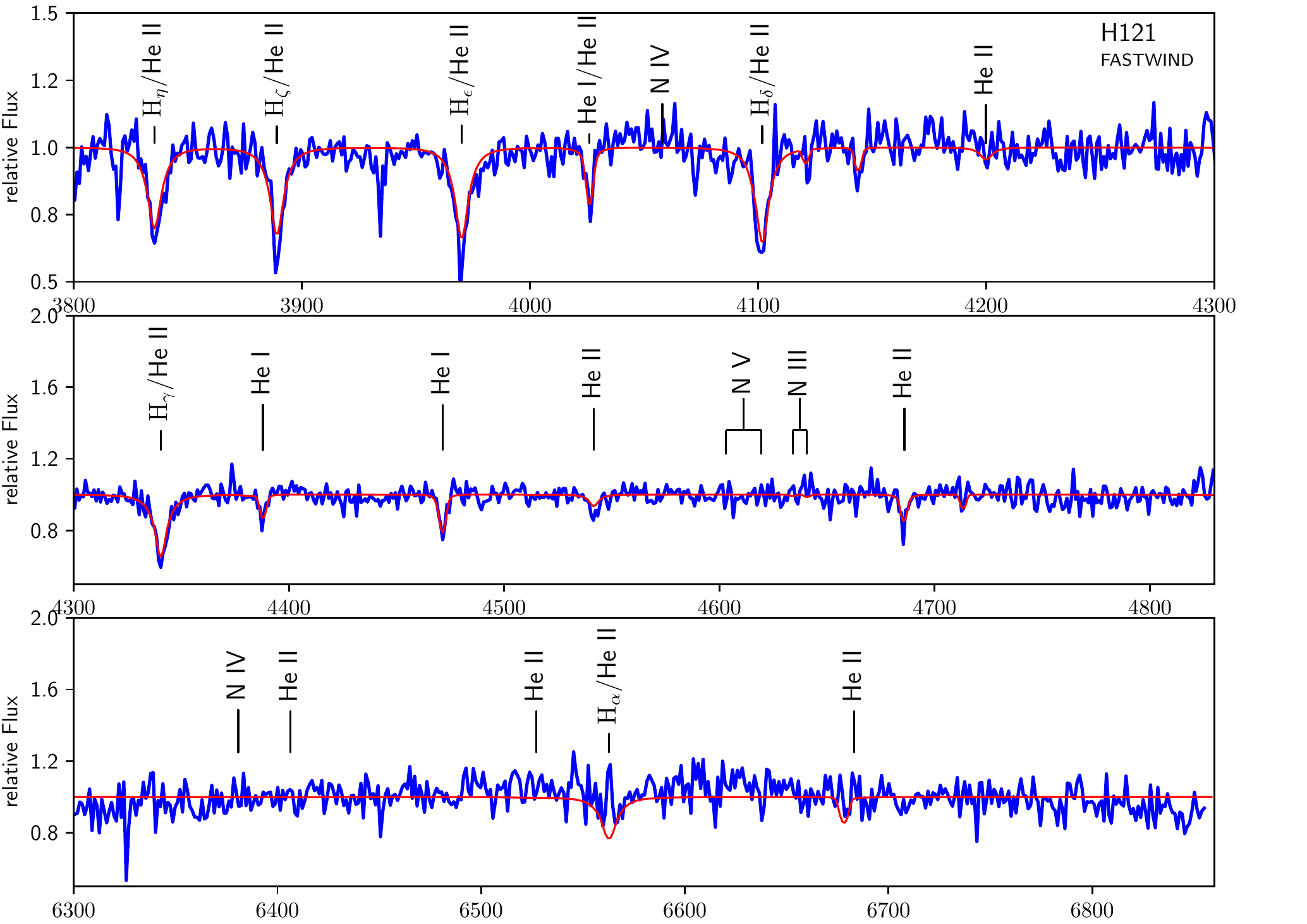}}
\end{center}
\caption{Spectroscopic fit to the data of H121. Blue solid line is the observed HST/STIS spectrum. Red solid line is the synthetic spectrum computed with FASTWIND. Stellar parameters are given in Table\,1.}
\end{figure*}

\begin{figure*}
\begin{center}
\resizebox{0.825\hsize}{!}{\includegraphics{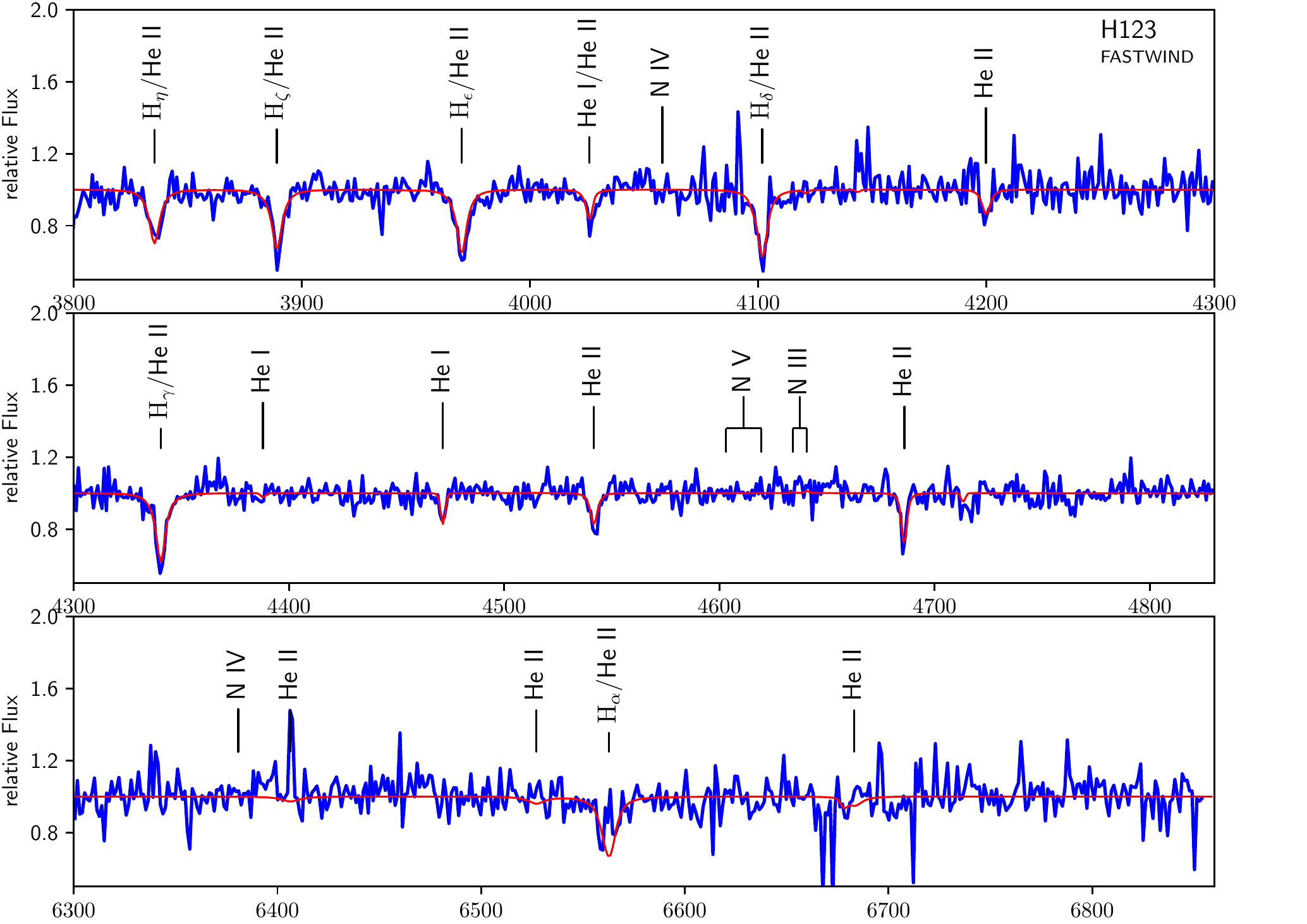}}
\end{center}
\caption{Spectroscopic fit to the data of H123. Blue solid line is the observed HST/STIS spectrum. Red solid line is the synthetic spectrum computed with FASTWIND. Stellar parameters are given in Table\,1.}
\end{figure*}
\begin{figure*}
\begin{center}
\resizebox{0.825\hsize}{!}{\includegraphics{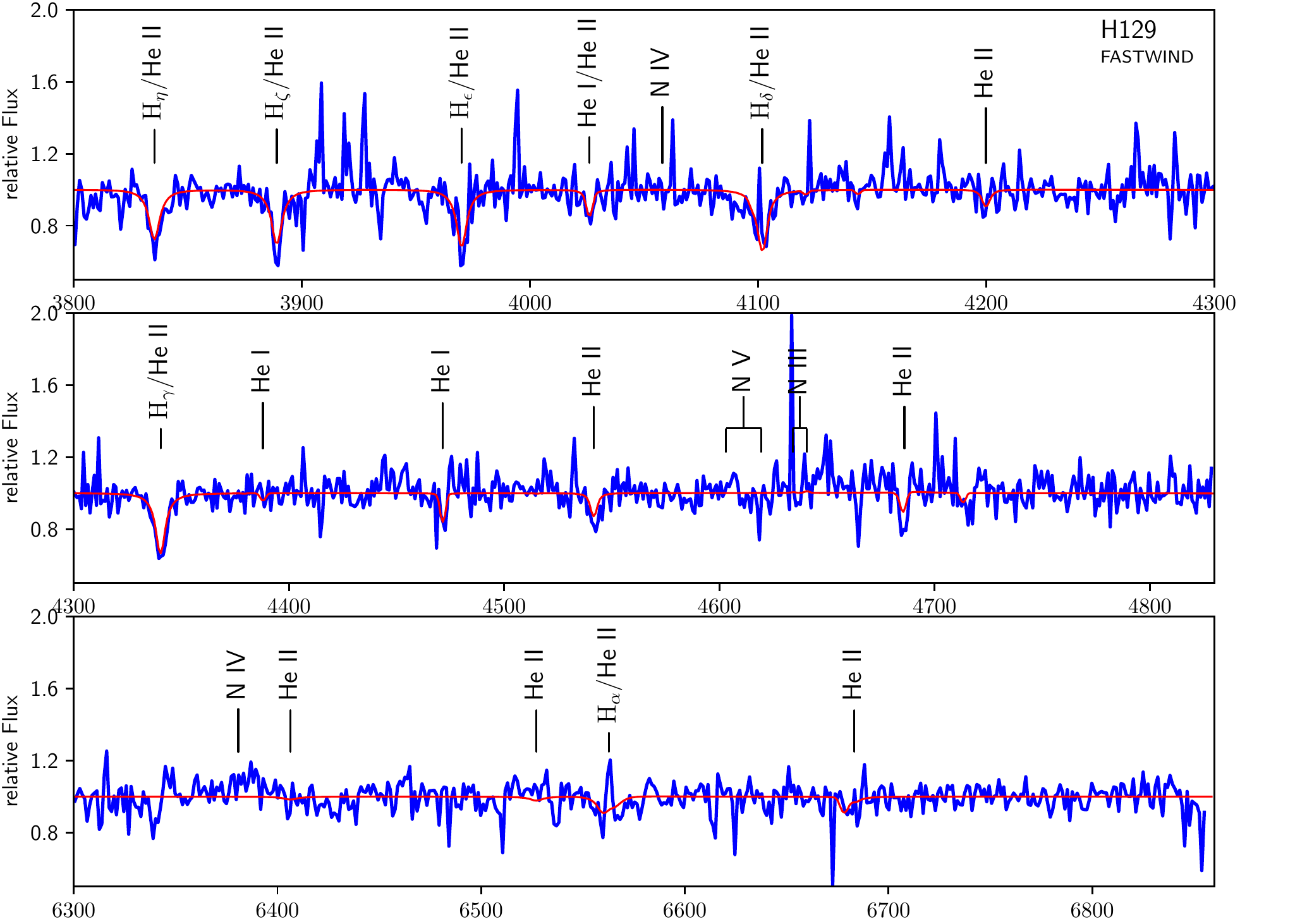}}
\end{center}
\caption{Spectroscopic fit to the data of H129. Blue solid line is the observed HST/STIS spectrum. Red solid line is the synthetic spectrum computed with FASTWIND. Stellar parameters are given in Table\,1.}
\end{figure*}

\begin{figure*}
\begin{center}
\resizebox{0.825\hsize}{!}{\includegraphics{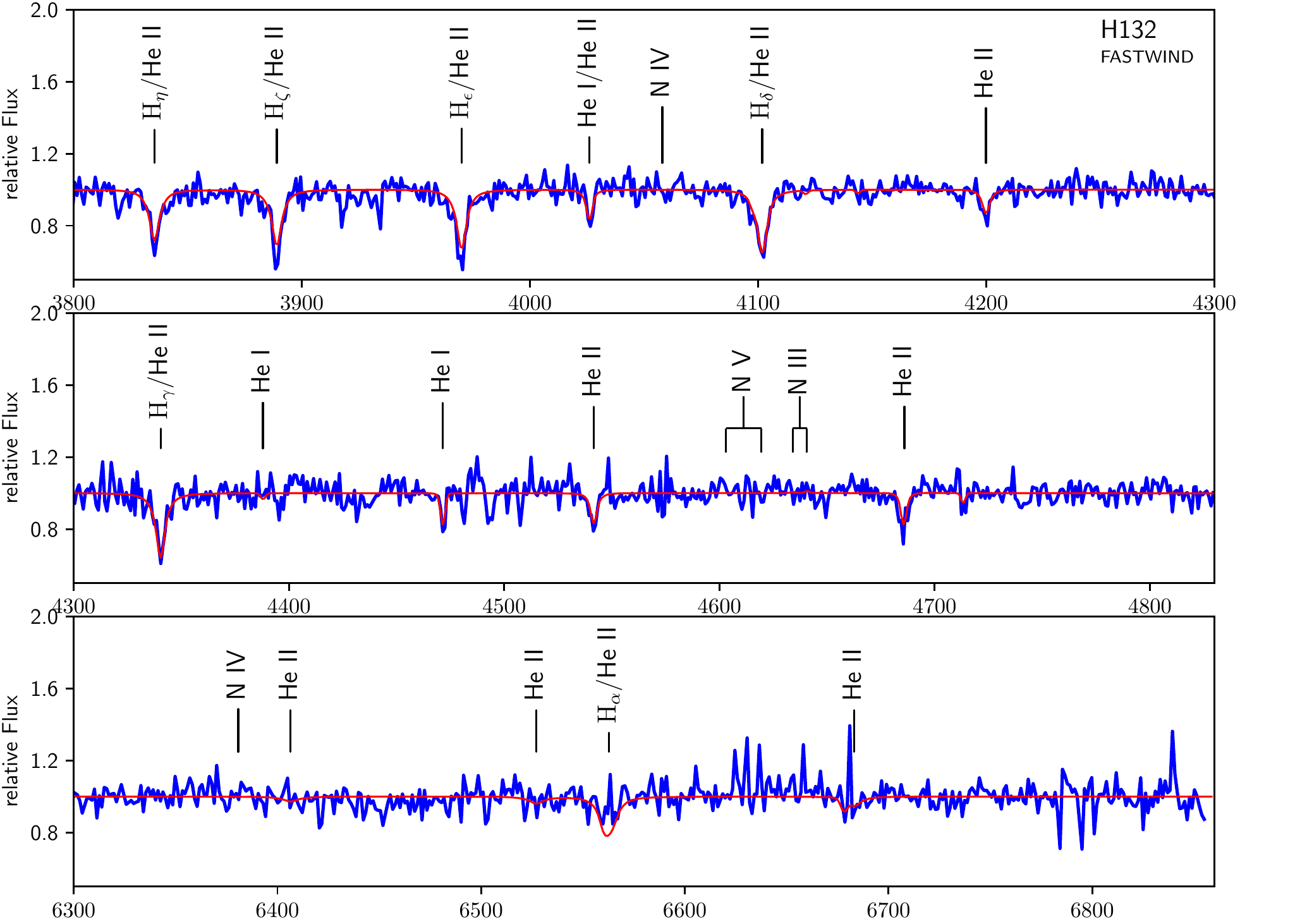}}
\end{center}
\caption{Spectroscopic fit to the data of H132. Blue solid line is the observed HST/STIS spectrum. Red solid line is the synthetic spectrum computed with FASTWIND. Stellar parameters are given in Table\,1.}
\end{figure*}
\begin{figure*}
\begin{center}
\resizebox{0.825\hsize}{!}{\includegraphics{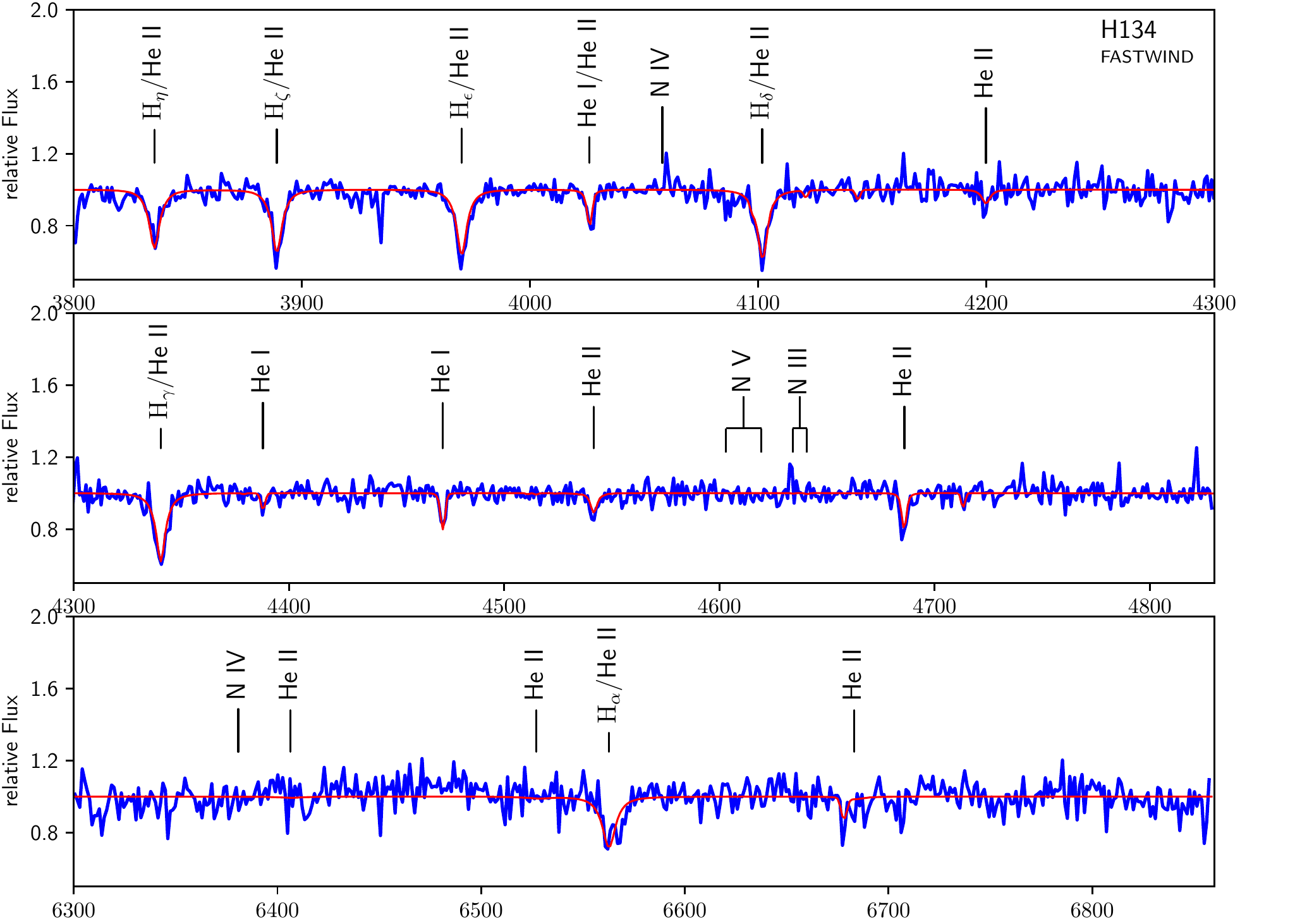}}
\end{center}
\caption{Spectroscopic fit to the data of H134. Blue solid line is the observed HST/STIS spectrum. Red solid line is the synthetic spectrum computed with FASTWIND. Stellar parameters are given in Table\,1.}
\end{figure*}

\begin{figure*}
\begin{center}
\resizebox{0.825\hsize}{!}{\includegraphics{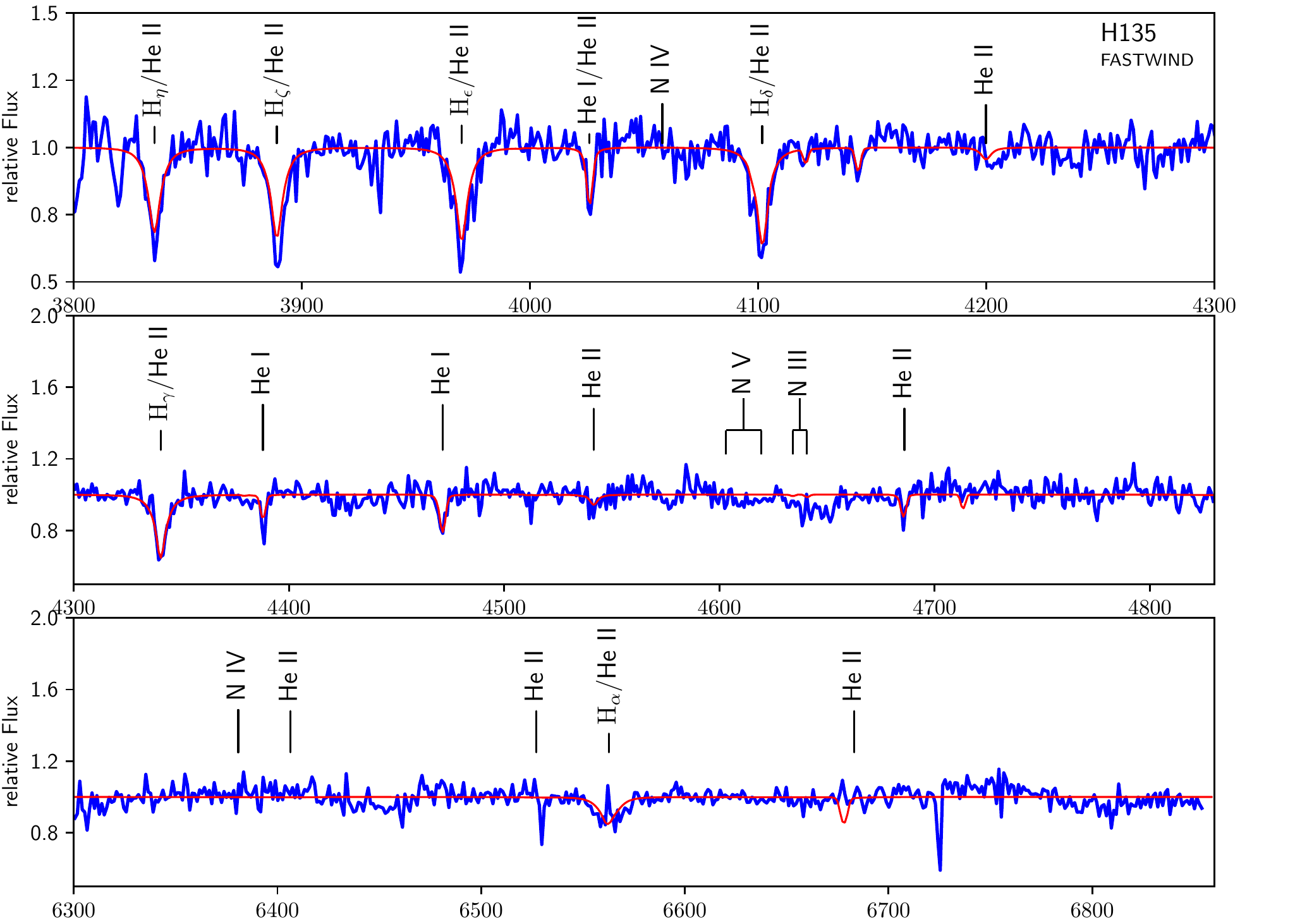}}
\end{center}
\caption{Spectroscopic fit to the data of H135. Blue solid line is the observed HST/STIS spectrum. Red solid line is the synthetic spectrum computed with FASTWIND. Stellar parameters are given in Table\,1.}
\end{figure*}
\begin{figure*}
\begin{center}
\resizebox{0.825\hsize}{!}{\includegraphics{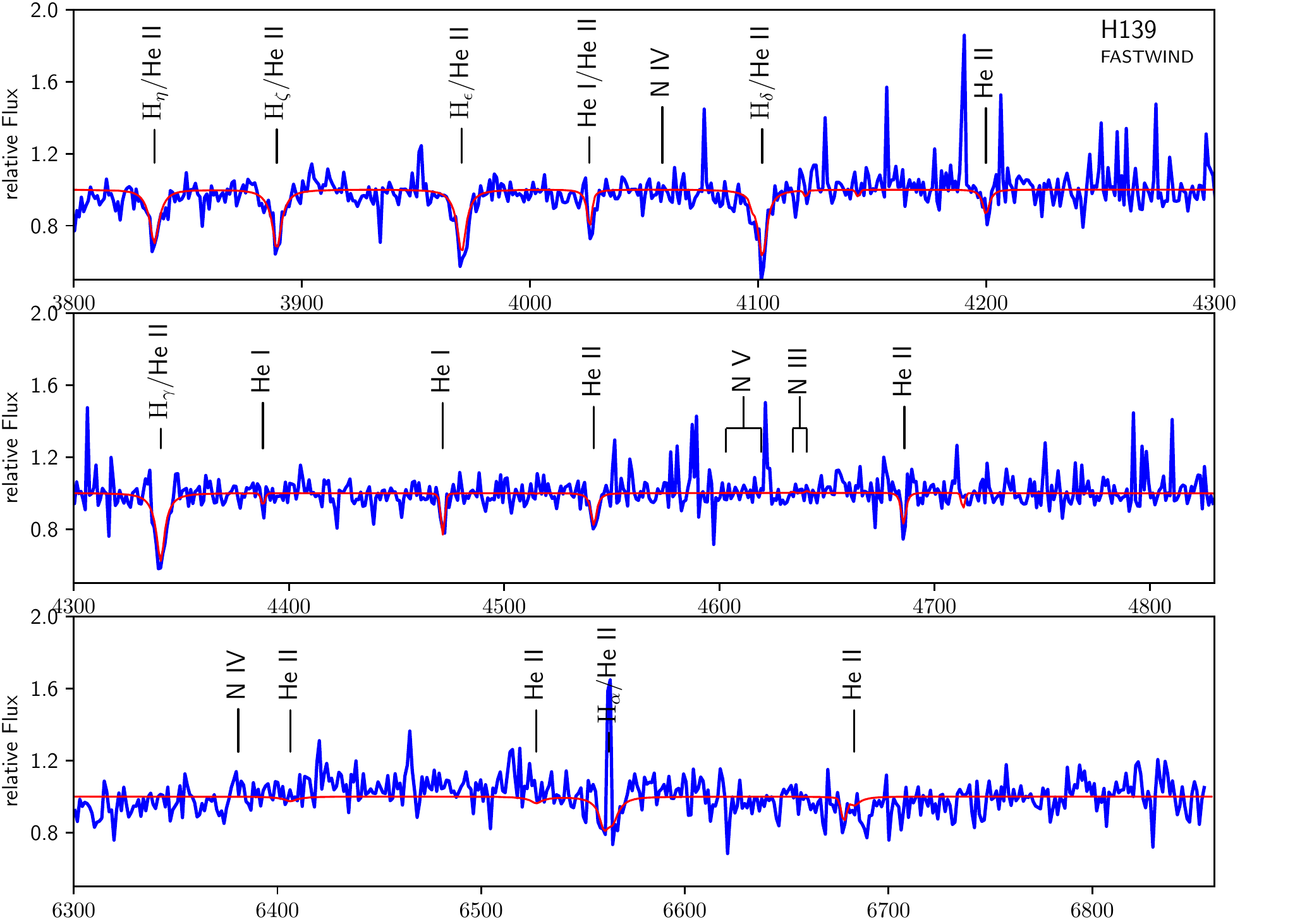}}
\end{center}
\caption{Spectroscopic fit to the data of H139. Blue solid line is the observed HST/STIS spectrum. Red solid line is the synthetic spectrum computed with FASTWIND. Stellar parameters are given in Table\,1.}
\end{figure*}

\clearpage
\begin{figure*}
\begin{center}
\resizebox{0.825\hsize}{!}{\includegraphics{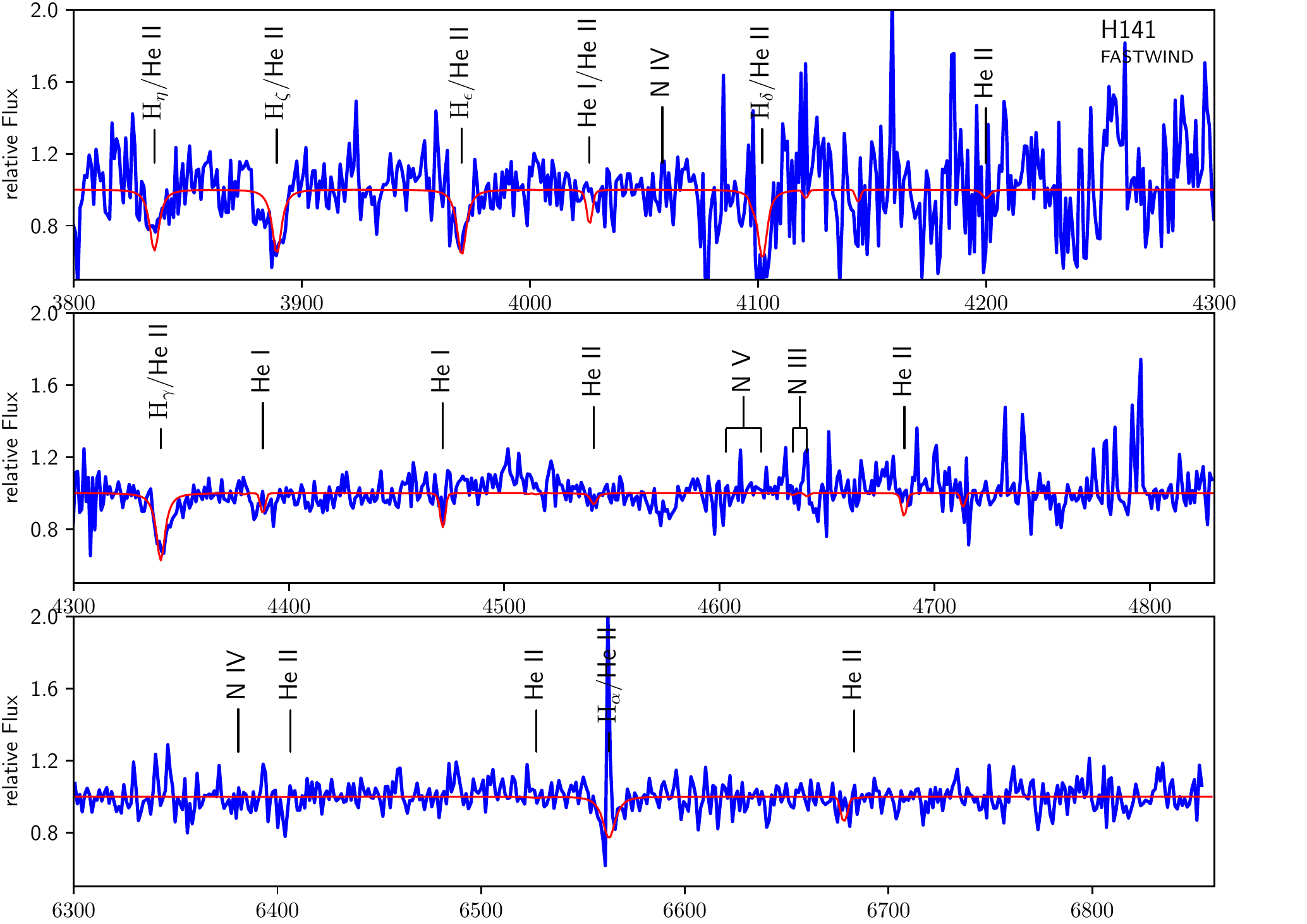}}
\end{center}
\caption{Spectroscopic fit to the data of H141. Blue solid line is the observed HST/STIS spectrum. Red solid line is the synthetic spectrum computed with FASTWIND. Stellar parameters are given in Table\,1.}
\end{figure*}
\clearpage
\begin{figure*}
\begin{center}
\resizebox{0.825\hsize}{!}{\includegraphics{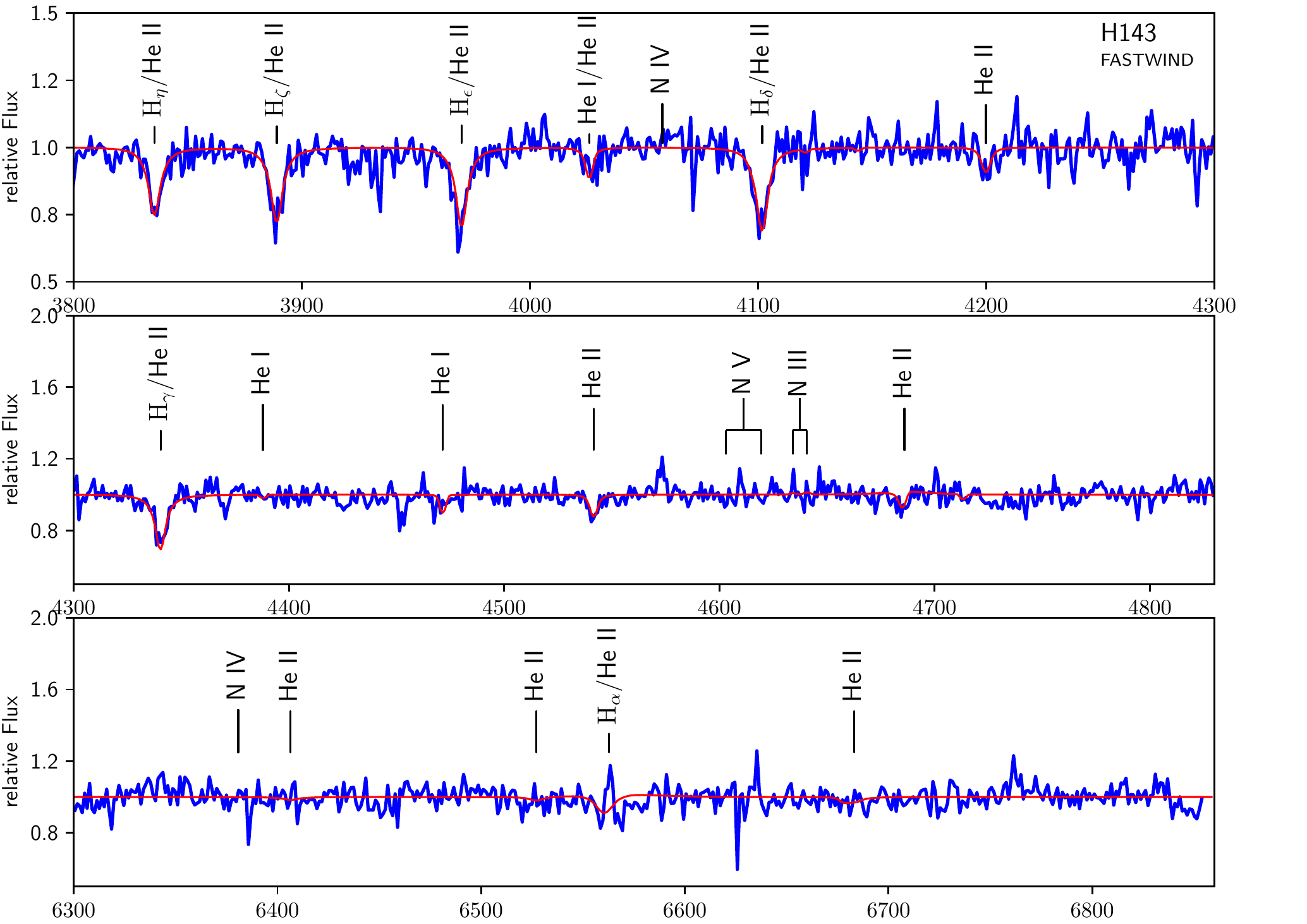}}
\end{center}
\caption{Spectroscopic fit to the data of H143. Blue solid line is the observed HST/STIS spectrum. Red solid line is the synthetic spectrum computed with FASTWIND. Stellar parameters are given in Table\,1.}
\end{figure*}

\begin{figure*}
\begin{center}
\resizebox{0.825\hsize}{!}{\includegraphics{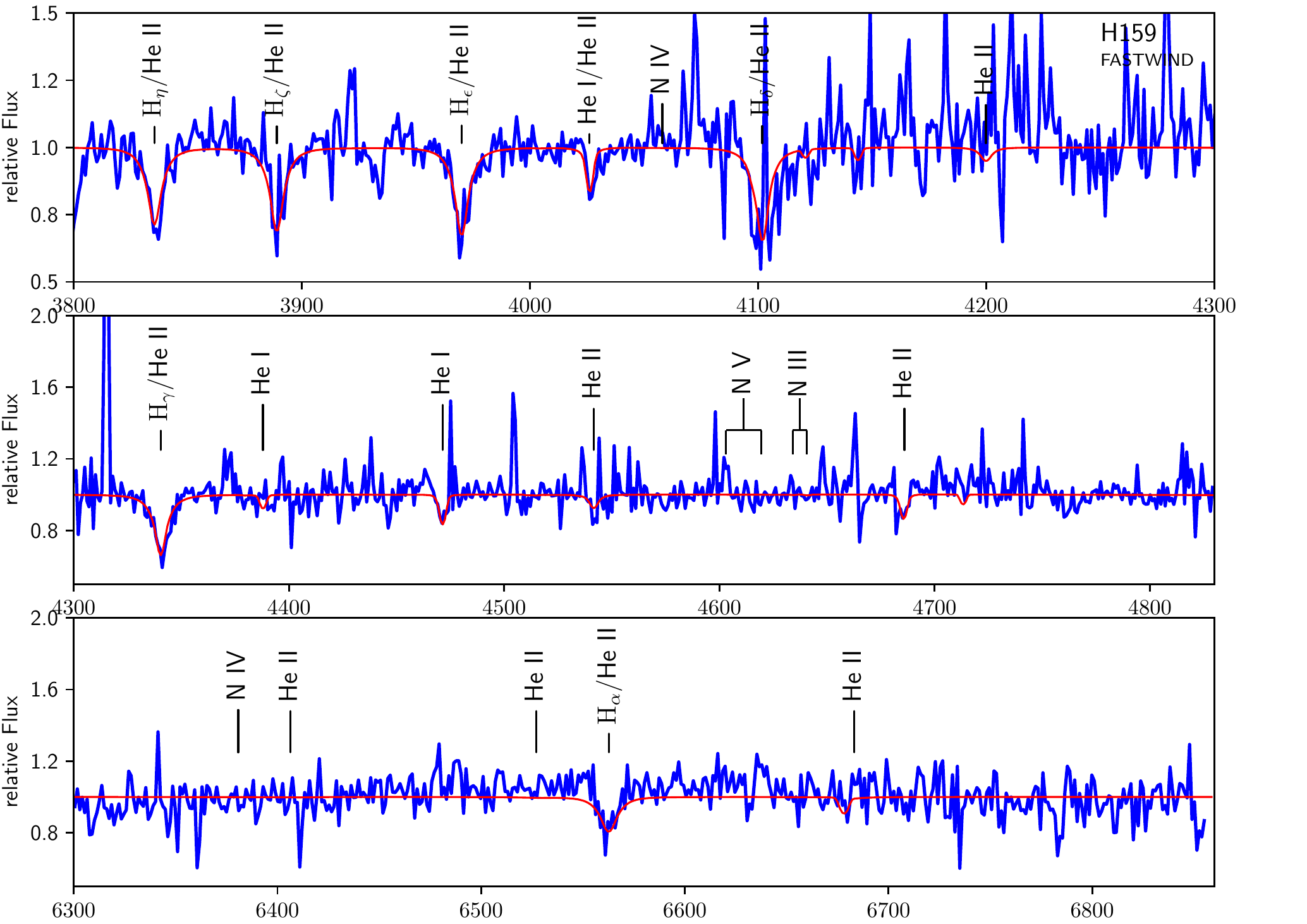}}
\end{center}
\caption{Spectroscopic fit to the data of H159. Blue solid line is the observed HST/STIS spectrum. Red solid line is the synthetic spectrum computed with FASTWIND. Stellar parameters are given in Table\,1.}
\end{figure*}
\begin{figure*}
\begin{center}
\resizebox{0.825\hsize}{!}{\includegraphics{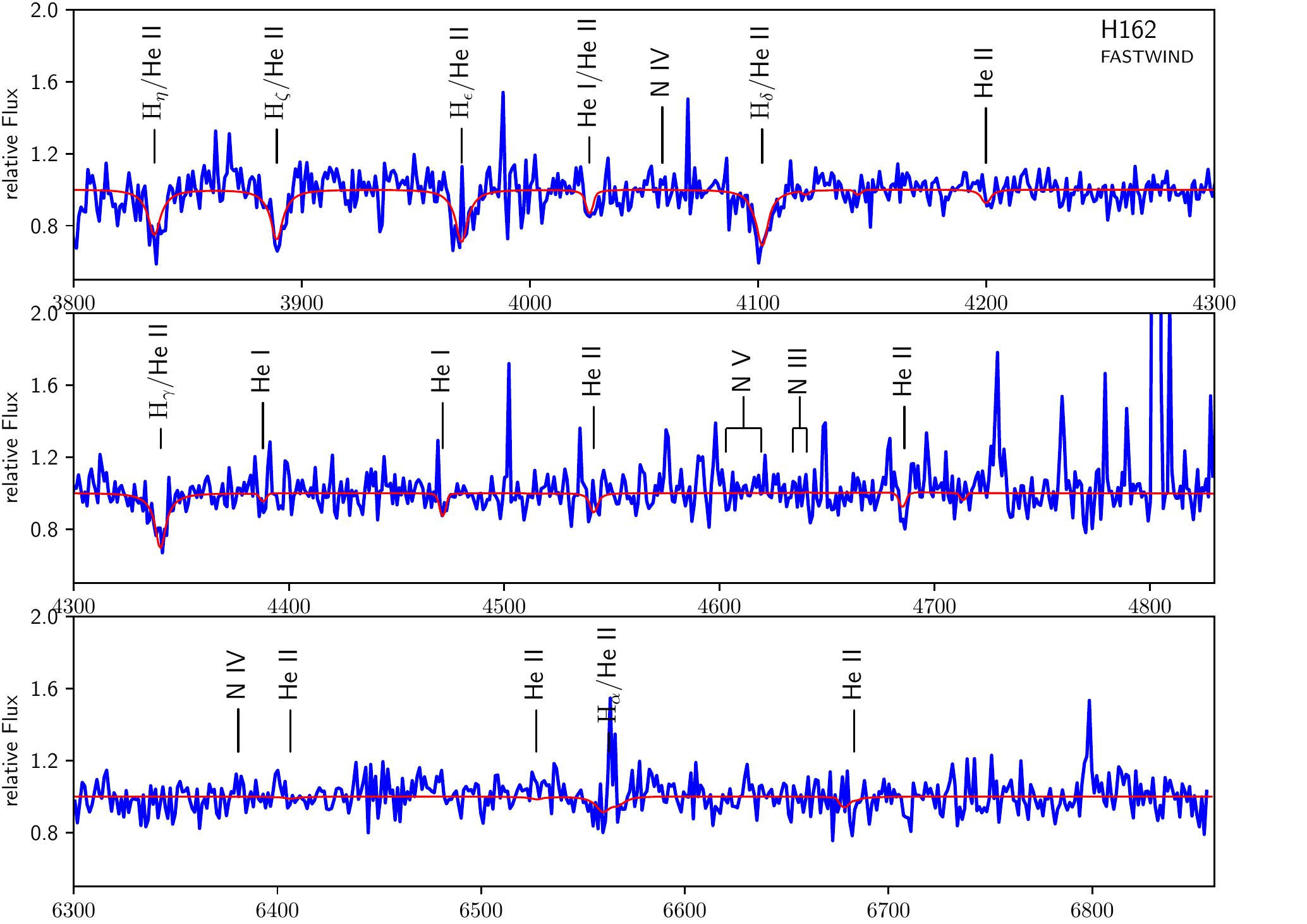}}
\end{center}
\caption{Spectroscopic fit to the data of H162. Blue solid line is the observed HST/STIS spectrum. Red solid line is the synthetic spectrum computed with FASTWIND. Stellar parameters are given in Table\,1.}
\end{figure*}

\begin{figure*}
\begin{center}
\resizebox{0.825\hsize}{!}{\includegraphics{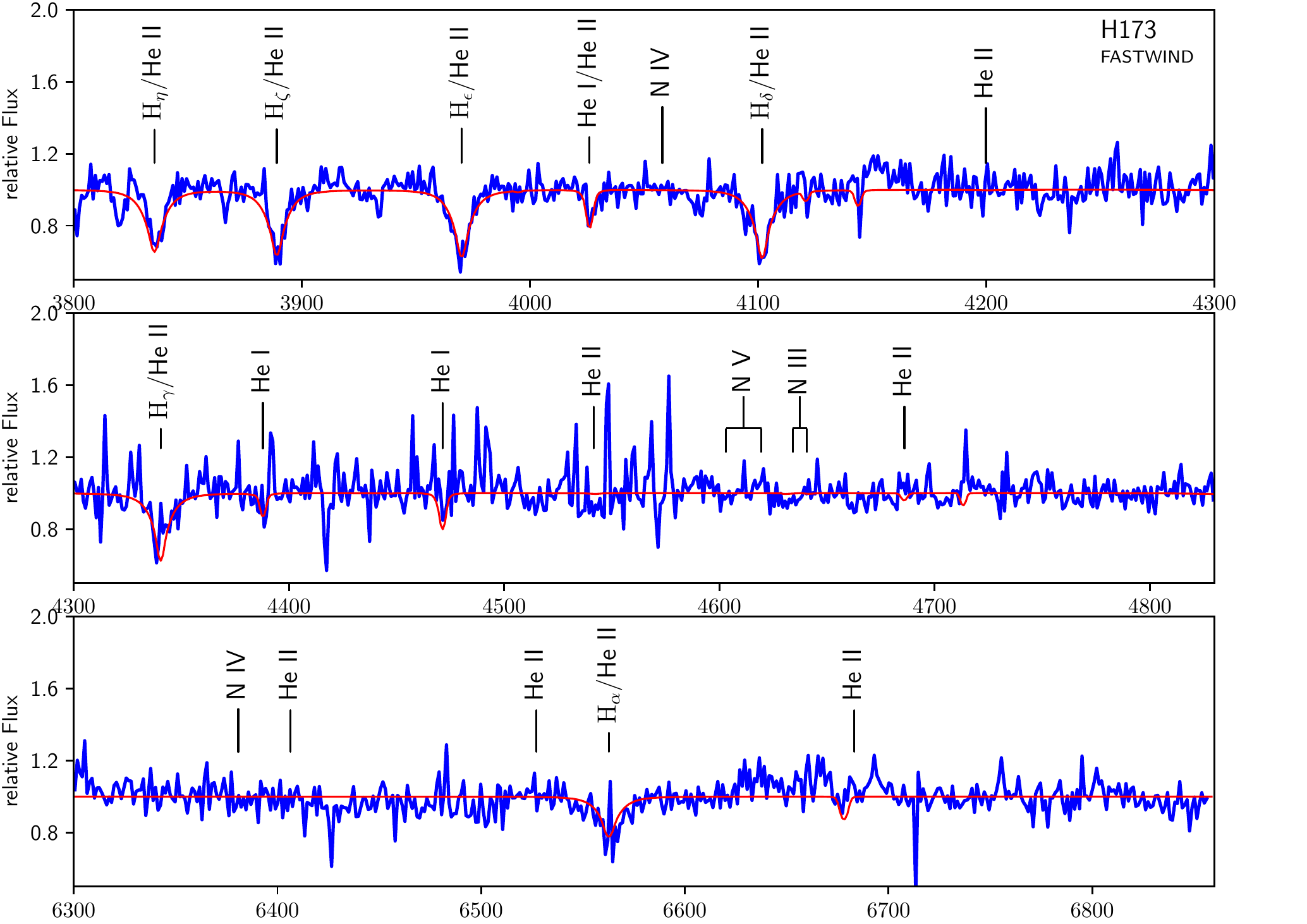}}
\end{center}
\caption{Spectroscopic fit to the data of H173. Blue solid line is the observed HST/STIS spectrum. Red solid line is the synthetic spectrum computed with FASTWIND. Stellar parameters are given in Table\,1.}
\end{figure*}

\end{document}